\documentclass[12pt,a4paper,twoside,openright,dvips]{report}

\usepackage{a4}
\usepackage{textcomp}
\usepackage{amsmath}
\usepackage{graphicx}
\usepackage{axodraw}
\usepackage{verbatim}
\usepackage[vietnam, english]{babel}
\usepackage[utf8]{inputenc}

\usepackage{amsfonts}
\usepackage{amssymb}
\usepackage{amsthm}
\usepackage{newlfont}
\usepackage{xthesis} 
\usepackage{xtocinc} 

\newlength{\defbaselineskip}
\newcommand{\setlinespacing}[1]%
           {\setlength{\baselineskip}{#1 \defbaselineskip}}

\newcommand{\A}{{\cal A}}
\newcommand{\OO}{{\cal O}}
\newcommand{\Q}{{\cal Q}}

\newcommand{\Rel}{\operatorname{Re}}
\newcommand{\Img}{\operatorname{Im}}

\newcommand{\Trace}{\operatorname{Tr}}
\newcommand{\Sp}{\operatorname{Sp}}
\newcommand{\sign}{\operatorname{sign}}
\newcommand{\GeV}{\textrm{GeV}}
\newcommand{\lum}{\textrm{cm}^{-2}\textrm{s}^{-1}}
\newcommand{\beq}{\begin{equation}}
\newcommand{\eeq}{\end{equation}}
\newcommand{\bea}{\begin{eqnarray}}
\newcommand{\eea}{\end{eqnarray}}
\newcommand{\ben}{\begin{enumerate}}
\newcommand{\een}{\end{enumerate}}
\newcommand{\bitm}{\begin{itemize}}
\newcommand{\eitm}{\end{itemize}}
\newcommand{\bc}{\begin{center}}
\newcommand{\ec}{\end{center}}
\newcommand{\beqn}{\begin{eqnarray}}
\newcommand{\eeqn}{\end{eqnarray}}
\newcommand{\nn}{\nonumber}
\newcommand{\crn}{\nonumber \\}

\newcommand{\noi}{\noindent}

\newcommand{\la}{\lambda}

\newcommand{\ga}{\gamma}

\newcommand{\fr}{\frac}

\newcommand{\eps}{\epsilon}
\newcommand{\varep}{\varepsilon}

\newcommand{\up}{\upsilon}

\newcommand{\vph}{\varphi}

\newcommand{\xiw}{\xi_W}
\newcommand{\xiz}{\xi_Z}

\newcommand{\vv}{\textbf{v}}
\newcommand{\pp}{\textbf{p}}
\newcommand{\Lar}{\mathcal{L}}

\newcommand{\ppbbH}{pp\to b\bar{b}H}
\newcommand{\ggbbH}{gg\to b\bar{b}H}
\newcommand{\qqbbH}{qq\to b\bar{b}H}

\newcommand{\alio}{{\cal A}(\hat\la)}
\newcommand{\aliot}{{\tilde{\cal A}}(\hat\la)}
\newcommand{\ali}{{\cal A}(\la_1,\la_2;\la_3,\la_4)}
\newcommand{\aliq}{{\cal A}(\la_1,\la_2;\la_3,\la_4;q_1,q_2)}
\newcommand{\aliqp}{{\cal A}(\la_1,\la_2;\la_3,\la_4;q^\prime_1,q^\prime_2)}
\def\slashepi{\epsilon_i\kern -.720em {/}}
\def\slashpi{p_i\kern -.600em {/}}

\def\slashc{c\kern -.400em {/}}
\def\slashp{p\kern -.400em {/}}
\def\slashq{q\kern -.450em {/}}
\def\slashL{L\kern -.450em {/}}
\def\slashcl{\cl\kern -.600em {/}}
\def\slashr{r\kern -.450em {/}}
\def\slashk{k\kern -.500em {/}}
\def\slashep{\epsilon\kern -.450em {/}}
\def\slashpbar{\bar{p}\kern -.450em {/}}
\def\slashD{D\kern -.650em {/}}

\newcommand{\hs}{\hspace*{3mm}}

\newcommand{\ie}{{\it i.e.\;}}
\newcommand{\eg}{{\it e.g.\;}}
\newcommand{\ra}{\rightarrow}

\newcommand{\sma}{S-matrix\ }

\let\originalcleardoublepage=\cleardoublepage
\renewcommand{\cleardoublepage}
             {\newpage{\pagestyle{empty}\originalcleardoublepage}}

\begin{document}
{
\typeout{Title page}
\pagestyle{empty}
\thispagestyle{empty}
\voffset-10pt
\vspace*{-2.5cm}\rightline{LAPTH-1261/08}
\begin{center}
\vspace*{-0.5cm}
{\sc{Universit\'e de Savoie},\\ 
\sc{Laboratoire d'Annecy-le-Vieux de Physique Th{\'e}orique}}\\[8mm]

{\bf{TH\`ESE}}\\[8mm]

{\it{pr\'esent\'ee pour obtenir le grade de}}\\[4mm]
DOCTEUR EN PHYSIQUE\\
DE L'UNIVERSIT\'E DE SAVOIE\\
Sp{\'e}cialit{\'e}: Physique Th{\'e}orique\\[4mm]

{\it {par}}\\[4mm]
{\bf\large{\selectlanguage{vietnam} {\sc LÊ} Đức Ninh}}\\[4mm]

{\it{Sujet :}}\\[4mm]

{\bf{\large{ONE-LOOP YUKAWA CORRECTIONS TO THE PROCESS $\ppbbH$ IN THE STANDARD MODEL AT THE LHC: LANDAU SINGULARITIES.}}}\\[8mm]

{\it{Soutenue le 22 Juillet 2008 apr\`es avis des
rapporteurs ~:}}\\[4mm]

\begin{tabular}{l}
       Mr. Guido ALTARELLI \\
       Mr. Ansgar DENNER  \\
\end{tabular}\\[4mm]
 {\it{devant la commission d'examen compos\'ee de :}}\\[4mm]
\begin{tabular}{ll}
       Mr. Guido ALTARELLI & Rapporteur \\
       Mr. Patrick AURENCHE &  \\
       Mr. Fawzi BOUDJEMA & Directeur de th\`ese \\
       Mr. Ansgar DENNER  & Rapporteur\\
       Mr. Luc FRAPPAT & Pr\'esident du jury\\
       Mr. {\selectlanguage{vietnam}HOÀNG Ngọc Long} & Co-directeur de th\`ese \\ 
       Mr. {\selectlanguage{vietnam}NGUYỄN Anh Kỳ} &  
\end{tabular}
\end{center}

\cleardoublepage
}


\setcounter{page}{0}
\pagenumbering{roman}
\tableofcontents

\prefacesection{Abstract}
The aim of this thesis is twofold. First, to study methods to calculate one-loop corrections in the context of perturbative theories. Second, to apply those methods to calculate the leading electroweak (EW) corrections to the important process of Higgs production associated with two bottom quarks at the CERN Large Hadron Collider (LHC). Our study is restricted to the Standard Model (SM). 

The first aim is of theoretical importance. Though the general method to calculate one-loop corrections in the SM is, in principle, well understood by means of renormalisation, it presents a number of technical difficulties. They are all related to loop integrals. The analytical method making use of various techniques to reduce all the tensorial integrals in terms of a basis of scalar integrals is most widely used nowadays. A problem with this method is that for processes with more than $4$ external particles the amplitude expressions are extremely cumbersome and very difficult to handle even with powerful computers. 
In this thesis, we have studied this problem and realised that the whole calculation can be easily optimised if one uses the helicity amplitude method. 
Another general problem is related to the analytic properties of the scalar loop integrals. An important part of this thesis is devoted to studying this by using Landau equations. We found significant effects due to Landau singularities in the process of Higgs production associated with two bottom quarks at the LHC. 

The second aim is of practical (experimental) importance. Higgs production associated with bottom quarks at the LHC is a very important process to understand the bottom-Higgs Yukawa coupling. If this coupling is strongly enhanced as predicted by the Minimal Supersymmetric Standard Model (MSSM) then this process can have a very large cross section. In this thesis, based on the theoretical study mentioned above, we have calculated the leading EW corrections to this process. The result is the following. If the Higgs mass is about 120GeV then the next-to-leading order (NLO) correction is small, about $-4\%$. If the Higgs mass is about $160$GeV then the EW correction is strongly enhanced by the Landau singularities, leading to  a significant  correction of about $50\%$. This important phenomenon is carefully studied  in this thesis. 
\prefacesection{Acknowledgements}
{\small
I would like to thank Fawzi BOUDJEMA, my friendly supervisor, for accepting me as his student, 
giving me an interesting topic, many useful suggestions and constant support during this research. In particular, 
he has suggested and encouraged me a lot to attack the difficult problem of Landau singularities. 
His enthusiasm for physics was always 
great and it inspired me a lot. By guiding me to finish this thesis, he has done so much to mature my approach to physics. 

I admire Patrick AURENCHE for his personal character and physical understanding. It was always a great pleasure for me to see and talk to him. 
In every physical discussion since the first time we met in Hanoi (2003), I have learnt something new from him. The way he attacks any physical problem 
is so simple and pedagogical. I thank him for bringing me to Annecy (the most beautiful city I have ever seen), filling my Ph.D 
years with so many beautiful weekends at his house. I will never forget the trips to Lamastre. He has carefully read the manuscript and given me a lot of suggestions. Without his help and continuous support I would not be the 
person I am today. Thanks, Patrick!     
 
I am deeply indebted to Guido ALTARELLI for his guidance, support and a lot of fruitful discussions during the one-year period I was at CERN. He 
has also spent time and effort to read the manuscript as a rapporteur. 

Ansgar DENNER, as a rapporteur, has carefully read the manuscript and given me many comments and suggestions which improved a lot the thesis. 
I greatly appreciate it and thank him so much.

I am grateful to {\selectlanguage{vietnam}HOÀNG Ngọc Long} for his continuous encouragement and support. He has read the manuscript and given me valuable comments. I thank {\selectlanguage{vietnam}NGUYỄN Anh Kỳ} for suggesting me to apply for the CERN Marie Curie fellowship and constant support. 
The help of the Institute of Physics in {\selectlanguage{vietnam}Hà Nội} is greatly acknowledged.     

For interesting discussions and help I would like to thank Nans BARO, James BEDFORD, Genevi{\`e}ve B{\'E}LANGER, Christophe BERNICOT, Thomas BINOTH, Noureddine BOUAYED, {\selectlanguage{vietnam}ĐÀO Thị Nhung}, Cedric DELAUNAY, 
Ansgar DENNER, Stefan DITTMAIER, {\selectlanguage{vietnam}ĐỖ Hoàng Sơn}, John ELLIS, Luc FRAPPAT, Junpei FUJIMOTO, Jean-Philippe GUILLET, Thomas HAHN, Wolfgang HOLLIK, Kiyoshi KATO, Yoshimasa KURIHARA, M{\"u}hlleitner MARGARETE, Zoltan NAGY, {\'E}ric PILON, Gr{\'e}gory SANGUINETTI, Pietro SLAVICH, Peter UWER, Jos VERMASEREN, {\selectlanguage{vietnam}VŨ Anh Tuấn}, John WARD and Fukuko YUASA. Special thanks go to {\'E}ric PILON for many fruitful discussions and explaining me useful mathematical tricks related to Landau singularities. Other special thanks go to YUASA-san for comparisons
between her numerical code and our code for the four-point
function with complex masses. 

I would like to thank Jean-Philippe GUILLET for his help with the computer system and his suggestion to use Perl. 

{\selectlanguage{vietnam}ĐỖ Hoàng Sơn} is very good at computer and Linux operating 
system. He has improved both my computer and my knowledge of it. Thanks, {\selectlanguage{vietnam}Sơn}!   

I acknowledge the financial
support of LAPTH, {\em Rencontres du Vietnam} sponsored by Odon VALLET and the {\em Marie Curie Early Stage Training Grant of the European Commission}. In particular, I am grateful to {\selectlanguage{vietnam}TRẦN Thanh Vân} for his support. 

Dominique TURC-POENCIER, V{\'e}ronique JONNERY, Virginie MALAVAL, 
Nanie PERRIN, Diana DE TOTH and Suzy VASCOTTO make LAPTH and CERN really special places and I thank them for their help. 
 
Last, but by no means least I owe a great debt to my parents {\selectlanguage{vietnam}NGUYỄN Thị Thắm} and {\selectlanguage{vietnam}LÊ Trần Phương}, my sister {\selectlanguage{vietnam}LÊ Thị Nam} and her husband {\selectlanguage{vietnam}LÊ Quang Đông}, and my wife {\selectlanguage{vietnam}ĐÀO Thị Nhung}, for their invaluable love.
}

\pagestyle{headings}
\let\MakeUppercase\relax
\setcounter{page}{0}
\pagenumbering{arabic}

\nonumchapter{Introduction}
In the realm of high energy physics, the Standard Model (SM) of particle physics \cite{glashow_sm, weinberg_sm, salam_sm, GellMann:1964nj, Fritzsch:1973pi, Gross:1973id, Politzer:1973fx} is the highest 
achievement to date. Almost all its predictions have been verified by various experiments \cite{PDBook, range_MH}.
The only prediction of the SM which has not been confirmed by any experiment is the existence of a scalar fundamental
particle called the Higgs boson. The fact that we have never observed any fundamental scalar particle in nature so far
makes this the truly greatest challenge faced by physicists today. For this greatest challenge we have the world
largest particle accelerator to date, the CERN Large Hadron Collider (LHC) \cite{lhc}. The LHC collides two proton beams with a center-of-mass energy up to $14$TeV and is expected to start this year. It is our belief that the Higgs boson will be found within a few years.

The prominent feature of the Higgs boson is that it couples mainly to heavy particles with large couplings. This makes the theoretical calculations
of the Higgs production rates as perturbative expansions in those large couplings complicated. The convergence rate of the perturbative expansion is
slow and one cannot rely merely on the leading order (LO) result. Loop calculations are therefore mandatory. The most famous example is the Higgs production mechanism via gluon fusion, the Higgs discovery channel. The LO contribution in this example is already at one-loop level.
The two-loop contribution, mainly due to the gluon radiation in the initial state and the QCD virtual corrections, increases the total cross section by about $60\%$ for a Higgs mass about $100$GeV at the LHC \cite{djouadi_H1}. 
Indeed, loop calculations are required in order to understand the structure of perturbative field theory and the uncertainties of the theoretical predictions. The only way to reduce the error of a theoretical prediction so that it can be comparable to the small error (say $10\%$) of precision measurements nowadays is to pick up higher order terms, \ie loop corrections.

There are two methods to calculate loop integrals: analytical and numerical methods. The traditional analytical method decomposes each
Feynman diagram's numerator into a sum of scalar and tensorial Passarino-Veltman functions. The advantage is that the whole calculation of cross sections involving the numerical integration over phase space is faster. The disadvantage is that the numerator decomposition usually results in huge algebraic expressions with various spurious singularities, among them the inverse of the Gram determinant (defined as $\det(G)=\det(2p_i.p_j)$ with $p_i$ are external momenta) which can vanish in some region of phase space. Recently, Denner and Dittmaier have developed a numerically stable method for reducing one-loop tensor integrals \cite{denner_5p,denner_1loop}, which has been
used in various electroweak processes including the $e^+e^-\to 4\ \text{fermions}$ process \cite{denner_ee_4fa,denner_ee_4fb}.
For the numerical method, the loop integration should be performed along with the integrations over the momenta of final state particles. In this method one should not decompose the various numerators but rather combine various terms in one common denominator. Thus the algebraic expression of the integrand is much simpler this way and no spurious singularities appear. The disadvantage is that the number of integration variables is large resulting in large integration errors. In both methods, the ultra-violet (UV)-, infrared (IR)- and collinear- divergences have to be subtracted before performing the numerical integration.

Recently, there has appeared {\em on-shell methods} to calculate
one-loop multi-leg QCD processes (see \cite{Bern:2007dw} for a
review). These methods are analytical but very different from
traditional methods based on Passarino-Veltman reduction
technique. On-shell methods have already led to a host of new
results at one loop, including the computation of non-trivial
amplitudes in QCD with an arbitrary number of external legs
\cite{Berger:2008sj, Berger:2006vq, Berger:2006ci}. These methods
work as follows. A generic one-loop amplitude can be expressed in
terms of a set of scalar master integrals multiplied by various
rational coefficients, along with the additional purely rational
terms. The relevant master integrals consist of box, triangle,
bubble and (for massive particles) tadpole integrals. All these
basic integrals are known analytically. The purely rational terms
have their origin in the difference between $D=4-2\varep$ and four
dimensions when using dimensional regularization. One way to
calculate the rational terms is to use on-shell recursion
\cite{Britto:2005fq, Britto:2004ap} to construct the rational
remainder from the loop amplitudes' factorization poles
\cite{Bern:2008ef, Berger:2008sj, Bern:2007dw}. The various
rational coefficients are determined by using generalized
unitarity cuts \cite{Britto:2004nc, Forde:2007mi}. The evaluation
is carried out in the context of the spinor formalism. Like the
traditional analytical method, spurious singularities occur in
intermediate steps. However, it is claimed in \cite{Bern:2007dw}
that they can be under control. More detailed studies on this
important issue are necessary to confirm this statement though.
On-shell methods can also deal with massive internal/external
particles \cite{Britto:2006fc} and hence can be used for electroweak
processes. It is not clear for us whether these on-shell methods
can be extended to include the case of internal unstable
particles.

Although the on-shell methods differ from the traditional analytical
methods in many respects, they have a common feature that one-loop
amplitudes are expressed in terms of a set of basic scalar loop
integrals. One may wonder if there is a method to express a one-loop
amplitude in terms of tree-level amplitudes? The answer was known
$45$ years ago by Feynman \cite{Feynman:1963, Feynman:1972}.
Feynman has proved that any diagram with closed loops can be
expressed in terms of sums (actually phase-space integrals) of tree diagrams.
This is called the Feynman Tree Theorem (FTT) whose very
simple proof can be found in \cite{Feynman:1972}. This theorem can
be used in several ways. The simplest application is to calculate
scalar loop integrals needed by other analytical methods described
above. The best application is to calculate loop corrections for
physical processes. Feynman has shown that this important
application can be realized for many processes. Let us explain
this a little bit more. After making use of the FTT, one has a lot
of tree diagrams obtained by cutting a $N$-point one-loop diagram with multiple cuts (single-cut, double-cut, $\ldots$, $N$-cut).
One can re-organize this result as a sum of sets
of tree diagrams, each set representing the complete set of tree
diagrams expected for some given physical process. In this way,
one obtains relations among the diagrams for various processes.
Surprisingly, no one has applied this FTT to calculate QCD/EW one-loop
corrections to important processes at colliders, to the best of our
knowledge. However, there is ongoing effort in this direction by
Catani, Gleisberg, Krauss, Rodrigo and Winter. They have very recently
proposed a method to numerically compute multi-leg one-loop cross sections in perturbative field theories  \cite{Catani:2008xa}.
The method relies
on the so-called duality relation between one-loop integrals and phase space integrals. This duality relation is very similar to
the FTT. The main difference is that the duality relation involves only single cuts of the one-loop diagrams. Interestingly,
the duality relation can be applied to one-loop diagrams with internal complex masses \cite{Catani:2008xa}.

In general, Higgs production processes involve unstable internal
particles. If these unstable particles can be on-shell then the
width effect can be relevant and therefore must be taken into account. In
particular, scalar box integrals with unstable internal particles
can develop a Landau singularity (to be discussed below) which is not integrable
at one-loop amplitude square level. In this case, the internal
widths are regulators as they move the singularity outside the
physical region. Thus, a good method to calculate one-loop corrections must be able to handle internal complex masses.

Independent of calculation methods, the analytic structure of \sma is intrinsic and is related to
fundamental properties like unitarity and causality \cite{book_eden}. Analytic properties of \sma can be studied by using 
Landau equations \cite{landau, book_eden} applied to an individual Feynman diagram. Landau equations are necessary and sufficient conditions
for the appearance of a pinch singularity of Feynman loop integrals \cite{Coleman:1965xm}.
Solutions of Landau equations are singularities
of the loop integral as a function of internal masses and external momenta, called Landau singularities. These singularities occur when
internal particles are on-shell. They can be finite
like the famous normal threshold in the case of one-loop two-point function. The normal thresholds are branch points \cite{book_eden}.
Landau singularities can be divergent like in the case of three-point and four-point functions. The former is integrable but the latter is
not at the level of one-loop amplitude squared. This four-point Landau divergence can be due to the presence of internal unstable particles and hence must be
regularized by taking into account their widths.
A detailed account on this topic is given in chapters~\ref{section_landau_introduction} and \ref{chapter_bbH2}. 

The main calculation of this thesis is to compute the leading electroweak one-loop correction to
Higgs production associated with two bottom quarks at the LHC in the SM. Our calculation involves $8$ tree-level diagrams and
$115$ one-loop diagrams with $8$ pentagons. The loop integrals include 2-point, 3-point, 4-point and 5-point functions which contain internal unstable particles,
namely the top-quark and the W gauge boson. Interestingly, Landau singularities occur in all those functions.
We follow the traditional analytical method of Veltman and Passarino \cite{pass_velt} to calculate the one-loop corrections. For the 5-point function part,
we have adapted the new reduction method of Denner and Dittmaier \cite{denner_5p}, which replaces the inverse of vanishing Gram determinant with the inverse
of the Landau determinant and hence replaces the spurious Gram singularities with the true Landau singularities of loop integrals.
In our opinion, this is one of the best ways to deal with those spurious Gram singularities. However, as will be explained in
chapter~\ref{section_landau_introduction}, the condition of vanishing Landau determinant is necessary but not sufficient for a Landau singularity to actually occur in the physical region.
Thus, spurious singularities can still be encountered but very rarely.
This new reduction method for 5-point functions has been implemented in the library LoopTools
\cite{looptools,looptools_5p} based on the library FF \cite{ff}. Our calculation has proved the efficiency of this method. The reason for us to choose this
traditional method is that our calculation involves massive internal particles. Furthermore, in order to deal with Landau singularities, our calculation
must include also complex masses.\\ 
Although the calculation method is well understood, the difficulty is that we have to handle very huge algebraic
expressions since we have to expand the numerator of each Feynman diagram. Thus, we cannot use the traditional amplitude squared method as it will
result in extremely enormous algebraic expressions of the total amplitude squared.
Fortunately, there is a very efficient way to organize the calculation based
on the helicity amplitude method (HAM) \cite{kleiss_stirling}. Using this HAM, one just needs to calculate all the independent helicity amplitudes
which are complex numbers. This way of calculating makes it very easy to divide the whole complicated computation into independent blocks
therefore factorizes out terms that occur several times in the calculation.

Our calculation consists of two parts. In the first part, we calculate the NLO corrections, \ie the interference terms between tree-level and one-loop
amplitudes. Although Landau singularities do appear in many one-loop diagrams, they are integrable hence do not cause any problem of
numerical instability. The bottom-quark mass is kept in this calculation. In the second part, we calculate the one-loop correction in the limit of
massless bottom-quark therefore the bottom-Higgs Yuakawa coupling vanishes. The process is loop induced and we have to calculate one-loop amplitude squared.
In this calculation, the Landau singularity of a scalar four-point function is not integrable and causes a severe problem of numerical instability
if $M_H\ge 2M_W$. This problem is solved by introducing a width for the top-quark and W gauge boson in the loop diagrams. It turns out that the
width effect is large if $M_H$ is around $2M_W$. 

Although the main calculation of this thesis is for a very specific process, we have gained several insights that can be equally used for other 
practical calculations. First of all, the method to optimise complicated loop calculations using the HAM is general. Second, the method to check the final/intermediate results by using QCD gauge invariance in the framework of the HAM can be used for any process with at least one gluon in the external states. Third, some general results related 
to Landau singularities are new and can be used for practical purposes. They are equations (\ref{landau_cond1}) and (\ref{TN0_LLS}). Finally, we 
have applied the loop calculation method of 't Hooft and Veltman to write down explicitly two formulae to calculate scalar box integrals with complex 
internal masses. They are equations (\ref{d0_y_13_sumij}) and (\ref{box_adj}). The restriction is that at least two external momenta are lightlike. We have implemented those two formulae into the library LoopTools.  

The outline of this thesis is as follows.
First, a short review of the SM including QCD is presented in chapter \ref{chapter_sm}. We pay special attention
to the one-loop renormalisation of the EW part and the Higgs sector.
We also give a short introduction to the Minimal Supersymmetric Standard Model and discuss its Higgs sector in the same chapter. 
In chapter \ref{higgs_production} we discuss the dominant mechanisms for SM Higgs production at the LHC and Higgs signatures at the colliders. 
In chapter \ref{chapter_bbH1} we present the main calculation of this thesis, one-loop Yukawa corrections to the SM process $\ppbbH$ at the LHC, for the 
case $M_H\le 150$GeV. There are two reasons to start with small values of the Higgs mass: it is preferred by the latest EW data and there is no problem of numerical instability related to Landau singularities. The framework of a one-loop calculation based on the helicity amplitude method is also given in this chapter. In chapter \ref{section_landau_introduction} we explain in detail the Landau singularities of a general one-loop Feynman diagram. We emphasize the conditions to have a Landau singularity and its nature. In chapter \ref{chapter_bbH2}, we complete the study of chapter \ref{chapter_bbH1} for larger values of $M_H$, up to $250$GeV. We show that the one-loop process $\ggbbH$ is an ideal example for understanding Landau singularities. It contains several 
types of Landau singularities related to two-point, three-point and four-point functions. The conclusions are given in chapter \ref{chapter_conclusions}.\\ This thesis includes several appendices. In appendix \ref{appendix-helicity} we explain 
the helicity amplitude method and how to check the correctness of the result by using QCD gauge invariance. In appendix \ref{optimisation} we show how to 
optimise the calculation of various one-loop helicity amplitudes and how that can be easily achieved by using FORM. Appendix \ref{appendix_integral_phase} concerns the phase space integral of $2\to 3$ process. We explain how to use the Fortran routine BASES \cite{bases} to do numerical integration. 
Appendix \ref{appendix_math} gives useful mathematical formulae related to loop integrals. In appendix \ref{appendix-box-integral} we explain the analytical 
calculation of scalar one-loop four-point integrals with complex internal masses. The restriction is that at least two external momenta are lightlike.

\chapter{The Standard Model and beyond}
\label{chapter_sm}
The Glashow-Salam-Weinberg (GSW) model of the electroweak
interaction was proposed by Glashow \cite{glashow_sm}, Weinberg
\cite{weinberg_sm} and Salam \cite{salam_sm} for leptons and
extended to the hadronic degrees of freedom by Glashow, Iliopoulos
and Maiani \cite{Glashow:1970gm}. The GSW model is a Yang-Mills
theory \cite{Yang:1954ek} based on the symmetry group
$SU(2)_L\times U(1)_Y$. It describes the electromagnetic and weak
interactions of the known $6$ leptons and $6$ quarks. The electromagnetic
interaction is mediated by a massless gauge boson, the photon
($\gamma$). The short-range weak interaction is carried by $2$
massive gauge bosons, $Z$ and $W$. The strong interaction,
mediated by the massless gluon, is also a Yang-Mills theory based
on the gauge group $SU(3)_C$. This is known as Quantum
chromodynamics (abbreviated as QCD) \cite{GellMann:1964nj,
Fritzsch:1973pi, Gross:1973id, Politzer:1973fx}. The Standard
Model of particle physics is just a trivial combination of GSW
model and QCD. The particle content of the SM is listed in Table.
\ref{tab_sm}. There is an additional scalar field called the Higgs
boson ($H$), the only remnant of the spontaneous symmetry breaking
(SSB) mechanism invented by Brout, Englert, Guralnik, Hagen, Higgs
and Kibble \cite{Higgs:1964ia, Higgs:1966ev, Englert:1964et,
Guralnik:1964eu, Kibble:1967sv}. The SSB mechanism is responsible
for explaining the mass spectrum of the SM.

To date, almost all experimental tests of the three forces
described by the Standard Model agree with its predictions
\cite{PDBook, range_MH, Altarelli:2008}. The measurements of $M_W$
and $M_Z$ together with the fact that their relation
$M_W^2=M_Z^2c_W^2$ (with $c_W^2\approx 0.77$ defined in Eq.
(\ref{def_cW_sW})) has been experimentally proven imply two things.
First, the existence of massive gauge bosons means that the local
gauge symmetry is broken. Second, the mass relation
indicates that the effective Higgs (be it fundamental or
composite) is isospin doublet \cite{Altarelli:2008}. Experiments
have also confirmed that couplings that are mass-independent like
the ones of quarks and leptons to the $W^\pm$ and $Z$ gauge bosons
or triple couplings among electroweak gauge bosons agree with
those described by the gauge symmetry \cite{Altarelli:2008}. It
means that the only sector which remains untested is the mass
couplings or in other words the nature of SSB mechanism.

The primary goal of the LHC is to find the scalar Higgs boson and
to understand its properties. The main drawback here is that we do
not know the value of the Higgs mass which uniquely defines the Higgs
profile. The LEP direct searches for the Higgs and precision
electroweak measurements lead to the conclusion that
$114\text{GeV}< M_H <190$GeV \cite{range_MH}. The most prominent
property of the Higgs is that it couples mainly to heavy
particles at tree level. This has two consequences at the LHC: the Higgs production cross
section is small and the Higgs decay product is very complicated
and usually suffers from huge QCD background. Thus, it is
completely understandable that searching for the Higgs is not an
easy task, even at the LHC.
\begin{table}[t]
\caption{Particle content of the standard model}
\begin{center}
\begin{tabular}{|c|c|c|c|}  \hline
 &Particles & Spin & Electric charge \\
\hline
 Leptons & $(e,\mu,\tau)$ &$1/2$ &$-1$\\
  & $(\nu_e,\nu_\mu,\nu_\tau)$ &$1/2$ &$0$\\
\hline
 Quarks & $(u,c,t)$ &$1/2$ &$2/3$\\
  & $(d,s,b)$ &$1/2$ &$-1/3$\\
\hline
 Gauge bosons & gluon ($g$) &$1$ &$0$\\
  & $(\gamma,Z)$ &$1$ &$0$\\
  & $W^\pm$ &$1$ &$\pm 1$\\
\hline
 Higgs & $H$ &$0$ &$0$\\
\hline
\end{tabular}\label{tab_sm}
\end{center}
\end{table}
\section{QCD}
\label{section_QCD}
The classical QCD Lagrangian reads
\bea
\Lar_{QCD}=\bar{\psi}(i\slashD-m)\psi-\fr{1}{2}\Trace F_{\mu\nu}F^{\mu\nu},
\label{Lar_qcd}
\eea
where
\bea
\slashD&=&\gamma^\mu D_\mu, \hs D_\mu=\partial_\mu-ig_s\A_\mu , \hs \A_\mu=A_\mu^aT_a,\crn
F_{\mu\nu}&=&\partial_\mu\A_\nu-\partial_\nu\A_\mu-ig_s[\A_\mu,\A_\nu],
\eea
with $a=1,\ldots,8$; $\psi$ is a fermion field belonging to the triplet representation of $SU(3)_C$ group;
$A$ the gauge boson field and $g_s$ is the strong coupling; $T_a$ are Gell-Mann generators. The corresponding Feynman rules in the 't Hooft-Feynman gauge read:
\begin{center}
\begin{minipage}[c]{4cm}
\begin{center}
\includegraphics[width=4cm]{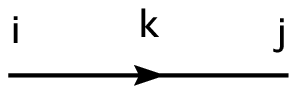}
\end{center}
\end{minipage}
\hspace*{3mm}
\begin{minipage}[c]{8cm}
\[\fr{-\delta_{ij}}{\slashk-m+i\eps} \]
\end{minipage}\\[7mm]
\begin{minipage}[c]{4cm}
\begin{center}
\includegraphics[width=4cm]{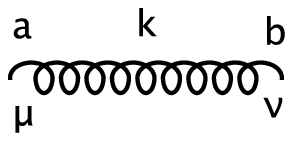}
\end{center}
\end{minipage}
\hspace*{3mm}
\begin{minipage}[c]{8cm}
\[\fr{\delta_{ab}g_{\mu\nu}}{k^2+i\eps} \]
\end{minipage}\\[7mm]
\begin{minipage}[c]{4cm}
\begin{center}
\includegraphics[width=4cm]{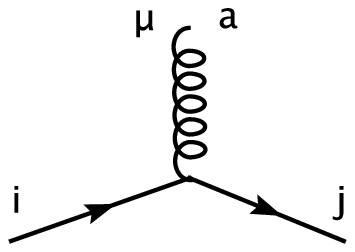}
\end{center}
\end{minipage}
\hspace*{3mm}
\begin{minipage}[c]{8cm}
\[g_s\gamma^\mu\left(T_a\right)_{ji}\]
\end{minipage}\\[7mm]
\begin{minipage}[c]{4cm}
\begin{center}
\includegraphics[width=4cm]{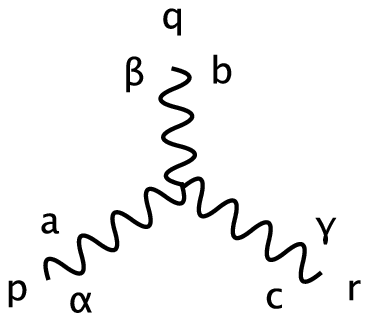}
\end{center}
\end{minipage}
\hspace*{3mm}
\begin{minipage}[c]{8cm}
\[-ig_sf^{abc}[(p-q)_\gamma g_{\alpha\beta}+(q-r)_{\alpha}g_{\beta\gamma}+(r-p)_{\beta}g_{\alpha\gamma}] \]
\end{minipage}\\[7mm]
\begin{minipage}[c]{4cm}
\begin{center}
\includegraphics[width=4cm]{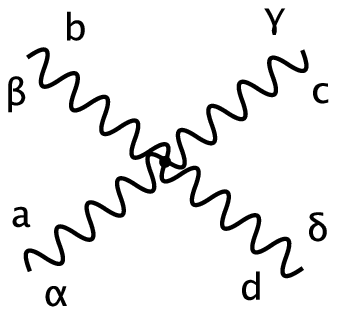}
\end{center}
\end{minipage}
\hspace*{3mm}
\begin{minipage}[c]{8cm}
\bea
&&g_s^2f^{abe}f^{cde}(g_{\alpha\gamma}g_{\beta\delta}-g_{\alpha\delta}g_{\beta\gamma})\crn
&+& g_s^2f^{ace}f^{bde}(g_{\alpha\beta}g_{\gamma\delta}-g_{\alpha\delta}g_{\beta\gamma})\crn
&+& g_s^2f^{ade}f^{bce}(g_{\alpha\beta}g_{\gamma\delta}-g_{\alpha\gamma}g_{\beta\delta})\nn
\eea
\end{minipage}
\end{center}
We have adopted the Feynman rules of \cite{aoki,grace} (derived by using $\Lar$) which differ from the normal Feynman rules (derived by using $i\Lar$) by a factor $i$. One can use those Feynman rules to calculate tree-level QCD processes or QED-like processes by keeping in mind that the gluon has only two transverse polarisation components. However, in a general situation where a loop calculation is involved one needs to quantize the classical Lagrangian (\ref{Lar_qcd}). The covariant quantization following the Faddeev-Popov method \cite{Faddeev:1967fc} introduces unphysical scalar Faddeev-Popov ghosts with additional Feynman rules:
\begin{center}
\begin{minipage}[c]{4cm}
\begin{center}
\includegraphics[width=4cm]{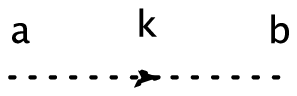}
\end{center}
\end{minipage}
\hspace*{3mm}
\begin{minipage}[c]{8cm}
\[\fr{-\delta_{ab}}{k^2} \]
\end{minipage}\\[7mm]
\begin{minipage}[c]{4cm}
\begin{center}
\includegraphics[width=4cm]{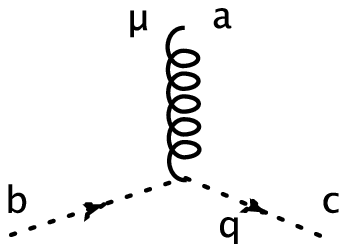}
\end{center}
\end{minipage}
\hspace*{3mm}
\begin{minipage}[c]{8cm}
\[-igf^{abc}q^\alpha\]
\end{minipage}
\end{center}
The main difference between QCD and QED is that the gluon couples to itself while the photon does not. In QED, only
the transverse photon can couple to the electron hence the unphysical components (longitudinal and scalar polarisations) decouple from
the theory and the Faddeev-Popov ghosts do not appear. The same thing happens for the gluon-quark coupling. However, an external transverse gluon can
couple to its unphysical states via its triple and quartic self couplings. Those unphysical states, in some situation, can propagate as internal particles without coupling to any quarks and give an unphysical contribution to the final result. In that situation, one has to take into account also the ghost contribution for compensation.

Indeed, there is another way to calculate QCD processes by taking into account only the physical contribution, \ie only the transverse gluon components involve and no ghosts appear. This is called the axial (non-covariant) gauge \cite{Bassetto:1991ue}. The main difference compared to the above covariant gauge is with the form of the gluon propagator. The covariant propagator includes the unphysical polarisation states via\footnote{See p.$511$ of \cite{Peskin}.}
\bea
g_{\mu\nu}=\eps_\mu^{-}\eps_\nu^{+*}+\eps_\mu^+\eps_\nu^{-*}-\sum_{i=1}^2\eps_{i\mu}\eps_{i\nu}^{*},
\eea
where $\eps_\mu^{\pm}$ are two unphysical polarisation states and $\eps_{i\mu}$ with $i=1,2$ are the two transverse polarisation states. In the axial gauge, the gluon propagator takes the form
\bea
P_{\mu\nu}&=&-\fr{\delta_{ab}}{k^2+i\eps}\sum_{i=1}^2\eps_{i\mu}\eps_{i\nu}^{*}\crn
&=&-\fr{\delta_{ab}}{k^2+i\eps}\left[-g_{\mu\nu}+\fr{k_\mu n_\nu +k_\nu n_\mu}{n.k}\right]
\label{gluon_axial}
\eea
with $n^2=0$ and $n.k\neq 0$, which includes only the transverse polarisation states. The main drawback of this axial gauge is that the propagator's numerator becomes very complicated.

The main calculation of this thesis is to compute the one-loop electroweak corrections to the process $\ggbbH$. Though the triple gluon coupling does appear in various Feynman diagrams, it always couples to a fermion line hence the virtually unphysical polarisation states cannot contribute and the ghosts do not show up. We will therefore use the covariant Feynman rules and take into account only the contribution of the transverse polarisation states of the initial gluons\footnote{If one follows the traditional amplitude squared method and wants to use the polarisation sum identity $\sum\eps_\mu\eps_\nu =-g_{\mu\nu}$ then one has to consider the Feynman diagrams with two ghosts in the initial state.}.
\section{The Glashow-Salam-Weinberg Model}
The classical Lagrangian of the GSW model is composed of a gauge, a Higgs, a fermion and a Yukawa part \footnote{For more technical details of the GSW model, its one-loop renormalisation prescription and Feynman rules, we refer to \cite{Denner:1991kt, aoki, grace}.}
\bea
\Lar_{C}=\Lar_{G}+\Lar_{H}+\Lar_{F}+\Lar_{Y}.
\eea
Each of them is separately gauge invariant and specified as follows:
\subsection{Gauge sector}
The Lagrangian of the gauge part of the group $SU(2)_L\times U(1)_Y$ reads
\bea
\Lar_{G}=-\fr{1}{4}(\partial_{\mu}W_\nu^a-\partial_\nu W_\mu^a+g\eps^{abc}W_\mu^bW_\nu^c)^2
-\fr{1}{4}(\partial_\mu B_\nu -\partial_\nu B_\mu)^2,
\eea
where $a,b,c\in \{1,2,3\}$, $W_\mu^a$ are the $3$ gauge fields of the $SU(2)$ group, $B_\mu$ is the
$U(1)$ gauge field, the $SU(2)$ gauge coupling $g$, the $U(1)$ gauge coupling $g^\prime$ and $\eps^{abc}$ are
the totally antisymmetric structure constants of $SU(2)$. The covariant derivative is given by
\bea
D_{\mu}=\partial_\mu-igT^aW_\mu^a-ig^\prime YB_\mu ,
\eea
where $T^a =\sigma^a/2$ with $\sigma^a$ are the usual Pauli matrices, the hypercharge according to
the Gell-Mann Nishijima relation
\bea
Q=T^3+Y.
\eea
The physical fields $W^\pm , Z, A$ relate to the $W^a$ and $B$ fields as
\bea
\left\{
\begin{array}{ll}
W_{\mu}^{\pm}& = \fr{W_\mu^1\mp iW_\mu^2}{\sqrt{2}}\\
Z_\mu& = c_WW_\mu^3-s_WW_\mu^0\\
A_\mu & = s_WW_\mu^3+c_WW_\mu^0,
\end{array}
\right. \eea with \bea c_W=\fr{g}{\sqrt{g^2+g^{\prime 2}}}, \hs
s_W=\fr{g^\prime}{\sqrt{g^2+g^{\prime 2}}}, \label{def_cW_sW} \eea
the electromagnetic coupling $e$ \bea
e=\fr{gg^\prime}{\sqrt{g^2+g^{\prime 2}}}, \hs g=\fr{e}{s_W}, \hs
g^\prime=\fr{e}{c_W}. \eea
\subsection{Fermionic gauge sector}
Left-handed fermions $L$ of each generation belong to $SU(2)_L$ doublets while right-handed fermions $R$ are in $SU(2)_L$
singlets. The fermionic gauge Lagrangian is just
\bea
\Lar_{F}=i\sum\bar{L}\gamma^\mu D_\mu L+i\sum\bar{R}\gamma^\mu D_\mu R,
\eea
where the sum is assumed over all doublets and singlets of the three generations. Note that in the covariant derivative $D_\mu$ acting on right-handed fermions the term involving $g$ is absent since they are $SU(2)_L$ singlets. Neutrinos are left-handed in the SM. Fermionic mass terms are forbidden by 
gauge invariance. They are introduced through the interaction with the scalar Higgs doublet.
\subsection{Higgs sector}
Mass terms for both the gauge bosons and fermions are generated in a gauge invariant way through the Higgs mechanism.
To that effect one introduces minimally a complex scalar $SU(2)$ doublet field with hypercharge $Y=1/2$
\bea
\Phi=\left(
\begin{array}{l}
\phi^+\\
\phi^0
\end{array}
\right)=\left(
\begin{array}{l}
\hs\hs\hs\hs i\chi^+\\
(\upsilon+H-i\chi_3)/\sqrt{2}
\end{array}
\right), \hs \langle 0\mid\Phi\mid 0\rangle =\upsilon/\sqrt{2},
\eea
where the electrically neutral component has been given a non-zero vacuum expectation value $\upsilon$ to break spontaneously the gauge symmetry $SU(2)_L\times U(1)_Y$ down to $U(1)_Q$. The scalar Lagrangian writes
\bea
\Lar_{H}=(D_\mu \Phi)^\dagger(D^\mu \Phi)-V(\Phi), \hs V(\Phi)=-\mu^2\Phi^\dagger\Phi+\lambda(\Phi^\dagger\Phi)^2.
\label{V_Higgs_sm}
\eea
After rewriting $\Lar_{H}$ in terms of $\chi^\pm$, $\chi_3$, $H$ and imposing the minimum condition on the potential
$V(\Phi)$ one sees that $\chi^\pm$ and $\chi_3$ are massless while the Higgs boson obtains a mass
\bea
M_H^2=2\mu^2, \hs \mu^2=\lambda\upsilon^2.
\eea
$\chi^\pm$, $\chi_3$ are called the Nambu-Goldstone bosons. They are unphysical degrees of freedom and get absorbed by
the $W^\pm$ and $Z$ to give the latter masses given by
\bea
M_W=\fr{e\upsilon}{2s_W}, \hs M_Z=\fr{e\upsilon}{2s_Wc_W}.
\eea
\subsection{Fermionic scalar sector}
Fermion masses require the introduction of Yukawa interactions of fermions and the scalar Higgs doublet
\bea
\Lar_{Y}=-\sum_{up}f^{ij}_{U}\bar{L}^{i}\tilde{\Phi}R^j_{U}-\sum_{down}f^{ij}_{D}\bar{L}^i\Phi R^j_{D}+(h.c.), \hs M_{U,D}^{ij}=\fr{f_{U,D}^{ij}\upsilon}{\sqrt{2}}, \label{sm_lar_yukawa}
\eea
where $f_{U,D}^{ij}$ with $i,j\in\{1,2,3\}$ the generation indices are Yukawa couplings, $\tilde{\Phi}=i\sigma_2\Phi^*$. Neutrinos, which are only right-handed, do not couple to the Higgs boson and thus are massless in the SM. The diagonalization of the fermion mass matrices $M_{U,D}^{ij}$ introduces a matrix into the quark-W-boson couplings, the unitary quark mixing matrix \cite{PDBook}
\bea
V=\left( \begin{array}{ccc}
V_{ud} & V_{us} & V_{ub}\\
V_{cd} & V_{cs} & V_{cb}\\
V_{td} & V_{ts} & V_{tb}
\end{array}\right)=
\left( \begin{array}{ccc}
0.97383 & 0.2272 & 0.00396\\
0.2271 & 0.97296 & 0.04221\\
0.00814 & 0.04161 & 0.9991
\end{array}\right),
\eea
which is well-known as Cabibbo-Kobayashi-Maskawa (CKM) matrix. There is no corresponding matrix in the lepton sector as the
neutrinos are massless in the SM. 

For later reference, we define $\la_f=\sqrt{2}m_f/\upsilon$ where $m_f$ is the physical mass of a fermion. 
\subsection{Quantisation: Gauge-fixing and Ghost Lagrangian}
The classical Lagrangian $\Lar_C$ has gauge freedom. A Lorentz invariant quantisation requires a gauge fixing (otherwise the propagators of gauge fields are not well-defined). The 't Hooft linear gauge fixing terms read
\bea
F^{A}&=&(\xi^{A})^{-1/2}\partial^\mu A_\mu,\crn
F^Z&=&(\xi_1^Z)^{-1/2}\partial^\mu Z_\mu-M_Z(\xi_2^Z)^{1/2}\chi_3,\crn
F^{\pm}&=&(\xi_1^W)^{-1/2}\partial^\mu W^{\pm}_\mu+M_W(\xi_2^W)^{1/2}\chi^{\pm}.
\label{gauge_linear}
\eea
This leads to a gauge fixing Lagrangian
\bea
\Lar_{fix}=-\fr{1}{2}[(F^A)^2+(F^Z)^2+2F^+F^-].
\eea
$\Lar_{fix}$ involves the unphysical components of the gauge fields, \ie field components with negative norm, which lead to a serious problem that the theory is not
gauge invariant and violates unitarity. In order to compensate their effects one introduces Faddeev Popov ghosts $u^\alpha(x)$, $\bar{u}^{\alpha}(x)$ ($\alpha=A,Z,W^\pm$) with the Lagrangian
\bea
\Lar_{ghost}=\bar{u}^\alpha(x)\fr{\delta F^\alpha}{\delta\theta^\beta(x)}u^\beta(x),
\eea
where $\fr{\delta F^\alpha}{\delta\theta^\beta(x)}$ is the variation of the gauge fixing operators $F^\alpha$ under infinitesimal gauge transformation parameter $\theta^\beta(x)$. An element of the $SU(2)_L\times U(1)_Y$ group has a typical form
$G=e^{-igT^\alpha \theta_\alpha (x)-ig^\prime Y\theta_Y (x)}$. Faddeev Popov ghosts are scalar fields following anticommutation rules and belonging to the adjoint representation of the gauge group.

In a practical calculation, the final result does not depend on gauge parameters. Thus one can choose for these parameters some special values to make the calculation simpler. For tree-level calculations, one can think of the unitary gauge $\xi^Z=\xi^W=\infty$ where the Nambu-Goldstone bosons and ghosts do not appear and the number of Feynman diagrams is minimized.
For general one-loop calculations, it is more convenient to use the 't Hooft Feynman gauge $\xi^A=\xi^Z=\xi^W=1$ where the numerator structure is simplest.

It is worth knowing that the 't Hooft linear gauge fixing terms defined in Eq. (\ref{gauge_linear}) can be generalised to include non-linear terms as follows \cite{Boudjema:1995cb, grace}
\bea
F^Z&=&(\xi_Z)^{-1/2}\left[\partial^\mu Z_\mu+M_Z\xi_Z^\prime\chi_3+\fr{g}{2c_W}\xi_Z^\prime\tilde{\eps}H\chi_3\right],\crn
F^{\pm}&=&(\xi_W)^{-1/2}\left[\partial^\mu W^{\pm}_\mu + M_W\xi_W^\prime\chi^{\pm}\right. \crn
&\mp & \left. (ie\tilde{\alpha}A_\mu+igc_W\tilde{\beta}Z_\mu)W^{\mu\pm}+\fr{g}{2}\xi_W^\prime
(\tilde{\delta}H \pm i\tilde{\kappa}\chi_3)\chi^{\pm}\right],
\eea
with the gauge fixing term for the photon $F^A$ remains unchanged. It is simplest to choose $\xi_{Z,W}^\prime=\xi_{Z,W}$.
Those non-linear fixing terms involve five extra arbitrary parameters
$\zeta=(\tilde{\alpha},\tilde{\beta},\tilde{\delta},\tilde{\kappa},\tilde{\eps})$. The advantage of this non-linear gauge is twofold. First, in an automatic calculation involving a lot of Feynman diagrams one can perform the gauge-parameter independence checks to find bugs. Second, for some specific calculations involving gauge and scalar fields one can kill some triple and quartic vertices by judiciously choosing some of those gauge parameters and thus reduce the number of Feynman diagrams. This is based on the fact that the new gauge parameters modify some vertices involving the gauge, scalar and ghost sector and at the same time introduce new quartic vertices \cite{grace}. In the most general case, the Feynman rules with non-linear gauge are much more complicated than those with 't Hooft linear gauge, however.

With $\Lar_{fix}$ and $\Lar_{ghost}$ the complete renormalisable Lagrangian of the GSW model reads
\bea
\Lar_{GSW}=\Lar_{C}+\Lar_{fix}+\Lar_{ghost}.
\label{lagrangian_full}
\eea
\subsection{One-loop renormalisation}
\label{subsection_one-loop-renorm}
Given the full Lagrangian $\Lar_{GSW}$ above, one proceeds to calculate the cross section of some physical process.
In the framework of perturbative theory this can be done order by order. At tree level, the cross section
is a function of a set of input parameters which appear in $\Lar_{GSW}$. These parameters can be chosen to be
$O=\{e,M_W,M_Z,M_H,M_{U,D}^{ij}\}$ which have to be determined experimentally. There are direct relations between these
parameters and physical observables at tree level. However, these direct relations are destroyed when one considers loop
corrections.
Let us look at the case of $M_W$ as an example. The tree-level $W$ mass is directly related to the Fermi constant
$G_\mu$ through
\bea s_W^2M_W^2=\fr{\pi\alpha}{\sqrt{2}G_\mu}.\eea
When one takes into account higher order corrections, this becomes \cite{Sirlin:1980nh, Marciano:1980pb, Sirlin:1981yz}
\bea
s_W^2M_W^2=\fr{\pi\alpha}{\sqrt{2}G_\mu}\fr{1}{1-\Delta r},
\eea
where $\Delta r$ containing all loop effect is a complicated function of $M_W$ and other input parameters. A question arises naturally, how to calculate $\Delta r$ or some cross section at one-loop level? The answer is the following.
If we just use the Lagrangian given in Eq. (\ref{lagrangian_full}), follow the corresponding Feynman rules to calculate
all the relevant one-loop Feynman diagrams then we will end up with something infinite. This is because there are
a lot of one-loop diagrams being UV-divergent.
This problem can be solved if $\Lar_{GSW}$ is renormalisable. The renormalisability of nonabelian gauge theories with spontaneous symmetry breaking and thus the GSW model was proven by 't Hooft \cite{Hooft:1971fh, Hooft:1971rn}.
The idea of renormalisation is that we have to get rid of all UV-divergence terms originating from one-loop diagrams
by redefining a finite number of fundamental input parameters $O$ in the original Lagrangian $\Lar_{GSW}$. This is done as follows
\bea
e&\to& (1+\delta Y)e,\crn
M&\to& M+\delta M,\crn
\psi &\to& (1+\delta Z^{1/2})\psi.
\label{rules_renorm}
\eea
The latter is called wave function renormalisation. The renormalisation constants $\delta Y$, $\delta M$ and $\delta Z^{1/2}$ are fixed by using renormalisation conditions to be discussed later. The one-loop renormalised Lagrangian writes
\bea
\Lar_{GSW}^{1-loop}=\Lar_{GSW}+\delta\Lar_{GSW}.
\label{Lar_one-loop}
\eea
The parameters $O$ in $\Lar_{GSW}^{1-loop}$ are now called the renormalised parameters determined from experiments. From this renormalised Lagrangian one can write down the corresponding Feynman rules and use them to calculate $\Delta r$ or any
cross section at one loop. The results are guaranteed to be finite by 't Hooft.

We now discuss the renormalisation conditions which define a renormalisation scheme. In this thesis, we stick with the on-shell scheme where all renormalisation conditions are formulated on mass shell external fields. To fix $\delta Y$, one imposes a condition on the $e^+ e^- A$ vertex as in QED. The condition reads
\bea
(e^+ e^- A\;\; \text{one-loop term}+e^+ e^- A\;\; \text{counterterm})\mid_{q=0,p_{\pm}^2=m_e^2}=0,
\eea
where $q$ is the photon momentum, $p_{\pm}$ are the momenta of $e^{\pm}$ respectively. All $\delta M$s are fixed by the requirement that the corresponding renormalised mass parameter is equal to the physical mass which is the single pole of the two-point Green function. This translates into the condition that the real part of the inverse of the corresponding propagator is zero. $\delta Z^{1/2}$s are found by requiring that the residue of the propagator at the pole is $1$. To be explicit we look at the cases of Higgs boson, fermions and gauge bosons, which will be useful for our main calculation of $\ppbbH$.
The Higgs one-particle irreducible two-point function is $\tilde{\Pi}^H(q^2)$ with $q$ the Higgs momentum. One calculates this function by using Eq. (\ref{Lar_one-loop})
\bea
\tilde{\Pi}^H(q^2)=\Pi^H(q^2)+\hat{\Pi}^H(q^2)
\eea
where the counterterm contribution is denoted by a caret, the full contribution is denoted by a tilde. The two renormalisation conditions read
\bea
\Rel\tilde{\Pi}^H(M_H^2)=0, \hs \fr{d}{dq^2}\Rel\tilde{\Pi}^H(q^2)\Big\vert_{q^2=M_H^2}=0.
\eea
This gives
\bea
\delta Z_H^{1/2}=-\fr{1}{2}\fr{d}{dq^2}\Rel\Pi^H(q^2)\Big\vert_{q^2=M_H^2}.
\label{dZH_PiH}
\eea
For a fermion with $\psi=\psi_L+\psi_R$ ($\psi_{L,R}=P_{L,R}\psi$ with $P_{L,R}=\fr{1}{2}(1\mp\gamma_5)$, respectively), the one-particle irreducible two-point function takes the form
\bea
\tilde\Sigma(q^2)&=&\Sigma(q^2)+\hat\Sigma(q^2),\crn
\Sigma(q^2)&=&K_1+K_\gamma\slashq+K_{5\gamma}\slashq\gamma_5,\crn
\hat\Sigma(q^2)&=&\hat K_1+\hat K_\gamma\slashq+\hat K_{5\gamma}\slashq\gamma_5,
\eea
with
\bea
\hat K_1&=&-m_f(\delta Z_{f_L}^{1/2}+\delta Z_{f_R}^{1/2})-\delta m_f,\crn
\hat K_\gamma&=&(\delta Z_{f_L}^{1/2}+\delta Z_{f_R}^{1/2}),\crn
\hat K_{5\gamma}&=&-(\delta Z_{f_L}^{1/2}-\delta Z_{f_R}^{1/2}).
\eea
The two renormalisation conditions become
\bea
\left\{
\begin{array}{ll}
m_f\Rel\tilde{K}_\gamma(m_f^2)+\Rel\tilde{K}_1(m_f^2)=0\hs \textrm{and} \hs \Rel\tilde{K}_{5\gamma}(m_f^2)=0\\
\fr{d}{d\slashq}\Rel\left[\slashq \tilde{K}_\gamma(q^2)+\tilde{K}_1(q^2) \right]_{\slashq=m_f}=0.
\end{array}
\right.
\eea
This gives
\bea
\delta m_f&=&\Rel\Big(m_f K_\gamma(m_f^2)+K_1(m_f^2)\Big),\crn \delta
Z_{f_L}^{1/2}&=&
\fr{1}{2}\Rel\Big(K_{5\gamma}(m_f^2)-K_\gamma(m_f^2)\Big)-m_f\fr{d}{dq^2}\Big(m_f\Rel K_\gamma(q^2)+\Rel K_1(q^2)\Big)\/\Big\vert_{q^2=m_f^2},\crn
\delta
Z_{f_R}^{1/2}&=&-\fr{1}{2}\Rel\Big(K_{5\gamma}(m_f^2)+K_\gamma(m_f^2)\Big)
-m_f\fr{d}{dq^2}\Big(m_f\Rel K_\gamma(q^2)+\Rel K_1(q^2)\Big)\/\Big\vert_{q^2=m_f^2}.\crn
\label{counter_vertex}
\eea
For gauge bosons, the one-particle irreducible two-point functions write\footnote{For massless gauge bosons like the photon, the longitudinal part $\Pi_L^V$ vanishes.}
\bea
\tilde\Pi_T^V&=&\Pi_T^V+\hat\Pi_T^V,\crn
\Pi_{\mu\nu}^V(q^2)&=&(g_{\mu\nu}-\fr{q_\mu q_\nu}{q^2})\Pi_T^V(q^2)+\fr{q_\mu q_\nu}{q^2}\Pi_L^V(q^2),\crn
\hat\Pi_{\mu\nu}^V(q^2)&=&(g_{\mu\nu}-\fr{q_\mu q_\nu}{q^2})\hat\Pi_T^V(q^2)+\fr{q_\mu q_\nu}{q^2}\hat\Pi_L^V(q^2),\crn
\hat\Pi_T^V&=&\delta M_V^2+2(M_V^2-q^2)\delta Z_V^{1/2}, \hs \hat\Pi_L^V=\delta M_V^2+2M_V^2\delta Z_V^{1/2},
\eea
where $V=W,Z$. We do not touch the photon\footnote{In a general case, one should keep in mind that there is mixing between the photon and the Z boson.} since it is irrelevant to the calculations in this thesis, which are only related to the Yukawa sector. It is sufficient to
impose the two renormalisation conditions (for the pole-position and residue) on the transverse part $\Pi_T^V(q^2)$ to determine $\delta M_V^2$ and
$\delta Z_V^{1/2}$. The longitudinal part is automatically renormalised when the transverse part is, if the theory is renormalisable. The two conditions write
\bea
\Rel\tilde\Pi_T^V(M_V^2)=0, \hs \fr{d}{dq^2}\Rel\tilde\Pi_T^V(q^2)\Big\vert_{q^2=M_V^2}=0,
\eea
which give
\bea
\delta M_V^2=-\Rel\Pi_T^V(M_V^2), \hs \delta Z_V^{1/2}=\fr{1}{2}\fr{d}{dq^2}\Rel\Pi_T^V(q^2)\Big\vert_{q^2=M_V^2}.
\label{renorm_gauge}
\eea
In practical calculations, one has to calculate $\Pi^H(q^2)$, $K_1(q^2)$, $K_\gamma(q^2)$, $K_{5\gamma}(q^2)$ and $\Pi_T^V(q^2)$ as sums of various
two-point functions. The full results in the SM can be found in \cite{aoki,grace,Denner:1991kt}.
\section{Higgs Feynman Rules}
\label{section_HFrules}
In order to understand the phenomenology of Higgs production, it is important to write down the relevant Feynman rules.

The Feynman rules listed here are taken from \cite{grace}. Their Feynman rules derived from $\Lar_{GSW}$ differs 
from the normal Feynman rules derived by using $i\Lar_{GSW}$ by a factor $i$\footnote{In the QCD section \ref{section_QCD} we have adapted the same 
rules of this section.}. A particle at the endpoint
\textit{enters} the vertex. For instance, if a line is denoted as
$W^+$, then the line shows either the incoming $W^+$ or the
outgoing $W^-$. The momentum assigned to a particle is defined as
\textit{inward}. The following Feynman rules are for the linear gauge.
\subsubsection{Propagators}
\begin{tabular}{cll}
\hline $ W^{\pm} $ \rule[-5mm]{0mm}{12mm}
 & $\displaystyle{ \frac{1}{k^2-M_W^2}
 \left( g_{\mu\nu}-(1-\xiw)\frac{k_{\mu}k_{\nu}}{k^2-\xiw M_W^2}\right) }$ \\
$ Z $ \rule[-5mm]{0mm}{12mm}
 & $\displaystyle{ \frac{1}{k^2-M_Z^2}
 \left( g_{\mu\nu}-(1-\xiz)\frac{k_{\mu}k_{\nu}}{k^2-\xiz M_Z^2}\right) }$ \\
\hline $ f $ \rule[-5mm]{0mm}{12mm}
& $\displaystyle{\frac{-1}{\slashk-m_f}}$ \\
\hline $ H $ \rule[-5mm]{0mm}{12mm}
 & $\displaystyle{\frac{-1}{k^2-M_H^2}}$ \\
\hline $ \chi^{\pm} $ \rule[-5mm]{0mm}{12mm}
 & $\displaystyle{\frac{-1}{k^2-\xiw M_W^2}}$ \\
\hline $ \chi_3 $ \rule[-5mm]{0mm}{12mm}
 & $\displaystyle{\frac{-1}{k^2-\xiz M_Z^2}}$ \\
\hline
\end{tabular}
\subsubsection{Vector-Vector-Scalar}
\begin{minipage}[c]{5cm}
\begin{center}
\includegraphics[width=5cm,height=5cm]{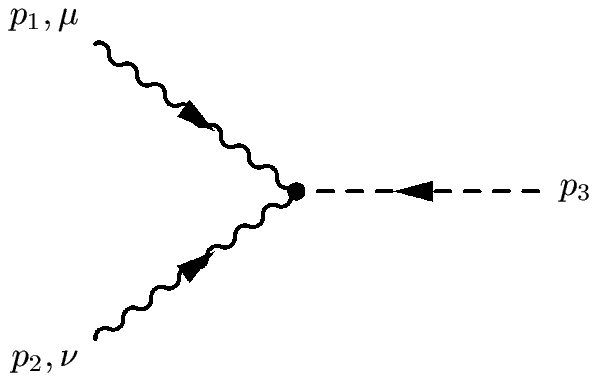}
\end{center}
\end{minipage}
\hspace*{3mm}
\begin{minipage}[c]{10cm}
\begin{tabular}{cccl}
\hline $p_1 \ (\mu)$ & $p_2 \ (\nu)$ &
$p_3$ & \\
\hline
& & & \\
$W^-$ & $W^+$ & $H$ &
$\displaystyle{e\frac{1}{s_W}M_W g^{\mu\nu}}$ \\
& & & \\
\hline
& & & \\
$Z$ & $Z$ & $H$ &
$\displaystyle{e\frac{1}{s_Wc_W^2}M_W g^{\mu\nu}}$ \\
& & & \\
\hline
\end{tabular}
\end{minipage}
\subsubsection{Scalar-Scalar-Vector}
\begin{minipage}[c]{5cm}
\begin{center}
\includegraphics[width=5cm,height=5cm]{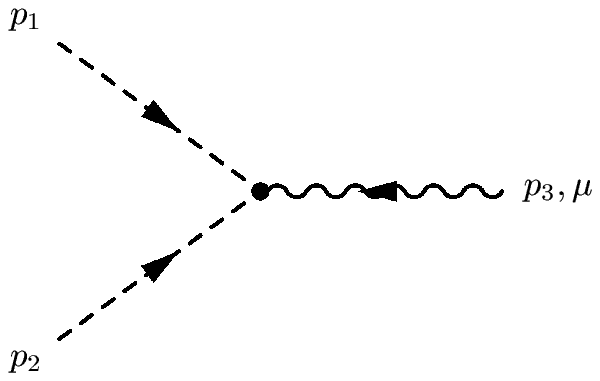}
\end{center}
\end{minipage}
\hspace*{3mm}
\begin{minipage}[c]{10cm}
\begin{tabular}{cccl}
\hline $p_1$ & $p_2$ &
$p_3 \ (\mu)$ & \\
\hline
& & & \\
$H$ & $\chi^{\mp}$ & $W^{\pm}$ & $\displaystyle{i
e\frac{1}{2s_W}\left( p_2^{\mu}-p_{1}^{\mu}
\right)}$ \\
& & & \\
\hline
& & & \\
$H$ & $\chi_3$ & $Z$ & $\displaystyle{i e \frac{1}{2s_Wc_W}\left(
p_2^{\mu}-p_{1}^{\mu}
\right)}$ \\
& & & \\
\hline
\end{tabular}
\end{minipage}
\subsubsection{Scalar-Scalar-Scalar}
\begin{minipage}[c]{5cm}
\begin{center}
\includegraphics[width=5cm,height=5cm]{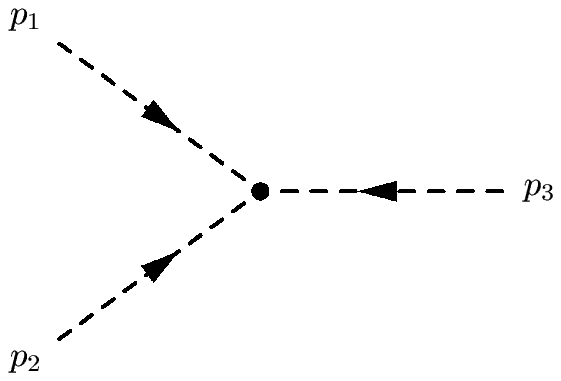}
\end{center}
\end{minipage}
\hspace*{3mm}
\begin{minipage}[c]{10cm}
\begin{tabular}{cccl}
\hline $p_1$ & $p_2$ &
$p_3$ & \\
\hline
& & & \\
$H$ & $H$ & $H$ &
$\displaystyle{-e\frac{3}{2s_W M_W}M_H^2}$ \\
& & & \\
\hline
& & & \\
$H$ & $\chi^-$ & $\chi^+$ &
$\displaystyle{-e\frac{M_H^2}{2s_W M_W}}$ \\
& & & \\
\hline
& & & \\
$H$ & $\chi_3$ & $\chi_3$ &
$\displaystyle{-e\frac{M_H^2}{2s_W M_W}}$ \\
& & & \\
\hline
\end{tabular}
\end{minipage}
\subsubsection{Fermion-Fermion-Scalar}
\begin{minipage}[c]{5cm}
\begin{center}
\includegraphics[width=5cm,height=5cm]{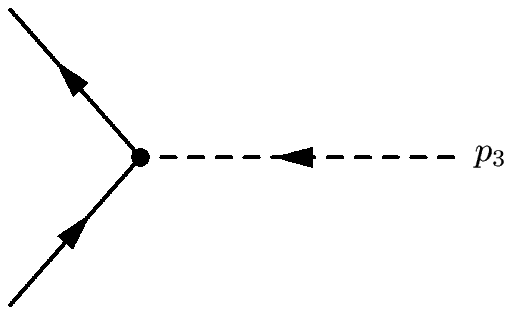}
\end{center}
\end{minipage}
\hspace*{3mm}
\begin{minipage}[c]{10cm}
\begin{tabular}{cccl}
\hline $p_1 $ & $p_2 $ &
$p_3 $ & \\
\hline \rule[-5mm]{0mm}{12mm} $\bar{f}$ & $f$ & $H$ &
$\displaystyle{-e\frac{1}{2s_W}\frac{m_f}{M_W} }$ \\
\hline \rule[-5mm]{0mm}{12mm} $\bar{U}/\bar{D}$ & $U/D$ & $\chi_3$
&
$\displaystyle{(-/+) i e\frac{1}{2s_W}\frac{m_f}{M_W}\ \gamma_5}$ \\
\hline \rule[-1mm]{0mm}{5mm}
$\bar{U}$ & $D$ & $\chi^+$ & \\
\multicolumn{4}{r}{ \rule[-5mm]{0mm}{12mm} $\displaystyle{-i
e\frac{1}{2\sqrt{2}s_W}\frac{1}{M_W}
\left[ (m_D-m_U)+(m_D+m_U)\gamma_5\right] }$} \\
\hline \rule[-1mm]{0mm}{5mm}
$\bar{D}$ & $U$ & $\chi^-$ & \\
\multicolumn{4}{r}{ \rule[-5mm]{0mm}{12mm} $\displaystyle{-i
e\frac{1}{2\sqrt{2}s_W}\frac{1}{M_W}
\left[ (m_U-m_D)+(m_U+m_D)\gamma_5\right] }$ } \\
\hline
\end{tabular}
\end{minipage}\\

We would like to make some connections between the underlying Feynman rules of the SM and the main calculation of 
this thesis, one-loop Yukawa corrections to the process $\ggbbH$. The relevant vertices will be "scalar-scalar-scalar" and "fermion-fermion-scalar". Of these, the vertex $\langle bb\chi_3 \rangle$ will be excluded as it will result in Feynman diagrams proportional to $\la_{bbH}^2$, which are neglected in our calculation.

\section{Problems of the Standard Model}
In spite of its great experimental success, the SM suffers from a
conceptual problem known as the hierarchy problem
\footnote{Indeed, there are other conceptual as well as
phenomenological problems of the SM such as those related to
gravity and dark matter. These discussions can be found in the
recent review of Altarelli \cite{Altarelli:2008} and references
therein. The discussion on the hierarchy problem can be found also
in \cite{Martin:1997ns, djouadi_H2}.}. This problem is related to
the quantum corrections to the Higgs mass. 
In the calculation of one-loop corrections to the Higgs mass, we see that
quadratic divergences appear. 
Of course, these UV-divergences have
to be canceled by the corresponding counter terms. The leading
correction is proportional to the largest mass squared, assumed
to be $m_t^2$. Since the value of $m_t\approx 174$GeV is not so
large, this correction is well under control in the SM. However,
the SM is just an effective theory of a more general theory with
heavy particles at some high energy scale, say the GUT scale
$\Lambda_{GUT}\sim 2\times 10^{16}$GeV where the three gauge
coupling constants unify. The masses of those heavy particles are
at the order of $\Lambda_{GUT}$. Those heavy particles must couple
to the SM Higgs boson and hence give enormous corrections to
$M_H$. The fact the $M_H/\Lambda_{GUT}\sim 10^{-14}$ means that an
extreme cancelation occurs among those huge corrections. This is
known as the naturalness or fine-tuning problem. A related
question, called the hierarchy problem, is why $\Lambda_{GUT}\gg
M_Z$. These problems can be solved if there is a symmetry to
explain that cancelation. There are a few options for such a
symmetry, among them supersymmetry is the most promising
candidate.
\section{Minimal Supersymmetric Standard Model}
\label{bbH_mssm}
The Minimal Supersymmetric Standard Model is a theory describing the interactions of all SM fundamental particles, their
superpartners and some additional Higgs particles. None of these superpartners and new Higgs bosons has been seen in experiment. The fundamental superpartners arise as a consequence of the so-called supersymmetry (SUSY) imposed on the Lagrangian of the theory. The SUSY generator $\Q$ transforms a fermion into a boson and vice versa:
\bea
\Q\vert \text{fermion}\rangle = \vert \text{boson} \rangle, \hs \Q\vert \text{boson}\rangle = \vert \text{fermion} \rangle.
\eea
It means that each SM particle has a corresponding superpartner. The superpartners of a fermion, a vector gauge boson, a scalar Higgs boson are called a sfermion, a gaugino, a higgsino respectively. SUSY requires that a superpartner has the same quantum numbers as its corresponding particle except for the spin. Sfermions are scalar while gauginos and higgsinos have spin $1/2$. One notices immediately that some mixings among gauginos and higgsinos are allowed. The MSSM Lagrangian has three symmetries: Lorentz symmetry, SM gauge symmetry and SUSY.

The fact that we have never observed a fundamental scalar selectron with the same mass as the electron means that SUSY is broken. To date there is no completely satisfactory dynamical way to break SUSY. In the MSSM, SUSY is broken by introducing extra terms that explicitly break SUSY into the Lagrangian \cite{Girardello:1981wz}. They are called soft-SUSY-breaking terms, all contained in $\Lar_{soft}$. The purpose of $\Lar_{soft}$ is to give (quite heavy) masses to superpartners \cite{djouadi_H2, Girardello:1981wz, Chamseddine:1982jx, Barbieri:1982eh}.

The SM particles obtain masses by the Higgs mechanism. In the SM, we just need one Higgs doublet $\Phi$ (and $\tilde{\Phi}=i\sigma_2\Phi^*$) to generate masses for down quarks (up quarks). However, the same trick cannot be used for the MSSM since it will break SUSY. Thus one needs two complex Higgs doublets with opposite hypercharges
\bea
H_1=\left(\begin{array}{c} H_1^0\\ H_1^- \end{array} \right)\hs \text{with} \hs Y_{H_1}=-1/2;\hs
H_2=\left(\begin{array}{c} H_2^+\\ H_2^0 \end{array} \right)\hs \text{with} \hs Y_{H_2}=1/2,
\eea
to give masses for down fermions and up fermions respectively. Before symmetry breaking, these two Higgs doublets have $8$ independent real fields. After symmetry breaking, $3$ vector gauge bosons $Z$, $W^{\pm}$ get masses by "eating" $3$ Goldstone bosons, so five real fields remain. The MSSM therefore predicts the existence of $3$ neutral Higgs bosons denoted $H$, $h$, $A$ and $2$ charged Higgs bosons denoted $H^{\pm}$.

In the unconstrained MSSM, $\Lar_{soft}$ introduces a huge number ($105$) of unknown parameters (\eg intergenerational mixing, complex phases), in addition to $19$ parameters of the SM \cite{djouadi_H2, Dimopoulos:1995ju}. This makes the phenomenology study of the MSSM extremely difficult if not impossible. There exists however the so-called contrained MSSMs with only a handful of parameters. Among them, mSUGRA is most well-known with the $5$ following parameters \cite{Girardello:1981wz, Chamseddine:1982jx, Barbieri:1982eh, Hall:1983iz, Ohta:1982wn}
\bea
\tan\beta, \hs m_{1/2}, \hs m_{0}, \hs A_{0}, \hs sign(\mu).
\eea
This is achieved by imposing some conditions on the soft-SUSY-breaking parameters. These parameters are required to be real and satisfy a set of boundary conditions at the GUT scale ($\Lambda_{GUT}\sim 2\times 10^{16}\GeV$) where the three gauge coupling constants unify. These boundary conditions say that: all gauginos have the same masses ($m_{1/2}$), all sfermions and Higgs bosons have the same mass ($m_0$) and all trilinear couplings in $\Lar_{soft}$ are equal at the GUT scale.
\subsection{The Higgs sector of the MSSM}
The scalar Higgs potential $V_H$ comes from three different sources \cite{djouadi_H2,Martin:1997ns,Gunion:1984yn}:
\bea
V_H&=&V_D+V_F+V_{soft},\crn
V_{D}&=&\fr{g^2}{8}\left[4\vert H_1^\dagger.H_2\vert^2-2\vert H_1\vert^2\vert H_2\vert^2+(\vert H_1\vert^2)^2+(\vert H_2\vert^2)^2\right]+\fr{g^{\prime 2}}{8}(\vert H_2\vert^2-\vert H_1\vert^2)^2,\crn
V_F&=&\mu^2(\vert H_1\vert^2+\vert H_2\vert^2),\crn
V_{soft}&=&m_{H_1}^2H_1^\dagger H_1+m_{H_2}^2H_2^\dagger H_2+B\mu(H_2.H_1+h.c.),
\label{V_Higgs_mssm}
\eea
where $g$, $g^\prime$ are the usual two couplings of the groups $SU(2)$ and $U(1)$ respectively; $\mu$ and $B\mu$ are bilinear couplings; $\vert H_1\vert^2=\vert H_1^0\vert^2+\vert H_1^-\vert^2$ and the same definition for $\vert H_2\vert^2$. The first two terms of $V_H$ are the so-called D- and F- terms. The last term $V_{soft}$ is just a part of $\Lar_{soft}$ discussed above. The MSSM Higgs potential contains the gauge couplings while the SM one given in Eq. (\ref{V_Higgs_sm}) does not.

The neutral components of the two Higgs fields develop vacuum expectations values
\beq
\langle H_1^0\rangle = \frac{\upsilon_1}{\sqrt 2} \ \ , \ \
\langle H_2^0 \rangle = \frac{\upsilon_2}{\sqrt 2}.
\eeq
One defines
\beq
\tan\beta=\fr{\upsilon_2}{\upsilon_1}.
\eeq
Comparing to the SM we have
\bea
\upsilon_1^2+\upsilon_2^2=\upsilon^2.
\eea
We now develop the
two doublet complex scalar fields $H_1$ and $H_2$ around the  vacuum, into
real and imaginary parts
\bea
H_1&=&(H_1^0,H_1^-)=  \frac{1}{\sqrt{2}} \left( \upsilon_1+ H_1^0+ i P_1^0 \ , \  H_1^- \right) \crn
H_2&=&(H_2^+,H_2^0)= \frac{1}{\sqrt{2}} \left( H_2^+ \ , \ \upsilon_2+ H_2^0+ i P_2^0 \right)
\eea
where the real parts correspond to the CP--even Higgs bosons and the imaginary
parts corresponds to the CP--odd Higgs and the Goldstone bosons. By looking at the Eq. \ref{V_Higgs_mssm}, 
we see that those fields mix. After diagonalizing the mass
matrices, one gets
\bea
\left( \begin{array}{c}   \chi_3 \\ A \end{array} \right)
&=& \left( \begin{array}{cc} \cos \beta & \sin \beta \\
- \sin \beta & \cos \beta \end{array} \right) \
\left( \begin{array}{c}   P_1^0 \\ P_2^0 \end{array} \right),\crn
\left( \begin{array}{c}   \chi^\pm \\ H^\pm \end{array} \right)
&=& \left( \begin{array}{cc} \cos \beta & \sin \beta \\
- \sin \beta & \cos \beta \end{array} \right) \
\left( \begin{array}{c}   H_1^\pm \\ H_2^\pm \end{array} \right),\crn
\left( \begin{array}{c}   H \\ h \end{array} \right)
&=& \left( \begin{array}{cc} \cos \alpha & \sin \alpha \\
- \sin \alpha & \cos \alpha \end{array} \right) \
\left( \begin{array}{c}   H_1^0 \\ H_2^0 \end{array} \right),
\eea
with the mixing angle $\alpha$ given by
\beq
\cos 2\alpha = -\cos2\beta \, \frac{M_A^2 - M_Z^2}{ M_H^2-M_h^2} \ , \
\sin 2\alpha = -\sin2\beta \, \frac{M_H^2 + M_h^2}{ M_H^2-M_h^2},
\label{alpha:tree}
\eeq
where $\chi_3$, $\chi^{\pm}$ are massless Goldstone bosons to be eaten by the $Z$, $W^{\pm}$ respectively; $A$, $H^\pm$, $H$ and $h$ are five physically massive Higgs bosons; the tree-level masses are given by
\bea
M_A^2&=& - B\mu (\tan\beta + \cot\beta) = - \frac{2 B\mu}{
\sin 2\beta},\crn
M_{H^\pm}^2&=& M_A^2 + M_W^2,\crn
M_{h,H}^2&=& \frac{1}{2} \left[ M_A^2+M_Z^2 \mp \sqrt{ (M_A^2+M_Z^2)^2 -4M_A^2
M_Z^2 \cos^2 2\beta } \right].
\label{mass_higgs_mssm}
\eea
We remark that $M_h\le M_Z$ at tree level. From Eqs. (\ref{alpha:tree}) and (\ref{mass_higgs_mssm}) we get
\bea
\cos^2(\beta-\alpha)=\fr{M_h^2(M_Z^2-M_h^2)}{M_A^2(M_H^2-M_h^2)},
\label{relat_cos_MA}
\eea
which will be useful later.
\subsection{Higgs couplings to gauge bosons and heavy quarks}
\label{section_HFrules_mssm}
\vspace*{-3mm}
Like in the SM, the Higgs boson couplings to the gauge bosons are obtained
from the kinetic terms with covariant derivatives of the Higgs fields $H_1$ and $H_2$. 
The Yukawa Higgs boson couplings to the fermions are obtained from the Yukawa Lagrangian. 
We list here some relevant couplings needed in this thesis. For a full account of Higgs couplings in the 
MSSM, we refer to \cite{djouadi_H2, Gunion:1989we, Gunion:1992hs}. With $\la_{WWH}$, $\la_{ZZH}$, $\la_{bbH}$, $\la_{ttH}$ are 
the SM couplings and using the same Feynman rules as in section~\ref{section_HFrules}, we have:\\[2mm]
\begin{minipage}[c]{5cm}
\begin{center}
\includegraphics[width=5cm,height=5cm]{psfig/vvs.eps}
\end{center}
\end{minipage}
\hspace*{3mm}
\begin{minipage}[c]{10cm}
\begin{tabular}{cccl}
\hline $p_1 \ (\mu)$ & $p_2 \ (\nu)$ &
$p_3$ & \\
\hline
$W^-$ & $W^+$ & $H$ &
$\la_{WWH}\cos(\beta-\alpha)$ \\
$W^-$ & $W^+$ & $h$ &
$\la_{WWH}\sin(\beta-\alpha)$ \\
\hline
$Z$ & $Z$ & $H$ &
$\la_{ZZH}\cos(\beta-\alpha)$ \\
$Z$ & $Z$ & $h$ &
$\la_{ZZH}\sin(\beta-\alpha)$ \\
\hline
\end{tabular}
\end{minipage}\\[4mm]
%
\begin{minipage}[c]{5cm}
\begin{center}
\includegraphics[width=5cm,height=5cm]{psfig/ffs.eps}
\end{center}
\end{minipage}
\hspace*{3mm}
\begin{minipage}[c]{10cm}
\begin{tabular}{cccl}
\hline $p_1 $ & $p_2 $ &
$p_3 $ & \\
\hline \rule[-5mm]{0mm}{12mm} $\bar{t}$ & $t$ & $H$ &
$-\la_{ttH}[\cos(\beta-\alpha)-\cot\beta\sin(\beta-\alpha)]$ \\
$\bar{b}$ & $b$ & $H$ &
$-\la_{bbH}[\cos(\beta-\alpha)-\tan\beta\sin(\beta-\alpha)] $\\
\hline \rule[-5mm]{0mm}{12mm} $\bar{t}$ & $t$ & $h$ &
$-\la_{ttH}[\sin(\beta-\alpha)+\cot\beta\cos(\beta-\alpha)]$ \\
$\bar{b}$ & $b$ & $h$ &
$-\la_{bbH}[\sin(\beta-\alpha)-\tan\beta\cos(\beta-\alpha)]$ \\
\hline
\end{tabular}
\end{minipage}\\[3mm]
We remark that the $bb$($tt$) coupling of either the $H$ or $h$ boson is enhanced(suppressed) by a factor $\tan\beta$ with the enhancement(suppression) magnitude 
depending on the value 
of $\sin(\beta-\alpha)$ or $\cos(\beta-\alpha)$. Thus one can have very large value of bottom-Higgs Yukawa coupling, leading to a large cross section if $\tan\beta$ is large. In the decoupling limit where $M_A\to\infty$, \ie $\cos(\beta-\alpha)\to 0$ (see Eq. (\ref{relat_cos_MA})), the $h$ is SM-like (the same couplings) while the $H$ Yukawa coupling to $bb$($tt$) is exactly enhanced(suppressed) by a factor $\tan\beta$.

\chapter{Standard Model Higgs production at the LHC}
\label{higgs_production}
\section{The Large Hadron Collider}
\label{section_lhc}
\begin{table}[tp]
\caption{LHC beam parameters relevant for the peak luminosity \cite{lhc_report}}
\begin{center}
\begin{tabular}{|c|c|}  
\hline
Number of particles per bunch ($N_b$) & $1.15\times 10^{11}$\\
\hline
Number of bunches per beam ($n_b$) & $2808$\\
\hline
Revolution frequency ($f_{rev}$) & $11245$Hz \\
\hline
 Relativistic gamma ($\gamma_r$) & $7461$ ($E=7$TeV)\\
\hline
Normalized transverse emittance ($\eps_{n}$) & $3.75\times 10^{-4}$cm \\
\hline
Full crossing angle at the IP ($\theta_c$) for ATLAS/CMS & $285$$\mu$rad $=0.0163^\circ$\\
\hline
RMS bunch length ($\sigma_z$) & $7.55$cm \\
\hline
Transverse RMS beam size ($\sigma^*$) at ATLAS/CMS & $16.7\mu$m \\
\hline
Geometric luminosity reduction factor ($F$) at ATLAS/CMS & $0.84$ \\
\hline
Optical beta function at ATLAS/CMS ($\beta^*$) & $55$cm \\
\hline
 \end{tabular}\label{beam_parameters_lhc}
\end{center}
\end{table}
The Large Hadron Collider (LHC) is the world largest particle accelerator to date \cite{lhc}. It collides two proton beams with the center-of-mass energy up to $14$TeV. It is expected to start this year. 
It has four main experiments: ATLAS, CMS, LHCb and ALICE. ATLAS and CMS are general-purpose detectors.
Their goals are to find the Higgs boson and discover new physics expected to be Supersymmetry. LHCb is for B-physics and CP violation. ALICE aim is to 
study the physics of strongly interacting matter at extreme energy densities, where the formation of a new phase of matter, the quark-gluon plasma, is expected. The number of events per second generated in the  LHC collisions given by 
\bea
N_{event}=L\sigma_{event},
\eea
where $\sigma_{event}$ is the cross section for the event under study and $L$ the machine luminosity. The machine luminosity depends only on the beam parameters and can be written for a Gaussian beam distribution as 
\bea
L =\fr{N^2_b n_b f_{rev}\gamma_r}{4\pi \eps_n\beta^*}F,
\label{formula_luminosity}
\eea  
where $N_b$ is the number of particles per bunch, $n_b$ number of bunches per beam, $f_{rev}$ the revolution frequency,
$\gamma_r$ the relativistic gamma factor, $\eps_n$ the normalized transverse beam emittance\footnote{The beam emittance ($\eps$), units of length, is the extent occupied by the particles of the beam in phase space. A low emittance particle beam is a beam where the particles are confined to a small distance and have nearly the same momentum. The normalized beam emittance 
$\eps_n=\gamma_r\beta_r\eps$.}, $\beta^*$ the optical beta function at
the collision point\footnote{The optical beta function $\beta(s)$ appears in the amplitude of the solution of the equation of a harmonic ossilator $x^{\prime\prime}(s)+Kx(s)=0$ describing the transverse beam dynamics (linear force in $x$ and $y$, $\textbf{s}$ is the beam direction).} and $F$ the geometric luminosity reduction factor due to the crossing angle at the interaction point (IP):
\bea
F={1}/{\sqrt{1+\left(\fr{\theta_c\sigma_z}{2\sigma^*}\right)^2}},
\label{formula_F}
\eea
where $\theta_c$ is the full crossing angle at the IP, $\sigma_z$ the root-mean-square (RMS) bunch length and $\sigma^*$ the transverse RMS beam size at
the IP. Note that $F<1$ since the angle between two beams at the collision point is greater than zero. The above expressions assumes equal beam parameters for both circulating beams. In order to calculate the peak luminosity, we need the LHC beam 
parameters given in Table~\ref{beam_parameters_lhc}. We then get the value of the peak luminosity at ATLAS/CMS $L=10^{34}\lum$. Note that in luminosity formula 
(\ref{formula_luminosity}) there are only two beam parameters $\beta^*$ and $F$ which depend on the collision point (experiment position), other beam parameters are the same for LHCb and ALICE. Other experiments LHCb and ALICE with $Pb-Pb$ collision have lower luminosity, $10^{32}\lum$ and 
$10^{27}\lum$ respectively. If the Higgs production cross section is $1$pb then one has $10^{-2}$ events per second at ATLAS/CMS.

The luminosity in the LHC is not constant over a physics run but decays due to the degradation of intensities
and emittances of the circulating beams. The LHC integrated luminosity ($L_{int}=\int_0^T L(t)dt$) expected per year is between $80\textrm{fb}^{-1}$ and $120\textrm{fb}^{-1}$ \cite{lhc_report}. 

Despite of a number of enormous advantages: the high values of luminosity and center-of-mass energy, the dedicated detectors ATLAS and CMS designed to discover the Higgs boson and Supersymmetry particles; to get something out of a huge amount of data produced by the LHC is not easy. The main challenge for LHC physics is at the way one analyses data. Let us make it clear. 
The total cross section for inelastic scatterings which can be seen by LHC detectors is about $60$mb \cite{lhc_event_rate}. It is translated to 
$6\times 10^8$ events per second (event rate). To be optimistic and forget about the background, 
we say that the typical total cross section for Higgs production at the LHC is $60$pb which corresponds to $6\times 10^{-1}$ event rate. Thus we have to know how to find out $6\times 10^{-1}$ events in $6\times 10^8$ events per second. The task becomes much more complicated when backgrounds are taken into account. The only way to distinguish signal from backgrounds is to study various distributions.  

\section{SM Higgs production at the LHC}
\label{section_higgs_sm_lhc_prod}
\begin{figure}[t]
\begin{center}
\begin{minipage}[c]{5cm}
\includegraphics[width=5cm,height=4.5cm]{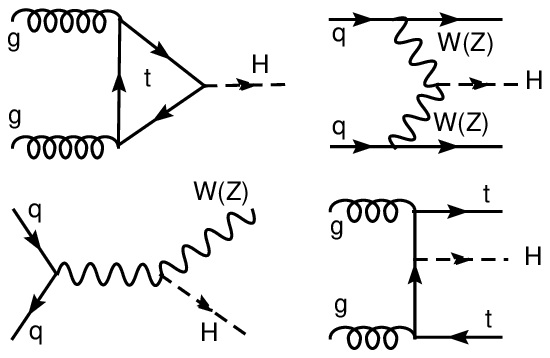}
\end{minipage}
\hs
\begin{minipage}[c]{8cm}
\includegraphics[width=8cm]{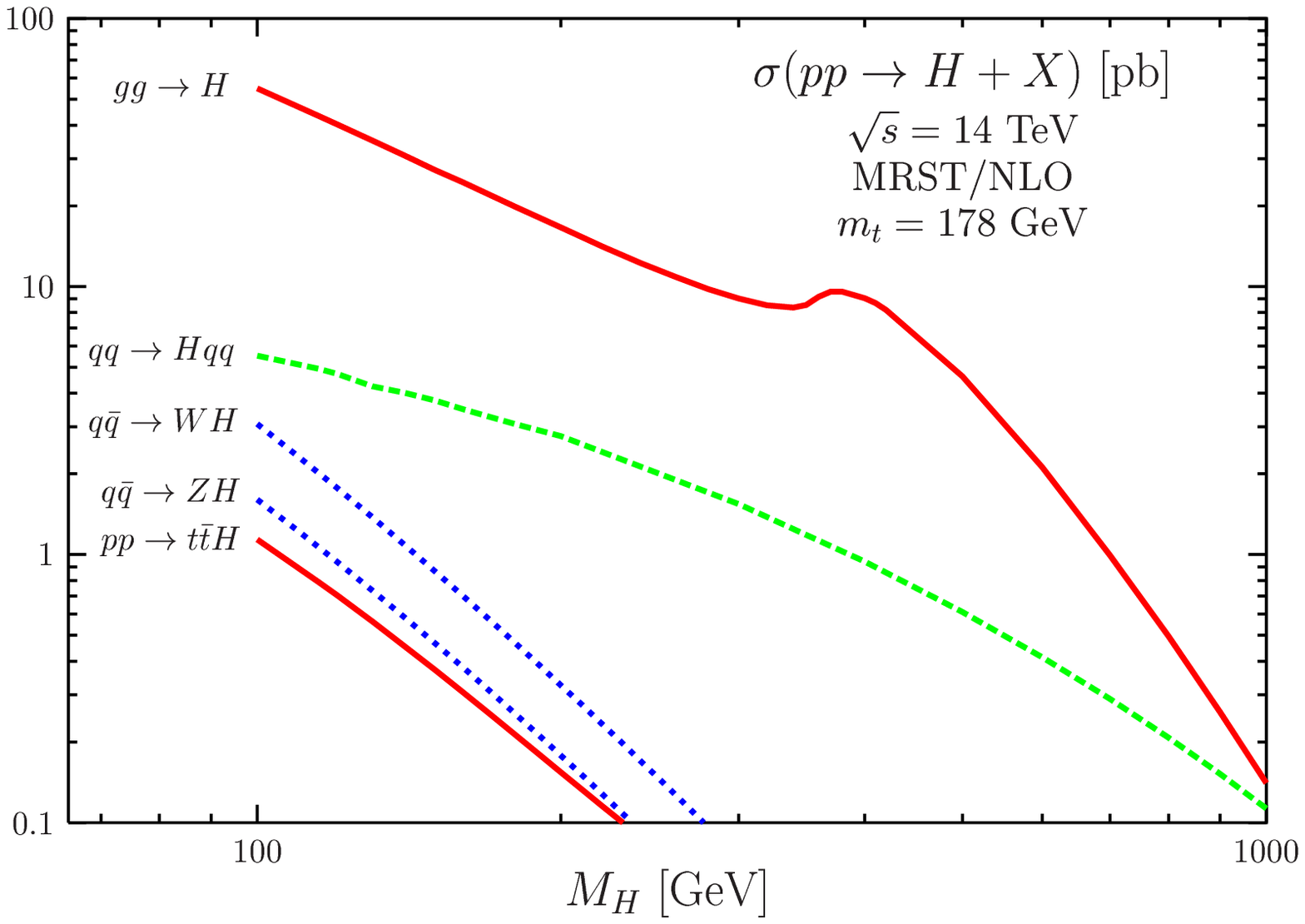}
\end{minipage}
\caption{\label{H_production_lhc}{\em Four mechanisms to produce the Higgs at the LHC. The right figure shows the cross sections as functions 
of $M_H$, which include the full NLO QCD corrections \cite{djouadi_H1}.}}
\end{center}
\end{figure}
The SM Higgs profile is: 
\begin{center}
\framebox[15cm][c]{
\begin{minipage}[c]{5cm}
\begin{description}
\item[($1$)] Electric charge: neutral 
\item[($2$)] Color charge: neutral
\item[($3$)] Spin: $0$
\item[($4$)] CP: even 
\end{description}
\end{minipage}
\hs \hs
\begin{minipage}[c]{8cm}
\begin{description}
\item[($5$)] Mass: $114\text{GeV} <M_H < 190$GeV \cite{range_MH}
\item[($6$)] Higgs couplings: $\lambda_{VVH}\propto M_V$, $\lambda_{ffH}\propto m_f/\upsilon$, $\lambda_{HHH}\propto \lambda\upsilon=\fr{M_H^2}{2\upsilon}$ with $V$ is a massive gauge boson, $f$ a fermion (see section~\ref{section_HFrules}).
\end{description}
\end{minipage}
}
\end{center}
In this section, we assume that the above Higgs profile is correct. The last property means that the Higgs boson couples mainly to heavy particles: $W$ and $Z$ gauge bosons, the top quark and, to lesser extent, the bottom quark. The four main mechanisms for single Higgs boson production are \cite{djouadi_H1}:
\begin{enumerate}
\item gluon-gluon fusion: $gg\to H$
\item associated production with heavy quarks: $gg\to Q\bar{Q}+H$ with $Q=b,t$
\item vector boson fusion: $q\bar{q}\to V^*V^*\to q\bar{q}+H$ with $q$ is a light quark 
\item associated production with $W$($Z$): $qq(q\bar{q})\to W(Z)+H$
\end{enumerate}
We will use the item number to refer to the corresponding process. 
The Feynman diagrams and total cross sections are shown in Fig.~\ref{H_production_lhc}. It is important to know that in the second mechanism, there is also a contribution from $q\bar{q}$ in the initial state. However, this contribution is very small compared to the $gg$ contribution and can be neglected. There are two reasons for this. First, the quark density in the proton is very small compared to the gluon density. Second, $q\bar{q}$ contribution contains only S-channel diagrams strongly suppressed at high energy while $gg$ contribution contains T and U channels. 
 
From the plot of total cross sections as functions of $M_H$ in Fig.~\ref{H_production_lhc} we see that for $M_H\in [114,190]$GeV the total cross section for $gg$ fusion process is largest and almost $10$ times bigger than the second one from vector boson fusion process. For $M_H=120$GeV, the total cross sections for the four processes are about: 
\bea
\left(
\begin{array}{cccc}
H & t\bar{t}H & q\bar{q}H & WH(ZH) \\
40\text{pb} & 0.7\text{pb} & 5\text{pb} & 2(1)\text{pb}
\end{array}
\right) 
\eea
We can conclude that: for Higgs production, the LHC is just the $gg$ collider. The Higgs discovery channel is clearly the $gg$ fusion process. 

The first important signal for a Higgs boson is a peak in the invariant mass distribution of its leptonic decay products \footnote{The Higgs hadronic decay products suffer from extreme QCD jet backgrounds.}. However, if we observe a new bump at around $120$GeV in some production channel it is not necessarily the SM Higgs boson. All we can say is that it is new physics, a new particle with mass around $120$GeV. Now we look back at the Higgs profile and assume that all the properties are confirmed except for the last one (the coupling property). We still cannot say that it is the SM Higgs boson. Remember that the Higgs originates from the Higgs potential defined by its self coupling ($\lambda$) and interacts with vector gauge bosons and heavy fermions in some special ways. The best way to check the SSB mechanism in the SM is to determine the three independent Higgs couplings: $\la_{VVH}$, $\la_{ffH}$ and $\la_{HHH}$. If all these three couplings are consistent with the SM prediction then we can definitely say that it is the SM Higgs boson. However, measuring accurately those three Higgs couplings is an extremely difficult task if not impossible at the LHC. The Higgs self-coupling is the most difficult one. For this, one can think of $gg\to HH$ but the total cross section is small and background is large. The two other couplings can be accessed at the LHC \cite{djouadi_H1, Rainwater:2007cp}.

\section{Experimental signatures of the SM Higgs}
\label{section_higgs_sm_lhc_decay}
\begin{figure}[htbp!]
\begin{center}
\begin{minipage}{0.8\textwidth}
\includegraphics[width=0.8\textwidth]{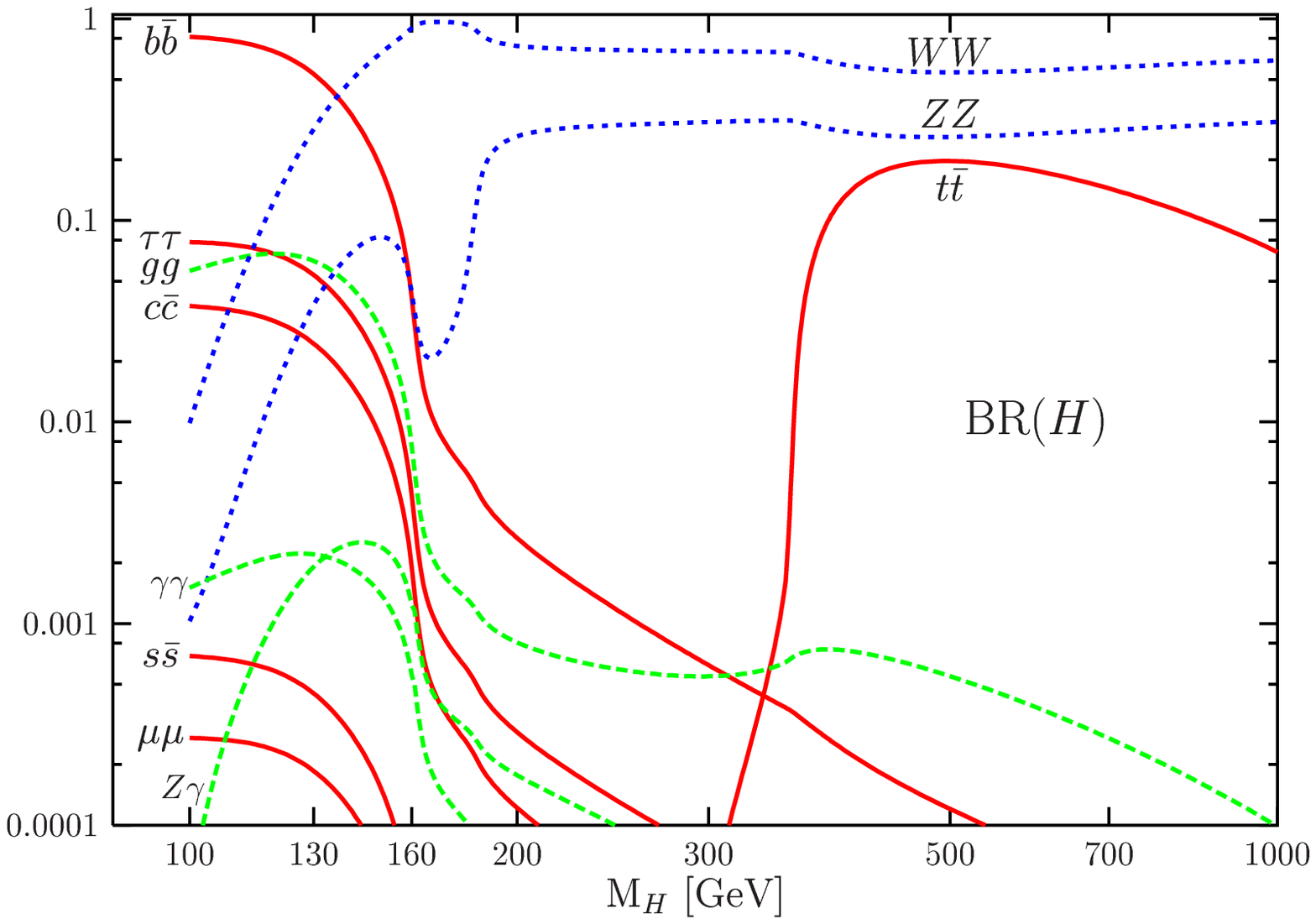}
\vspace*{1cm}
\end{minipage}
\begin{minipage}{0.8\textwidth}
\includegraphics[width=0.8\textwidth]{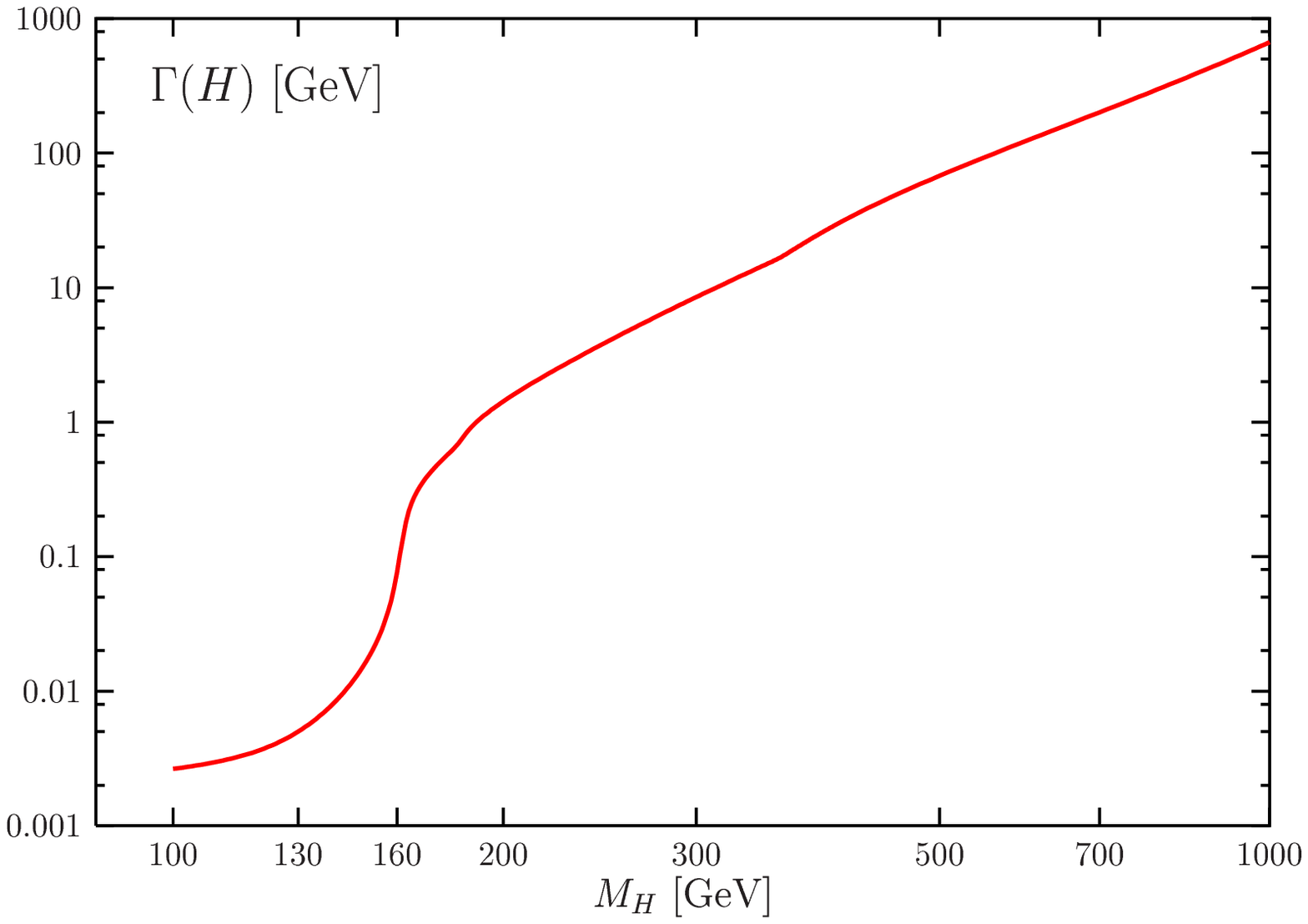}
\end{minipage}
\caption{\label{H_branch_width}{\em Upper: The SM Higgs boson decay branching ratios as a function of $M_H$. Lower: The SM Higgs boson total decay width as a function of $M_H$. Ref. \cite{djouadi_H1}.}}
\end{center}
\end{figure}
The SM Higgs boson is a heavy particle and decays. Its decay branching ratios ($\Gamma_i/\Gamma_H$) and total width ($\Gamma_H$) are shown in Fig.~\ref{H_branch_width}. We first look at the plot for the Higgs decay branching ratios ($BR$) and impose the following conditions: $BR>10^{-3}$, leptonic (photonic) decay product or heavy quarks in the decay product. These are the conditions for observing the Higgs in experiment. With the Higgs profile in hand ($M_H\in [114,182]$GeV) we are left with $7$ branching ratios: $b\bar{b}$, $W^+W^-$, $ZZ$, $\tau^+\tau^-$, $c\bar{c}$, $\gamma\gamma$, $Z\gamma$. 
Among these, $c\bar{c}$ and $Z\gamma$ can be discarded. For $c\bar{c}$ there are two reasons: $BR(c\bar{c})$ is $10$ times 
smaller than $BR(b\bar{b})$ and it is more difficult to tag a charm-quark in experiment than a bottom-quark. For $Z\gamma$ there are also two reasons: $BR(Z\gamma)$ is at the same order as $BR(\gamma\gamma)$ and the $Z$ is heavy and decays dominantly into hadrons and neutrinos and thus the combined branching ratio for this channel is very small. Thus we now have $5$ 
potential Higgs signatures
\bea
\left(
\begin{array}{ccccc}
 b\bar{b} & \gamma\gamma & \tau^+\tau^- &  W^+W^- &  ZZ \\
 0.68 & 2.16\times 10^{-3} & 6.78\times 10^{-2} & 0.13 & 1.49\times 10^{-2}
\end{array}
\right)  
\eea
where the second row are branching ratios for $M_H=120$GeV. Of these $b\bar{b}$ and $\gamma\gamma$ can be observed "directly" in experiments. The three other decay modes cannot be directly seen in experiments and more branching ratios must be taken into account. $\gamma\gamma$ is the most beautiful signal in experiment but its branching ratio is smallest. This is unlikely to be a discovery channel. The $gg\to H\to b\bar{b}$ suffers from huge QCD backgrounds 
\footnote{In a small window about the Higgs mass, the QCD background $pp\to b\bar{b}$ is still very large.} \cite{Gianotti:2004qs} and cannot be realised in LHC experiments \cite{djouadi_H1}. For $b\bar{b}$ signal, we have to use the mechanism of Higgs production associated with massive gauge bosons or heavy quarks. If we stick with the $gg$ fusion mechanism then the Higgs discovery signal can be $\tau^+\tau^-$, $W^+W^-$ or $ZZ$ depending on the value of $M_H$. $\tau^+\tau^-$ signal is important for small Higgs mass. $ZZ$ can give a beautiful $4$-lepton signal but only for very large Higgs mass ($M_H>2M_Z$). For the large range of $M_H$ left, $W^+W^-$ decay mode is the best signal to pursue. 

The total Higgs decay width is shown in the Fig.~\ref{H_branch_width}. We first notice that $\Gamma_H\le 0.63$GeV for $M_H\le 180$GeV. The SM Higgs boson is a heavy long-lived particle. This makes it easy to determine the Higgs mass accurately by using $H\to \gamma\gamma$ for small $M_H$ and $H\to ZZ\to l^+l^-l^+l^-$ for large $M_H$ \cite{Gianotti:2000tz, Drollinger:2001bc}. However, it will be difficult to measure $\Gamma_H$ precisely.                

\section{Summary and outlook}
\label{higgs_lhc_summary}
If there exists only one Higgs boson in nature as predicted by the SM, it should be found at 
the LHC via the gluon-gluon fusion channel. However, if nature prefers having more than a unique Higgs boson, say 
five Higgs bosons as anticipated by the MSSM, then the situation of the SM-like Higgs 
(with the same Higgs profile given in section~\ref{section_higgs_sm_lhc_prod} except for the last coupling property) production at the LHC will be very different. 
From the results of subsection~\ref{section_HFrules_mssm} we know that the SM-like Higgs coupling to the bottom quark can be strongly 
enhanced for large value of $\tan\beta$ while the same coupling with the top quark is much suppressed 
by the same factor. Thus the SM-like Higgs production associated with two bottom quarks can be the dominant mechanism at the LHC \cite{djouadi_H2}. 
This channel with $H\to W^+W^-$-signature is also very valuable for the determination of the bottom-Higgs Yukawa coupling. 
In order to identify those new physics characters or just to confirm the SM Higgs properties, the study of SM Higgs production associated with two bottom quarks at the LHC is very important. Indeed, it is the topic of the next chapter.


\chapter{Standard Model $b\bar{b}H$ production at the LHC}
\label{chapter_bbH1}
The content of this chapter is based on our publications \cite{fawzi_bbH, ninh_pic_2007, ninh_moriond_2008}.
\section{Motivation}
The study of the Higgs
properties such as its self-couplings and couplings to the other
particles of the standard model (SM) will be crucial in order to
establish the nature of the scalar component of the model. In this
respect most prominent couplings, in the SM, are the Higgs ($\lambda_{HHH}$), the
top ($\lambda_{ttH}$), and to a much lesser degree the bottom ($\lambda_{bbH}$), Yukawa couplings. The
top Yukawa coupling is after all of the order of the strong QCD
coupling and plays a crucial role in a variety of Higgs related
issues. In our main calculation of EW corrections to $b\bar{b}H$, the leading contribution includes terms with 
largest powers of $\la_{ttH}$. 

The next-to-leading order (NLO) QCD correction to $pp\to
b\bar{b}H$ has been calculated by different groups relying on
different formalisms. In a nut-shell, in the five-flavour scheme
(5FNS)\cite{Barnet-Haber-Soper,Willenbrock_bbh_original}, use is
made of the bottom distribution function so that the process is
approximated (at leading order, LO) by the fusion $b \bar b \ra H$.
This gives an approximation to the inclusive cross section
dominated by the untagged low $p_T$ outgoing $b$ jets. If only one
final $b$ is tagged, the cross section is approximated by $g b \ra
bH$. The four flavour scheme (4FNS) has no $b$ parton initiated
process but is induced by gluon fusion $gg \ra b \bar b H$, with a
very small contribution from the light quark initiated process
$q\bar{q}\to b\bar{b}H$\footnote{In fact $q\bar{q}\to b\bar{b}H$
is dominated by $q\bar{q}\to HZ^*\to b\bar{b}H$ and does not
vanish for vanishing bottom Yukawa coupling. However this
contribution should be counted as $ZH$ production and can be
excluded by imposing an appropriate cut on the invariant mass of
the $b\bar{b}$ pair.}. Here again the largest contribution is due
to low $p_T$ outgoing $b$'s which can be accounted for by gluon
splitting into $b\bar b$. The latter needs to be resummed and
hence one recovers most of the 5FNS calculation while retaining
the full kinematics of the reaction. QCD NLO corrections have been
performed in both
schemes\cite{Willenbrock_bbh_original,dittmaier_bbH,dawson_bbH,dawson_bH_bbH}
and one has now reached a quite good agreement\cite{LH05-Higgs}.\\
The 5FNS approach, which at leading order is a two-to-one process
has allowed the computation of the NNLO QCD
correction\cite{WillenbrockNNLObbh,HarlanderKilgore} and very
recently the electroweak/SUSY (supersymmetry)
correction\cite{Dittmaier_bbH_susy} to $b \bar b \ra \phi$, $\phi$
any of the neutral Higgs boson in the MSSM. SUSY QCD corrections
have also been performed for $gg \ra b \bar b
h$\cite{Deltamb-bbH,Hollik_susy_QQh} where $h$ is the lightest
Higgs in the MSSM as well as to  $g b \ra b
\phi$\cite{Dawsonbbh_susy}.

In order to exploit this production mechanism to study
the Higgs couplings to $b$'s, one must identify the process and
therefore one needs to tag both $b$'s, requiring somewhat large
$p_T$ $b$. This reduces the cross section but gives much better
signal over background ratio. For large $p_T$ outgoing quarks one
needs to rely on the 4FNS to properly reproduce the hight $p_T$
$b$ quarks. The aim of this and next chapters is to report on the calculation
of the leading electroweak corrections to the exclusive $bbH$
final state, meaning two $b$'s are detected. These leading
electroweak corrections are triggered by top-charged Goldstone
loops whereby, in effect, an external $b$ quark turns into a top.
This transition has a specific chiral structure whose dominant
part is given by the top mass or, in terms of couplings, to the
top Yukawa coupling. Considering that the latter is of the order
of the QCD coupling constant, the corrections might be large. In
fact, as we shall see, such type of transitions can trigger $g g
\ra b \bar b H$ even with {\em vanishing} $\lambda_{bbH}$ in which
case the process is generated solely at one-loop. We will quantify
the effect of such contributions.

\section{General considerations}
Before discussing the details of the calculation it is educative
to expose some key features that appear when one considers the
electroweak corrections at one-loop compared to the structure we
have at tree-level or even the structure that emerges from QCD
loop calculations. In particular the helicity structure is quite
telling. So let us set our definition first. The process we
consider is
$g(p_1,\la_1)+g(p_2,\la_2)\rightarrow
b(p_3,\la_3)+\bar{b}(p_4,\la_4)+H(p_5)$. $\la_i=\pm$ with
$i=1,2,3,4$ are the helicities of the gluons, the bottom and
anti-bottom while $p_i$ are the momenta of particles 
\footnote{The cross section for $q\bar{q}\to g^*\to b\bar{b}H$ (with input parameters given in section~\ref{section_bbH_result1} and $q=u,d,s$) is about $0.7\%$ of the $\sigma(\ggbbH)$ and thus can be totally neglected. 
There are two reasons for this. First the density of gluon at the LHC is much bigger the the density of quarks. 
Second the $\qqbbH$ which is S-channel is rather suppressed at high energy while the $\ggbbH$ contains also T and U channels which give the dominant contribution.}. The
corresponding helicity amplitude will be denoted by ${\cal
A}(\la_1,\la_2;\la_3,\la_4)$.

\subsection{Leading order considerations}
\begin{figure}[h]
\begin{center}
\includegraphics[width=12cm]{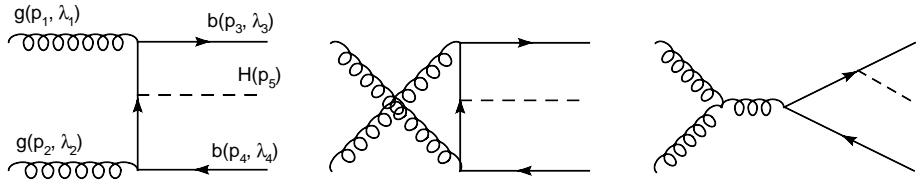}
\caption{\label{diag_gg_LO}{\em All the eight Feynman diagrams can
be obtained by inserting the Higgs line to all possible positions
in the bottom line.}}
\end{center}
\end{figure}

At tree-level,  see Fig. \ref{diag_gg_LO} for the contributing
diagrams, the Higgs can only attach to the $b$-quark and therefore
each diagram, and hence the total amplitude, is proportional to
the Higgs coupling to $b\bar b$, $\lambda_{bbH}$. Compared to the
gluon coupling this scalar coupling breaks chirality. These
features remain unchanged when we consider QCD corrections.
Moreover the QCD coupling and the Higgs coupling are parity
conserving which allows to relate the state with helicities
$(\la_1,\la_2;\la_3,\la_4)$ to the one with
$(-\la_1,-\la_2;-\la_3,-\la_4)$ therefore cutting by half the
number of helicity amplitudes to calculate. With our conventions
for the definition of the helicity states, see
Appendix~\ref{appendix-helicity},  parity conservation for the
tree-level helicity amplitude gives
\beqn
{\cal A}_0(-\la_1,-\la_2;-\la_3,-\la_4)=\la_3 \la_4 {\cal
A}_0(\la_1,\la_2;\la_3,\la_4)^\star.
\eeqn
This can be generalised at higher order in QCD with due care of
possible absorptive parts in taking complex conjugation.

The number of contributing helicity amplitudes can be reduced even
further at the leading order, in fact halved again, in the limit
where one neglects the mass of the $b$-quark that originates from
the $b$-quark spinors and therefore from the $b$ quark
propagators. We should in this case consider the $\lambda_{bbH}$
as an independent coupling, intimately related to the model of
symmetry breaking. In this case chirality and helicity arguments
are the same, the $b$ and $\bar b$ must have opposite helicities
for the leading order amplitudes and hence only ${\cal
A}_0(\la_1,\la_2;\la,-\la)$ remain non zero. In this limit, this
means that only a string containing an even number of Dirac
$\gamma$ matrices, which we will label in general as $\Gamma^{\textrm
even}$ as opposed to $\Gamma^{\textrm odd}$ for a string with an odd
number of $\gamma$'s, can contribute.

\noi In the general case and reinstating the $b$ mass, we may write the  helicity amplitudes  as
\bea \ali&=&\bar u(\lambda_3) \left( \Gamma^{\textrm
even}_{\la_1,\la_2} + \Gamma^{\textrm odd}_{\la_1,\la_2} \right)
v(\la_4) \crn &=&\delta_{\la_3, -\la_4} \left({\cal A}^{\textrm
even}+m_b \tilde{{\cal A}}^{\textrm odd} \right) \;+\;
\delta_{\la_3,\la_4} \left({\cal A}^{\textrm odd}+m_b \tilde{{\cal
A}}^{\textrm even} \right). \label{amp_form2}
\eea

The label $^{\textrm even}$  in ${\cal A}^{\textrm even}$ and
$\tilde{{\cal A}}^{\textrm even}$ are the contributions of
$\Gamma^{\textrm even}$ to the amplitude and likewise for $^{\textrm
odd}$. This way of writing shows that $m_b$ originates from the
mass insertion coming from the massive spinors and are responsible
for chirality flip. In the limit $m_b \ra 0$, $\Gamma^{\textrm
even}_{\la_1,\la_2}$ and $\Gamma^{\textrm odd}_{\la_1,\la_2}$
contribute to different independent helicity amplitudes. In
general $\Gamma^{\textrm even}$ and $\Gamma^{\textrm odd}$  differ by a
(fermion) mass insertion. In fact $\Gamma^{\textrm odd}$ is
proportional to a fermion mass insertion from a propagator. At
leading order the mass insertion is naturally $m_b$, such that
$\Gamma^{\textrm odd}$ is ${\cal O}(m_b)$. This shows that at leading
order, corrections from $m_b=0$ to the total cross section are of
order ${\cal O}(m_b^2)$. Of course there might be some enhancement
of the ${\cal O}(m_b^2)$ terms if one remembers that the cross
section can bring about terms of order  $m_b^2/{(p_{T}^{b})}^2$.
However,  in our calculation where we require the $b$'s to be
observed hence requiring a $p_T^b$ cut, the effect will be
minimal. With $m_b=4.62$GeV, the effect of neglecting $m_b$
is that the cross section is increased by $3.7\%$ for
$|\textbf{p}_T^{b,\bar{b}}|>20$GeV and $1.1\%$ for
$|\textbf{p}_T^{b,\bar{b}}|>50$GeV. At one-loop, the chiral
structure of the weak interaction and the contribution of the top
change many of the characteristics that we have just discussed for
the tree-level.

\subsection{Electroweak Yukawa-type contributions, novel characteristics}
\begin{figure}[h]
\begin{center}
\includegraphics[width=9cm]{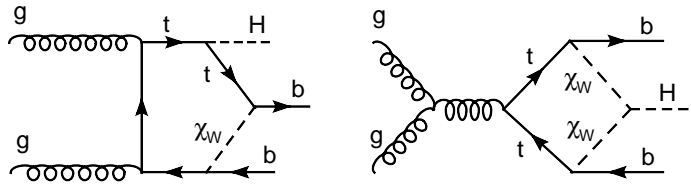}
\caption{\label{diag_gg_ew}{\em Sample of one-loop diagrams
related to the Yukawa interaction in the SM. $\chi_W$ represents
the charged Goldstone boson.}}
\end{center}
\end{figure}
\begin{figure}[htbp]
\begin{center}
\includegraphics[width=14cm]{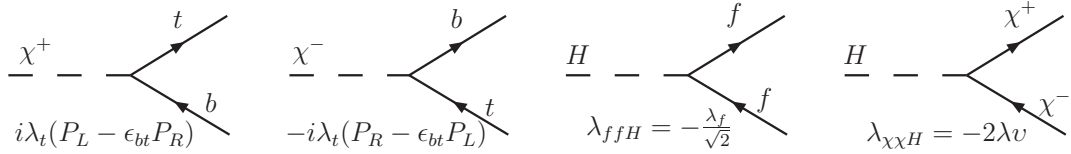}
\caption{\label{vertex_NLO}{\it Relevant vertices appearing at one
loop. $\varep_{bt}=\la_b/\la_t$ and $\la$ is the Higgs self-coupling, related to
the Higgs mass in the Standard Model. The relations to the gauge couplings can be obtained by 
comparing to the SM Feynman rules given in section~\ref{section_HFrules}.}}
\end{center}
\end{figure}

Indeed, look at the two contributions arising from the one loop
electroweak corrections given in Fig.~\ref{diag_gg_ew}. Now the
Higgs can attach to the top or to the $W$. Therefore these
contributions do not vanish in the limit $\lambda_{bbH}=0$.
The mass insertion in
what we called $\Gamma^{\textrm odd}$ is proportional to the top mass
and is not negligible. In fact the diagrams in
Fig.~\ref{diag_gg_ew} show the charged Goldstone boson in the
loop. The latter triggers a $ t \ra b \chi_W$ transition whose
dominant coupling is proportional to the Yukawa coupling of the
top. We will in fact be working in the approximation of keeping
only the Yukawa couplings. This reduces the number of diagrams and
if working in the Feynman gauge as we do in this computation, only
the Goldstone contributions survive. The neutral Goldstone bosons
can only contribute corrections of order $\lambda_b^2$ (see section~\ref{section_HFrules} for the 
Feynman rules). We will
neglect these ${\cal O}(\la_b^2)$ contributions at the amplitude
level. However the order ${\cal O}(\la_b)$ corrections will be kept.
All the corrections are then triggered by $t \ra b
\chi_W$ and apart from the QCD vertices, only the
Yukawa vertices shown in Fig.~\ref{vertex_NLO} below are needed to
build up the full set of electroweak corrections.
Note that in the MSSM, the Higgs coupling
to the fermion $f$, $\la_{ffH}$, can involve other parameters
beside the corresponding Yukawa coupling $\lambda_f$, as shown in subsection~\ref{section_HFrules_mssm}. The Higgs
coupling to the charged Goldstone involves the Higgs self-coupling
or Yukawa coupling of the Higgs, $\la=M_H^2/2 v^2$ proportional to
the square of the Higgs mass. The latter can be large for large
Higgs masses. These considerations allow to classify the
contributions into three gauge invariant classes.

\subsection{Three classes of diagrams and the chiral structure at one-loop}
\label{section_3classes}
\begin{figure}[hbt]
\begin{center}
\includegraphics[width=16cm]{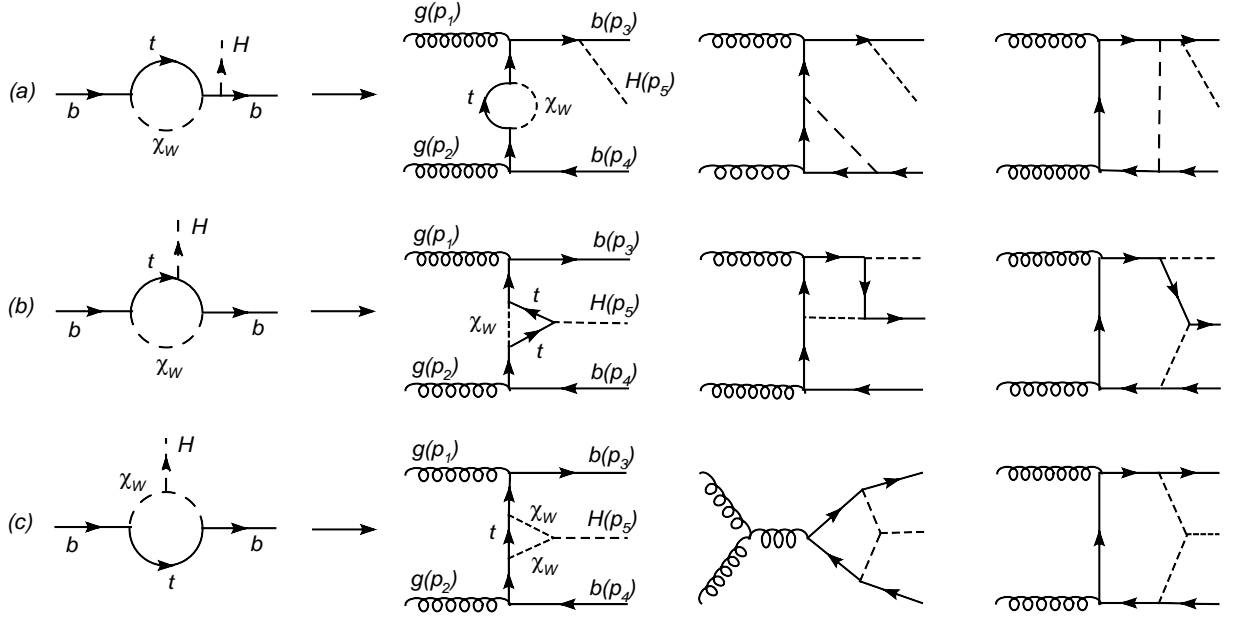}
\caption{\label{diag_3group}{\em All the diagrams in each group
can be obtained by inserting the two gluon lines or one triple
gluon vertex (not shown) to all possible positions in the generic
bottom line, which is the first diagram on the left. We have checked the number of diagrams through Grace-loop\cite{grace}.}}
\end{center}
\end{figure}
All the one-loop diagrams are classified into three gauge
invariant groups as displayed in Fig.~\ref{diag_3group}. The Higgs
couples to the bottom quark in the first group (Fig.
\ref{diag_3group}a), to the top quark in the second group (Fig.
\ref{diag_3group}b) and to the charged Goldstone boson in the
third group (Fig. \ref{diag_3group}c). As shown in
Fig.~\ref{diag_3group} each class can be efficiently reconstructed
from the one-loop vertex $b \bar b H$, depending on which leg one
attaches the Higgs, by then grafting the gluons in all possible
ways. We have also checked explicitly that each class with its
counterterms, see below, constitutes a QCD gauge invariant subset.
Note that these three contributions depend on different
combinations of independent couplings and therefore constitute
independent sets.

The chiral structure $t \ra b \chi_W$ impacts directly on the
structure of the helicity amplitudes at one-loop. The split of
each contribution according to $\Gamma^{\textrm even}$ and
$\Gamma^{\textrm odd}$, see Eq.~(\ref{amp_form2}) will turn out to be
useful and will indicate which helicity amplitude can be enhanced
by which Yukawa coupling at one-loop. We show only one example in
class $(b)$ of Fig.~\ref{diag_3group}. It is straight forward to
carry the same analysis for all other diagrams. We choose the
first diagram in group (b) in Fig. \ref{diag_3group}. For clarity
we will here take $m_b=0$, we have already shown how $m_b$
insertions are taken into account, see Eq.~(\ref{amp_form2}).
Leaving aside the colour part which can always be factorised out
(see Appendix~\ref{optimisation}) and the strong coupling
constant, we write explicitly the contribution of this diagram as
\bea
{\cal A}_{b1}(\la_1,\la_2;\la_3,\la_4)
=\la_{ttH}\la_t^2\bar{u}(\la_3,p_3)\slashep(\la_1,p_1)\fr{\slashpbar_{13}}{\bar{p}_{13}^2}C_{b1}\fr{\slashpbar_{24}}{\bar{p}_{24}^2}
\slashep(\la_2,p_2)u(\la_4,p_4) \label{amp_vertex}.
\eea
$C_{b1}$ is the Yukawa vertex correction. In $D$-dimension, with
$q$ the integration variable, the momenta as defined in
Fig.~\ref{diag_gg_LO} with $p_{ij}=p_i+p_j$ and $\bar{p}_{ij}=
p_j-p_i$ we have
\bea
C_{b1}=\int\fr{d^Dq}{(2\pi)^Di}\fr{(P_R-\varepsilon_{bt}P_L)(m_t+\slashq+\slashpbar_{13})(m_t+\slashq-\slashpbar_{24})
(P_L-\varepsilon_{bt}P_R)
}{(M_W^2-q^2)[m_t^2-(q+\bar{p}_{13})^2][m_t^2-(q-\bar{p}_{24})^2]},\label{cb1}
\eea
where $\varep_{bt}=\la_b/\la_t$ as defined in Fig.~\ref{vertex_NLO}.
The numerator of the integrand of Eq. (\ref{cb1}), neglecting terms of
${\cal O}(\la_b^2)$, can be re-arranged such as
\bea
{\cal A}_{b1}(\la_1,\la_2;\la_3,\la_4) \stackrel{\textrm
numerator}{\longrightarrow} &-&\underbrace{\varepsilon_{bt}
\left(m_t^2+(\slashq+\slashpbar_{13})(\slashq-\slashpbar_{24})\right)}_{\Gamma^{\textrm
even}}\crn 
&+& \underbrace{m_t P_R \left(
2\slashq+\slashpbar_{13}-\slashpbar_{24}\right)}_{\Gamma^{\textrm
odd}}.\label{C_numer}
\eea
This shows explicitly that ${\Gamma^{\textrm odd}}$ structures with a
specific chirality, $P_R$, can indeed be generated. They do not
vanish as $\la_{bbh} \ra 0$. The even one-loop structures on the
other hand are ${\cal O}(\la_b)$. The structure in class $(c)$,
Higgs radiation off the charged Goldstones, is the same. For class
$(a)$, radiation off the $b$-quark, the structure of the
correction is different, the odd part is suppressed and receives
an  ${\cal O}(\la_b)$ correction. To summarise, with $m_b=0$,
 making explicit the Yukawa couplings and the chiral structure if any, for example $P_R$, that characterise
each class and comparing to the leading order, one has
\begin{center}
\begin{tabular}{|c|c|c|}
\cline{2-3}
\multicolumn{1}{c|}{ }   & ${\Gamma^{\textrm even}}$ & ${\Gamma^{\textrm odd}}$ \\
  \hline
  tree-level &$\la_{bbH}$ & 0 \\
  $(a)$ & $\la_t^2 \la_{bbH}$&  $\la_{b}\la_t \la_{bbH}$\\
  $(b)$ & $\la_{b} \la_{t} \la_{ttH}$ &  $\la_t^2 \la_{ttH}$, ($P_R$)\\
  $(c)$ & $\la_{b} \la_{t} \la_{\chi \chi H}$  & $\la_t^2 \la_{\chi \chi H}$, ($P_R$) \\
  \hline
\end{tabular}
\end{center}
Despite the existence of the simple relation $\la_{ffH}=-\la_f/\sqrt{2}$ in the SM, we have kept $\la_{ffH}$ and $\la_f$ separate to distinguish their different origins ($\la_f$ comes from 
the Goldstone couplings). As discussed in subsection~\ref{section_HFrules_mssm}, in the MSSM $\la_{bbH}$ is enhanced by 
$\tan\beta$ but not $\la_b$. We clearly see that  all one-loop $\Gamma^{\textrm even}$
contributions vanish in the limit $\la_{b}=0$ and $\la_{bbH}=0$. On
the other hand this is not the case for the one-loop $\Gamma^{\textrm
odd}$ contribution belonging to class $(b)$ and $(c)$. However for
these contributions to interfere with the tree-level LO
contribution requires a chirality flip through a $m_b$ insertion.
Therefore in the SM for example, the NLO cross section is
necessarily of order $m_b^2$, like the LO, with corrections
proportional to the top Yukawa coupling for example. On the other
hand, in the limit of $\la_{bbH}=0$, the tree level vanishes
but $g g \ra b \bar b H$ still goes with an amplitude of order
$g_s^2 \la_t^2 \la_{ttH}$ or $g_s^2\la_t^2 \la_{\chi \chi H}$. For
$\la_{bbH} \neq 0$ these contributions should be considered as
part of the NNLO ``corrections" however they do not vanish in the
limit $m_b \ra 0$ (or $\la_{bbH}=0$) while the tree level does. These
contributions can be important and we will therefore study their
effects. For these  contributions at the ``NNLO" we can set
$m_b=0$.

The classification in terms of structures as we have done makes
clear also that the novel one-loop induced ${\Gamma^{\textrm odd}}$
contributions must be ultraviolet finite. This is not necessarily
the case of the ${\Gamma^{\textrm even}}$  structures where
counterterms to the tree-level structures are needed through
renormalisation to which we now turn.

\section{Renormalisation}\label{sec:renorm}
We use an on-shell (OS) renormalisation scheme exactly along the
lines described in subsection~\ref{subsection_one-loop-renorm}. Ultraviolet divergences are
regularised through dimensional regularisation. In our
approximation we only need to renormalise the vertices $b \bar b
g$ and $b \bar b H$ as well as the bottom mass, $m_b$. For the $b
\bar b g$ vertex, its counterterm reads
\bea
\delta^\mu_{bbg}=2g_s\gamma^\mu(\delta Z_{b_{L}}^{1/2}P_L+\delta Z_{b_{R}}^{1/2}P_R).
\label{renorm_bbg}
\eea 
$\delta Z_{b_{L,R}}^{1/2}$ are calculated by using Eq.~(\ref{counter_vertex}). For this one needs to know the 
coefficients $K_{1,\gamma,5\gamma}$ which are very simple in our approximation:
\begin{center}
\begin{minipage}[c]{5cm}
\begin{center}
\includegraphics[width=5cm]{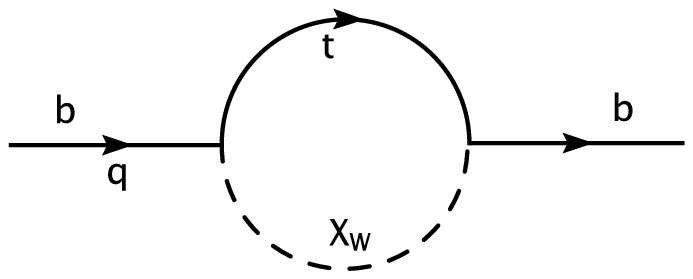}
\end{center}
\end{minipage}
\begin{minipage}[c]{8cm}
\[\Longrightarrow\hs \Sigma(q^2)=K_1+K_\gamma\slashq+K_{5\gamma}\slashq\gamma_5\]
\end{minipage}
\end{center}
We get
\bea
K_1(q^2)&=&-\fr{\la_t^2}{16\pi^2} \;\Big(
C_{UV}-F_0(t,W)\Big),\crn
K_{\gamma}(q^2)&=&-K_{5\gamma}(q^2)= \fr{\la_t^2}{64\pi^2} \;
\Big( C_{UV}-2 F_1(t,W) \Big)\crn {\textrm with} \;\;
C_{UV}&=&\fr{1}{\eps}-\gamma_E+\ln 4\pi\,\,,D=4-2\eps,\crn
F_n(A,B)&=&\int_0^1dxx^n\ln
\Big((1-x)m_A^2+xm_B^2-x(1-x)q^2 \Big). \label{KFG}
\eea

The reason we get $K_{\gamma}(q^2)=-K_{5\gamma}(q^2)$ is due to
the particular chiral structure of the $t \ra b \chi_W$ loop
insertion. In particular for $m_b=0$, one recovers that these
corrections only contribute to $\delta Z_{b_L}^{1/2}$ and not
$\delta Z_{b_R}^{1/2}$.

For the $b \bar b H$ vertex with the bare Lagrangian term $\Lar_{bbH}=-\fr{m_b}{\upsilon}\bar{\psi}_b\psi_b \vph_H$, 
one needs to renormalise $m_b$, $\upsilon$, $\psi_{b_L(R)}$ and $\vph_H$. With $\upsilon\to \upsilon(1+\delta\upsilon)$ and the rules 
given in Eq.~(\ref{rules_renorm}) we get the counterterm
\bea
\delta_{bbH}=\la_{bbH}\left[\fr{\delta m_b}{m_b}
+\delta Z_{b_L}^{1/2} +\delta Z_{b_R}^{1/2} +(\delta
Z_H^{1/2}-\delta\upsilon)\right].
\label{renorm_bbH}
\eea
Again $\delta Z_{b_{L,R}}^{1/2}$ and $\delta m_b$ are calculated by using Eqs. (\ref{counter_vertex}) and (\ref{KFG}). 
$\delta Z_{H}^{1/2}$ is calculated through Eq.~(\ref{dZH_PiH}) where $\Pi^H(q^2)$ comes from the diagrams with heavy particles in the loop. 
As we are interested in the Yukawa corrections, these particles are the top quark, the Higgs boson, the W-Goldstone bosons and the Z-Goldstone boson. 
Indeed, those corrections include all the leading Higgs couplings: $\la_{ttH}$ and $\la_{HHH}$. We get
\bea
\Pi^H(q^2)&=&\fr{M_H^4}{16\pi^2\upsilon^2}\left[3C_{UV}-\fr{9}{2}F_0(H,H)-F_0(W,W)-\fr{1}{2}F_0(Z,Z)-3(1-\ln M_H^2)\right]\crn
&+&\fr{3m_t^2}{8\pi^2\upsilon^2}\left\{q^2[C_{UV}-F_0(t,t)]+4[1-\ln m_t^2+F_0(t,t)]\right\}.
\eea  
From this and Eq.~(\ref{dZH_PiH}) we get
\bea
\delta Z_H^{1/2}&=&-\fr{1}{8\pi^2}\Rel\left\{\fr{3\la_{t}^2}{4}
\Big[C_{UV}-F_0(t,t)- M_H^2G_0(t,t)+4m_t^2G_0(t,t)\Big]
\right.\crn &-&\left.\fr{\la}{4}\Big[9 G_0(H,H)+2
G_0(W,W)+G_0(Z,Z)\Big]\right\}\Big\vert_{q^2=M_H^2},\crn
G_n(A,B)&=&q^2\fr{d}{dq^2}F_n(A,B)=q^2\int_0^1dx\fr{-x^nx(1-x)}{(1-x)m_A^2+xm_B^2-x(1-x)q^2}.
\label{delta_dZH}
\eea
For $\delta\upsilon$, we use the relation
\bea
\upsilon=\fr{2s_WM_W}{e}, \hs s_W=\sqrt{1-\fr{M_W^2}{M_Z^2}}
\eea
to write $\delta\upsilon$ in terms of $\delta M_W^2$, $\delta M_Z^2$ 
\bea
\delta\up=-\fr{c_W^2}{s_W^2}\left(\fr{\delta M_W^2}{2M_W^2}-\fr{\delta M_Z^2}{2M_Z^2}\right)+\fr{\delta M_W^2}{2M_W^2}.
\eea
We do not need to renormalise $e$ since we are interested only in the Yukawa 
sector hence do not touch the photon. We now use Eq.~(\ref{renorm_gauge}) to calculate $\delta M_W^2$ and $\delta M_Z^2$. 
Like in the case of $\Pi^H$, $\Pi_T^W$ and $\Pi_T^Z$ include all the leading contributions related to 
the Goldstone bosons: 
\bea
\fr{\Pi_T^W(M_W^2)}{M_W^2}&=&\fr{M_H^2}{8\pi^2\up^2}\left\{\left[F_1(H,W)-F_0(H,W)\right]\Big\vert_{q^2=M_W^2}+\fr{\ln M_H^2}{2}\right\}\crn
&+&\fr{m_t^2}{8\pi^2\up^2}\left[3C_{UV}-6F_1(b,t)\right]\Big\vert_{q^2=M_W^2},\crn
\fr{\Pi_T^Z(M_Z^2)}{M_Z^2}&=&\fr{M_H^2}{8\pi^2\up^2}\left\{\left[F_1(H,Z)-F_0(H,Z)\right]\Big\vert_{q^2=M_Z^2}+\fr{\ln M_H^2}{2} \right\}\crn
&+&\fr{m_t^2}{8\pi^2\up^2}\left[3C_{UV}-3F_0(t,t)\right]\Big\vert_{q^2=M_Z^2}.
\eea
Then we get
\bea
\delta\upsilon &=&-\fr{1}{8\pi^2}\Rel\Big\{\fr{3\la_t^2}{4}
\Big[C_{UV}-2F_1(b,t)\big\vert_{q^2=M_W^2}\Big] \crn &-& \la
\Big[F_0(H,W)-F_1(H,W)-\fr{1}{2}\ln
M_H^2\Big]\Big\vert_{q^2=M_W^2}\crn
&-&\fr{c_W^2}{s_W^2}\Big[-\fr{3\la_t^2}{2}F_1(b,t)
+\la \big(- F_0(H,W)+F_1(H,W)\big)\Big]\Big\vert_{q^2=M_W^2}\crn
&-&\fr{c_W^2}{s_W^2}\Big[\fr{3\la_t^2}{4}F_0(t,t)
+ \la \big(F_0(H,Z)-F_1(H,Z)\big)\Big]\Big\vert_{q^2=M_Z^2}\Big\}.
\label{delta_v}
\eea
From Eqs. (\ref{delta_dZH}) and (\ref{delta_v}), one clearly sees that $(\delta Z_H^{1/2}-\delta\upsilon)$ is UV-finite. 
Therefore, by looking at Eqs. (\ref{renorm_bbg}) and (\ref{renorm_bbH}), we conclude that to make all the contributions of diagrams in Fig.~\ref{diag_3group} 
UV-finite, it is sufficient to renormalise the mass and wave function of bottom-quark as done above. On
the other hand $(\delta Z_H^{1/2}-\delta\upsilon)$ can be seen as
a universal correction to Higgs production processes. We will
include this correction as it has potentially large contributions
scaling like $\la_t^2$ and $\la$ which fall into the category of
the corrections we are seeking.

In the actual calculation, the counter term $\delta_{bbg}^\mu$
belongs to class $(a)$ in the classification of
Fig~\ref{diag_3group}. This makes class $(a)$ finite. The
counterterm we associate to class $(b)$ is the part of
$\delta_{bbH}$ from the $t \ra b \chi_W$ loops and therefore does
not include what we termed the universal Higgs correction, {\it
i.e} does not include the contribution $(\delta
Z_H^{1/2}-\delta\upsilon)$. This is sufficient to make class $(b)$
finite. In our approach $(c)$ is finite without the addition of a
counterterm. We will keep the $(\delta Z_H^{1/2}-\delta\upsilon)$
contribution separate from the  contributions in  classes
$(a),(b),(c)$. We will of course include it in the final result.

\section{Calculation details}
\label{bbH1_cal}
We have written two independent codes. In the first one we set
$m_b=0$ in all propagators and other spinors that emerge from the
helicity formalism we follow. In this limit, the helicity
formalism is very much simplified and the expression quite
compact. This code is in fact subdivided in two separate
sub-codes. One sub-code is generated for the ``even" part
(constituted by the $\Gamma^{\textrm even}$ contributions, see
~Eq.~(\ref{amp_form2})) and the other by the ``odd" part. We also
generate a completely independent code for the case $m_b \neq 0$
where in particular we use the helicity formalism with massive
fermions. Details of the helicity formalism that we use are given
in Appendix~\ref{appendix-helicity}.

The steps that go into writing these codes are the following. In
the first stage, we use FORM\cite{form} to generate expressions
for the tree level and one loop helicity amplitudes. Each helicity
amplitude is written in terms of Lorentz invariants, scalar spinor
functions $(A,B,C)_{\la_i\la_j}$ defined in
Appendix~\ref{appendix-helicity} and the
Passarino-Veltman\cite{pass_velt} tensor functions $T^{N}_{M}$ for
a tensor of rank $M$ for $N$-point function. We have also sought
to write the contribution of each amplitude as a product of
different structures or blocks that reappear for different graphs
and contributions. For example colour factorisation is
implemented, this further allows to rearrange the amplitude into
an Abelian part and a non-Abelian part which will not interfere
with each other at the matrix element squared level. The helicity
information is contained in a set of basic blocks for further
optimisation. Another set of blocks pertains to the loop integrals
and other elements. The factorisation of the full
amplitude in terms of independent building blocks is easily
processed within FORM. These building blocks can still consist of
long algebraic expressions which can be efficiently abbreviated
into compact variables with the help of a Perl script which also
allows to convert the output of FORM into the Fortran code ready
for a numerical evaluation. More details on the FORM code as well as the optimisation we implemented 
can be found in Appendix~\ref{optimisation}.

\subsection{Loop integrals, Gram determinants and phase space integrals}
\label{sec:gram}
The highest rank $M$ of the Passarino-Veltman
tensor functions $T^{N}_{M}$ with $M \leq N$ that we encounter in
our calculation is $M=4$ and is associated to a pentagon graph,
$N=5$. We use the library LoopTools\cite{looptools, ff} to
calculate all the tensorial one loop integrals as well as the
scalar integrals, this means that we leave it completely to LoopTools to perform the reduction of the tensor integrals to the
basis of the scalar integrals. In order to obtain the cross
section one needs to perform the phase-space integration and
convolution over the gluon distribution function (GDF), $g(x,Q)$
with $Q$ representing the factorisation scale. We have
\bea
\sigma(p p\to b\bar{b}H
)&=&\fr{1}{256}\int_0^1dx_1g(x_1,Q)\int_0^1dx_2g(x_2,Q)\crn
&\times &\fr{1}{\hat
F}\int\fr{d^3\textbf{p}_3}{2e_3}\fr{d^3\textbf{p}_4}{2e_4}\fr{d^3\textbf{p}_5}{2e_5}|{\cal
A}(gg\to b\bar{b}H)|^2\delta^4(p_1+p_2-p_3-p_4-p_5)
\,,\crn\label{sigma_pp}
\eea
where $\fr{1}{256}=\fr{1}{4}\times\fr{1}{8}\times\fr{1}{8}$ is the
spin and colour average factor and  the flux factor is
$1/\hat{F}=1/\Big(2\pi)^52\hat{s}\Big)$ with $\hat{s}=x_1x_2s\ge
(2m_b+M_H)^2$.\\
\noi The integration over the three body phase space and momentum
fractions of the two initial gluons is done by using two
``integrators": {BASES}\cite{bases} and {DADMUL}\cite{dadmul}. {BASES} is a Monte Carlo that uses the
importance sampling technique while  {DADMUL} is based on the
adaptive quadrature algorithm. The use of two different phase
space integration routines helps control the accuracy of the
results and helps detect possible instabilities. In fact some
numerical instabilities in the phase space integration do occur
when we use {DADMUL} but not when we use {BASES} which
gives very stable results with small integration error, typically
$0.08\%$ for $10^5$ Monte Carlo points per iteration (see section~\ref{section_bases} for more details). 
For the range of Higgs
masses we are studying in this chapter, the instabilities that are
detected with {DADMUL} were identified as spurious
singularities having to do with vanishing Gram determinants for
the three and four point tensorial functions calculated in
LoopTools by using the Passarino-Veltman reduction
method\footnote{The reduction of the five point function using the
method of  Denner and Dittmaier~\cite{denner_5p,looptools_5p}
which avoids the Gram determinant at this stage as implemented in
LoopTools gives very stable results.}. Because this problem
always happens at the boundary of phase space, we can avoid it by
imposing appropriate kinematic cuts in the final state. In our
calculation, almost all zero Gram determinants disappear when we
apply the  cuts on the transverse momenta of the bottom quarks
relevant for our situation, see section~\ref{results} for the
choice of cuts. The remaining zero Gram determinants occur when
the two bottom quarks or one bottom quark and the Higgs are
produced in the same direction. Our solution, once identified as
spurious, was to discard these points by imposing some tiny cuts
on the polar, $\theta$, and relative azimuthal angles, $\phi$ of
the outgoing $b$-quarks, the value of the cuts is $\theta_{\textrm
cut}^{b,\bar b}=\vert\sin\phi^{\bar b}\vert_{\textrm cut}=10^{-6}$. {DADMUL}
then produces the same result as {BASES} within the
integration error.
\subsection{Checks on the results}
i) Ultraviolet finiteness: \\
\noi The final results must be ultraviolet (UV) finite. It means
that they should be independent of the parameter $C_{UV}$ defined
in Eq.~(\ref{KFG}). In our code this parameter is treated as a
variable.The cancellation of $C_{UV}$ has been carefully checked
in our code. Upon varying the value of the parameter $C_{UV}$ from
$C_{UV}=0$ to $C_{UV}=10^{5}$, the results is stable within more
than 9 digits using double precision. This check makes sure that the divergent part of
the calculation is correct. The correctness of the finite part is
also well checked in our code by confirming that each helicity
configuration is QCD gauge invariant.\\
\noi ii) QCD gauge invariance: \\
\noi In the physical gauge we use, the
QCD gauge invariance reflects the fact that the gluon is massless
and has only two transverse polarisation components. In the
helicity formalism that we use, the polarisation vector of the
gluon of momentum $p$ and helicity $\la$ is constructed with the
help of a reference vector $q$, see Appendix~\ref{appendix-helicity} for details. 
The polarisation vector is
then labelled as $\eps^{\mu}(p,\la;q)$. A change of reference
vector from $q$ to $q^\prime$ amounts essentially to a gauge
transformation (up to a phase)
\bea
\eps^\mu(p,\la;q^\prime)=e^{i\phi(q^\prime,q)}\eps^\mu(p,\la;q)+\beta(q^\prime,q)p^\mu
.
\eea
QCD gauge invariance in our case amounts to independence of the
cross section in the choice of the reference vector, $q$. We have
carefully checked that the numerical result for the norm of each
helicity amplitude at various points in  phase space is
independent of the reference vectors say $q_{1,2}$ for gluon 1 and
2, up to 12 digits using double precision. By default, our
numerical evaluation is based on the use of $q_{1,2}=(p_2,p_1)$.
For the checks in the case of massive $b$ quarks the result with
the default choice $q_{1,2}=(p_2,p_1)$ is compared with a random
choice of $q_{1,2}$, keeping away from vectors with excessively
too small or too large components, see
Appendix~\ref{appendix-helicity} for more details.\\
\noi iii) As stated earlier, the result based on the use of  the
massive quark helicity amplitude are checked against those with
the independent code using the massless helicity amplitude by
setting the mass of the $b$ quark to zero. This is though just a
consistency check.\\
\noi iv) At the level of integration over phase space and density
functions we have used two integration routines and made sure that
we obtain the same result once we have properly dealt with the
spurious Gram determinant as we explained in
section~\ref{sec:gram}.\\
\noi v) Moreover, our tree level results have been successfully
checked against the results of CalcHEP\cite{calchep}.

\section{Results: $M_H<2M_W$}
\label{section_bbH_result1}
\subsection{Input parameters and kinematical cuts}
\label{results} Our input parameters are $\alpha(0)=1/137.03599911$,
$M_W=80.3766$GeV, $M_Z=91.1876$GeV, $\alpha_s(M_Z)=0.118$,
$m_b=4.62$GeV, $m_t=174.0$GeV with $c_W\equiv M_W/M_Z$. The CKM parameter $V_{tb}$ is set to be $1$.
We consider the case at the LHC where the center of
mass energy of the two initial protons is $\sqrt{s}=14$TeV.
Neglecting the small light quark initiated contribution, we use CTEQ6L\cite{cteq6,cteq6_1,cteq6_2,cteq6_3}
for the gluon distribution function (GDF) in the proton. The
factorisation scale for the GDF and energy scale for the strong
coupling constant are chosen to be $Q=M_Z$ for simplicity.

As has been done in previous analyses~\cite{dawson_bbH,LH03_bbh},
for the exclusive $b\bar{b}H$ final state, we require the outgoing
$b$ and $\bar{b}$ to have high transverse momenta
$|\textbf{p}_{T}^{b,\bar{b}}|\ge 20$GeV and pseudo-rapidity
$|\eta^{b,\bar{b}}|<2.5$. These kinematical cuts reduce the total
rate of the signal but also greatly reduce the QCD background. As
pointed in~\cite{dittmaier_bbH} these cuts also stabilise the
scale dependence of the QCD NLO corrections compared to the
case where no cut is applied. In the following, these kinematical
cuts are always applied unless otherwise stated.

Talking of the NLO QCD scale uncertainty and before presenting our
results, let us remind the reader of the size of the QCD
corrections. Taking a renormalisation/factorisation scale as we
take here at $M_Z$, the QCD corrections in a scheme where the
bottom Yukawa coupling is taken on-shell amount to $\sim -22\%$
for a Higgs mass of $120$GeV.

\subsection{NLO EW correction with $\la_{bbH}\neq 0$ }
\label{bbH_result_nlo_1}
\begin{figure}[h]
\begin{center}
\mbox{\includegraphics[width=0.45\textwidth]{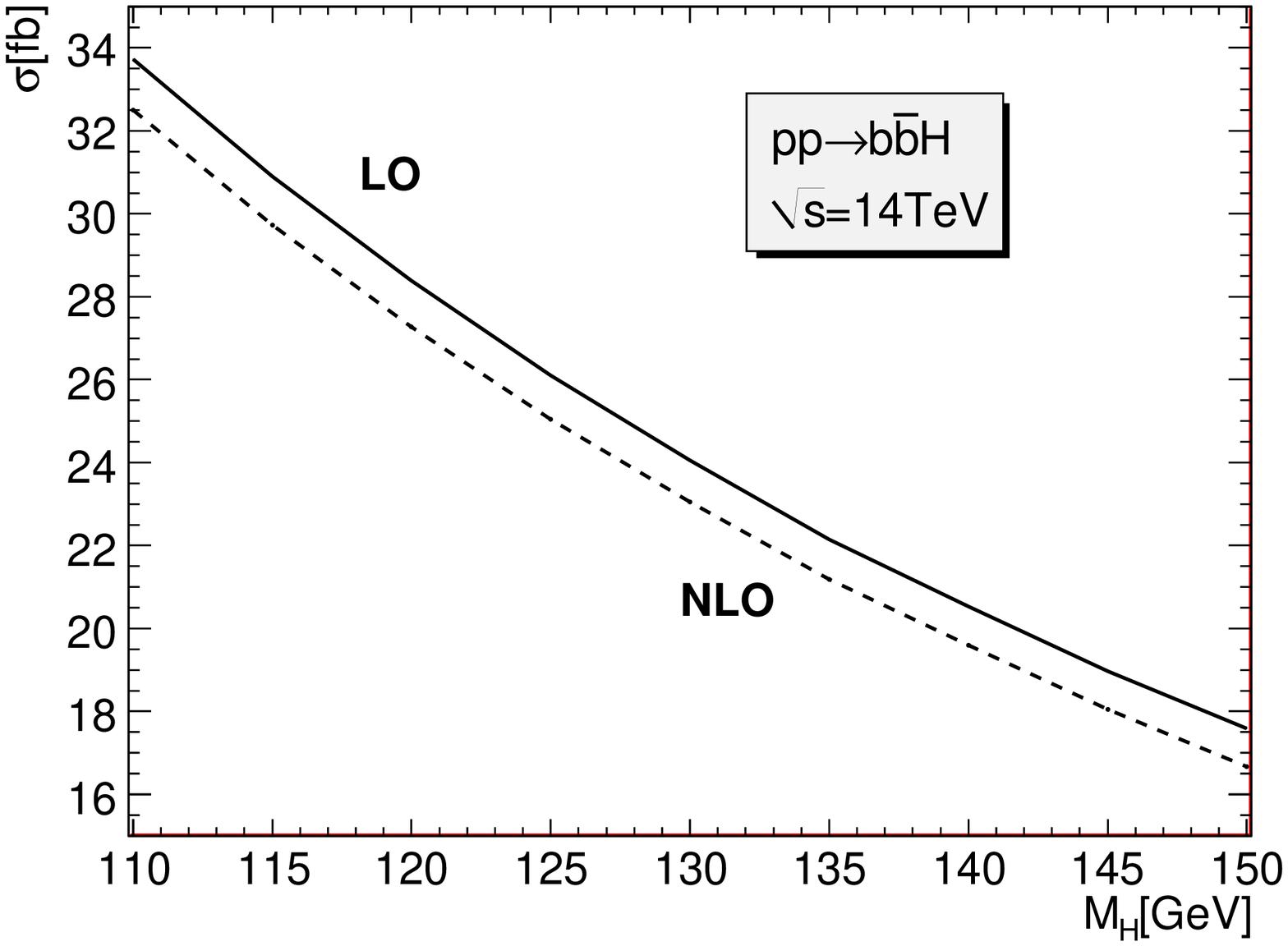}
\hspace*{0.075\textwidth}
\includegraphics[width=0.45\textwidth]{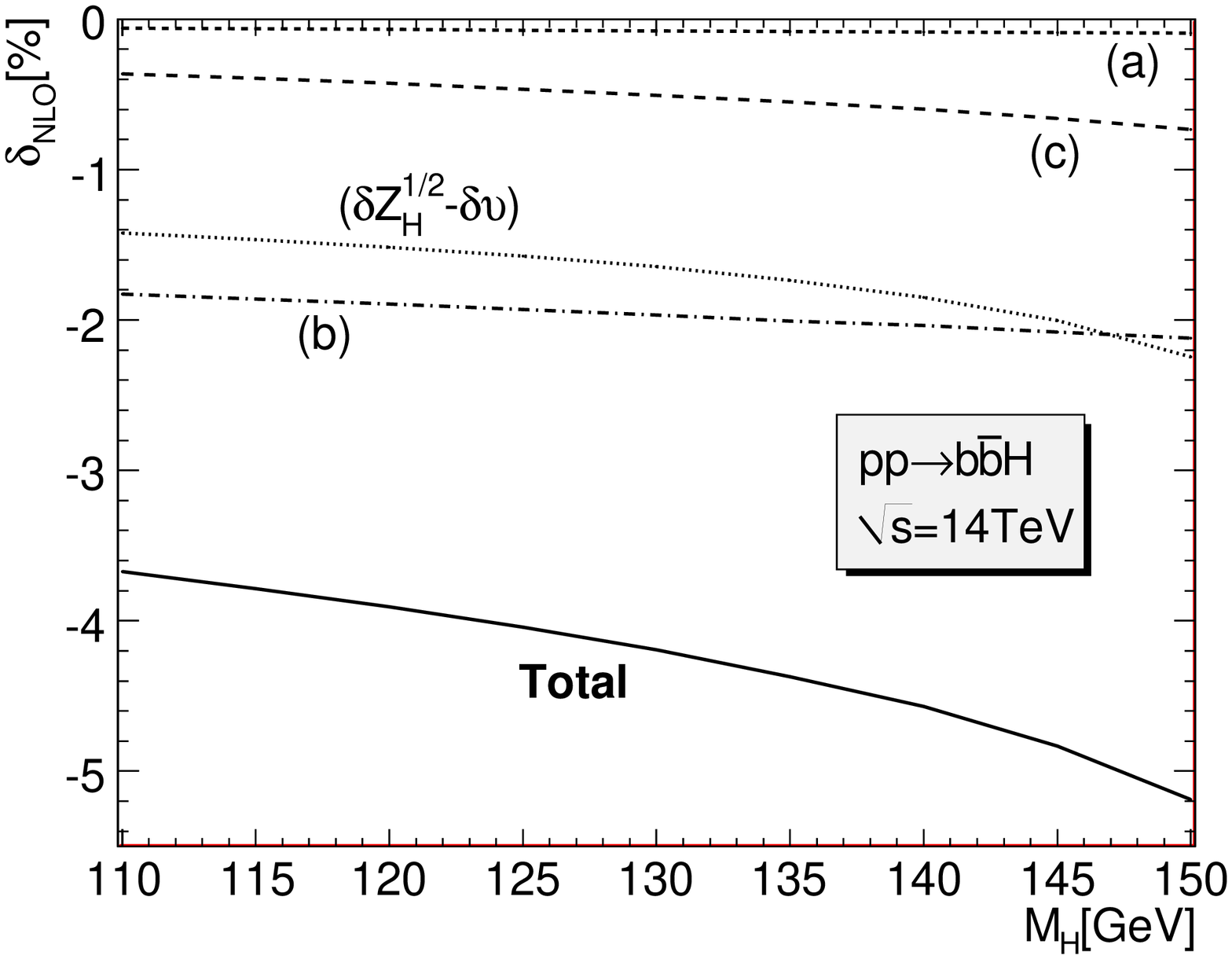}}
\caption{\label{p_LT_mH}{\em Left: the LO and NLO cross sections
as functions of $M_H$. Right: the relative NLO EW correction
normalized to tree level $\sigma_{LO}$. $(a)$, $(b)$, $(c)$ correspond to
the three classes of diagrams as displayed in Fig.
\ref{diag_3group} to which counterterms are added (see
section~\ref{sec:renorm}). $(\delta Z_{H}^{1/2}-\delta\upsilon)$
is the correction due to the  universal correction contained in
the renormalisation of the $b\bar bH$ vertex. ``Total'' refers to
the total electroweak correction, of Yukawa type, at one-loop.}}
\end{center}
\end{figure}

The cross sections with two high-$p_T$ bottom quarks at LO  and
NLO at the LHC are displayed in Fig. \ref{p_LT_mH} as a function
of the Higgs mass. The NLO EW correction reduces the cross section
by about $4\%$ to $5\%$ as the Higgs mass is varied from  $110$GeV
to $150$GeV. The first conclusion to draw is that this correction
is small if we compare it to the QCD correction or  even to the
QCD scale uncertainty. Considering that we have pointed to the
fact that the contributions could be grouped into three gauge
invariant classes that reflect the strengths of the Higgs coupling
to the $b$, the $t$ or its self-coupling, one can ask whether this
is the result of some cancellation. It turns out not to be the
case. All contributions are below $3\%$, see~Fig.~\ref{p_LT_mH}.
Class $(a)$ with a Higgs radiated from the bottom line is totally
negligible ranging from $-0.09\%$ to $-0.06\%$. We have failed in
finding a good reason for the smallness of this contribution
compared to the others. Those  due to the Higgs self-coupling are
below $1\%$. Radiation from the top contributes about $-2\%$ and
is of the same order as the contribution of the universal
correction. We had argued that the Yukawa corrections brought
about by the top might be large. It seems that the mass of the top
introduces also a large scale which can not be neglected compared
to the effective energy of the hard process even for LHC
energies.\\
\begin{figure}[htb]
\begin{center}
\mbox{\includegraphics[width=0.45\textwidth]{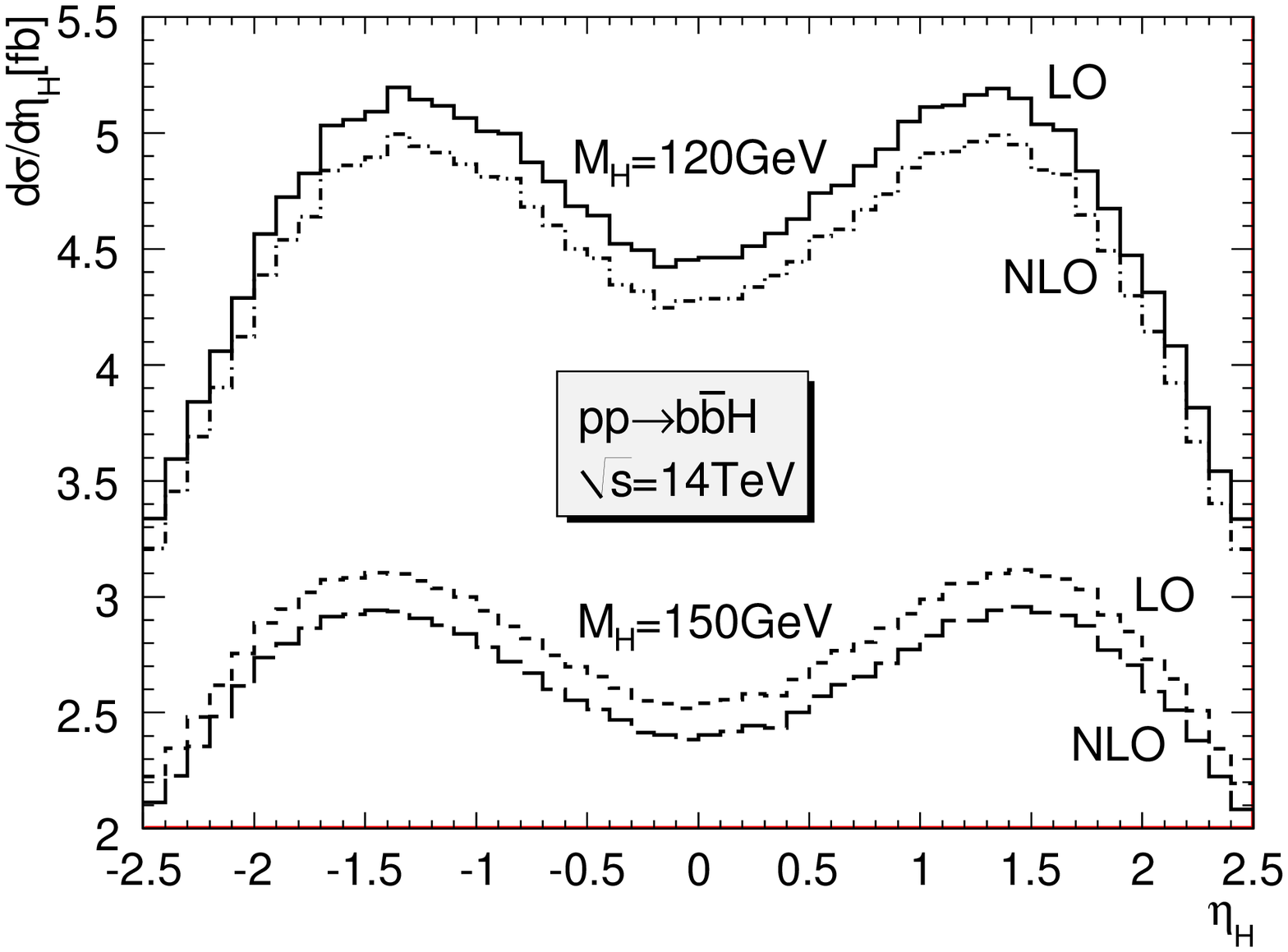}
\hspace*{0.075\textwidth}
\includegraphics[width=0.45\textwidth]{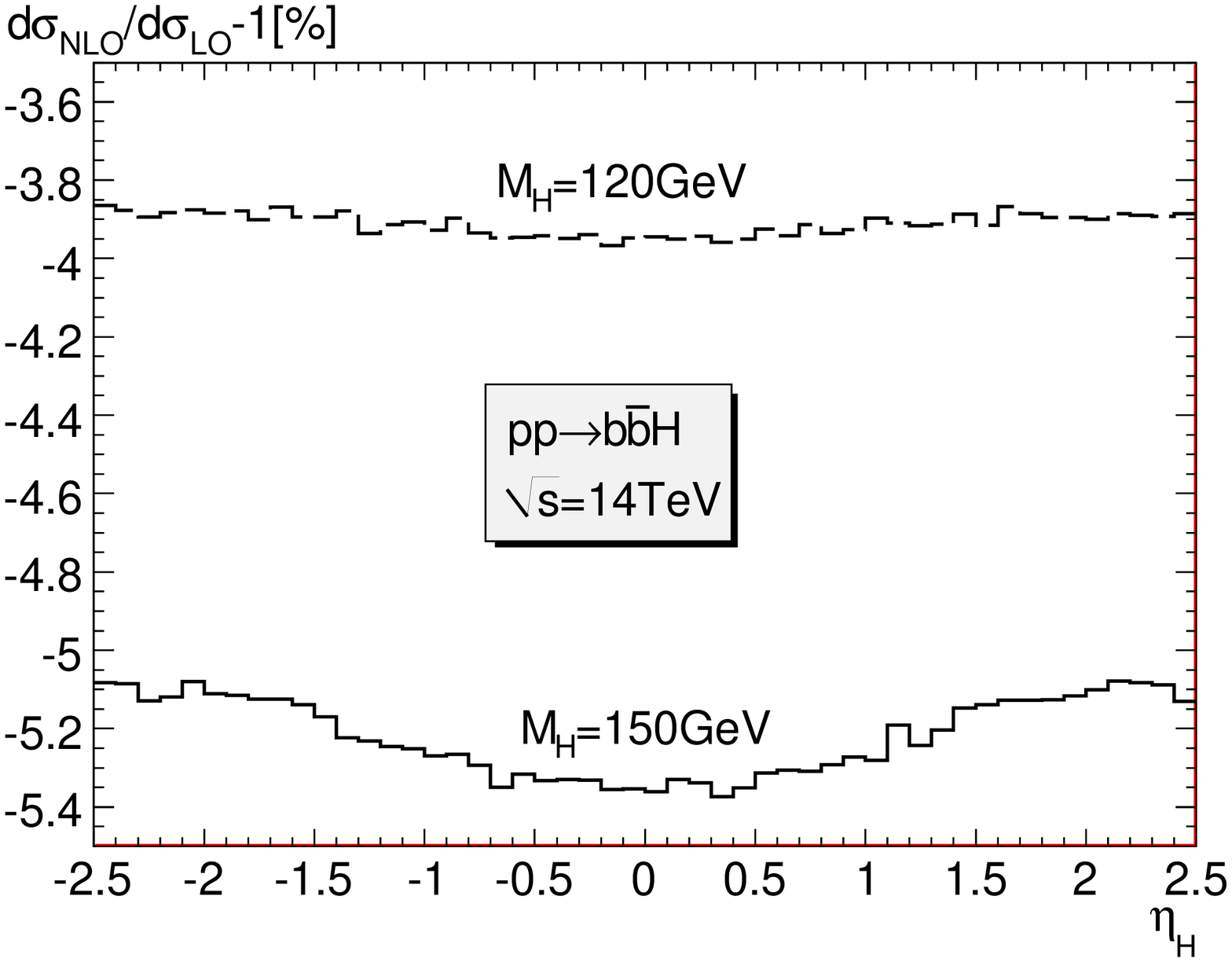}}
\mbox{\includegraphics[width=0.45\textwidth]{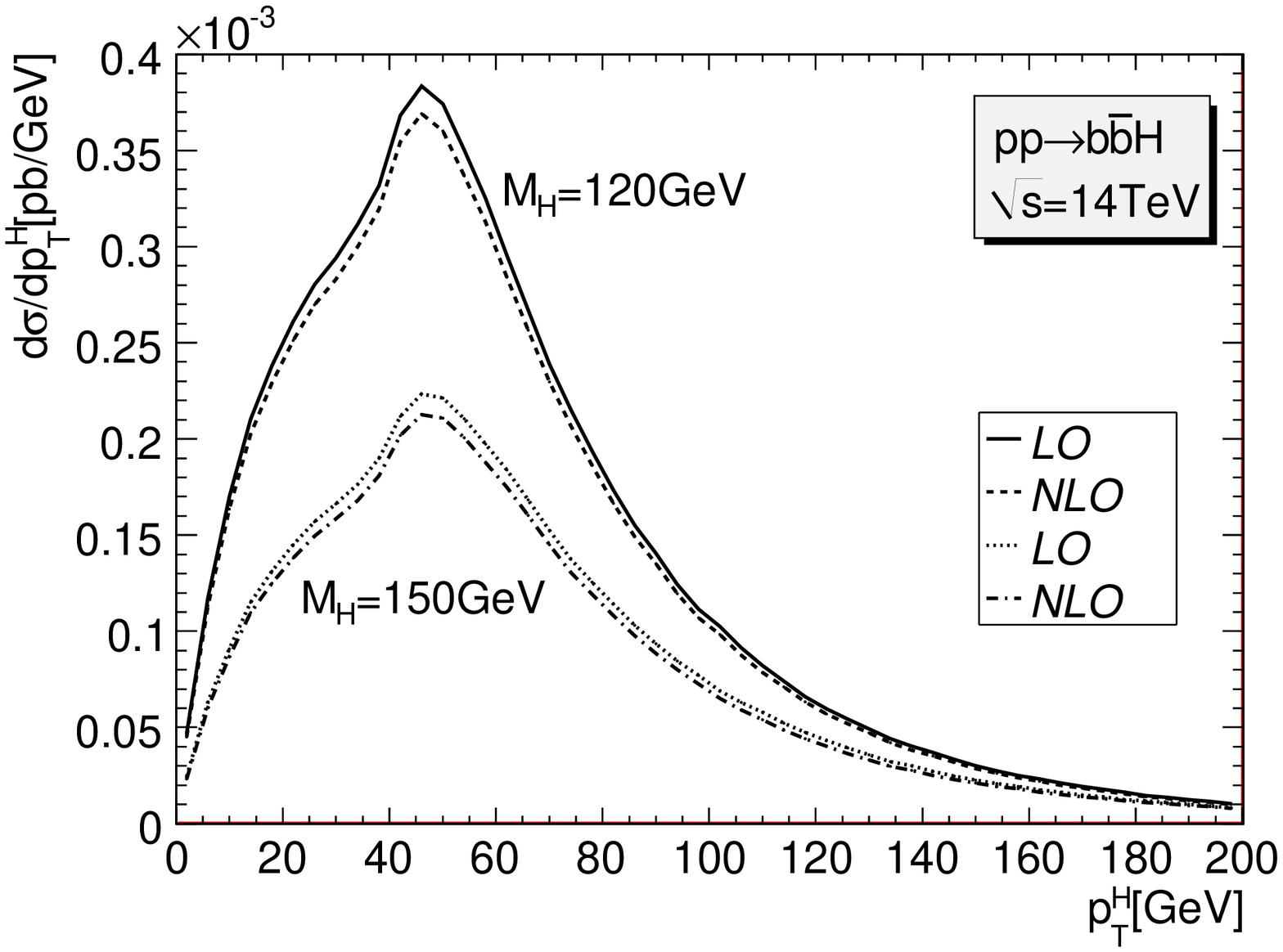}
\hspace*{0.075\textwidth}
\includegraphics[width=0.45\textwidth]{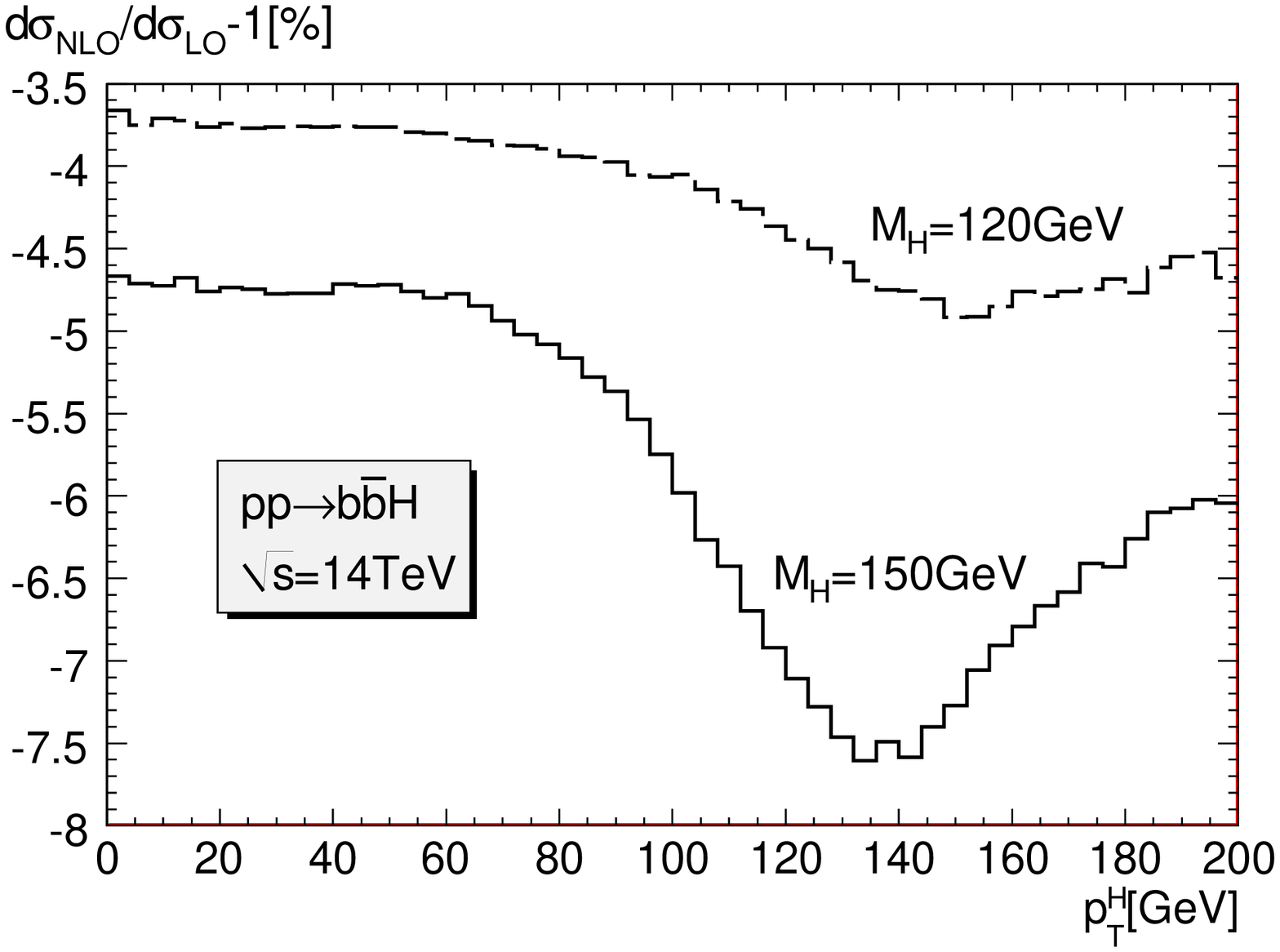}}
\caption{\label{fig:dist-nlo}{\em Effect of the NLO electroweak
corrections on the pseudo-rapidity and transverse momentum
distributions of the Higgs for $M_H=120,150$GeV. The relative
corrections $d\sigma_{NLO}/d\sigma_{LO}-1$ is also shown.}}
\end{center}
\end{figure}

The NLO corrections are spread rather uniformly on all the
distributions we have looked at. We have chosen to show in
Fig.~\ref{fig:dist-nlo} the effect on pseudo-rapidity and
transverse momentum distributions of the Higgs for two cases
$M_H=120$GeV and $M_H=150$GeV.  As  Fig.~ \ref{fig:dist-nlo} shows
the relative change in these two distributions is sensibly
constant especially for $M_H=120$GeV. For $M_H=150$GeV, the
corrections are largest for $p_T^H$ around $140$GeV, however this
is where the cross section is very small. A similar pattern, {\it
i.e.} a constant change in the distributions, is observed for the
bottom variables.

\subsection{EW correction in the limit of vanishing $\la_{bbH}$}
\begin{figure}[hbtp]
\begin{center}
\mbox{\includegraphics[width=0.45\textwidth]{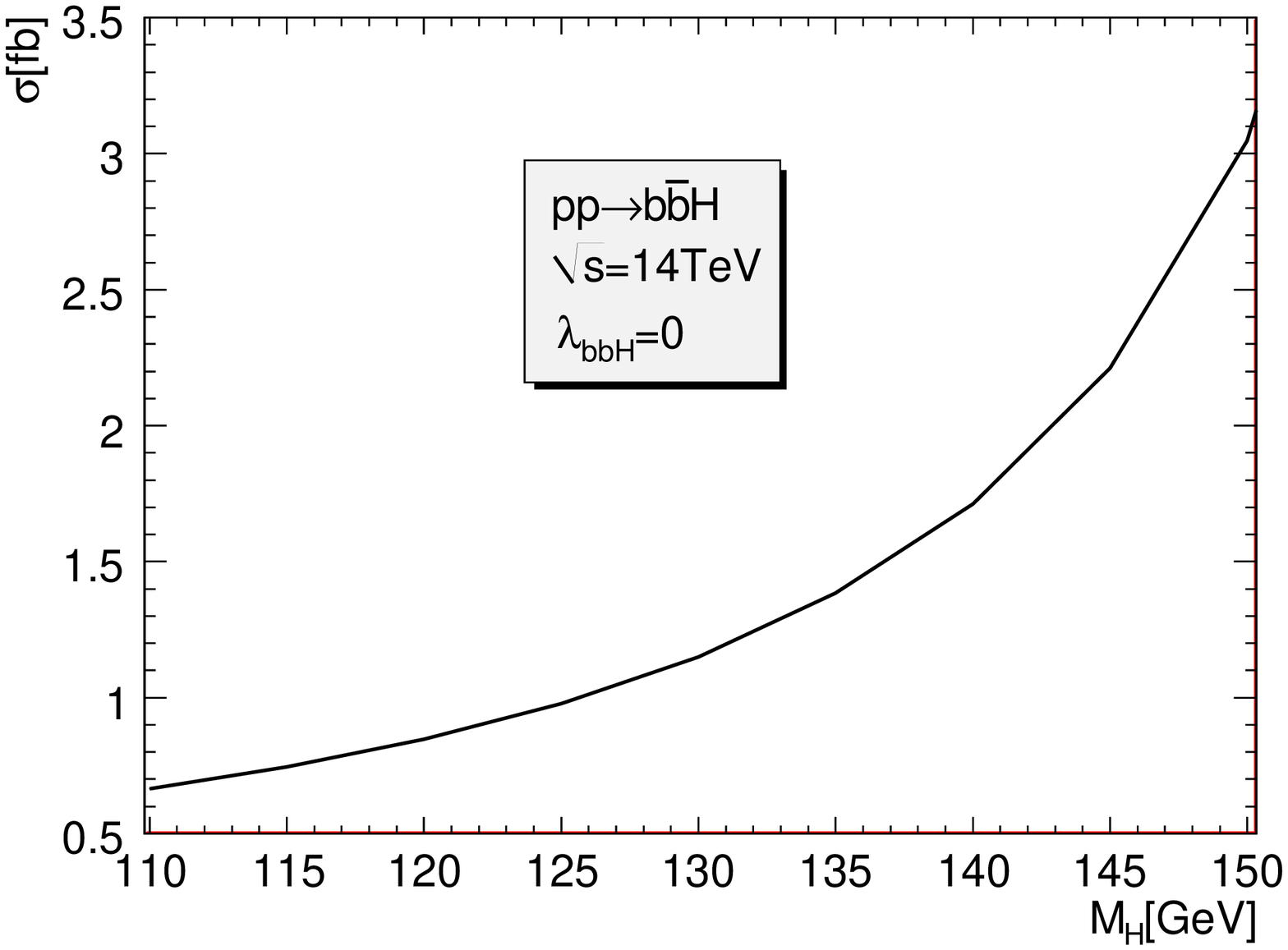}
\hspace*{0.075\textwidth}
\includegraphics[width=0.45\textwidth]{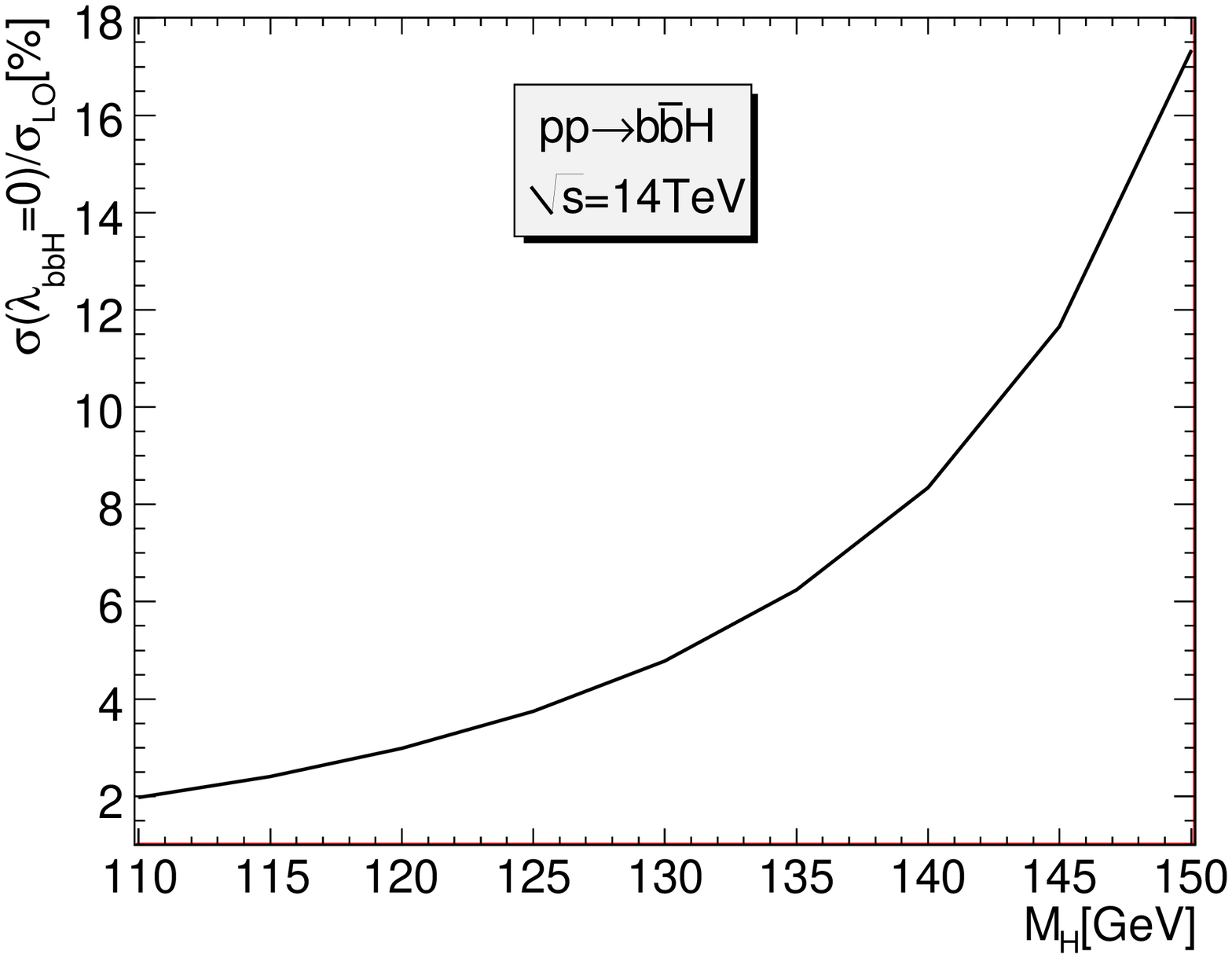}}
\caption{\label{p_LL_mH}{\em The one-loop induced cross section as
a function of $M_H$ in the limit of vanishing bottom-Higgs Yukawa
coupling. The right panel shows the percentage contribution of
this contribution relative to the tree level cross section calculated with
$\la_{bbH}\neq 0$.}}
\end{center}
\end{figure}
The cross section for $\la_{bbH}=0$ can be induced at one-loop
through the top loop. This ``NNLO" contribution rises rather
quickly as the Higgs mass increases even in the narrow range
$M_H=110-150$GeV as can be seen in Fig.~\ref{p_LL_mH}. Indeed
relative to the tree level,  the  cross section with $M_H=120$GeV amounts
to $3\%$ while for $M_H=150$GeV it has increased to as much as
$17\%$. Going past $M_H\geq 2M_W$ we encounter a Landau
singularity\cite{landau} (a pinch singularity in the loop
integral) from diagrams like the one depicted in
Fig.~\ref{diag_gg_ew} (right) with the Higgs being attached to the $W$'s
or their Goldstone counterpart. It corresponds to a situation
where all particles in the loop are resonating and can be
interpreted as the production and decay of the tops into
(longitudinal) $W$'s with the later fusing to produce the Higgs.
This leading Landau singularity is not integrable, at the level of
the loop amplitude squared and must be regulated by the
introduction of  a width for the unstable particles. This
issue together with a general discussion of Landau singularities
will be considered in the next chapters.
\begin{figure}[htp]
\begin{center}
\mbox{\includegraphics[width=0.45\textwidth]{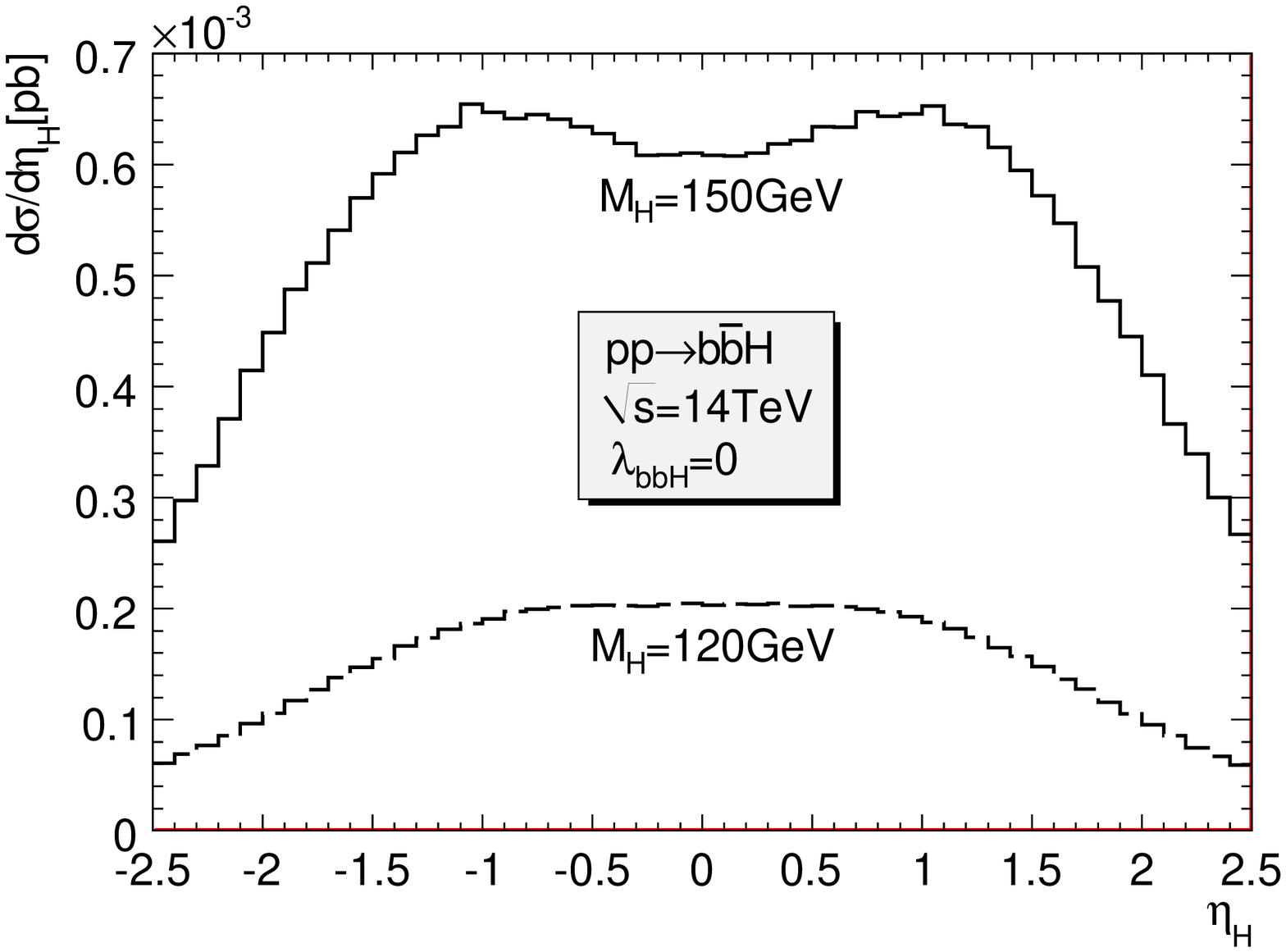}
\hspace*{0.075\textwidth}
\includegraphics[width=0.45\textwidth]{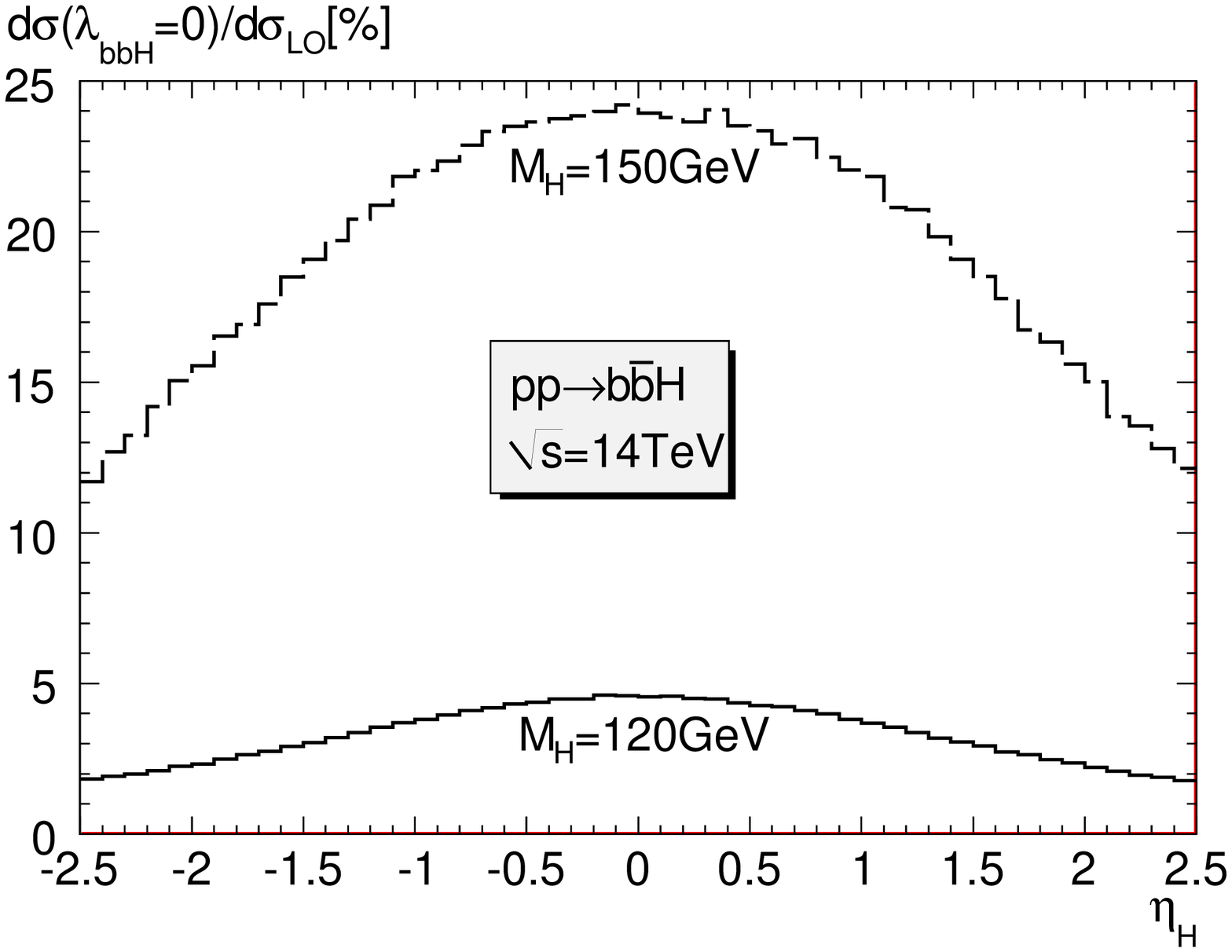}}
\mbox{\includegraphics[width=0.45\textwidth]{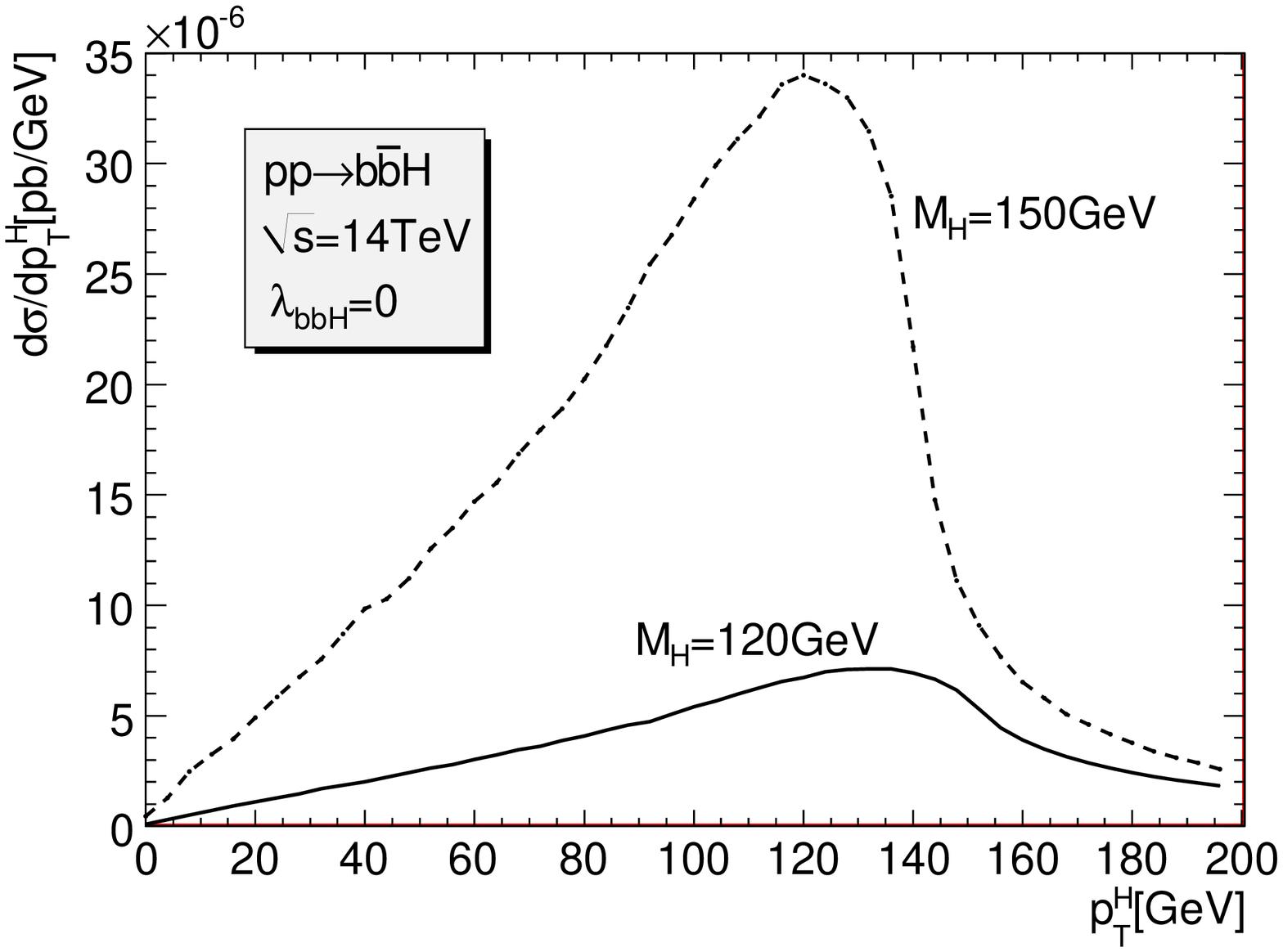}
\hspace*{0.075\textwidth}
\includegraphics[width=0.45\textwidth]{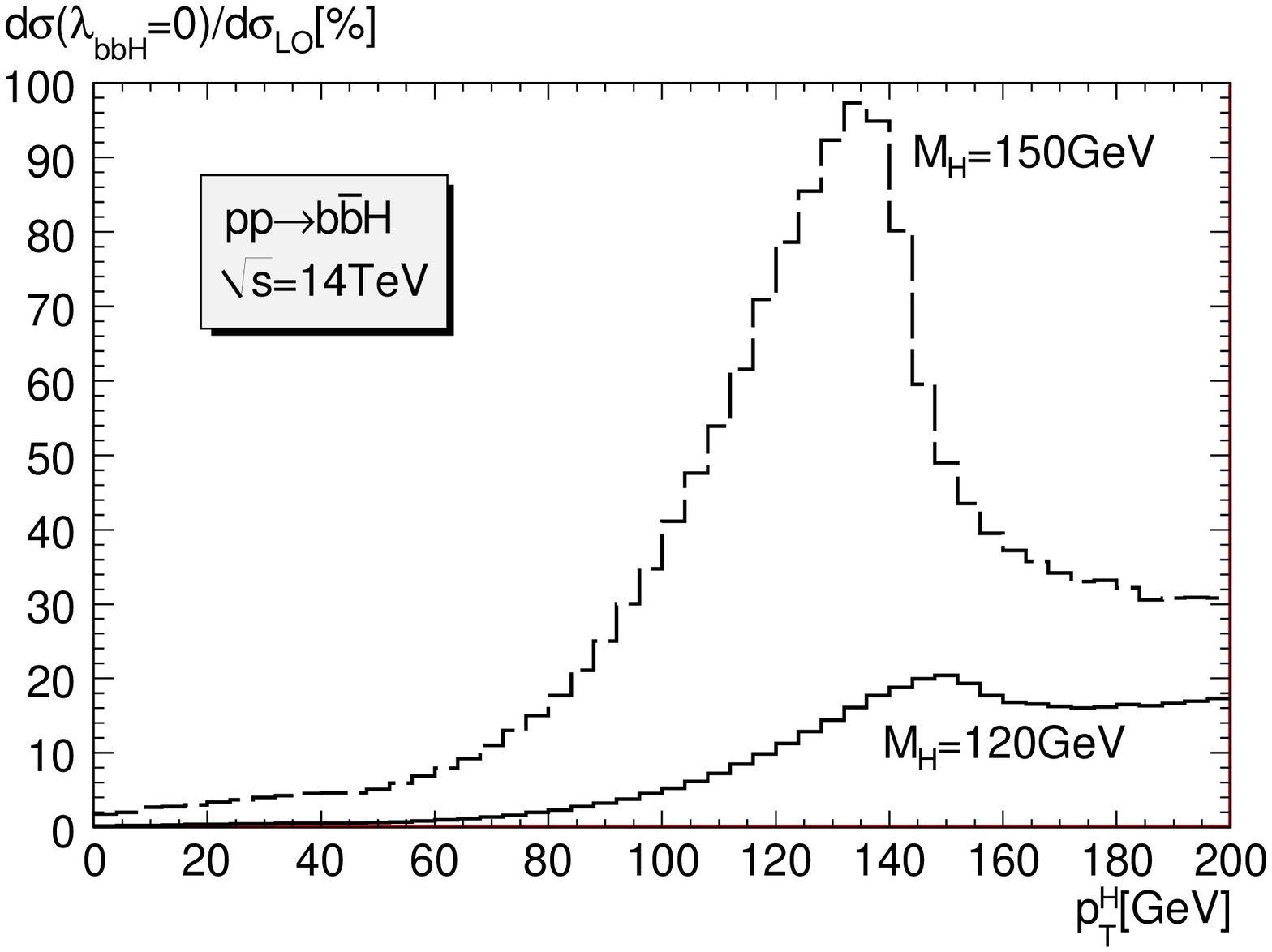}}
\mbox{\includegraphics[width=0.45\textwidth]{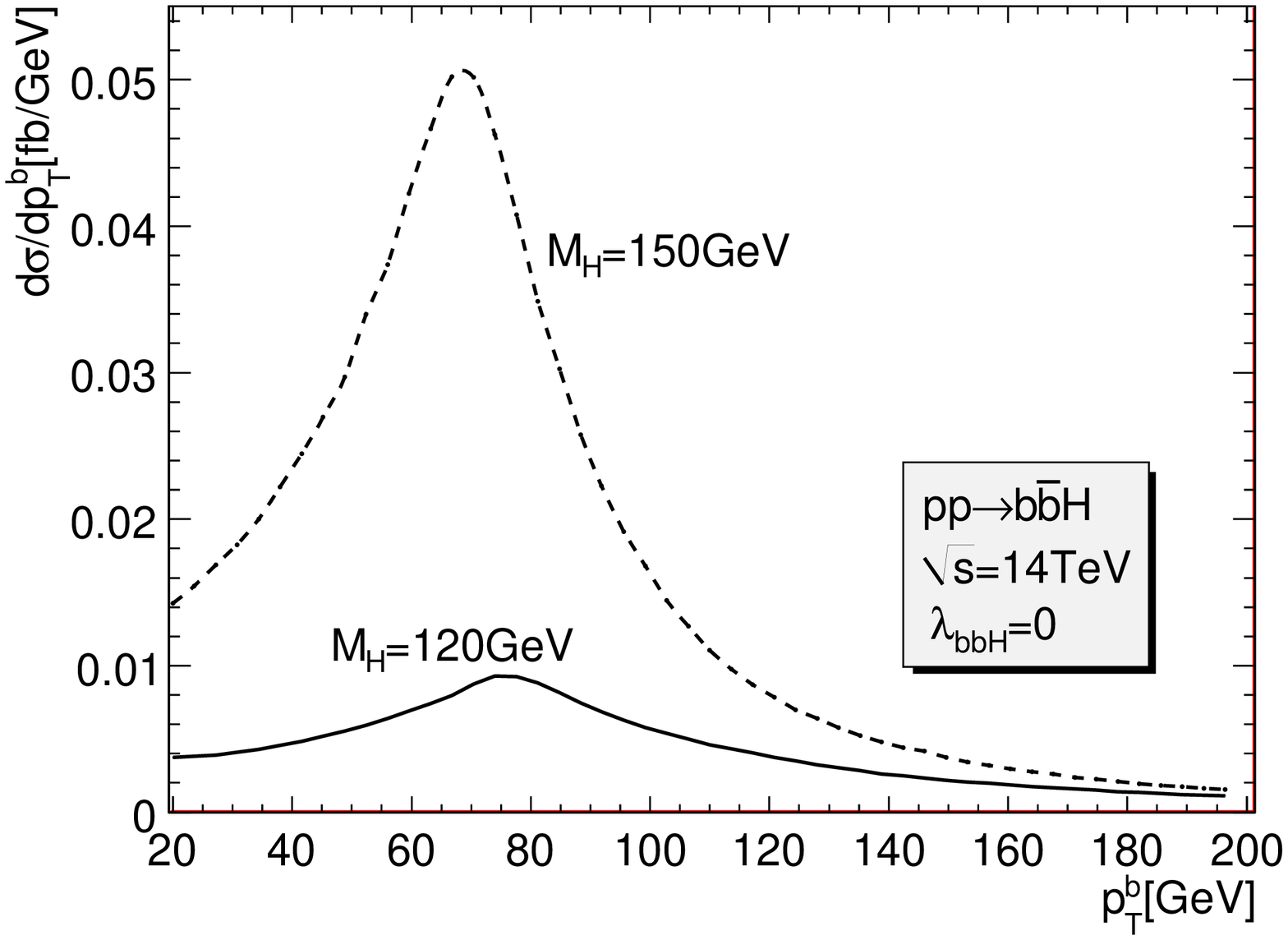}
\hspace*{0.075\textwidth}
\includegraphics[width=0.45\textwidth]{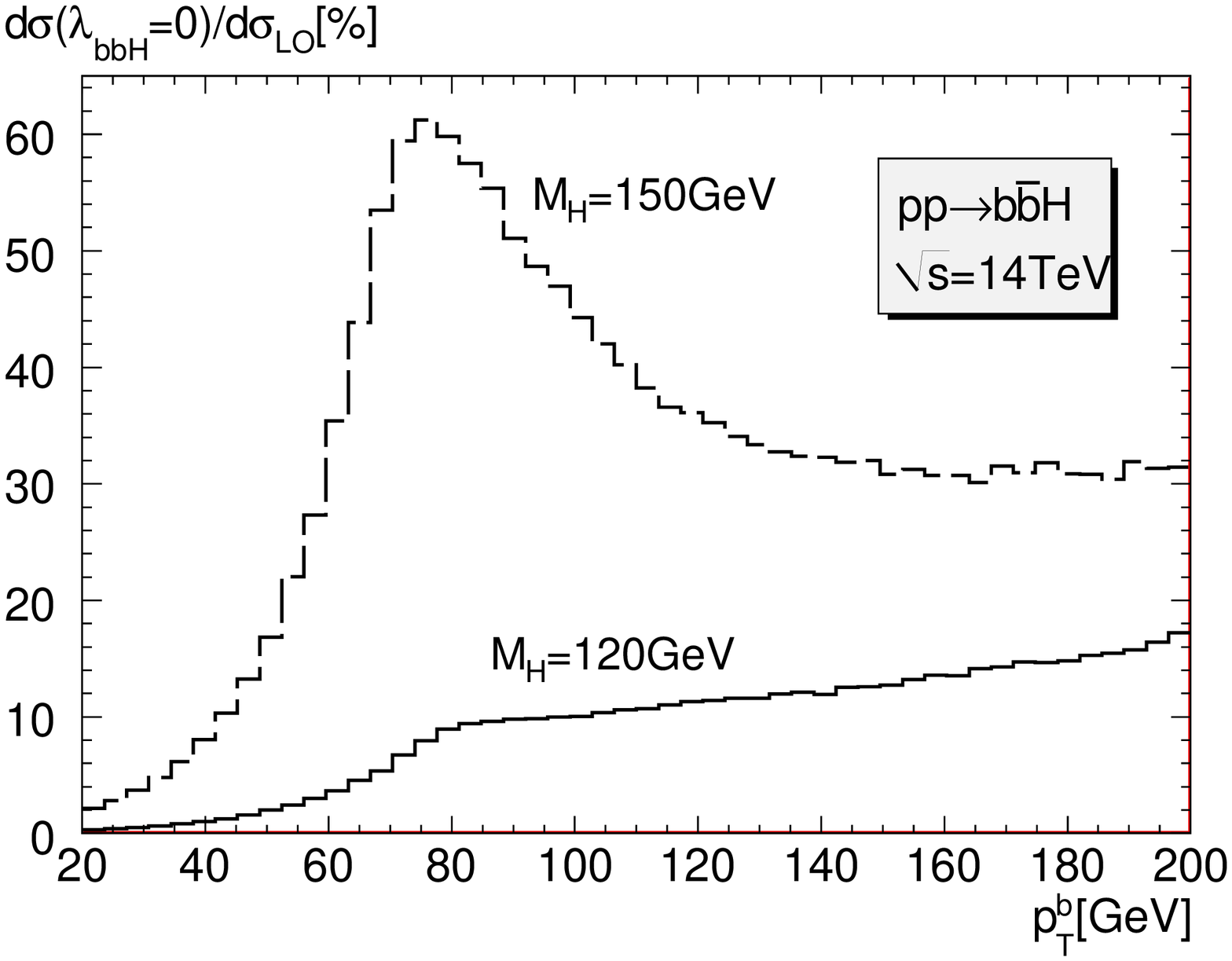}}
\caption{\label{p_LL_eta}{\em The pseudo-rapidity of the Higgs and
transverse momentum distributions of the Higgs and the bottom for
$M_H=120,150$GeV arising from the purely one-loop contribution in
the limit of vanishing LO ($\la_{bbH}=0$). Its relative percentage
contribution $d\sigma(\lambda_{bbH}=0)/d\sigma_{LO}$ is also
shown}}
\end{center}
\end{figure}
Fig. \ref{p_LL_eta}  shows the pseudo-rapidity and transverse
momentum distributions of the Higgs as well as the the $p_T$ of
the bottom for two cases $M_H=120$GeV and $M_H=150$GeV in the
limit of vanishing bottom-Higgs Yukawa coupling. These
distributions are significantly different from the ones we
observed at tree-level (and with the electroweak NLO corrections),
see Fig.~\ref{fig:dist-nlo}. The Higgs prefers being produced at
high value of transverse momentum, about $130$GeV. In the case of
a Higgs with $M_H=150$GeV this contribution can  significantly
distort the shape of the $p_T^H$ distribution for hight $p_T^H$
with a "correction" of more than $70\%$ over a rather large range.
The distribution in the $p_T$ of the bottom is also very telling.
The new contributions do not produce the bottom preferentially
with low $p_T^b$ as the case of the LO contribution.

\section{Summary}
We have calculated the EW radiative corrections triggered by the
Yukawa coupling of the top to the process $p p \to b\bar{b}H$ at
the LHC through gluon fusion in the Standard Model. This process
is triggered through Higgs radiation of the bottom quark with a
small coupling proportional to the mass of the bottom. Yet in
order to analyse this coupling, precision calculations that
include both the QCD and electroweak corrections are needed. In
this perspective, to identify the process  one needs to tag both
$b$-jets. Our calculation is therefore conducted in this
kinematical configuration. 

Inserting a top quark loop with a
Yukawa transition of the type $t \ra b \chi_W$, $\chi_W$ is the
charged Goldstone, allows now the Higgs  to be radiated from the
top or from the Goldstone boson. The latter coupling represents
the Higgs self-coupling and increases with the Higgs mass. The
former, the top Yukawa coupling, is also large. As a consequence,
the one-loop  amplitude $g g  \to b\bar{b}H$ no longer vanishes as
the Higgs coupling to $b$'s does, like what occurs  at leading
order. 

\begin{figure}[htb]
\begin{center}
\includegraphics[width=8cm]{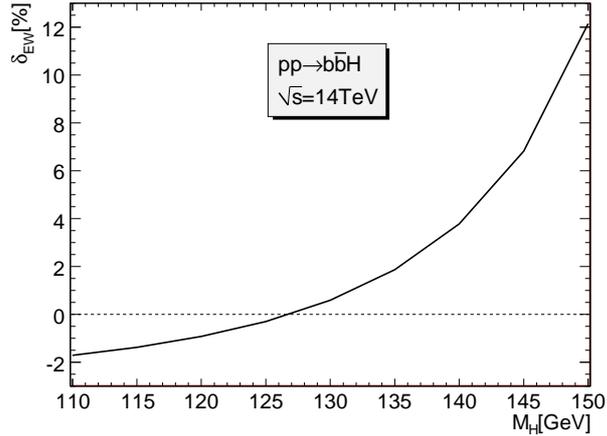}
\caption{\label{p_LandLL_mH}{\em
$\delta_{EW}=\delta_{NLO}+\fr{\sigma(\la_{bbH}=0)}{\sigma_0}$ as a
function of $M_H$.}}
\end{center}
\end{figure}

We find that in the limit of vanishing $\la_{bbH}$, the
one-loop induced electroweak process should be taken into account
for Higgs masses larger than $140$GeV or so. Indeed, though this
contribution is quite modest for a Higgs mass of $110$GeV it
increases quite rapidly as the Higgs mass increases, reaching
about $17\%$ of the leading order value, calculated with
$m_b=4.62$GeV, for $M_H=150$GeV. For these new corrections to
interfere with the leading order requires helicity flip. Therefore
at next-to-leading order in the Yukawa electroweak corrections,
all corrections involve either a bottom mass insertion or a bottom
Yukawa coupling. At the end the total Yukawa electroweak NLO
contribution brings in a correction which is within the range
$-4\%$ to $-5\%$ for Higgs masses in the range $110{\textrm GeV} < M_H
< 150$GeV. They are therefore negligible compared to the NLO QCD
correction and even the remaining QCD scale uncertainty. This
modest effect translates also as an uniform rescaling of the
distributions in the most interesting kinematical variables we
have looked at (pseudo-rapidities and $p_T$ of both $b$-quarks and
the Higgs). This is not the case of the one-loop induced
contributions which survive in the limit of $m_b \ra 0$ (and
$\la_{bbH} \ra 0$). Here the distributions for the Higgs masses
where the corrections for the total cross section is large are
drastically different from the LO distributions. A summary for the
corrections including the NLO with $\la_{bbH} \neq 0 $ and the
part of the NNLO counted as loop induced in the limit $\la_{bbH}
\ra 0$ is shown in Fig.~\ref{p_LandLL_mH}.

The analysis we have performed in this chapter does not cover Higgs
masses over $150$GeV and rests within the range of Higgs masses
preferred by indirect precision measurements. In fact as the
threshold for $ H \ra WW$ opens up, important phenomena take
place. Foremost Landau singularities, or pinch singularities in
some loop integrals, develop. This important issue will be addressed in the next two chapters.

There is another contribution which does not vanish for
vanishing $\lambda_{bbH}$ and which contributes to $gg \ra b \bar
b H$ through a closed top quark loop. This contribution represents
$ g g \ra H g^* \ra H b \bar b$. We have not included this
contribution in the present work as we do not consider it to be a
{\em genuine} $b \bar b H$ final state. This correction can be counted
as belonging to the {\em inclusive} $gg \ra H$ process. The same line
of reasoning has been argued in \cite{LH05-Higgs}. Nonetheless
from the experimental point of view it would be interesting to
include all these effects together with the NLO QCD corrections
and the electroweak corrections that we have studied here.

\chapter{Landau singularities}
\label{section_landau_introduction}
Let us start this chapter by quoting a paragraph in p.$30$ of the important book {\em The Analytic S-matrix} 
written by Eden, Landshoff, Olive and Polkinghorne ($1966$) \cite{book_eden}:
\begin{quote} 
For an individual Feynman integral, the singularities, and hence the analytic structure of the integral, arise 
from singularities in the integrand. Those that arise from ultraviolet divergences are removed by renormalisation and will not concern us. The singularities that do concern us come from zeros of the denominator factors. These denominator factors are the same for scalar particles as they are for particles with spin, namely of the general form
\bea
p^2-m_r^2+i\eps
\eea
for mass $m_r$ and four-momentum $p$.
\end{quote}
The issue of loop-integral singularities is also discussed in other text books \cite{Itzykson:1980rh,Sterman:1994ce,
Bjorken:1979dk}, see also the recent lecture notes of Denner for a practical review \cite{denner_landau}.  

In this chapter we will restrict ourselves to {\em one-loop} Feynman integrals. To set the mathematical background, we first consider the case of 
some general complex integrals. 
\section{Singularities of complex integrals}
We would like to consider two simple examples taken from \cite{book_eden}. 
\begin{description}
\item[(i)] With $w$ complex
\bea
f(w)=\int_a^b\fr{dz}{z-w}=\ln(b-w)-\ln(a-w). 
\label{landau_eg1}
\eea
\item[(ii)] With $w$, $a$ complex and real $x$
\bea
f(w)&=&\int_{0}^1\fr{dx}{(x-w)(x-a)}\crn
&=&\fr{[\ln(1-w)-\ln(1-a)]-[\ln(-w)-\ln(-a)]}{w-a}.
\label{landau_eg2}
\eea
\end{description}
In the first example we see that the integrand $1/(z-w)$ is singular at the end-points $a$, $b$ if $w=a,b$, corresponding 
to the singularity of the logarithm at these points. We see clearly that $f(w)$ is always singular if 
$w=a,b$ whatever the integration contour is. $w=a,b$ are called end-point singularities. We notice also that if $w$ lies somewhere on the 
integration contour but not at the end points then we can always use Cauchy's integral theorem to deform the contour to avoid the pole. 

The second example is more interesting. Again we see that $w=0,1$ are end-point singularities. One notices also that there may be a problem when $w=a$. In order to see the point, we consider the case where $a=a-i\eps$ with $\eps$ is infinitesimal positive and $w=w-i\rho$ with $\rho$ is also infinitesimal but can be positive or negative. $a$ and $w$ now should be considered as real numbers. We distinguish the three following cases:
\begin{enumerate}
\item For $0<a<1$ and $w\to a$ we have $\ln(-a+i\eps)=\ln(a)+i\pi$, $\ln(-w+i\rho)=\ln(w)+i\pi \sign(\rho)$ and 
\bea
f(w\to a)=-\fr{1}{1-a}-\fr{1}{a}-i\pi[\sign(\rho)-1]\lim_{w\to a}\fr{1}{w-a}.
\label{landau_eg2_c1}
\eea
\item For $a<0$ then all logarithmic arguments in Eq.~(\ref{landau_eg2}) are strictly positive. We then can forget about $\eps$ and 
$\rho$ to get
\bea
f(w\to a)=-\fr{1}{1-a}-\fr{1}{a}.
\eea
\item For $a>1$ then all logarithmic arguments in Eq.~(\ref{landau_eg2}) are negative. We have
\bea
f(w\to a)&=&-\fr{1}{1-a}-\fr{1}{a}+\left\{i\pi[\sign(\rho)-1]-i\pi[\sign(\rho)-1]\right\}\lim_{w\to a}\fr{1}{w-a}\crn
&=&-\fr{1}{1-a}-\fr{1}{a}.
\eea
\end{enumerate}
\begin{figure}[hptb!]
\begin{center}
\includegraphics[width=0.8\textwidth]{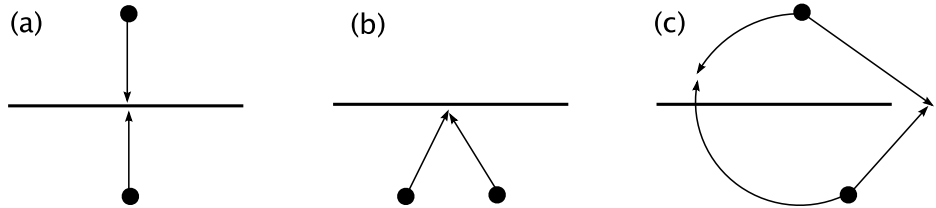}
\caption{\label{fig_landau_eg2}{\em
(a) Two singularities in the complex plane coming together and pinching the contour of integration, $[0,1]$ along the real axis. 
(b) and (c) Examples of coincidence of two singularities that do not pinch the integration contour.}}
\end{center}
\end{figure}
We remark immediately that $w=a$ is a singular point if and only if two conditions are satisfied: 
\bea 0<a<1\hs \text{and}\hs \sign(\rho)=-1,
\label{landau_eg2_cond}\eea 
which simply mean that the real integration contour $[0,1]$ is trapped when $w\to a$ (see Fig.~\ref{fig_landau_eg2}a). This is a pinch singularity. We learn also from the first case that if $\sign(\rho)=1$, \ie both $w$ and $a$ approach the real axis from the same side, then the contour is not trapped and there is no singularity (see Fig.~\ref{fig_landau_eg2}b). In the second and third cases there is no singularity because $w$ and $a$ come together nowhere near the contour (see Fig.~\ref{fig_landau_eg2}c). In this simple example, it is easy to find out the situation where the integration contour is trapped. However, in a general case with more than one integration variables finding whether the contour is actually trapped is very difficult. The ``$\eps$-prescription'' and multivalued function play crucial roles in the second example. These two elements will appear again when one considers loop integrals \footnote{In fact, we will see later in this thesis that one way to deal with pinch singularities of loop integrals, named as Landau singularities, is to kill the $\eps$-prescription by introducing the widths of internal unstable particles. The width moves the singularities away from the real axis, so they do not occur in the physical region (the real axis).}.    

Those two examples will guide us all the way. For now they lead us to two important statements for simple complex integrals. For this we first define the context \cite{book_eden}. Let g(w,z) be an analytic function of two complex variables and let $C$ be some finite contour in the complex $z$-plane. Define a function $f(w)$ by
\bea
f(w)=\int_C g(w,z)dz.
\eea
It is supposed that the singularities of the integrand $g$ are known and that their locations in the $z$-plane are
\bea
z=z_r(w), \hs r=1,2,\ldots .
\eea
$f(w)$ is analytic as long as $C$ does not intersect with any $z_r(w)$ or can be deformed accordingly. From the above examples, we see that $f(w)$ can be singular at $w_1$ for one of two reasons:
\begin{description}
\item[(i)] End-point singularities: One of the singularities $z_r$ reaches one of the end-points of the contour $C$ when 
$w\to w_1$. No deformation of $C$ can avoid them and  $w_1$ is a singularity of $f(w)$.
\item[(ii)] pinch singularities: If two (or more) singularities approach the contour from opposite sides and coincide, the contour $C$ will be trapped between them and no deformation can avoid them. $w_1$ is a singularity of $f(w)$.         
\end{description}

In order to deal with loop integrals, we have to generalise the above consideration to the case of multiple integrals. Consider
\bea
f(w)=\int_H\prod dz_ig(w,z_i),
\eea
where $H$ is a hypercontour in the multi-dimensional complex $z_i$-space. The singularities of the integrand $g(w,z_i)$ are given by various equations
\bea
S_r(w,z_i)=0, \hs r=1,2,\ldots\hs .
\eea
The boundary of $H$ is specified by 
\bea
\tilde{S}_s(w,z_i)=0, \hs s=1,2,\ldots\hs .
\eea 
Singularities occur when a surface of singularity meets a boundary surface or when the hypercontour $H$ is pinched between two or more surfaces of singularities. More precisely, it may be shown that a necessary condition of singularity is that there exists a set of complex parameters $\alpha_r$, $\tilde{\alpha}_s$ not all equal to zero such that \cite{book_eden, Itzykson:1980rh}
\bea
\alpha_rS_r=0,\hs \text{for each $r$},\nn
\eea
(so that either $\alpha_r$ or $S_r=0$),
\bea
\tilde{\alpha}_s\tilde{S}_s=0,\hs \text{for each $s$},\nn
\eea
and, for each integration variable $z_i$,
\bea
\fr{\partial}{\partial z_i}\left[\sum_{r}\alpha_rS_r+\sum_{s}\tilde{\alpha}_s\tilde{S}_s\right]=0.
\label{equation_pinch}
\eea 
The last condition expresses that the hypersurfaces are tangent. This is only a necessary condition and does not guarantee that the hypercontour $H$ is pinched. This is easy to understand if we look back at the example two. If the hypersurfaces come to be tangent from one side of the hypercontour or they are tangent at nowhere on the hypercontour then they are harmless. 

To find necessary and sufficient conditions for a pinch singularity of multiple integrals is very difficult and requires homology theory. This is obviously beyond the scope of this thesis and we refer to \cite{0173.09301, 0208.27905} for more study.  
\section{Landau equations for one-loop integrals}
\begin{figure}[htbp!]
\begin{center}
\includegraphics[width=0.4\textwidth]{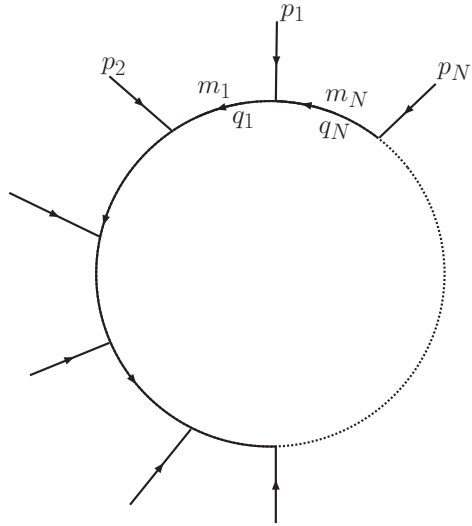}
\caption{\label{diagram_Npt}{\em One-loop Feynman diagram with $N$ external particles.}}
\end{center}
\end{figure}
We would like to clarify the important terminology used in this and the next chapters. 
The terminology "phase space" means the 
kinematical allowed region obtained by using the energy-momentum conservation on external momenta. 
The terminology "physical region" means the physical 
integration contour of the one-loop integral defined in Eq.~(\ref{landau_rep3}). It means $\left\{x_i=x_i^*,x_i\ge 0,q_i=q_i^*\right\}$ where $x_i$ are Feynman parameters and $q_i$ are loop momenta. The real condition on $x_i$ and $q_i$ accords with the $i\eps$ prescription.   

We are now in the position to apply the previous analysis for one-loop Feynman integrals. 
Consider the one-loop process $F_1(p_1)+F_2(p_2)+\ldots
F_N(p_N)\to 0,$ where $F_i$ stands for either a scalar, fermion or
vector field with momentum $p_i$ as in the figure opposite. The
internal momentum for each propagator is  $q_i$ with $i=1,\ldots
N$. The scalar one-loop integral reads
\begin{itemize}
\item In momentum space
\bea
T^{N}_{0}&\equiv & \int\fr{d^Dq}{(2\pi)^Di}\fr{1}{\prod_{i=1}^ND_i},\crn
D_i&=&q_i^2-m_i^2+i\epsilon, \hs q_i=q+r_i, \hs r_i=\sum_{j=1}^{i}p_j,
\label{landau_rep1}
\eea
where $m_i$ are the masses of internal particles, all the external particle are 
outgoing, $p_i=q_{i+1}-q_i$ are the external momenta.
\item Introducing Feynman parameters and integrating over $q$ one gets
\bea
T^{N}_{0}=\fr{(-1)^N\Gamma(N-D/2)}{(4\pi)^{D/2}}\int_0^\infty dx_1\cdots dx_N\fr{\delta(\sum_{i=1}^{N}x_i-1)}{\Delta^{N-D/2}}
\label{landau_rep2}
\eea
with
\bea
\Delta=\fr{1}{2}\sum_{i,j=1}^{N}x_ix_jQ_{ij}-i\eps , \hs Q_{ij}&=&m_i^2+m_j^2-(r_i-r_j)^2.
\label{delta_Q} 
\eea   
\end{itemize}
In representation (\ref{landau_rep1}) the {\em physical} hypercontour is $[-\infty,\infty]^D$ along the {\em real} axes, by definition. Each factor in the denominator is protected from zero by $i\eps$ with $\eps>0$. Thus the integral can be singular only when $\eps\to 0$. However, even in the case when $\eps\to 0$ and the singularities of the integrand reach the real hypercontour one can deform the real hypercontour to be complex to avoid the singularities. Indeed, the generalised definition of Feynman integral is Eq.~(\ref{landau_rep1}) and its analytical continuation \footnote{Based on this definition, a method called contour deformation technique to calculate Feynman integrals numerically has been realised in practice \cite{Nagy:2006xy,Kurihara:2005ja}. This method works as long as the contour can be deformed to avoid singularities, \ie when the singularities approach the contour from the same side like in Fig.~\ref{fig_landau_eg2}b.}. As already pointed out in the second example of the previous subsection, a singularity arises when the real hypercontour is pinched and therefore cannot be deformed. The pinch singularity of a Feynman integral is called Landau singularity. Necessary condition for this singularity is that equations (\ref{equation_pinch}) have solution with {\em real} $q$. 
The boundaries are at infinity so we can set all $\tilde{\alpha}_s=0$. 
With $S_i=q_i^2-m_i^2$ equations (\ref{equation_pinch}) become
\bea
\left\{
\begin{array}{ll}
\forall i \,\,\, \alpha_i(q_i^2-m_i^2)=0,\\
\sum_{i=1}^N\alpha_iq_i=0.\label{eq_landau_rep1}
\end{array}
\right.
\eea
These are the Landau equations corresponding to representation (\ref{landau_rep1}).       
If equations (\ref{eq_landau_rep1}) have a solution with some complex $\alpha_i\neq 0$ and {\em real} $q_i$ then one can say that integral (\ref{landau_rep1}) may have a Landau singularity. 

The drawback of this representation is that there are not any constraints on the range of $\alpha_i$. By introducing Feynman parameters and requiring them to be real (this is the purpose of $i\eps$ prescription), one imposes more constraint on the physical region. It will be clear later on that for Landau singularities to be in that physical region, $\alpha_i$ must be real and positive.

For representation (\ref{landau_rep2}), the {\em physical} hypercontour is $[0,\infty]^N$ along the {\em real} axes by definition \footnote{One might think that the hypercontour should be $[0,1]^N$. However, it can be expanded to infinity because of the Dirac delta function in the integrand.}. Thus the boundaries are $\tilde{S}_i=x_i=0$ for all $i$. The hypersurface of singularities of the integrand is $S=\Delta=0$. One might think that the hypercontour cannot be pinched now because there is only one hypersurface of singularities and the singularities are just end-point singularities if they occur. Indeed, a hypersurface can be very complicated and contains several sub-hypersurfaces which can trap the hypercontour. This is very difficult to imagine but the condition for that to happen is very simple, just like the extreme condition \cite{book_eden}: 
\bea
S=0=\fr{\partial S}{\partial x_i}.
\label{pinch_x1}
\eea
Other way to get the condition is to use equations (\ref{equation_pinch}). We can set $\alpha=1$ since there is only one singular hypersurface. One gets
\bea
\left\{
\begin{array}{ll}
\Delta=0,\\
\forall i \hs x_i\fr{\partial \Delta}{\partial x_i}=0.
\label{eq_landau_rep2}
\end{array}
\right.
\eea
The first condition comes from the second equation of (\ref{equation_pinch}). The second condition comes from 
the first and third equations of (\ref{equation_pinch}). If one picks up the solution $x_i=0$ from the second condition 
of (\ref{eq_landau_rep2}) then that is for end-point singularities. Otherwise, one has the condition for pinch singularities, which is the same as Eq.~(\ref{pinch_x1}). So everything is consistent. 
Since $\Delta$ is a homogeneous function of $x_i$, the first equation in (\ref{eq_landau_rep2}) is automatically 
satisfied when the second is. 
Necessary condition for integral (\ref{landau_rep2}) to have Landau singularities is that equations (\ref{eq_landau_rep2}) have solution with some $x_i>0$, $x_i$ are {\em real}. The drawback of this representation is that equations (\ref{eq_landau_rep2}) do not tell us anything about $q_i$.

There exists a representation which contains all the advantages of Eqs. (\ref{eq_landau_rep1}) and (\ref{eq_landau_rep2}). 
That is the mixed representation of Feynman integrals in the space of real momentum and real Feynman parameters \footnote{Landau has used this representation to devise the condition for singularities \cite{landau}.}: 
\bea
T^{N}_{0}=\Gamma(N)\int_0^\infty dx_1\cdots dx_N\delta(\sum_{i=1}^{N}x_i-1)\int\fr{d^Dq}{(2\pi)^Di}
\fr{1}{[\sum_{i=1}^Nx_i(q_i^2-m_i^2+i\eps)]^N}.
\label{landau_rep3}
\eea
The {\em physical} hypercontours are {\em real} $[-\infty,\infty]^D$ for $q$ and {\em real} $[0,\infty]^N$ for $x_i$. The boundaries are $\tilde{S}_i=x_i=0$ for all $i$. The hypersurface of singularities of the integrand is $S=\sum_{i=1}^Nx_i(q_i^2-m_i^2)=0$. 
From Eq.~(\ref{equation_pinch}) one gets
\bea
\left\{
\begin{array}{ll}
\forall i \,\,\, x_i(q_i^2-m_i^2)=0,\\
\sum_{i=1}^Nx_iq_i=0,\label{eq_landau_rep3}
\end{array}
\right.
\eea
together with 
\bea
\left\{
\begin{array}{ll}
x_i\ge 0,\\
q_i=q_i^*.\label{cond_landau_rep3}
\end{array}
\right.
\eea
The first condition of (\ref{eq_landau_rep3}) comes from the second equation and third equation (with $z_i=x_i$) of (\ref{equation_pinch}). 
The second condition of (\ref{eq_landau_rep3}) comes from the third equation of (\ref{equation_pinch}) with $z_i=q$. One notices immediately that 
Eq.~(\ref{eq_landau_rep3}) can be obtained from Eq.~(\ref{eq_landau_rep1}) by replacing complex $\alpha_i$ with real $x_i$. 
Conditions (\ref{cond_landau_rep3}) come from the definition of the physical hypercontours. 
The Landau singularities {\em may} occur in the physical region if equations (\ref{eq_landau_rep3}) and (\ref{cond_landau_rep3}) are satisfied. These are necessary conditions since we cannot be sure that the real hypercontours are pinched when $\eps\to 0$ (remember the second example of the previous subsection).

If all $x_i$ are strictly positive then we have a leading Landau singularity (LLS). Otherwise one has conditions for sub-leading Landau singularities (sub-LLS).      
 
\section{Necessary and sufficient conditions for Landau singularities}
\label{subsection_landau_condition}
It may be shown that if Landau matrix $Q_{ij}$ (defined in Eq.~(\ref{delta_Q}), see also Eq.~(\ref{def_Qij})) has {\em only one} zero eigenvalue then the 
necessary and sufficient conditions for the appearance of a singularity in the physical region are equations (\ref{eq_landau_rep3}) and (\ref{cond_landau_rep3}). 

This important conclusion has been pointed out in the paper of Coleman and Norton \cite{Coleman:1965xm} \footnote{See aslo Itzykson and Zuber \cite{Itzykson:1980rh} in p.$306$.}. The proof is very simple and will be given in the next subsection (after equation (\ref{TN0_zi_pinch})). 
It is based on the underlying fact that the Landau matrix $Q_{ij}$ is real and symmetric hence can be diagonalized by a real orthogonal co-ordinate transformation. It means that for unstable particles with complex masses the argument fails and the conditions based on Landau equations are no longer sufficient. 

We seek conditions for Eq.~(\ref{eq_landau_rep3}) to have a solution 
$x_i=0$ for $i=M+1,\ldots,N$ with $1\le M\le N$ and $x_i>0$ for every $i\in\{1,\ldots ,M\}$. 
The Eq.~(\ref{eq_landau_rep3}) becomes
\bea
\left\{
\begin{array}{lll}
x_i=0\,\,\, \text{for}\,\, i=M+1,\ldots,N,\\
q_i^2=m_i^2 \,\,\, \text{for}\,\, i=1,\ldots,M,\\
\sum_{i=1}^Mx_iq_i=0.\label{landau_eqM} 
\end{array}
\right.
\eea 
For $M=N$ one has a leading singularity, otherwise if $M<N$ this
is a sub-leading singularity.
Multiplying the third equation in (\ref{landau_eqM}) by $q_j$ leads to a system of $M$ equations
\bea
\left\{
\begin{array}{ll}
Q_{11}x_1 + Q_{12}x_2 + \cdots Q_{1M}x_M& =0,\\
Q_{21}x_1 + Q_{22}x_2 + \cdots Q_{2M}x_M& =0,\\
\vdots \\
Q_{M1}x_1 + Q_{M2}x_2 + \cdots Q_{MM}x_M& =0,\label{landau_Meqs}
\end{array}
\right.
\eea
where the $Q$ matrix is defined as  
\bea
Q_{ij}=2q_i.q_j=m_i^2+m_j^2-(q_i-q_j)^2=m_i^2+m_j^2-(r_i-r_j)^2,\,\,\, i,j=1,\ldots,M,\label{def_Qij}
\eea
which agrees with Eq.~(\ref{delta_Q}). The necessary and sufficient conditions for the appearance of a singularity in the physical region now become
\bea
\left\{
\begin{array}{ll}
\det(Q)=0\\
x_i>0\\
q_i^2=m_i^2\\
q_i=q_i^* 
\end{array}
\right.
\label{landau_cond0}
\eea
for $i=1,\ldots,M$.  

The condition $\det(Q)=0$ defines a singular surface or a Landau curve.
  
If some internal (external)
particles are massless like in the case of six photon scattering,
then some $Q_{ij}$ are zero, the above conditions can be easily
checked. However, if the internal particles are massive then it is
difficult to check the second condition explicitly, especially if $M$
is large. In this case, we can rewrite the second condition as
following
\bea
x_j=\det(\hat{Q}_{jM})/\det(\hat{Q}_{MM})>0,\,\,\,
j=1,\ldots,M-1,\label{landau_cond1}
\eea
where $\hat{Q}_{ij}$ is obtained from $Q$ by discarding row $i$
and column $j$ from $Q$ and $\det(\hat{Q}_{jM})=d[\det(Q)]/(2dQ_{jM})$, 
$\det(\hat{Q}_{MM})=d[\det(Q)]/dQ_{MM}$. 
The proof for Eq.~(\ref{landau_cond1}) is the following.
It is obvious that when the condition $\det(Q)=0$ is satisfied 
one can set $x_M=1$ and discard the last equation in (\ref{landau_Meqs}). After moving the $M$-column from 
the left hand side to the right hand side, one obtains a system of $M-1$ equations with $M-1$ variables. The solution of this is 
clearly equation (\ref{landau_cond1}). 
If $\det(\hat{Q}_{MM})=0$
then condition (\ref{landau_cond1}) becomes
$\det(\hat{Q}_{jM})=0$ with $j=1,\ldots,M-1$.

\begin{figure}[htbp]
\begin{center}
\includegraphics[width=0.8\textwidth]{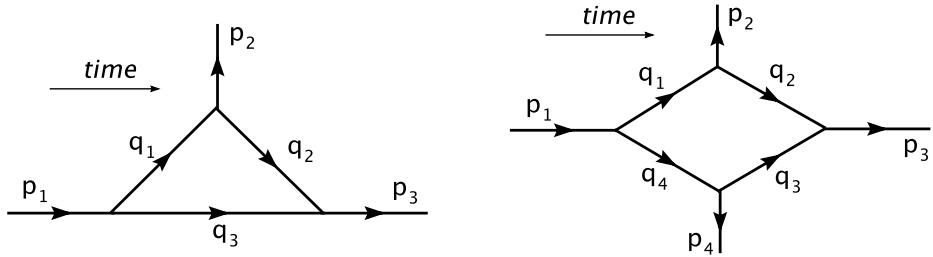}
\caption{\label{landau_diagrams}{\em Typical triangle and box Feynman diagrams.}}
\end{center}
\end{figure}

Conditions (\ref{eq_landau_rep3}) and (\ref{cond_landau_rep3}) admit a beautiful physical interpretation. This was discovered by Coleman and Norton \cite{Coleman:1965xm}. Consider the case where all $x_i$ are strictly positive. All the internal loop particles are therefore on-shell and have real momenta. An internal particle \footnote{One can regard different particles running in a loop as different states of one particle. Each vertex is associated with an external "force''.} has a real four-momentum: $q_i=m_iu_i$ (for each $i$) with $u_i$ is a four-velocity. Each vertex can be regarded as a real space-time point. The space-time separation between two vertices reads
\bea
dX_i=d\tau_i u_i=\fr{d\tau_i}{m_i}q_i,\hs \text{for each}\hs i,
\eea
where $d\tau_i$ is the proper time. Following a closed loop, one has $\sum_i dX_i=0$. Comparing this to the second equation of (\ref{eq_landau_rep3}) we get the correspondence: $d\tau_i=m_ix_i>0$ for each $i$. It means that the loop particle is moving forward in time. $dX_i$ can be positive or negative depending on the sign of $q_i$. 
If one chooses a reference frame where vertices are ordered in time, {\it i.e.} $dX_i^0>0$, then $q_i^0>0$ in that frame. This information can be very useful in practice. Let us illustrate this point. Consider two important Feynman diagrams in 
Fig.~\ref{landau_diagrams}.
We choose a reference frame where the arrows of the internal lines follow the time direction. 
We look at vertex $2$ (connected to $p_2$) of the triangle diagram and 
choose a co-ordinate system such that $q_1=(m_1,0,0,0)$, $q_2=(e_2,q_{2x},0,0)$. With $p_i^2=M_i^2$, from the energy-momentum conservation 
$p_2=q_1-q_2$ we get
\bea
e_2&=&\fr{m_1^2+m_2^2-M_2^2}{2m_1},\crn 
q_{2x}^2&=&\fr{\lambda(M_2^2,m_1^2,m_2^2)}{4m_1^2}=\fr{[M_2^2-(m_1-m_2)^2][M_2^2-(m_1+m_2)^2]}{2m_1}.
\eea 
From the conditions $q_{2}=q_{2}^*$ and $e_2>0$ we get 
\bea
M_2^2\le (m_1-m_2)^2.
\eea
Similarly, for other vertices of the triangle diagram we have
\bea
M_1^2\ge (m_1+m_4)^2, \hs M_3^2\ge (m_2+m_3)^2. 
\eea
By using the same trick one can easily see that necessary conditions to have a leading Landau singularity in the box diagram are
\bea
\left\{
\begin{array}{ll}
M_1^2\ge (m_1+m_4)^2, \hs M_3^2\ge (m_2+m_3)^2\\
M_2^2\le (m_1-m_2)^2, \hs M_4^2\le (m_3-m_4)^2.
\end{array}
\right.
\eea
Similarly, with $t=(p_2+p_3)^2$ and $u=(p_3+p_4)^2$ and using the energy-momentum conservation, we get the constraint 
\bea
t\ge (m_1+m_3)^2\hs \text{and}\hs u\ge (m_2+m_4)^2.
\eea
Thus a necessary condition for any diagram to have a leading Landau singularity is that it has at least two cuts which can produce physical on-shell particles \footnote{In the case of two-point function, the two cuts coincide due to energy-momentum conservation.}. Other external particle which does not correspond to those cuts must have mass smaller than the difference of the two internal masses associated with it.
\section{Nature of Landau singularities}
\label{subsection_landau_nature}
\subsection{Nature of leading Landau singularities}
\label{nature_LLS}
Our purpose is to extract the LLS by using Feynman parameter representation (\ref{landau_rep2}). The matrix $Q$ which 
appears in the denominator is real and symmetric hence can be diagonalized by a real orthogonal co-ordinate transformation. 
In general, $Q$ has $N$ real eigenvalues called 
$\la_1$, \ldots, $\la_N$. 
The characteristic equation of $Q$ is given by
\bea
f(\la)&=&\la^N+(-1)a_{N-1}\la^{N-1}+(-1)^2a_{N-2}\la^{N-2}-\ldots (-1)^{N-1}a_1\la+(-1)^{N}a_0\crn
&=&(\la-\la_1)(\la-\la_2)\ldots (\la-\la_n)=0.
\eea
For the case $N=4$ we have 
\bea
a_0&=&\la_1\la_2\la_3\la_4=\det(Q_4),\crn
a_1&=&\la_1\la_2\la_3+\la_1\la_2\la_4+\la_1\la_3\la_4+\la_2\la_3\la_4,\crn
a_2&=&\la_1\la_2+\la_1\la_3+\la_1\la_4+\la_2\la_3+\la_2\la_4+\la_3\la_4=\fr{1}{2}[\Trace(Q_{4})^2-\Trace(Q_{4}^2)],\crn
a_3&=&\la_1+\la_2+\la_3+\la_4=\Trace(Q_4),
\eea
Consider the case where $Q$ has only one very small eigenvalue $\la_N\ll 1$. Then, to 
leading order 
\bea
\la_N=\fr{a_0}{a_1},\,\,\, a_{1}=\la_1\la_2\ldots \la_{N-1}\neq 0.
\eea
Let $V=\{x_1^0,x_2^0,\ldots,x_N^0\}$ be the eigenvector corresponding to $\la_N$. $V$ is normalised to 
\bea 
\sum_{i=1}^Nx_i^0=1.
\eea
For later use, we define
\bea
\upsilon^2=V.V.
\eea
The expansion of $\Delta$ around $V$ reads
\bea
\Delta&=&\fr{1}{2}\sum_{i,j=1}^NQ_{ij}y_iy_j+\la_{N}\sum_{i=1}^Nx_i^0y_i+\fr{1}{2}\la_N\upsilon^2-i\eps,\crn
&\approx&\fr{1}{2}\sum_{i,j=1}^NQ_{ij}y_iy_j+\fr{1}{2}\la_N\upsilon^2-i\eps,
\label{Delta_DetQ}
\eea
where $y_i=x_i-x_i^0$. For the leading part of the singularity it is sufficient to neglect the 
linear terms. The $Q$-matrix can be diagonalised by rotating the $y$-vector
\bea
y_i=\sum_{j=1}^NA_{ij}z_j,
\eea
where $A$ is an orthogonal matrix whose columns are the normalised eigenvectors of $Q$. Thus we have
\bea 
\det(A)=1,\,\,\, \sum_{j=1}^NA_{Nj}=\fr{\sum_{i=1}^Nx_i^0}{\sqrt{V.V}}=\fr{1}{\upsilon},\crn
\Delta=\fr{1}{2}\sum_{i=1}^{N-1}\la_iz_i^2+\fr{1}{2}\la_N\upsilon^2-i\eps.
\eea
Note that the term $\la_Nz_N^2$ in the rhs has been neglected as this term would give a contribution of the order 
$\mathcal{O}(\la_N^2)$ to the final result. 
Equation (\ref{landau_rep2}) can now be re-written in the form 
\bea
T^{N}_{0}=\fr{(-1)^N\Gamma(N-D/2)}{\pi^{D/2}2^{3D/2-N}}\int_{-\infty}^{+\infty} dz_1\cdots dz_N\fr{\delta(\sum_{i,j=1}^{N}A_{ij}z_j)}
{(\sum_{i=1}^{N-1}\la_iz_i^2+\la_N\upsilon^2-i\eps)^{N-D/2}}.\label{eq_TN0_2a}
\eea
Although the original integration contour is some segment around the singular point $z_i=0$ with $i=1,\ldots,N$, the singular part 
will not be changed if we extend the integration contour to infinity, provided the power $(N-D/2)$ of the denominator in 
Eq.~(\ref{eq_TN0_2a}) is sufficiently large. Integrating over $z_N$ gives
\bea
T^{N}_{0}=\fr{(-1)^N\Gamma(N-D/2)\upsilon}{\pi^{D/2}2^{3D/2-N}}\int_{-\infty}^{+\infty} dz_1\cdots dz_{N-1}\fr{1}
{(\sum_{i=1}^{N-1}\la_iz_i^2+\la_N\upsilon^2-i\eps)^{N-D/2}},\;\;
\label{TN0_zi_pinch}
\eea
where the factor $\upsilon$ comes from the $\delta$-function. One sees clearly that each integration contour is pinched when $\eps\to 0$ if 
all $\la_i\neq 0$ with $i=1,\ldots,N-1$. 

Asumming that $\la_i>0$ for $i=1,\ldots,K$ and 
$\la_j<0$ for $j=K+1,\ldots,N-1$ with $0\le K\le N-1$. We then change the integration variables as follows
\bea
\left\{
\begin{array}{ll}
t_i=\sqrt{\la_i}z_i\,\,\, \text{for} \,\,\, i=1,\ldots K,\\
t_j=\sqrt{-\la_j}z_j\,\,\, \text{for} \,\,\, j=K+1,\ldots N-1. 
\end{array}
\right.
\eea
This makes sure that all $t_i$ are real. We get 
\bea
T^{N}_{0}&=&\fr{(-1)^N\Gamma(N-D/2)\upsilon}{\pi^{D/2}2^{3D/2-N}\sqrt{(-1)^{N-K-1}a_1}}\crn
&\times&\int_{-\infty}^{+\infty} dt_1\cdots dt_{K}\int_{-\infty}^{+\infty} dt_{K+1}\cdots dt_{N-1}\fr{1}
{(-\sum_{i=K+1}^{N-1}t_i^2+b^2)^{N-D/2}},
\eea
where 
\bea
b^2=\sum_{i=1}^{K}t_i^2+\la_N\upsilon^2-i\eps,\,\,\, \Rel(b^2)>0.
\eea
Changing to spherical coordinates by using formulae (\ref{int_spherical}) we get
\bea
T^{N}_{0}&=&\fr{(-1)^N\Gamma(N-D/2)\upsilon}{\pi^{D/2}2^{3D/2-N}\sqrt{(-1)^{N-K-1}a_1}}
\fr{2\pi^{(N-K-1)/2}}{\Gamma((N-K-1)/2)}\crn
&\times&\int_{-\infty}^{+\infty} dt_1\cdots dt_{K}\int_0^{\infty}dr\fr{r^{N-K-2}}
{(b^2-r^2)^{N-D/2}}.
\eea
Note that $(b^2-r^2)^{N-D/2}=e^{-i\pi(N-D/2)}(r^2-b^2)^{N-D/2}$ due to the fact that $\eps>0$.
Then by using formula (\ref{int_master_z}) we have
\bea
T^{N}_{0}&=&\fr{(-1)^Ne^{i\pi(N-K-1)/2}\upsilon}{\pi^{D/2}2^{3D/2-N}\sqrt{(-1)^{N-K-1}a_1}}
\pi^{(N-K-1)/2}\Gamma((N-D+K+1)/2)\crn
&\times&\int_{-\infty}^{+\infty} dt_1\cdots dt_{K}\fr{1}{(\sum_{i=1}^{K}t_i^2+\la_N\upsilon^2-i\eps)^{(N-D+K+1)/2}}.
\eea
Repeat the above steps to get
\bea
T^{N}_{0}&=&\fr{(-1)^Ne^{i\pi(N-K-1)/2}\upsilon}{2^{(3+N)/2}\pi\sqrt{(-1)^{N-K-1}a_1}}
\fr{(4\pi)^{(N-D+1)/2}\Gamma((N-D+1)/2)}{\left(\fr{\la_N\upsilon^2}{2}-i\eps\right)^{(N-D+1)/2}}.
\label{TN0_LLS}
\eea
This result holds provided 
\bea
a_1\neq 0 \,\,\, \text{and} \,\,\, N-D+1>0.
\eea
In the case where $N-D+1\le 0$ one can write $D=4-2\varep$ ($\varep >0$) and do the expansion in $\varep$. Apart from a divergent term 
of the form $1/\varep$ related to the artificial infinite boundary, the other terms give the nature of singularities.\\ 
A similar result for the nature of the singularity has
been derived in \cite{MR0127825} in the general case of a
multi-loop diagram including the behaviour of the sub-LLS. 
The extraction of the overall, regular, factor  which
is the $K$-dependent part in Eq.~(\ref{TN0_LLS}) (see also Eq.~(\ref{TN0_sub-LLS}) for the sub-LLS) is more
transparent in our derivation.\\  
For $N=4$, $D=4$  
\bea
(T_0^4)_{div}&=&e^{i\pi(3-K)/2}\fr{1}{4\sqrt{(-1)^{3-K}a_1}}\fr{1}{\sqrt{\la_4-i\eps}}\crn
&=&e^{i\pi(3-K)/2}\fr{1}{4\sqrt{(-1)^{3-K}a_0-i\eps}}.
\label{eq_T04h}
\eea
For $N=3$, $D=4-2\varep$, we use $\Gamma(\varep)=(1/\varep)-\gamma_E$ to get
\bea
(T^{3}_{0})_{div}=\fr{e^{i\pi(2-K)/2}\upsilon}{8\pi\sqrt{(-1)^{2-K}\la_1\la_2}}\ln(\la_3\upsilon^2-i\eps).
\label{T30_div_final}
\eea
For $N=2$, $D=4-2\varep$, we use $\Gamma(-1/2)=-2\sqrt{\pi}$ to get
\bea
(T^{2}_{0})_{div}=-\fr{\upsilon}{8\pi\sqrt{\la_1}}(\la_2\upsilon^2-i\eps)^{1/2}.
\eea 
For $N=1$, $D=4-2\varep$, we have
\bea
T^{1}_{0}&=&\fr{-\Gamma(-1+\varep)(4\pi)^\varep}{16\pi^2\la_1^\varep}\la_1\crn
&=&\fr{\la_1}{16\pi^2}\left(\fr{1}{\varep}-\gamma_E+\ln(4\pi)-\ln(\la_1)\right),
\eea
with $\la_1=m^2$.
 
Remarks: The leading Landau singularity appears when $\lambda_N\to 0$. The nature of the leading singularities for the scalar one-, two-, three-, four- functions are 
$1$, $1/2$, log, $-1/2$ respectively. One remarks that in the case $N=4,3$ the LLS is divergent, \ie becomes infinite. The LLS is finite but singular, \ie the derivative is divergent at the singular point, in the case $N=2$ and is regular in the case $N=1$. 
The scalar three-point function and its square are 
integrable at the LLS point. The scalar four-point function is also integrable at the LLS point but its square is not.

One may wonder if we can use the general result in Eq.~(\ref{TN0_LLS}) for the case $N\ge 5$. The answer is YES as long as $a_1\neq 0$. 
As will be proved in the next subsection, $a_1$ is proportional to the Gram determinant $\det(G)$ at the singular point. Since $\det(G)=0$ 
for $N\ge 6$ in four dimensional space, we conclude that Eq.~(\ref{TN0_LLS}) cannot be used for the case $N\ge 6$. However, the LLS can occur 
for $N=5$. If this happens, we will have five on-shell equations $q_i^2=m_i^2$ with $i=1,\ldots, 5$ to solve for $q^\mu$. We just need four 
equations to find $q^\mu$, the rest is a $\delta$-function to give some constraint on the internal masses and external momenta. Thus the nature 
of $5$-point function LLS is a pole \cite{book_eden}. Indeed, it is highly nontrivial to find a physical process 
which contains a $5$-point function LLS.   
\subsection{Nature of sub-LLS}
In order to understand the nature of sub-leading Landau singularities, one should integrate over $x_N$ from Eq.~(\ref{landau_rep2}). 
This gives
\bea
T^{N}_{0}=\fr{(-1)^N\Gamma(N-D/2)}{(4\pi)^{D/2}}\int_0^1dx_1\cdots dx_{N-1}\fr{\eta(1-\sum_{i=1}^{N-1}x_i)}{[\hat{\Delta}(x_1,\ldots,x_{N-1})]^{N-D/2}},\label{eq_TN0_3}
\eea 
where $\eta$ is Heaviside step function and
\bea
\hat\Delta(x_1,\ldots,x_{N-1})&\equiv&\Delta(x_1,\ldots,x_{N-1},1-x_1-\ldots-x_{N-1})\crn
&=&\fr{1}{2}\sum_{i,j=1}^{N-1}x_ix_jG_{ij}-\sum_{i=1}^{N-1}x_i\beta_i+\fr{1}{2}Q_{NN}-i\eps,\label{delta_prime}
\eea
where
\bea
G_{ij}=Q_{ij}-Q_{iN}-Q_{jN}+Q_{NN}=2r_i.r_j, \,\,\, \beta_i=Q_{NN}-Q_{iN}=m_N^2-m_i^2+r_i^2.\label{eq_Q_G}
\label{defs_G_beta}
\eea
Thus $\det(G)$ is just the Gram determinant. From Eq.~(\ref{eq_Q_G}) we get 
\bea
\det(Q)=Q_{NN}\det(G)-\sum_{i,j=1}^{N-1}\beta_i\beta_j\hat{G}_{ij}.\label{det_Q_G}
\label{detQ_detG}
\eea
The Landau equations for representation (\ref{eq_TN0_3}) are\footnote{It is important to notice that when one performs the $x_N$-integration the boundaries become: $x_i=0$ for $i=1,\ldots N-1$ and $1-\sum_{i=1}^{N-1}x_i=0$. 
In the mean time $\hat\Delta$ becomes inhomogeneous, so that the first equation of (\ref{landau_eq_rep4}) is not automatically satisfied when the others are.} 
\bea
\left\{
\begin{array}{ll}
\hat\Delta=0,\\
x_i=0, \hs i=1,2,\ldots \nu,\\
\fr{\partial \hat{\Delta}}{\partial x_i}=0, \hs i=\nu+1,\ldots N-1. 
\end{array}
\right.
\label{landau_eq_rep4} 
\eea
The third equation of (\ref{landau_eq_rep4}) gives
\bea
\sum_{j=\nu+1}^{N-1}G_{ij}x_j=\beta_{i}, \hs i=\nu+1,\ldots N-1. 
\eea
If $\det(G)\neq 0$ then the solution reads
\bea
\bar{x}_i=\sum_{j=\nu+1}^{N-1}\beta_j G^{-1}_{ij}=\fr{1}{\det(G)}\sum_{j=\nu+1}^{N-1}\beta_j \hat{G}_{ij},  \hs i=\nu+1,\ldots N-1. 
\label{solution_x_bar}
\eea
Thus the solution of the second and third equations of (\ref{landau_eq_rep4}) is $x_i=\bar{x}_i$ with  
\bea
\bar{x}=(\underbrace{0,\ldots,0}_{\nu},\bar{x}_{\nu+1},\ldots,\bar{x}_{N-1}).
\eea
The first equation of (\ref{landau_eq_rep4}) gives the equation of the surface of singularity \cite{Eric}
\bea
\hat\Delta(\bar{x})&=&\fr{Q_{NN}}{2}-\fr{1}{2}\sum_{i=\nu+1}^{N-1}\bar{x}_i\beta_i\crn
&=&\fr{1}{2}\fr{\det(Q)}{\det(G)}=0,
\eea
where we have used Eqs. (\ref{solution_x_bar}) and (\ref{detQ_detG}). Not surprisingly, one obtains again $\det(Q)=0$. 
In the case $\det(G)=0$, the condition for the second and third equations of (\ref{landau_eq_rep4}) to have solution is 
$\sum_{j=\nu+1}^{N-1}\beta_j \hat{G}_{ij}=0$ (see Eq.~(\ref{solution_x_bar})). This together with Eq.~(\ref{detQ_detG}) give $\det(Q)=0$. 
 
In the neighbourhood of a point $\bar{x}$ that lies on the surface of singularity, we expand $\hat\Delta$, keeping only the lowest terms:
\bea
\hat\Delta(x)&=&\hat\Delta(\bar{x})+\sum_{i=1}^\nu x_i\fr{\partial \hat\Delta}{\partial x_i}\Big{\vert}_{x=\bar{x}}
+\fr{1}{2}\sum_{i,j=\nu+1}^{N-1}(x_i-\bar{x}_i)(x_j-\bar{x}_j)\fr{\partial^2\hat\Delta}{\partial x_i\partial x_j}\Big{\vert}_{x=\bar{x}}\crn
&=&\fr{1}{2}\left[C(x_i)+\sum_{i,j=\nu+1}^{N-1}y_iy_jG_{ij}\right],
\label{Delta_DetG}
\eea
with 
\bea 
C(x_i)&=&2\hat\Delta(\bar{x})+2\sum_{i=1}^\nu x_i\fr{\partial \hat\Delta}{\partial x_i}\Big{\vert}_{x=\bar{x}}-i\eps,\crn
y_i&=&x_i-\bar{x}_i, \hs i=\nu+1,\ldots, N-1.
\eea
Integral (\ref{eq_TN0_3}) becomes
\bea
T^{N}_{0}&=&\fr{(-1)^N\Gamma(N-D/2)}{\pi^{D/2}2^{3D/2-N}}\int_0^1\prod_{i=1}^{\nu}dx_i\crn
&\times&\int_{-\infty}^{+\infty}\prod_{i=\nu+1}^{N-1}dy_i
\fr{1}{[\sum_{i,j=\nu+1}^{N-1}y_iy_jG_{ij}+C(x_i)]^{N-D/2}},
\label{eq_TN0_4}
\eea
where, similar to the calculation in the previous subsection, we have let each $y_i$-integration contour run from $-\infty$ to $+\infty$, provided the power $(N-D/2)$ of the denominator is sufficiently large. To understand the difference between the LLS and sub-LLS we should compare Eq.~(\ref{Delta_DetQ}) to Eq.~(\ref{Delta_DetG}). One remarks that the linear terms only appear in the case of sub-LLS. The $y_i$-integration is exactly the same for the two cases. $G_{ij}$ is a real symmetric matrix hence can be 
diagonalized by a real orthogonal co-ordinate transformation. Using the same method described in the previous subsection we integrate over $y_i$ to get
\bea
T^{N}_{0}&=&\fr{(-1)^Ne^{i\pi(N-\nu-K-1)}}{2^{3D/2-N}\sqrt{(-1)^{N-\nu-1-K}\det(G)}}\pi^{(N-D-\nu-1)/2}\crn
&\times&\Gamma((N-D+\nu+1)/2)\int_0^1\prod_{i=1}^{\nu}dx_i
[C(x_i)]^{-(N-D+\nu+1)/2},
\label{TN0_subLLS}
\eea 
where $K$ is the number of positive eigenvalues of Gram matrix $G_{ij}$. Of course, one can recover Eq.~(\ref{TN0_LLS}) by setting $\nu=0$. Asumming that
$b_i=\fr{\partial \hat\Delta}{\partial x_i}\Big{\vert}_{x=\bar{x}}\neq 0$, for the leading part of the singularity occurred when $x_i\to 0^+$ we have
\bea
\int_0^1\prod_{i=1}^{\nu}dx_i
[C(x_i)]^\alpha&=&2^\alpha\int_0^1\prod_{i=1}^{\nu}dx_i
[b_ix_i+\hat\Delta(\bar{x})]^\alpha,\crn
&\sim&2^\alpha\fr{(-1)^\nu}{\prod_{i=1}^\nu b_i}\fr{\Gamma(\alpha+1)}{\Gamma(\alpha+\nu+1)}[\hat\Delta(\bar{x})-i\eps]^{\alpha+\nu}.
\eea
where $\alpha=-(N-D+\nu+1)/2$. With $\gamma=-\alpha-\nu=(N-\nu-D+1)/2$ we get
\bea
(T^{N}_{0})_{div}&=&\fr{(-1)^{N}e^{i\pi(N-\nu-K-1)}}{2^{3D/2-N}\sqrt{(-1)^{N-\nu-1-K}\det(G)}}\pi^{(N-D-\nu-1)/2}\crn
&\times&\fr{2^{-\nu}}{\prod_{i=1}^\nu b_i}\fr{(-1)^\nu\Gamma(\gamma+\nu)\Gamma(-\gamma-\nu+1)}{\Gamma(-\gamma+1)}[2\hat\Delta(\bar{x})-i\eps]^{-\gamma}.
\eea 
We then make use of the following identity
\bea
\Gamma(1-z)\Gamma(z)=\fr{\pi}{\sin(z\pi)}
\eea
to get
\bea
\fr{(-1)^\nu\Gamma(\gamma+\nu)\Gamma(-\gamma-\nu+1)}{\Gamma(-\gamma+1)}=\Gamma(\gamma).
\eea
The final result reads\footnote{Similar result has been obtained by Polkinghorne and Screaton \cite{MR0127825}.}
\bea
(T^{N}_{0})_{div}&=&\fr{(-1)^{N}e^{i\pi(N-\nu-K-1)}}{2^{(3+N-\nu)/2}\pi\sqrt{(-1)^{N-\nu-K-1}\det(G)}}
\fr{(4\pi)^{\gamma}\Gamma(\gamma)}{\prod_{i=1}^\nu b_i}\left[\fr{\det(Q)}{2\det(G)}-i\eps\right]^{-\gamma}
\label{TN0_sub-LLS}
\eea
which holds provided 
\bea
\det(G)\neq 0,\hs b_i=\fr{\partial \hat\Delta}{\partial x_i}\Big{\vert}_{x=\bar{x}}\neq 0 \hs \text{and} \hs \gamma=(N-\nu-D+1)/2>0.
\eea 
If some $b_i=0$ then expansion (\ref{Delta_DetG}) must be modified to take into account the higher order terms $x_i(x_j-\bar{x}_j)\fr{\partial^2\hat\Delta}{\partial x_i\partial x_j}\Big{\vert}_{x=\bar{x}}$ with $i=1,\ldots,\nu$. If $\gamma\le 0$ one can write $D=4-2\varep$ ($\varep>0$) and do the expansion in $\varep$ to get the divergent term independent of $\varep$ as in the case of LLS.\\
Comparing Eq.~(\ref{TN0_LLS}) to Eq.~(\ref{TN0_sub-LLS}) we see that, apart from the finite factor related to $b_i$, the nature of the singlarities given by the latter can be obtained from the former by simply replacing $N$ by $N-\nu$ which are the number of on-shell internal particles. We remark also that there must be some relation between $a_1$ and $\det(G)$ at the singular point, namely $\det(G)\propto a_1$ when $\det(Q)=0$. As a consequence, if $a_1=\det(Q)=0$ (Landau matrix $Q$ has at least two zero-eigenvalues) then $\det(G)=0$ (Gram matrix $G$ has at least one zero eigenvalue). There is a beautiful way to prove this mathematically \footnote{We have learnt this trick from Eric Pilon, many thanks!}. Let $V_a=\{x_i^{(a)}\}$ with $a=1,2$ and $i=1,N$ be two linearly independent vectors of the degenerate zero-eigenvalue \footnote{In the case of the real symmetric matrix $Q$, 
the degree of degeneracy of one zero-eigenvalue is equal to the number of zero-eigenvalues.}. We can always normalise $V_a$ such that
\bea
\sum_{i=1}^Nx_i^{(a)}=1, \hs a=1,2.
\eea 
One has
\bea
0&=&\sum_{j=1}^{N} Q_{ij}x_j^{(a)}=\sum_{j=1}^{N-1} Q_{ij}x_j^{(a)}+Q_{iN}(1-\sum_{j=1}^{N-1}x_j^{(a)})\crn
&=&\sum_{j=1}^{N-1} (Q_{ij}-Q_{iN})x_j^{(a)}+Q_{iN}\crn
&=&\sum_{j=1}^{N-1} (Q_{ij}-Q_{iN}-Q_{jN}+Q_{NN})x_j^{(a)}+\sum_{j=1}^{N-1}Q_{Nj}x_j^{(a)}-Q_{NN}\sum_{j=1}^{N-1}x_j^{(a)}+Q_{iN}\crn
&=&\sum_{j=1}^{N-1} G_{ij}x_j^{(a)}-\beta_i\, ,
\eea
where relations (\ref{defs_G_beta}) have been used.
Thus one gets
\bea
\sum_{j=1}^{N-1} G_{ij}x_j^{(a)}=\beta_i, \hs a=1,2.
\eea
With $V_0=V_1-V_2\neq 0$ this gives
\bea
G.V_0=0.
\eea
This means that the Gram matrix has at least one zero-eigenvalue hence $\det(G)=0$.   
\section{Conditions for leading Landau singularities to terminate}
\label{LLS_terminate}
\begin{figure}[htb]
\begin{center}
\includegraphics[width=0.8\textwidth]{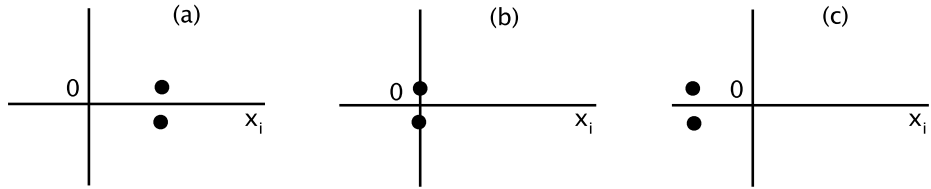}
\caption{\label{LLS_terminate}{\em Mechanism for termination of LLS in $x_i$-plane.}}
\end{center}
\end{figure}
This section concerns the termination of LLS as we vary the external parameters denoted by $M_i$ 
(without any loss of generality, we assume that the internal masses are fixed for the sake of simplicity).

It is obvious that the position of LLS and its properties depend on the values of $M_i$.  If we vary $M_i$ continuously, while maintaining the pinch conditions, the only mechanism for the termination of a LLS is the following\footnote{In the case of more than one loop, there is another mechanism. The LLS can terminate if two pinches meet on the integration contour. If this happens, the singularity may somehow leave the physical region \cite{Eden_notes, cunningham}.}.
The LLS moves to the end of the integration contour ($x_i=0$). Thus the LLS will coincide with a sub-LLS and move off the physical region afterwards \cite{Eden_notes, cunningham, tarski}. This is illustrated in Fig.~\ref{LLS_terminate}. 

The following remark will be useful later. The $4$-point LLS terminates when it coincides with a $3$-point sub-LLS which in turn will terminate when 
coinciding with a normal threshold. The normal threshold coincides with itself, \ie it occurs at one point.

A good question to ask is "How does a LLS terminate?" This is a mathematically complicated question and 
we do not attempt to give a complete answer. What we understand is as follows.  
When moving from Fig.~\ref{LLS_terminate}a to Fig.~\ref{LLS_terminate}b, the leading Landau 
curve ($\det(Q_N)=0$) changes continuously until it makes an {\em effective intersection} with a sub-leading Landau curve ($\det(Q_{N-1})=0$) \cite{Eden_notes, tarski, cunningham}. At the point of contact, both curves 
have the same slope and both corresponding Landau equations have the same solution of $x_i$ \cite{Eden_notes, cunningham}. At effective intersections the nature of the Landau singularity may change \cite{tarski, cunningham}. In Fig.~\ref{LLS_terminate}a we have a $N$-point LLS whose nature is given by 
\bea
T_a(M_i)=A_a(M_i)+B_a(M_i)f_a(\det Q_{N}),
\label{nature_a}
\eea 
where the functions $A$ and $B$ are analytic in a neighborhood of the singular point $M_i$, the function $f_a(\det Q_{N})$ is singular at this point. $f_a(z)$ can be 
$z^{1/2}$, $\ln z$ or $z^{-1/2}$ if $N$ is $2$, $3$ or $4$ respectively. For Fig.~\ref{LLS_terminate}c we have a similar equation
\bea
T_c(M_i)=A_c(M_i)+B_c(M_i)f_c(\det Q_{N-1}),
\label{nature_c}
\eea 
where we have assumed that Fig.~\ref{LLS_terminate}c has a $(N-1)$-point Landau singularity. The nature of the coincident singularity in Fig.~\ref{LLS_terminate}b is a product of two factors which are similar to $T_a$ and $T_c$ \cite{tarski}. Thus, we have
\bea
T_b(M_i)=A+Bf_a(\det Q_{N})+Cf_c(\det Q_{N-1})+Df_a(\det Q_{N})f_c(\det Q_{N-1}).
\eea   
If $D\neq 0$ then the leading singularity is given by the last term which means that the Landau singularity can be enhanced at termination point. This kind of enhancement can be somehow understood if we look at some formulae in this thesis: from Eq.~(\ref{TN0_sub-LLS}) we see that if a leading $N$-point singularity coincides with a sub-LLS then $b_i=\fr{\partial \hat\Delta}{\partial x_i}\Big{\vert}_{x=\bar{x}}=0$ leading to an enhancement from the prefactor; from Eq.~(\ref{nature_DPS}) we observe a product of two singularities (a leading Landau singularity and a collinear divergence); from Eq.~(\ref{d0_y_13_sumij}) we see the possibility that a product of two singularities can occur if the integrals related to $3$-point functions develop a pinch singularity at the same position where the prefactor related to the leading Landau determinant ($1/\det(Q_4)$) vanishes. 

If a four-point LLS coincides with a three-point sub-LLS, the nature of this singularity can be $z^{-1/2}\ln z$ which is integrable but its square is not.
\section{Special solutions of Landau equations}
\label{section_landau_special}
\subsection{Infrared and collinear divergences}
In this section, we will show that infrared and collinear singularities are solutions of Landau equations. 
However, in order for them to become divergent additional conditions must be satisfied. As one might anticipate from Eq.~(\ref{TN0_sub-LLS}) sub-leading Landau singularities can be enhanced by various factors.
\subsubsection{Infrared divergence}
We consider the case of a sub-LLS where $x_1=\ldots =x_{N-1}=0$ and $x_N>0$.
The Landau equations become
\bea q_N^2=m_N^2\hs \text{and} \hs q_N=0.\eea
We get $m_N=0$. As remarked in subsection~\ref{subsection_landau_nature}, for the case $N=1,2$ the Landau singularities are finite hence there is no 
infrared divergence in those cases. We thus consider the case $N=3$. With $\nu=2$, equation (\ref{eq_TN0_4}) becomes
\bea
T^{3}_{0}\sim\int_{0^+} dx_1dx_2
\fr{1}{C(x_i)},
\eea 
with 
\bea
C(x_i)=2m_3^2+2\sum_{i=1}^{2}x_i\beta_i+\sum_{i,j=1}^{2}G_{ij}x_ix_j -i\eps.
\eea
If $\beta_i=m_3^2-m_i^2+r_i^2\neq 0$ then one can neglect the quadratic terms in $C(x_i)$ to get
\bea
(T^{3}_{0})_{div}\sim\int_{0^+} dx_1dx_2
\fr{1}{2m_3^2+2(\beta_1x_1+\beta_2x_2)}\sim \fr{1}{\beta_1\beta_2}m_3^2\to 0.
\eea 
If $\beta_i=m_3^2-m_i^2+r_i^2=0$ then we get
\bea
(T^{3}_{0})_{div}\sim\int_{0^+} dx_1dx_2
\fr{1}{2m_3^2+\sum_{i,j=1}^{2}G_{ij}x_ix_j}\sim \ln(m_3^2)\to \infty.
\eea
The nature of Landau singularity is $m_3^2$ if $\beta_i\neq 0$ and is enhanced to $\ln(m_3^2)$ if $\beta_1=\beta_2=0$. 
The condtions to have an infrared divergence for the case of three-point function therefore are
\bea
m_3=0, \hs m_1^2=r_1^2=p_1^2, \hs m_2^2=r_2^2=p_3^2.
\label{cond_ir}
\eea
For the cases $N>3$ we can always reduce them to three-point functions hence we get the same conclusion. Physically, one sees that conditions 
(\ref{cond_ir}) can be satisfied only by the photon or gluon. For the electroweak theory, if we take the limit $M_W\to 0$ then the one-loop 
diagrams involving the W-gauge boson as an internal particle have no infrared divergence since it couples to particles with different masses. 
\subsubsection{Collinear divergence}
For $M=2$, \ie $x_{1,2}>0$ and $x_3=\ldots =x_N=0$ equations (\ref{landau_eqM}) become
\bea
\left\{
\begin{array}{lll}
x_{1,2}>0,\\
q_1^2=m_1^2, \hs q_2=m_2^2,\\
x_1q_1+x_2q_2=0.\label{landau_eq2} 
\end{array}
\right.
\eea
One gets $x_1m_1=x_2m_2$. If $m_{1,2}\neq 0$ then $x_1(q_1+\fr{m_1}{m_2}q_2)=0$ whose solution corresponds to the normal 
threshold $p^2=(q_1-q_2)^2=(m_1+m_2)^2$. If $m_{1,2}=0$ one gets
\bea
x_1m_1^2+x_2(q_1.q_2)=0,
\eea 
which gives $q_1.q_2=0$ or $p^2=(q_1-q_2)^2=0$. This solution corresponds to a collinear divergence whose nature is also logarithmic \cite{Kinoshita:1962ur}. Clearly, this collinear divergence cannot occur if $N=1$.  

It is important to remark that the solutions for infrared and collinear divergences appear in the limit of massless internal particles. These solutions 
require no constraint on relevant Feynman parameters, even the positive condition is not necessary. 
\subsection{Double parton scattering singularity}
\begin{figure}[hptb!]
\begin{center}
\includegraphics[width=6cm]{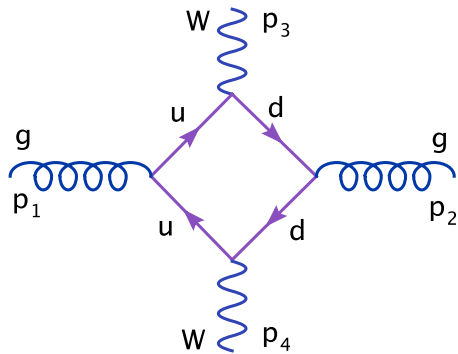}
\caption{\label{fig_ggWW}{\em A typical box Feynman diagram which has a double parton scattering singularity.
}}
\end{center}
\end{figure}
There exists a special case of Landau singularity called double parton scattering (DPS) singularity \cite{Nagy:2006xy, Bern:2008ef} which appears also in the massless limit. 
Unlike the sub-leading infrared and collinear divergences, the DPS singularity is a LLS and its solution requires some sort of constraint on relevant Feynman parameters (the positive sign condition is important). 

Let us consider the case of $g(p_1)g(p_2)\to W(p_3)W(p_4)$ box diagram displayed in Fig. \ref{fig_ggWW}.
The internal particles $u$-quark and $d$-quark are massless.
The $Q$-matrix is given by
\bea Q_4=\left( \begin{array}{cccc}
0 & 0 & -t & -M_W^2\\
0 & 0 & -M_W^2 & -u\\
-t & -M_W^2 & 0 & 0\\
-M_W^2 & -u & 0 & 0\\ 
\end{array}\right)\eea
where $t=(p_1-p_3)^2$, $u=(p_2-p_3)^2$. The Landau determinant, characteristic polynomial and Gram determinant read
\bea
\det(Q_4)&=&(tu-M_W^4)^2=0,\crn
f(\la)&=&\la^4-(t^2+u^2+2M_W^4)\la^2+\det(Q_4),\crn
\det(G_3)&=&2s(tu-M_W^4),
\eea
with $s=(p_1+p_2)^2$.
One sees that $a_1=0$ and $\det(Q_4)=\det(G_3)=0$ at the boundary of phase space where the Landau matrix has two zero-eigenvalues. 
The phase space is defined as 
\bea
s\ge 4M_W^2,\hspace*{3mm} tu-M_W^4\ge 0\hspace*{3mm} \text{with} \hspace*{3mm} t+u=2M_W^2-s\le -2M_W^2.
\label{ggWW_phase}
\eea
In this special case where the Landau matrix has two zero-eigenvalues at the singular point, the conditions given in subsection~\ref{subsection_landau_condition} (see Eq.~(\ref{landau_cond0})) are necessary but no longer sufficient. One should keep in mind that this box diagram always has a collinear divergence associated with the reduced two-point functions as discussed in the previous subsection. 
The necessary conditions for a LLS read
\bea
\left\{
\begin{array}{ll}
tu=M_W^4,\\
t<0,\,\,\, u<0, \label{landau_cond_ggWW} 
\end{array}
\right.
\eea
which are compatible with Eq.~(\ref{ggWW_phase}). We now have two questions to be answered: whether they are sufficient conditions for a pinch singularity? If Yes then what is the nature of the singularity? 

For these questions, we come back to equation (\ref{landau_rep2}) and perform the following nonlinear change of variables \cite{Eric,Nagy:2006xy}
\bea
x_1=\sigma\alpha,\hs x_2=\sigma(1-\alpha), \hs x_3=\tau\beta, \hs x_4=\tau(1-\beta).
\eea
The inverse solution gives us the range of new variables
\bea
0\le \sigma , \tau <\infty; \hs 0\le \alpha, \beta \le 1.
\eea
The Jacobian is simply $\sigma\tau$. With $D=4+2\varep$ ($\varep>0$) equation (\ref{landau_rep2}) factorizes
\bea
T_0^4&=&\fr{1}{(4\pi)^2}\int_0^\infty d\sigma d\tau\fr{\delta(\sigma+\tau-1)}{(\sigma\tau)^{1-\varep}}\crn
&\times&\int_0^1 d\alpha d\beta\fr{1}{[s\alpha\beta+(u-M_W^2)(\alpha+\beta)-u-i\eps]^{2-\varep}}.
\eea
The first integral is just the Beta function $B(\varep,\varep)$ producing a collinear divergence. We use
\bea
B(\varep,\varep)=\fr{\Gamma(\varep)\Gamma(\varep)}{\Gamma(2\varep)}=\fr{2}{\varep}+\OO(\varep)
\eea
to get
\bea
T_0^4&=&\fr{1}{8\pi^2\varep}\int_0^1 d\alpha d\beta\fr{1}{[s\alpha\beta+(u-M_W^2)(\alpha+\beta)-u-i\eps]^{2}}\crn
&+&\fr{1}{8\pi^2}\int_0^1 d\alpha d\beta\fr{\ln[s\alpha\beta+(u-M_W^2)(\alpha+\beta)-u]}{[s\alpha\beta+(u-M_W^2)(\alpha+\beta)-u-i\eps]^{2}}.
\eea
We are interested in the collinear divergent term. The relevant integral reads
\bea
I_1=-\fr{1}{(M_W^2-t)(M_W^2-u)}\int_0^1 d\alpha\fr{1}{(\alpha-w)(\alpha-a)},
\eea
where we have integrated over $\beta$ and 
\bea
a&=&\fr{M_W^2}{M_W^2-t}-i\eps_a, \hs \eps_a=\fr{\eps}{M_W^2-t},\crn 
w&=&\fr{-u}{M_W^2-u}+i\eps_w, \hs \eps_w=\fr{\eps}{M_W^2-u}.
\eea
This is nothing but the second example (\ref{landau_eg2}). From Eq.~(\ref{landau_eg2_cond}) 
we have the necessary and sufficient conditions for a pinch singularity:
\bea
0<w=a<1 \hs \text{and}\hs (M_W^2-t)(M_W^2-u)>0,
\eea
which are completely equivalent to Eq.~(\ref{landau_cond_ggWW}). From Eq.~(\ref{landau_eg2_c1}) we get the following result 
at the singular point $tu=M_W^4$
\bea
(I_1)_{div}&=&\fr{1}{M_W^4}+\fr{2\pi i}{M_W^4-tu},\crn
(T_0^4)_{div}&=&\fr{i}{4\pi (M_W^4-tu)}\fr{1}{\varep}+\ldots =-\fr{i}{4\pi \sqrt{\det(Q_4)}}\fr{1}{\varep}+\ldots \hs .
\label{nature_DPS}
\eea
We conclude that in this special case where the Landau matrix has two zero-eigenvalues, conditions (\ref{landau_cond_ggWW}) 
given by the Landau equations are necessary and sufficient for the appearance of a LLS. 
The nature of this singularity is $1/\sqrt{\det(Q_4)}$ which is consistent 
with Eq.~(\ref{TN0_LLS}). The LLS goes together with the collinear divergence (see also section \ref{LLS_terminate}). 
This LLS is called double parton scattering singularity 
first pointed out in \cite{Nagy:2006xy}.
 
We notice that Eq.~(\ref{nature_DPS}) disagrees with the result of Ellis and Zanderighi given in Eq.~(4.21) of \cite{Ellis:2007qk}\footnote{Many thanks to Denner for pointing out this reference to us.} where they claimed that, apart 
from the collinear divergence, a finite result can be obtained by doing an expansion around the LLS point $tu=M_W^4$.  

This DPS singularity is not integrable. 
In a practical calculation, the cross section must be finite. This singularity must somehow disappear order 
by order. For the process $gg\to WW$ there is no tree level diagram hence there is no real radiation at the leading order. One infers that the numerator must vanish at the singular point. The actual calculations \cite{Binoth:2006mf,Bernicot:2007hs} have confirmed this. 
We have conjectured it as a consequence of the gauge dynamics \cite{Bern:2008ef}. To understand this dynamical cancellation, a physical investigation is necessary. The same phenomenon of DPS singularity happens in the case of six photon scattering whose numerical analysis around the singular point is 
given in \cite{Bern:2008ef}.

\chapter{SM $b\bar{b}H$ production at the LHC: $M_H\ge 2M_W$}
\label{chapter_bbH2}
The content of this chapter is based on our publications \cite{fawzi_bbH2, ninh_moriond_2008}. 

An important notation used in this chapter should be clarified. The notation of center-of-mass energy, $\sqrt{s}$, can be used for the 
sub-process $\ggbbH$ or $\ppbbH$. Which use will depend on the context, as you will notice very easily.
\section{Motivation}
The aim of this chapter is to extend the study we made in chapter \ref{chapter_bbH1} to higher Higgs masses. 

When trying to do so for the case $M_H\ge 2M_W$ we have encountered severe numerical instabilities for the cross section involving 
the one-loop amplitude squared. We have tried to understand this problem and found the following facts. At the level of NLO which involves 
the interference term between the tree-level and
one-loop amplitudes no instability was present. On close inspection it was found that
the instabilities were only due to the box and pentagon
diagrams of class (c) of Fig.~\ref{diag_3group}. The contribution of various triangle diagrams does not show any numerical instabilities. 
At the partonic gluon-gluon level it was found that there is no instability
for $\sqrt{s} < 2m_t$ and that independently of $M_H$ and $\sqrt{s}$ the result was completely
stable for $m_t = M_W$. These threshold conditions were a sign for the possible existence of
a leading Landau singularity for the box diagrams whose square is not integrable. The five point functions are reduced to four point 
functions hence should have the same problem. Indeed, some triangle diagrams of class (c) of Fig.~\ref{diag_3group} have also LLS whose nature 
is integrable. That's why they do not show any numerical instability. 

In order to solve this problem of Landau singularities, we first have to understand them in detail by applying the general analysis of the 
previous chapter to the specific problem of $\ggbbH$.

\section{Landau singularities in $\ggbbH$}
In this section, we discuss all Landau singularities that occur in the Feynman diagrams of class (c) of Fig.~\ref{diag_3group}. 
It is 
pedagogical to start with the three point function. 
\subsection{Three point function}
\label{appendix_3pt}
\begin{figure}[h]
\begin{center}
\mbox{\includegraphics[width=0.45\textwidth]{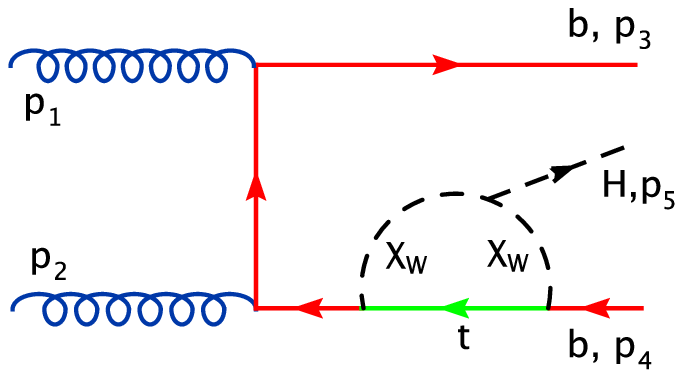}
\hspace*{0.075\textwidth}
\includegraphics[width=0.45\textwidth]{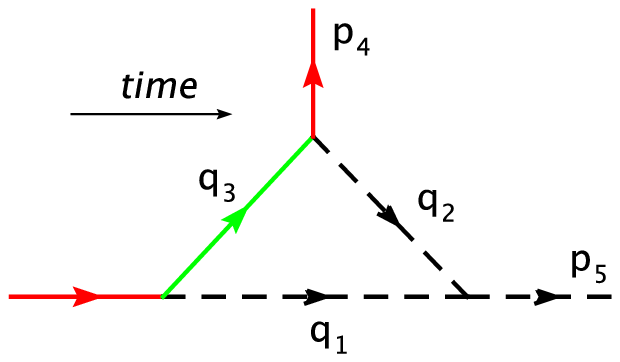}}
\caption{\label{3pt-LLS}{\em Left: A triangle diagram contributing to
$gg \ra b \bar b H$  that can develop a leading Landau singularity for
$M_H\ge 2M_W$ and $\sqrt{s_2}\ge m_t+M_W$, i.e. all the three particles
in the loop can be simultaneously on-shell. Right: the three point function of the left diagram where the arrows indicate 
momentum flow.}}
\end{center}
\end{figure}
As concluded in section~\ref{subsection_landau_condition}, 
a necessary condition for a three point function to have a LLS is that it has two cuts which 
can produce physical on-shell particles. The other external particle which does not correspond to those cuts must have mass smaller than the difference of the two internal masses associated with it. The diagram in Fig.~\ref{3pt-LLS} satisfies those conditions. The scalar three point function of this diagram reads
\bea
T_0^3(s_2)=C_0(s_2,M_H^2,0,m_t^2,M_W^2,M_W^2)
\eea
with $s_2=(p_4+p_5)^2$ and the bottom-quark mass has been neglected. Necessary conditions to have LLS are
\bea
M_H\ge 2M_W\hs \textrm{and}\hs \sqrt{s_2}\ge m_t+M_W.
\label{cond_mass_3pt}
\eea
The phase space is defined by
\bea
M_H^2\le s_2\le s.
\eea
The characteristic equation writes
\bea
-\la^3+a_2\la^2-a_1\la+a_0=0
\eea
with
\bea
a_0&=&\det(Q_3)=-2M_W^2s_2^2+2M_H^2(m_t^2 + M_W^2)s_2-2M_H^2[M_H^2m_t^2+(m_t^2-M_W^2)^2],\crn
a_1&=&-\left[M_H^2(M_H^2-4M_W^2)+(s_2-m_t^2-M_W^2)^2+m_t^4+M_W^4-6m_t^2M_W^2 \right]\crn
&\le& -\left[M_H^2(M_H^2-4M_W^2)+(m_t^2-M_W^2)^2\right]< 0,\crn
a_2&=&2(m_t^2+2M_W^2),
\label{def_ai_3pt}
\eea
where we have used conditions (\ref{cond_mass_3pt}) to prove that $a_1$ is negative. The equation $\det(Q_3)=0$ has two roots
\bea
s_2^{\pm}=\fr{1}{2M_W^2}\left[M_H^2(M_W^2+m_t^2)\pm (m_t^2-M_W^2)M_H\sqrt{M_H^2-4M_W^2}\right].
\label{root_s2_3pt}
\eea
The sign condition ($x_i>0$), Eq.~(\ref{landau_cond1}), is very simple. For this, we need 
\bea
\det(\hat{Q}_{33})&=&-M_H^2(M_H^2-4M_W^2)\le 0,\crn
\det(\hat{Q}_{13})&=&-M_H^2(m_t^2+M_W^2)+2s_2M_W^2\le 0,\crn
\det(\hat{Q}_{23})&=&-M_H^2(m_t^2+M_W^2)+s_2(M_H^2-2M_W^2)\le 0,
\eea
which together with Eq.~(\ref{cond_mass_3pt}) give
\bea
s_2\le \fr{M_W^2+m_t^2}{M_H^2-2M_W^2}M_H^2\le 2(m_t^2+M_W^2).
\label{s2_max}
\eea
Only the minus solution given in Eq.~(\ref{root_s2_3pt}) satisfies this, thus 
\bea
s_2^{LLS}=s_2^{-}=\fr{1}{2M_W^2}\left[M_H^2(M_W^2+m_t^2)-(m_t^2-M_W^2)M_H\sqrt{M_H^2-4M_W^2}\right].
\label{s2_3pt}
\eea
This result tells us many things. First, if $M_H<2M_W$ there is no LLS. Second, if $M_H=2M_W$ the LLS 
occurs at $s_2^{LLS}=2(m_t^2+M_W^2)$ which is the maximum value of $s_2^{LLS}$ given by Eq.~(\ref{s2_max}). If $M_H$ increases then 
$s_2^{LLS}$ becomes smaller and smaller. The maximum value of $M_H$ is determined when $s_2^{LLS}$ reaches the normal threshold 
(this condition for the termination of a LLS was discussed in section~\ref{LLS_terminate}): 
\bea
&&4M_W^2\le M_H^2\le 4M_W^2+\fr{M_W}{m_t}(m_t-M_W)^2,\crn
&&(m_t+M_W)^2\le s_2\le 2(m_t^2+M_W^2).
\label{range_MH_s2_3pt}
\eea
Numerically, we have
\bea 
&&160.7532\text{GeV}\le M_H\le 172.889\text{GeV},\crn
&&254.3766\text{GeV}\le \sqrt{s_2}\le 271.059\text{GeV}.
\eea
If one keeps increasing $M_H>172.889\text{GeV}$, $s_2$ will increase from its minimum value $254.3766$GeV and become larger 
than the limit $\fr{M_W^2+m_t^2}{M_H^2-2M_W^2}M_H^2$ (see Eq.~(\ref{s2_max})) required by the sign condition ($x_i>0$) hence be outside the 
physical region. 
\begin{figure}[ht]
\begin{center}
\mbox{\includegraphics[width=0.45\textwidth,height=5.5cm]{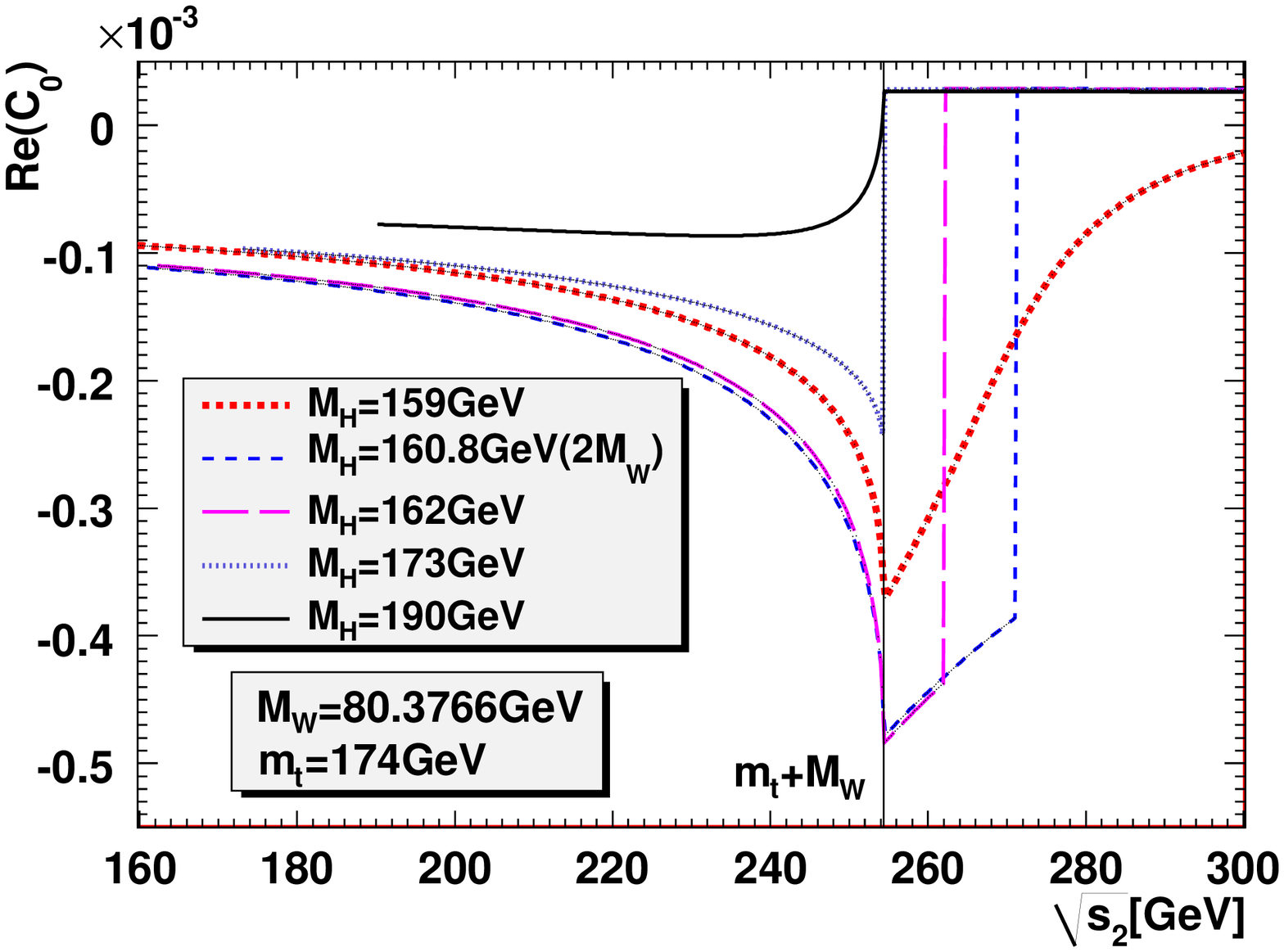}
\hspace*{0.075\textwidth}
\includegraphics[width=0.45\textwidth,height=5.5cm]{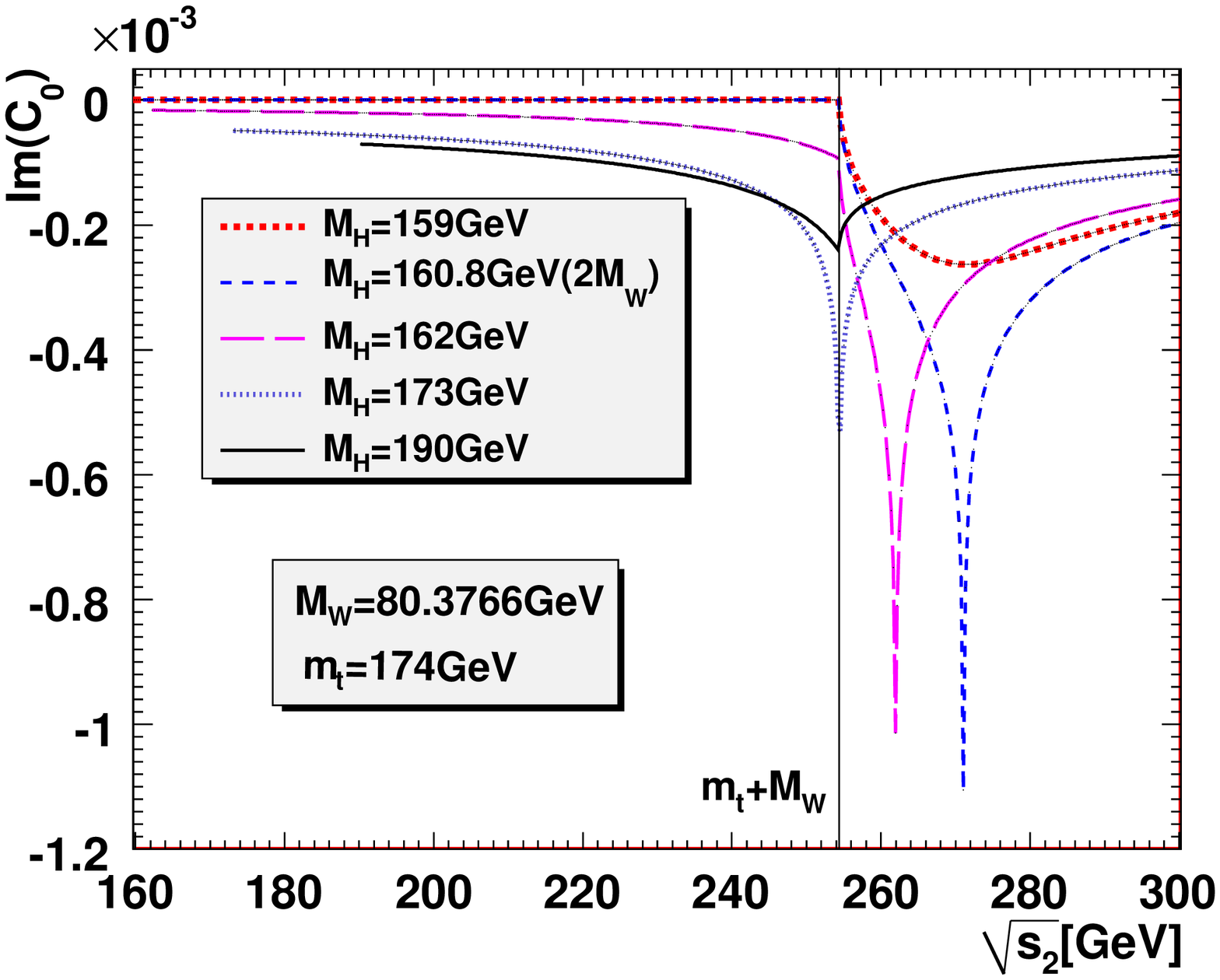}}
\caption{\label{fig_c0_width0}{\em Left: the real part of $C_0$ as functions of $\sqrt{s_2}$ with various values of $M_H$. Right: the same plots for the imaginary part.}}
\end{center}
\end{figure}
In the mean time, one should notice that the normal threshold is moving towards the left boundary of phase space as $M_H$ increases. When $M_H=m_t+M_W=254.3766$GeV, the normal threshold is at the boundary and disappears if $M_H$ passes that value. The function $T_0^3$ then has no structure. All this phenomenon is shown in Fig.~\ref{fig_c0_width0}.

We would like to explain the behaviour of the real and imaginary parts of $C_0$ at the LLS point. 
Since $\la_3\to 0$, we have $\la_1\la_2=a_1<0$ as proved in Eq.~(\ref{def_ai_3pt}). Thus there is one positive eigenvalue, \ie $K=1$. 
From Eq.~(\ref{T30_div_final}) we get
\bea
(C_0)_{div}=\fr{i\upsilon}{8\pi\sqrt{\vert a_1\vert}}\ln(\la_3\upsilon^2-i\eps)
=\fr{\upsilon}{8\pi\sqrt{\vert a_1\vert}}(i\ln\vert\la_3\upsilon^2\vert +\pi).
\label{nature_LLS_3pt}
\eea 
Thus, one observes a positive jump in the real part and a logarithmic singularity in the imaginary part as described in Fig.~\ref{fig_c0_width0}. This singularity is integrable at the NLO as well as one-loop amplitude square level hence does not cause any numerical instability.

\subsection{Four point function}
\label{bbH_landau_4pt}
\begin{figure}[htb]
\begin{center}
\includegraphics[width=0.5\textwidth]{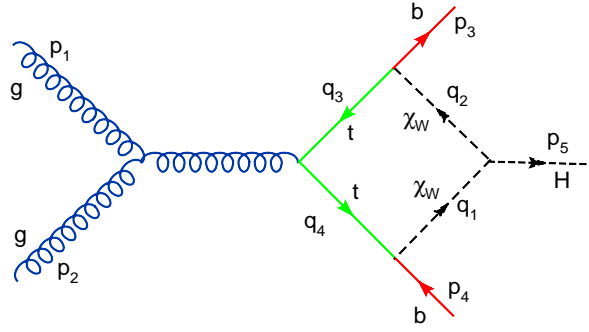}
\caption{\label{landau_box}{\em A box diagram contributing to
$gg \ra b \bar b H$  that can develop a Landau singularity for
$M_H\ge 2M_W$ and $\sqrt{s}\ge 2m_t$, i.e. all the four particles
in the loop can be simultaneously on-shell. The arrows indicate momentum flow. 
Internal momenta $q_i$ are real, on-shell and have positive energies.}}
\end{center}
\end{figure}

The $5$-point functions have no LLS but contain a sub-leading Landau singularity which is exactly the same as the LLS of the box diagram in Fig.~\ref{landau_box}. It is thus enough to study the singularity structure of this box diagram. 
With $s_1=(p_3+p_5)^2,\,\, s_2=(p_4+p_5)^2$, and the
on-shell conditions $p_1^2=p_2^2=0$, $p_3^2=p_4^2=m_b^2=0$,
$p_5^2=M_H^2$, fixing $s$ and $M_H$, the scalar box integral is a
function of two variables $s_{1,2}$
\bea
T_0^4(s_1,s_2)=D_0(M_H^2,0,s,0,s_1,s_2,M_W^2,M_W^2,m_t^2,m_t^2).
\label{def_d0}
\eea
The kinematically allowed region (phase space) is
\bea
M_H^2\le  s_{1} \le s,\;\; M_H^2\fr{s}{s_1}\le s_2  \le
M_H^2+s-s_1,\label{phys_region_ggbbH}
\eea
where the latter is obtained by writing $s_2=(p_4+p_5)^2=$ $M_H^2+2e_4(e_5-\vert \pp_5\vert \cos\theta_{45})$ in 
the center-of-mass system of $(p_3+p_5)$. 
\subsubsection{Conditions for the opening of normal thresholds}
The condition $\sum x_iq_i=0$ with $x_i>0$ of Eq.~(\ref{landau_eqM}) can be re-written in the form
\bea
x_2q_2+x_3q_3=x_1q_1+x_4q_4
\label{sum_xq_new}
\eea
with all $q_i^0>0$ as shown in Fig.~\ref{landau_box}. Indeed, there are other possibilities like $x_3q_3=x_1q_1+x_4q_4+x_2q_2$ with all 
$q_i^0>0$ but this will require $m_b\ge (m_t+M_W)$ which is impossible in our case. Thus equation (\ref{sum_xq_new}) is the unique possibility for a LLS. 
This condition of positive energy together with the real and on-shell condition ($q_i=q_i^*$, $q_i^2=m_i^2$) give a beautiful physical picture 
as shown in Fig.~\ref{landau_box} where internal and external momenta share the same physical properties. 

As already discussed in section~\ref{subsection_landau_condition}, the above picture gives the following necessary conditions for the appearance of a LLS: 
\bea
&&M_H\ge 2M_W \hs \text{and} \hs \sqrt{s}\ge 2m_t, \label{cond_MH_s_0}\\
&&s_1\ge (m_t+M_W)^2\hs \textrm{and}\hs s_2\ge (m_t+M_W)^2, \label{cond_s1_s2_0}\\
&&m_t>M_W, \label{cond_mt_MW}
\eea
where we have used the fact that the momenta of the bottom-quarks and the Higgs boson flow in the same positive direction to get the last constraint 
(if we consider the inverse process $H\to b\bar{b}gg$ where momenta of the bottom-quarks and the Higgs boson are in opposite directions then we 
get $M_W>m_t$ which cannot be satisfied by experimental data.). They are conditions for the opening of $4$ normal thresholds.
\subsubsection{Landau determinant}
The reduced  matrix, $S^{(4)}$, which is equivalent in this case
to the $Q$ matrix for studying the Landau singularity, is given by
\bea S_{4}=\left( \begin{array}{cccc}
1 & \fr{2M_W^2-M_H^2}{2M_W^2} & \fr{m_t^2+M_W^2-s_1}{2M_Wm_t} & \fr{M_W^2+m_t^2}{2M_Wm_t}\\
\fr{2M_W^2-M_H^2}{2M_W^2} & 1 & \fr{M_W^2+m_t^2}{2M_Wm_t} & \fr{m_t^2+M_W^2-s_2}{2M_Wm_t}\\
\fr{m_t^2+M_W^2-s_1}{2M_Wm_t} & \fr{M_W^2+m_t^2}{2M_Wm_t} & 1 & \fr{2m_t^2-s}{2m_t^2}\\
\fr{M_W^2+m_t^2}{2M_Wm_t} & \fr{m_t^2+M_W^2-s_2}{2M_Wm_t} & \fr{2m_t^2-s}{2m_t^2} & 1\\
\end{array}\right), \;\; S_{4}^{ij}=\frac{Q_4^{ij}}{2m_i m_j}.\eea
The determinant reads
\bea
\det(Q_4)&=&16M_W^4m_t^4\det(S_4)=as_2^2+bs_2+c,\crn
a&=&\la(s_1,m_t^2,M_W^2)=[s_1-(m_t+M_W)^2][s_1-(m_t-M_W)^2],\crn
b&=&2\left\{-s_1^2(m_t^2+M_W^2)+s_1[(m_t^2+M_W^2)^2-(s-2m_t^2)(M_H^2-2M_W^2)]\right.\crn
&+&\left. sM_H^2(m_t^2+M_W^2)\right\},\crn
c&=&s_1^2(m_t^2-M_W^2)^2+2M_H^2s(m_t^2+M_W^2)s_1\crn
&+&sM_H^2[(s-4m_t^2)(M_H^2-4M_W^2)-4(m_t^2+M_W^2)^2].\label{det_abc}
\eea
We remark that $a=0$ corresponds to a normal threshold and defines the asymptotes of the Landau curve.

The Landau determinant can be written in the following beautiful form 
\bea
\det(Q_4)&=&a(s_2-s_2^\prime)^2-\fr{\Delta}{4a},\hs s_2^\prime=-\fr{b}{2a},\crn
\Delta&=&\det Q_3(s,s_1)\det Q_3(M_H^2,s_1),\crn
\fr{\det Q_3(M_H^2,s_1)}{2M_W^2}&=&-\left[(s_1-\fr{m_t^2+M_W^2}{2M_W^2}M_H^2)^2-\fr{M_H^2(M_H^2-4M_W^2)(m_t^2-M_W^2)^2}{4M_W^4}\right],\crn
\fr{\det Q_3(s,s_1)}{2m_t^2}&=&-\left[(s_1-\fr{m_t^2+M_W^2}{2m_t^2}s)^2-\fr{s(s-4m_t^2)(m_t^2-M_W^2)^2}{4m_t^4}\right],
\eea
which is very useful when one knows that the LLS coincides with a three-point sub-LLS. If that happens the Eq. $\det(Q_4)=0$ has only one 
root $s_2=s_2^\prime$. The fact that the discriminant of a Landau determinant can be written as a product of lower-order Landau determinants 
is known as the Jacobi ratio theorem for determinants \cite{tarski, book_turnbull}.  
\subsubsection{Discussion of singular structure}
We would like to understand the singular structure of the scalar $4$-point function defined in Eq.~(\ref{def_d0}). Namely, we will look 
at the behaviour of its real and imaginary parts as functions of $s_1$ and $s_2$ while other parameters are fixed. 
We take $m_t=174\;$GeV, $M_W=80.3766\;$GeV, $\sqrt{s}=353\;$GeV and $M_H=165\;$GeV. 

\begin{figure}[htb]
\begin{center}
\mbox{
\includegraphics[width=0.45\textwidth]{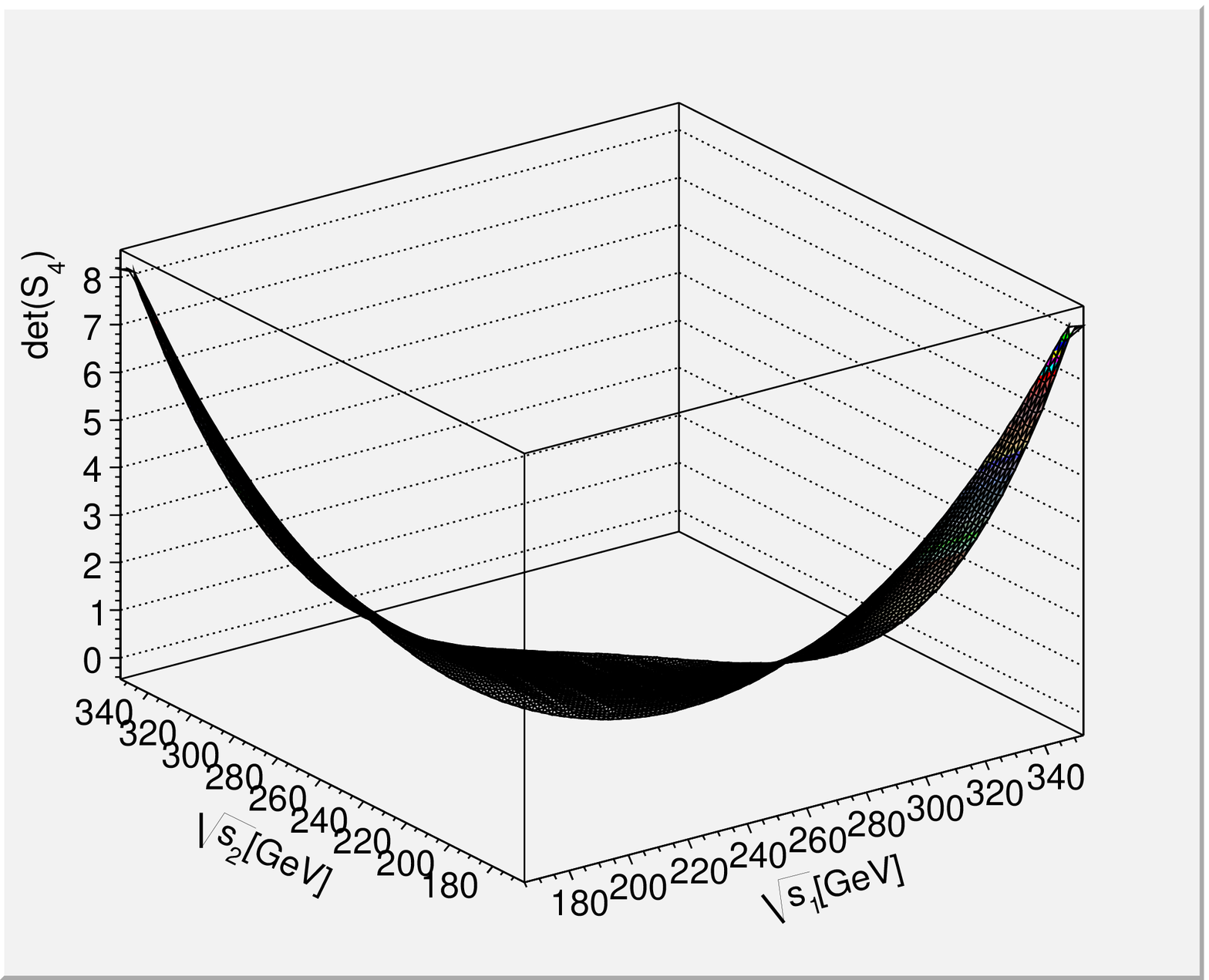}}
\mbox{\includegraphics[width=0.45\textwidth]{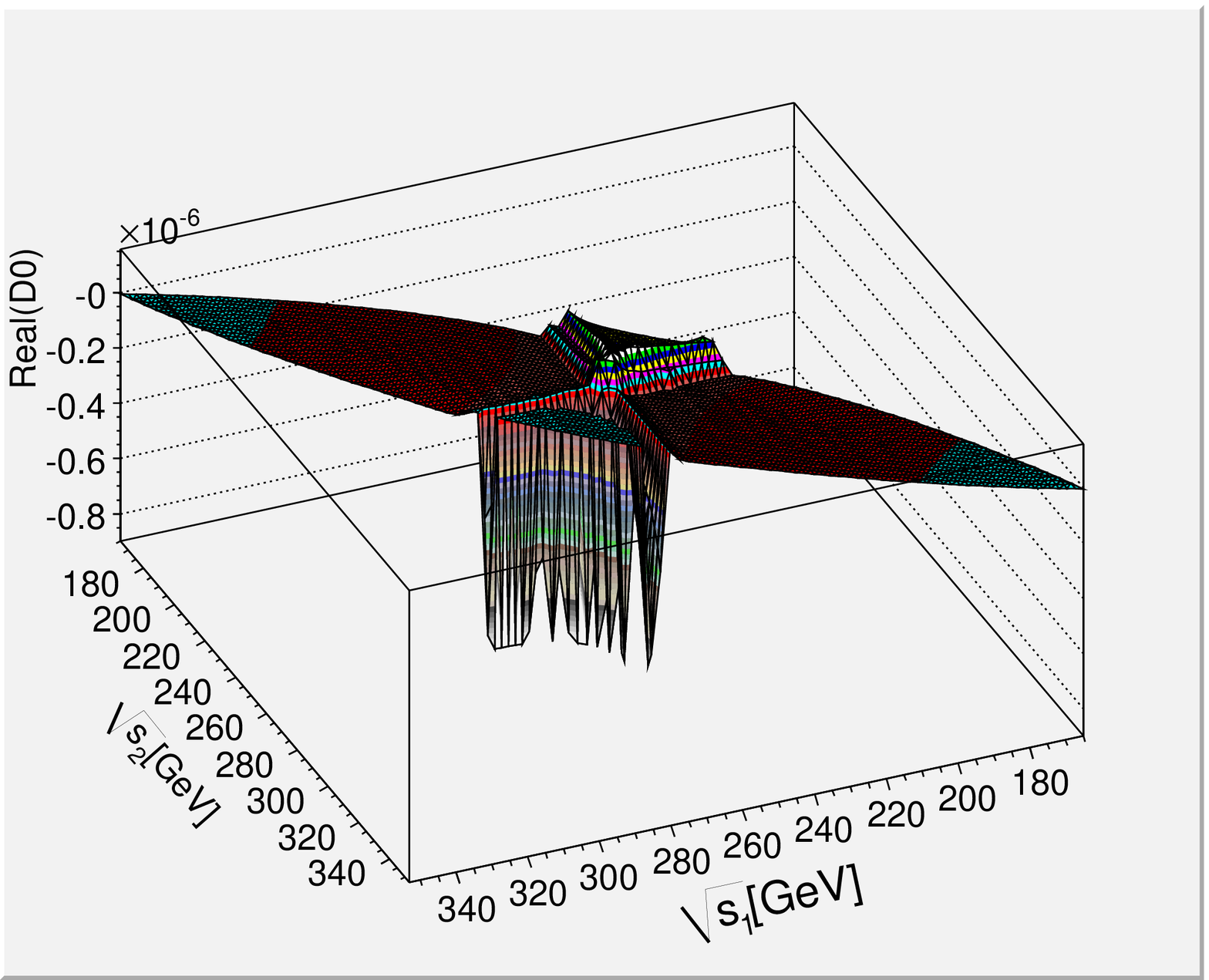}
\hspace*{0.075\textwidth}
\includegraphics[width=0.45\textwidth]{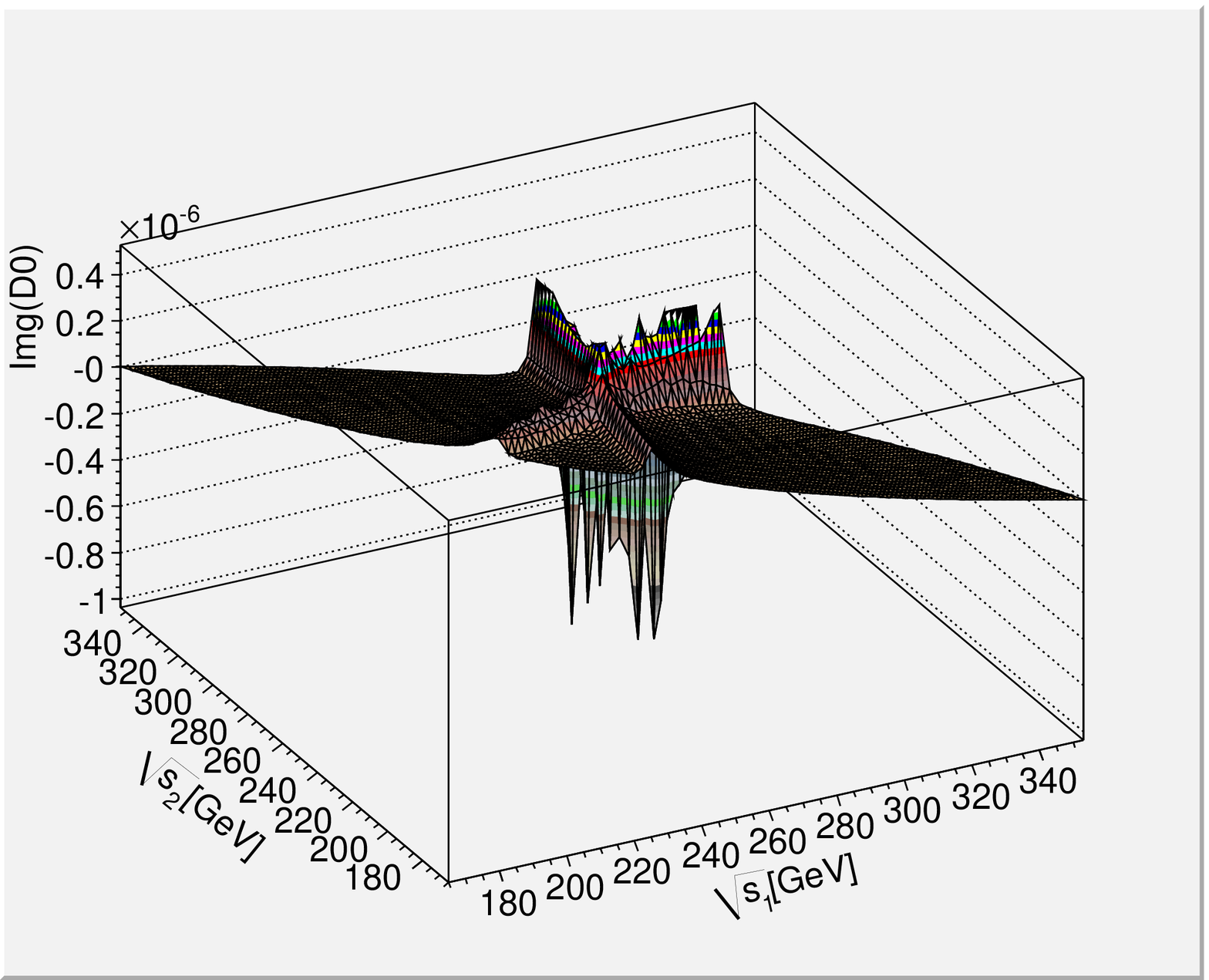}}
\caption{\label{box_diag_3D_plots}{\em The Landau determinant as a
function of $s_1$ and $s_2$ (upper figure). The real and imaginary
parts of $D_0$ as a function of $s_1$ and $s_2$.}}
\end{center}
\end{figure}

The behaviour of the Landau determinant, the real
and imaginary parts of the $4-$point function $T_0^4$ are
displayed in Fig.~\ref{box_diag_3D_plots} as function of $s_1$,
$s_2$ within the phase space defined by Eq.~(\ref{phys_region_ggbbH}). 
We clearly see that the Landau
determinant vanishes inside the phase space and leads to regions
of severe instability in both the real and imaginary parts of the
scalar integral. We notice that this region of Landau singularities is localised at the center of phase space. 
The boundary of singular region will be explained later, see Eqs. (\ref{eq_s1_s2}) and (\ref{bounds_s1_s2}).  

\begin{figure}[h]
\begin{center}
\includegraphics[width=0.7\textwidth]{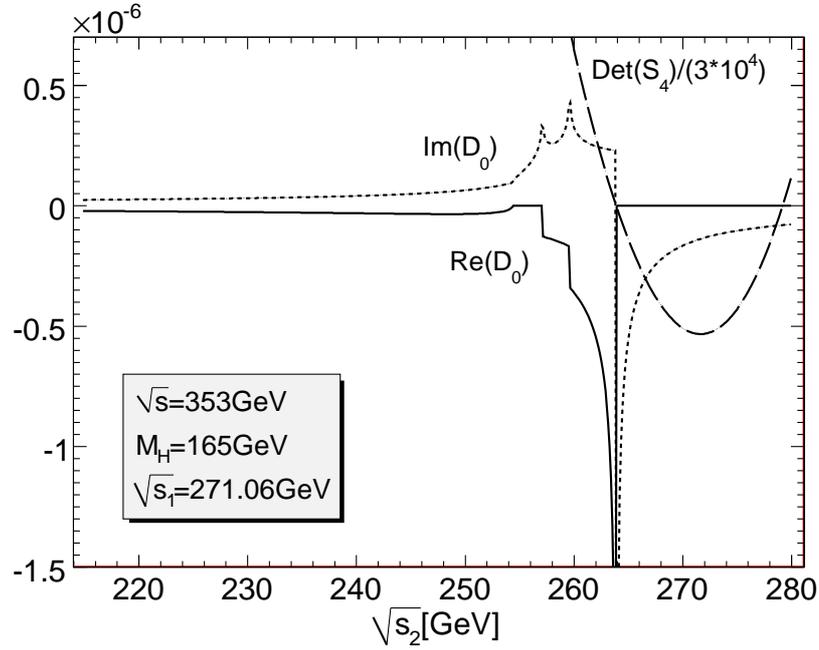}
\caption{\label{gbbH_img_real_det0_2D}{\em The imaginary, real
parts of $D_0$ and the Landau determinant as functions of $\sqrt{s_2}$.}}
\end{center}
\end{figure}

To investigate the structure of the singularities in more detail
let us fix $\sqrt{s_1}=\sqrt{2(m_t^2+M_W^2)}\approx 271.06\;$GeV,
so that the properties are studied for the single variable $s_2$.
{In order to do so, we have to find all the reduced diagrams containing $s_2$ as an external momentum squared.
There are $3$ diagrams shown in Fig.~\ref{fig_gbbH_sub-LLS}: one self-energy and two triangle diagrams.  
\begin{figure}[htb]
\begin{center}
\includegraphics[width=0.9\textwidth]{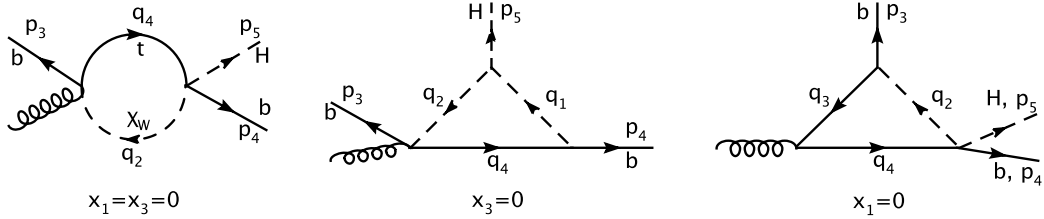}
\caption{\label{fig_gbbH_sub-LLS}{\em Three reduced diagrams of the box diagram in Fig.~\ref{landau_box}, that contain $s_2$ as an external momentum squared. The self-energy diagram has a normal threshold. 
The two triangle diagrams contain anomalous thresholds. We refer to subsection~\ref{appendix_3pt} for a detailed account of the $3$-point Landau singularity.}}
\end{center}
\end{figure}
}
The plots for the real and imaginary parts are shown in Fig.~\ref{gbbH_img_real_det0_2D}.
This figure is very educative. We see that
there are four discontinuities in the function representing the
real part of the scalar integral in the variable $\sqrt{s_2}$: 
\begin{itemize} 
\item As
$s_2$ increases we first encounter a discontinuity  at the normal
threshold $\sqrt{s_2}=m_t+M_W=254.38\;$GeV, representing $H b \ra
W t$. This corresponds to the solution (for the Feynman
parameters) $x_{1,3}=0$ and $x_{2,4}>0$ of the Landau equations {(see Fig.~\ref{fig_gbbH_sub-LLS}-left)}.
\item The second discontinuity occurs at the anomalous threshold
$\sqrt{s_2}=257.09\;$GeV of a reduced triangle diagram. This
corresponds to the solution $x_{3}=0$ and $x_{1,2,4}>0$ of the
Landau equations {(see Fig.~\ref{fig_gbbH_sub-LLS}-middle)}.  
{As discussed in subsection~\ref{appendix_3pt}, the singular 
position is given by
\bea
s_2^H=\fr{1}{2M_W^2} \left( M_H^2(m_t^2+M_W^2)-
M_H\sqrt{M_H^2-4M_W^2}(m_t^2-M_W^2) \right)\label{landaupole3_x3}
\eea
which gives $\sqrt{s_2}=257.09\;$GeV. 
\item The third discontinuity corresponds to the diagram of Fig.~\ref{fig_gbbH_sub-LLS}-right. The position of this Landau singularity is given by
\bea
s_2^s=\fr{1}{2m_t^2} \left( s(m_t^2+M_W^2)-
\sqrt{s}\sqrt{s-4m_t^2}(m_t^2-M_W^2) \right),\label{landaupole3_x3b}
\eea
which gives $\sqrt{s_2}=259.58\;$GeV.}
\item The last singular discontinuity is the
leading Landau singularity. The condition $\det(S_4)=0$ for the
box has two solutions which numerically correspond to
$\sqrt{s_2}=263.88\;$GeV or $\sqrt{s_2}=279.18\;$GeV. Both values
are inside the phase space, see Fig.~\ref{gbbH_img_real_det0_2D}.
However after inspection of the corresponding sign condition, only
$\sqrt{s_2}=263.88\;$GeV (with  $x_1 \approx 0.53, x_2 \approx
0.75, x_3 \approx 0.77$)  qualifies as a leading Landau
singularity. $\sqrt{s_2}=279.18\;$GeV has $x_1\approx -0.74,
x_2 \approx -0.75, x_3 \approx 1.07$ and is outside the physical region.
\end{itemize}
The nature of the LLS in Fig.~\ref{gbbH_img_real_det0_2D} can be
extracted  by using general formula (\ref{eq_T04h}). With the
input parameters given above, the Landau matrix has only one
positive eigenvalue at the leading singular point, {\it i.e.}
$K=1$. The leading singularity behaves as
\bea D_0^{div}=-\fr{1}{16M_W^2m_t^2\sqrt{\det(S_4)-i\eps}}.\label{d0_detS}
\eea
When approaching the singularity from the left, $\det(S_4)>0$,
the real part turns singular. When we cross the leading
singularity from the right, $\det(S_4)<0$, the imaginary part of
the singularity switches on, while the real part vanishes. In this
example, both the real and imaginary parts are singular because
$\det(S_4)$ changes  sign when  the leading singular point is
crossed. 

At the position of the two $3$-point sub-LLSs, 
the imaginary part shows logarithmic divergences while the real part has 
two finite negative jumps. This is similar to that shown in Fig.~\ref{fig_c0_width0} whose explanation 
is given in Eq.~(\ref{nature_LLS_3pt}). However, there is an important difference between Fig.~\ref{gbbH_img_real_det0_2D} and 
Fig.~\ref{fig_c0_width0} in the sign of the singularities. This is because the sign of the $3$-point LLS is $(-1)^3$ (see Eq.~(\ref{TN0_LLS}) for $N=3$) 
while the sign of the $3$-point sub-LLS is $(-1)^4$ (see Eq.~(\ref{TN0_sub-LLS}) for $N=4$).  
\subsection{Conditions on external parameters to have LLS}
\label{sect_landau_range}
\begin{figure}[htbp!]
\begin{center}
\includegraphics[width=0.6\textwidth]{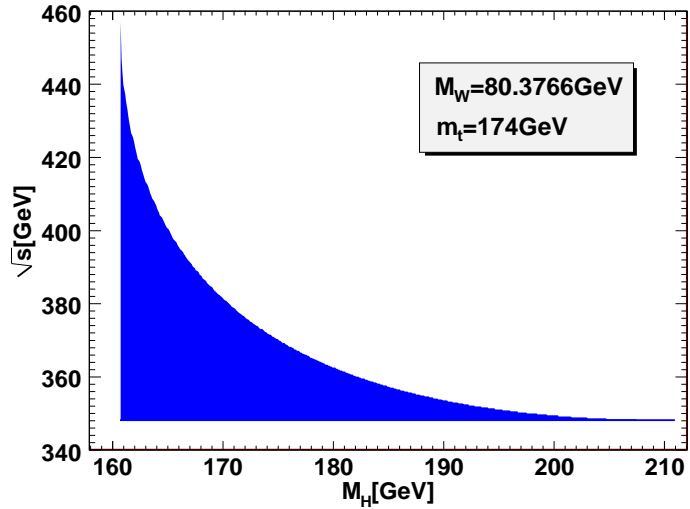}
\caption{\textit{The region of leading landau singularity.}} \label{landau_range}
\end{center}
\end{figure}
The conditions for the opening of normal thresholds give the lower bounds on external parameters $M_H^2$, $s$, $s_{1,2}$ 
as given in Eqs. (\ref{cond_MH_s_0}) and (\ref{cond_s1_s2_0}). However, we have learnt from section~\ref{LLS_terminate} that the LLS can terminate 
as those external parameters increase. The conditions for the termination of LLS define the upper bounds of those parameters, as illustrated in 
subsection~\ref{appendix_3pt} for the case of $3$-point function. 

However, the situation becomes much more complicated in the case of $4$-point function since there are $4$ variables ($M_H$, $s$, $s_1$, $s_2$) and $2$ parameters ($M_W$, $m_t$) involved. We will show that there are two ways to find out the upper bounds by using numerical and analytical methods. 
%
%

We first explain the numerical method to find the upper bounds of $M_H^2$ and $s$. This is done by a very simple Fortran 
code which includes the following steps. For each $M_H^2$, what is the condition on $s$ to have a LLS in the phase space? 
The Landau determinant takes the form $\det(Q_4)=as_2^2+bs_2+c$ as given in Eq.~(\ref{det_abc}). 
If $\Delta=b^2-4ac<0$ then there is no LLS. If $\Delta=b^2-4ac\ge 0$ then the Landau determinant can 
vanish at $2$ points
\bea
s_2^\mp=\fr{-b\mp\sqrt{\Delta}}{2a}.
\label{LLS_s2}
\eea 
If $s_2^\mp$ are not in the phase space defined by Eq.~(\ref{phys_region_ggbbH}) then there is no LLS. 
If at least one root is in the phase space then we have to check condition (\ref{landau_cond1}). If 
this condition is satisfied then there is a LLS. The result is shown in Fig.~\ref{landau_range}. 
We conclude that the LLS occurs when $2M_W\le M_H< 211$GeV and $2m_t\le \sqrt{s}< 457$GeV. 
The range of LLS region 
depends on $M_W$ and $m_t$. If $m_t/M_W\le 1$ then the first two conditions in Eq.~(\ref{landau_cond0}) can never be satisfied. In 
particular, if $m_t/M_W=1$ then the Landau determinant can vanish but the sign condition cannot be realised. 
When $M_H>211$GeV or $\sqrt{s}>457$GeV the Landau determinant $\det(Q_4)$ can vanish inside the phase space but the sign condition
$x_i > 0$ cannot be fulfilled.
\begin{figure}[h]
\begin{center}
\includegraphics[width=0.8\textwidth]{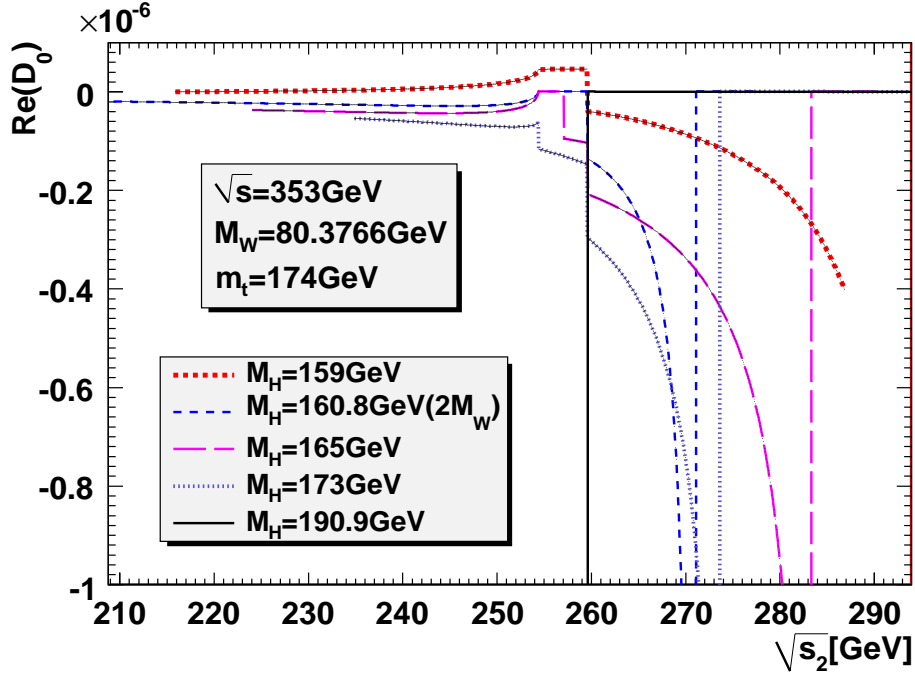}
\caption{\label{ggbbH_d0_width0}{\em The real part of $D_0$ as a function of $\sqrt{s_2}$ for various values of $M_H$. 
For $M_H=2M_W$ we have taken $s_1=2(m_t^2+M_W^2)$. For the other cases, we take 
$s_1=260$GeV.}}
\end{center}
\end{figure}

Before explaining the analytical method, we would like to show pictorially how the LLS moves and terminates as $M_H$ increases. 
We fix $\sqrt{s}=353$GeV as in Fig.~\ref{gbbH_img_real_det0_2D}. We will increase $M_H$ and look at the behaviour of the scalar $4$-point 
function $D_0$ as a function of $s_2$. We will explain what value of $s_1$ should be chosen. 
The result is shown in Fig.~\ref{ggbbH_d0_width0} which 
is just an extension of Fig.~\ref{gbbH_img_real_det0_2D}. The key points to understand this picture are as follows. 
At most, there are four discontinuities in the real part as a function of $s_2$ as already explained (see Fig.~\ref{gbbH_img_real_det0_2D}). 
When we fix $s$ and increase $M_H$, two of them are fixed and the other two move. The normal threshold is fixed at $\sqrt{s_2^{tW}}=m_t+M_W=254.3766$GeV, 
one $3$-point sub-LLS is fixed at $\sqrt{s_2^s}=259.576$GeV as given in Eq.~(\ref{landaupole3_x3b}). The position of the other $3$-point sub-LLS depends 
only on $M_H$ as given in Eq.~(\ref{landaupole3_x3}). The position of LLS depends on $M_H$ and $s_1$ as given in Eq.~(\ref{LLS_s2}). 
As $M_H$ increases, the Fig.~\ref{ggbbH_d0_width0} shows:
\begin{itemize}  
\item For $M_H=159\text{GeV}<2M_W$, only the normal threshold and the three-point sub-LLS $\sqrt{s_2^s}$ show up. 
\item For $M_H=2M_W$ the second three-point sub-LLS appears at 
$\sqrt{s_2^H}=271.059$GeV. One has to change $s_1$ in the range defined by Eq.~(\ref{phys_region_ggbbH}) with the condition $s_1\ge (m_t+M_W)^2$ to make a LLS appeared. 
It is easy to find out that the LLS only occurs when $\sqrt{s_1}=\sqrt{2(m_t^2+M_W^2)}=271.059$GeV and the LLS position coincides with the position of the three-point singularity $\sqrt{s_2^H}$. At this LLS point, the sign condition has the form $x_i=0/0$ for $i=1,2,3$. We have the ordering 
$\sqrt{s_2^{tW}}<\sqrt{s_2^{s}}<\sqrt{s_2^{H}}=\sqrt{s_2^{LLS}}$.
\item For $M_H=165$GeV then $\sqrt{s_2^H}=257.088$GeV and we see that the LLS starts showing up at $\sqrt{s_2}\approx 283.5$GeV when $\sqrt{s_1}\approx 260$GeV (before this value there is no LLS) and moves to the left as $\sqrt{s_1}$ increases. 
We have the ordering 
$\sqrt{s_2^{tW}}<\sqrt{s_2^{H}}<\sqrt{s_2^{s}}<\sqrt{s_2^{LLS}}$.
\item For $M_H\approx 173$GeV then $\sqrt{s_2^H}=254.3766$GeV coinciding with the normal threshold and we see that the LLS starts showing up at $\sqrt{s_2}\approx 274$GeV when $\sqrt{s_1}\approx 260$GeV and moves to the left as $\sqrt{s_1}$ increases. After this value of $M_H$, the $\sqrt{s_2^H}$-three-point singularity disappears from the physical region and the LLS continues moving to the left as $M_H$ increases. 
We have the ordering 
$\sqrt{s_2^{tW}}=\sqrt{s_2^{H}}<\sqrt{s_2^{s}}<\sqrt{s_2^{LLS}}$.
\item For the special value $M_H=190.877$GeV, the LLS starts showing up and coincides with the fixed three-point singularity at $\sqrt{s_2}=\sqrt{s_2^s}=259.576$GeV when $\sqrt{s_1}\approx 260$GeV $\approx \sqrt{s_2^{s}}$ and moves off the physical region as $\sqrt{s_1}$ increases. If $M_H>190.877$GeV then the LLS disappears from the physical region. 
\end{itemize}
We conclude that for $\sqrt{s}=353$GeV, the upper bound of $M_H$ to have a LLS is $190.877$GeV which is consistent with Fig.~\ref{landau_range}. The above picture also leads to the conclusion that the upper bound is determined when the LLS coincides with a sub-LLS. This important remark agrees with the 
explanation on the termination of LLS given in section~\ref{LLS_terminate}.

We now are in the position to explain the equation of the upper bounds of $M_H$ and $\sqrt{s}$, which is shown in Fig.~\ref{landau_range}, by using 
analytical method. The key points are as follows. The termination of LLS occurs when {\em all} LLSs coincides with a three-point sub-LLS. One should keep 
in mind that for each value of $(M_H,\sqrt{s})$ there may be a lot of LLSs corresponding to different values of $(s_1,s_2)$ which make a $s_{1,2}$-LLS-range. At the termination of LLS, this $s_{1,2}$-LLS-range must become a point in order to coincide with a three-point sub-LLS. 

If we express $\det(Q_4)$ as a quadratic polynomial of $s_2$, there are $2$ three-point sub-LLSs whose positions are given in Eqs. (\ref{landaupole3_x3}) and 
(\ref{landaupole3_x3b}). Without losing any generality, we assume that the former coincides with the LLS. Thus, at the termination point we have 
\bea
s_2=s_2^H\hs \text{and} \hs \det(Q_4)=0.
\label{eq_s2_s2s}
\eea
However, since our problem is completely symmetric under exchange of $s_1$ and $s_2$, one should observe the same thing when expressing $\det(Q_4)$ 
as a quadratic polynomial of $s_1$. The LLS therefore coincides with the $s_2^H$ three-point sub-LLS when 
\bea
s_1=s_2^H \hs \text{and} \hs \det(Q_4)=0.
\eea
We conclude that the LLS terminates when $s_1=s_2$. With this information, the Eq. $\det(Q_4)=0$ gives\footnote{Other roots do not satisfy the positive 
sign condition of the three-point sub-LLS.}
\bea
s_1=s_2=2(m_t^2+M_W^2)-\sqrt{(s-4m_t^2)(M_H^2-4M_W^2)}.
\label{s2-root_LLS}
\eea
Equating the Eq.~(\ref{s2-root_LLS}) with Eq.~(\ref{landaupole3_x3}), we get
\bea
\sqrt{(s-4m_t^2)}&=&\frac{1}{2 M_W^2} \bigg(M_H
(m_t^2-M_W^2)-(m_t^2+M_W^2)\sqrt{(M_H^2-4M_W^2)} \biggr).\hs\hs
\label{curve_s_MH_max}
\eea 
We observe that this equation shows, in a very transparent way, that
all thresholds:
 $$m_t > M_W, M_H\ge 2 M_W, \sqrt{s} \ge 2 m_t$$  need
to be open simultaneously. We can invert
Eq.~(\ref{curve_s_MH_max}) to write the solution in terms of
$M_H$. To arrive at the same result, it is more judicious however
to go through exactly the same steps but choosing $s_2^s$
instead of $s_2^H$. We derive
\bea
\sqrt{(M_H^2-4 M_W^2)}&=&\frac{1}{2 m_t^2} \bigg(\sqrt{s}
(m_t^2-M_W^2)-(m_t^2+M_W^2)\sqrt{(s-4 m_t^2)} \biggr).
\label{curve_MH_s_max}
\eea
The maximum value of $M_H$ ($\sqrt{s}$) is obtained by setting $\sqrt{s}=2m_t$ ($M_H=2M_W$), \ie when the LLS, the two $3$-point sub-LLSs and the normal threshold coincide. We have
\bea
&&4M_W^2\le M_H^2\le 4M_W^2+\fr{(m_t^2-M_W^2)^2}{m_t^2},\crn
&&4m_t^2\le s\le 4m_t^2+\fr{(m_t^2-M_W^2)^2}{M_W^2}.
\label{bounds_MH_s}
\eea
Numerically, it means
\bea
348\text{GeV}\le \sqrt{s}\le 457.053\text{GeV}\hs \text{and}\hs 160.7532\text{GeV}\le M_H\le 211.129\text{GeV}.
\eea
Of course, those analytical formulae agree with the curve obtained by numerical method in Fig.~\ref{landau_range}. 

There are many other ways to derive Eq.~(\ref{curve_s_MH_max}). Let us explain a very practical way. In order to obtain Eq.~(\ref{curve_s_MH_max}), we 
have assumed that $s_{1,2}=s_2^H$ as in Eq.~(\ref{eq_s2_s2s}). However, we can also equally assume that $s_{1,2}=s_2^s$ \ie the LLS coincides with the 
other three-point sub-LLS. Of course, this assumption does lead to the same result\footnote{This will fail if the equation of the upper bounds of $s$ and $M_H^2$ (the maximum curve) is not analytic. If it happens, each assumption will give a part of the maximum curve.}. But this also means that, in order to 
have a unique result for the equation of upper bounds, at the termination of LLS one must have
\bea
s_2^s=s_2^H.
\label{eq_s2s_s2H}
\eea
Thus, a very practical way to quickly obtain the result Eq.~(\ref{curve_s_MH_max}) is equating two equations (\ref{landaupole3_x3}) and (\ref{landaupole3_x3b}), 
without caring about Eq. $\det(Q_4)=0$.

The same argument can be repeated for the two parameters $s_{1,2}$ to get their upper bounds. Namely, we express $\det(Q_4)$ as a quadratic polynomial of 
$M_H^2$. The two 3-point singularities are found by considering the two reduced triangle diagrams containing $p_5$ as an external momentum. The equation of the upper bounds is obtained by requiring that $M_H^2(s_1)=M_H^2(s_2)$ (similar to Eq.~(\ref{eq_s2s_s2H})):
\bea
M_H^2&=&\fr{s_2(m_t^2+M_W^2)-(m_t^2-M_W^2)^2-(m_t^2-M_W^2)\sqrt{\lambda(s_2,m_t^2,M_W^2)}}{2m_t^2}\crn
&=&\fr{s_1(m_t^2+M_W^2)-(m_t^2-M_W^2)^2-(m_t^2-M_W^2)\sqrt{\lambda(s_1,m_t^2,M_W^2)}}{2m_t^2},
\eea
which is compatible with the sign condition $M_H^2\le [s_{1,2}(m_t^2+M_W^2)-(m_t^2-M_W^2)^2]/(2m_t^2)$ (the "plus" solution does not satisfies this); the equation writes
\bea
s_1-s_2=\frac{m_t^2-M_W^2}{m_t^2+M_W^2} \left[
\sqrt{\lambda(s_1,m_t^2,M_W^2)}  - \sqrt{\lambda(s_2,m_t^2,M_W^2)}
\right].
\label{eq_s1_s2}
\eea
The maximum of $s_2$ is got from this equation by setting $s_1=(m_t+M_W)^2$. We get
\bea
(m_t+M_W)^2\le s_{1,2}\le (m_t+M_W)^2+\fr{(m_t^2-M_W^2)^2}{m_tM_W},
\label{bounds_s1_s2}
\eea
which gives
\bea
254.3766\text{GeV}\le \sqrt{s_{1,2}}\le 324.442\text{GeV}.
\eea
From Eq.~\ref{eq_s1_s2}, one can make a very similar plot like the one in Fig.~\ref{landau_range}. We remark, by looking at Eqs. (\ref{bounds_MH_s}) and 
(\ref{bounds_s1_s2}), that there is no LLS if $m_t=M_W$ (because $\sqrt{s}>M_H$). 

A good question to ask is "What does the termination of the LLS mean physically?" The answer is very simple if one uses the physical interpretation of 
Coleman and Norton discussed in section \ref{subsection_landau_condition}. The relation between the Feynman parameter $x_i$ and the proper time 
$d\tau_i=m_ix_i$ where $m_i=M_W,m_t$ means that at the termination of the LLS (when $M_H$ or/and $\sqrt{s}$ are large enough) at least one internal particle has zero proper time, \ie it reaches the velocity of light.

\section{The width as a regulator of Landau singularities}
We will argue that the LLS which is not integrable at the level of one-loop amplitude squared can be tamed by introducing a width for 
unstable particles in the loops. 
\begin{figure}[htbp!]
\centering \mbox{
\includegraphics[width=0.45\textwidth]{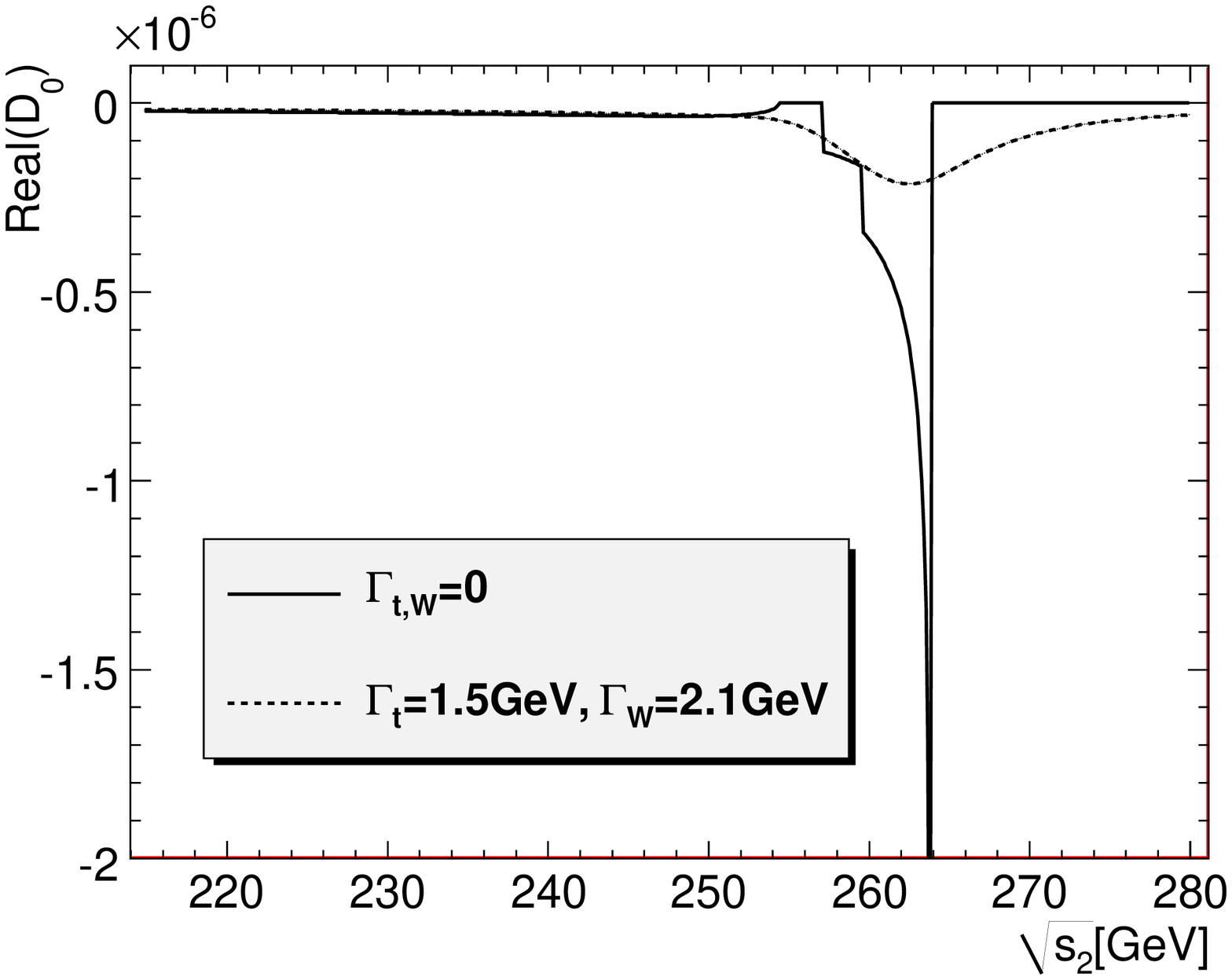}
\includegraphics[width=0.45\textwidth]{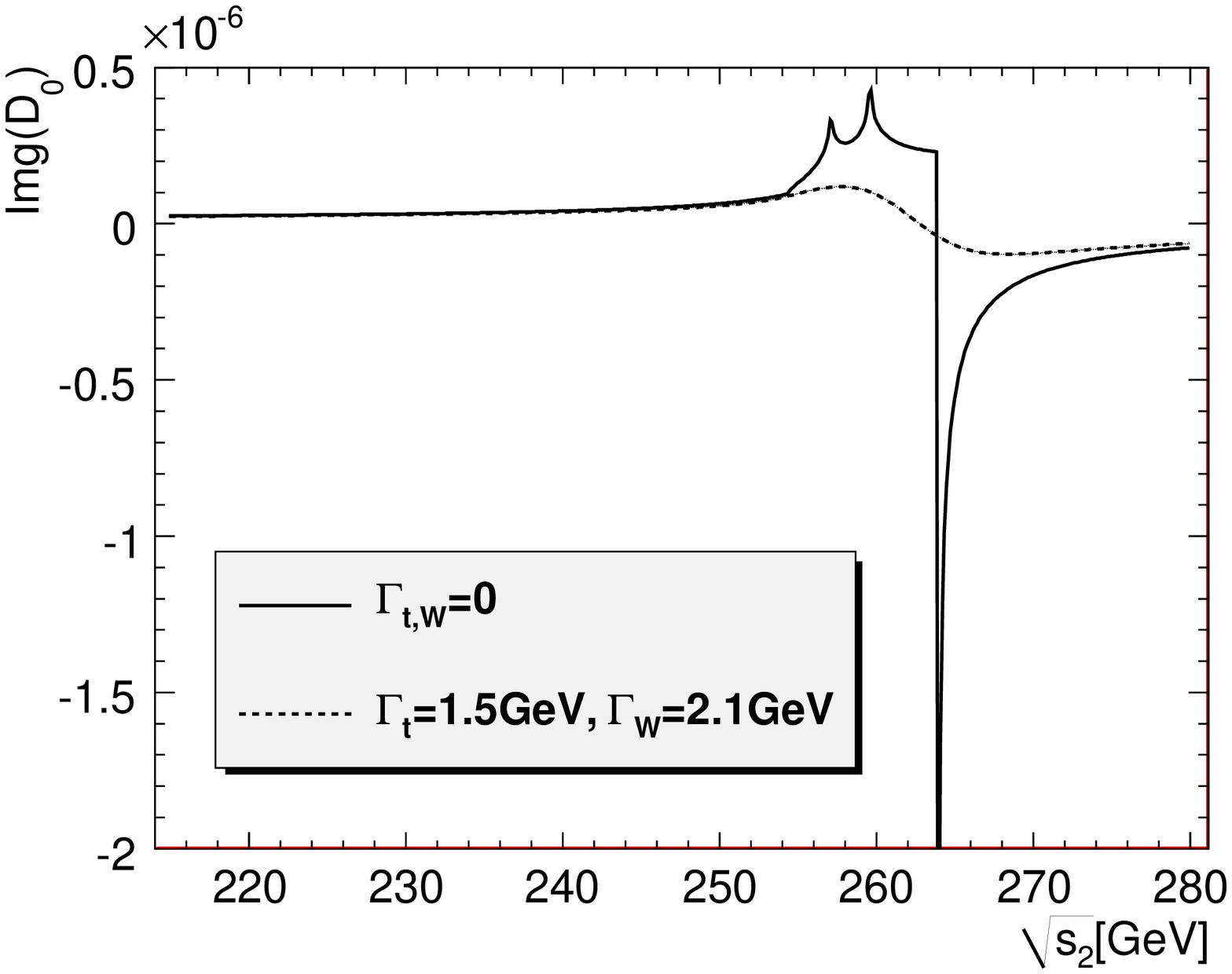}}
\caption{\textit{Effect of the width of the
$W$, $\Gamma_W$ and the top, $\Gamma_t$, on the real and imaginary
parts of the scalar four-point function.}} \label{dessin-width-smooth}
\end{figure}

It is well-known that if internal unstable particles are on their mass shell, 
then one, in general, has to introduce a width for those particles. 
For example in the process $e^+e^-\to f\bar{f}$ at tree level, there is 
a pole in a S-channel diagram associated with the $Z$ boson. 
It is obvious that one has to include the $Z$-width to solve this problem. 
On the other hand, there are several processes with unstable internal particles 
where the width effect is very small hence can be safely neglected. A famous example is a T-channel 
diagram. In that case, the internal particle is forced to be far away from the on-shell region. 

In our case at hand, the LLS occurs when all loop internal particles (the top-quark and W Goldstone boson) are on-shell. Thus the width 
effect can be important. We take the simple prescription of a fixed width and make the substitution
\bea
m_t^2\to m_t^2-im_t\Gamma_t,\hs M_W^2\to M_W^2-iM_W\Gamma_W.\label{mass_real_complex}
\eea
Mathematically, the width effect is to move Landau singularities into the complex plane, so they do not occur in the physical region (the real axis). 

Applying rules (\ref{mass_real_complex}) to the case of the scalar four-point function defined by Eq. (\ref{def_d0}) one sees in Fig. \ref{dessin-width-smooth} that indeed the width regulates the
LLS and gives a smooth result that nicely interpolates with the
result at zero width away from the singularity. The normal
threshold and the 3-point sub-leading singularity are also
softened. The real part of the 4-point function still shows a
smooth valley at the location of the LLS after regularisation. For
the imaginary part we note that after introducing the width the
LLS singularity is drastically reduced with a contribution of the
order of the sub-leading singularity. \\
\noi As we will explain in the next section and in more detail in
Appendix~\ref{appendix-box-integral} the introduction of the width in a four-point function
requires careful extension of the usual 4-point function
libraries.

In our calculation of Yukawa corrections where all the relevant couplings 
depend only on the top-quark mass and the vacuum expectation value $\upsilon$, 
we will keep $m_t$ and $\upsilon$ real while applying rules (\ref{mass_real_complex}) to all the loop 
integrals\footnote{In a general calculation of full electroweak correction and if one chooses the input parameters 
to be $\{\alpha(0), M_W, M_Z, m_t, \ldots\}$ then rules (\ref{mass_real_complex}) should be applied everywhere in the calculation so that  
the EW gauge invariance can be preserved. This idea of doing analytical continuation on gauge boson masses is the philosophy of the complex-mass scheme \cite{Denner:1999gp, denner_ee_4fb}. For a practical discussion of methods to deal with unstable particles, see Denner's lecture notes at \cite{denner_unstable}.}. 

\section{Calculation and checks}
The one-loop calculation of this chapter is exactly the same as in the real mass case (see section \ref{bbH1_cal}) except the fact that we now have to consider the tensorial and scalar loop integrals with complex masses. 

LoopTools \cite{looptools} can handle the complex masses up to $3$ point functions. The $5$ point functions are reduced to 
$4$ point functions \cite{denner_5p, looptools_5p}. The tensorial $4$ point functions are reduced to the scalar $4$ point function and $3$ point functions. We therefore have to calculate only the scalar $4$ point function. The analytical calculation of scalar $4$-point function
with complex masses in the most general case is practically intractable. The standard technique of 't Hooft and Veltman \cite{hooft_velt} has some restriction on the values of external momenta. In particular, the method works if at least one of the external momenta is lightlike. In that case, the result is written in terms of 
$72$ Spence functions. In our present calculation, there are at least $2$ lightlike external momenta in all boxes. If the 
positions of two lightlike momenta are opposite then we can write the result in terms of $32$ Spence functions. If the two 
lightlike momenta are adjacent, the result contains $60$ Spence functions. 
The detailed derivation and results are given in the Appendix~\ref{appendix-box-integral}. We have implemented those analytical formulae for the case of two massless external momenta into a code and added this into LoopTools \footnote{The implementation for the case of one massles external momentum is straight forward. However, we have not done this yet since it is not necessary for our present calculation.}. All the five point functions which have problem with Landau singularities in our calculation have $4$ external massless momenta (two gluon and two bottom quarks). Thus they can always be reduced to a set of $4$- point functions with at least $2$ massless external momenta. 

Checks on the loop integrals with complex internal masses: for all the tensorial and scalar loop integrals ($4$- and $5$- point functions), we have performed a consistency check where we have made sure that the results with complex internal masses become asymptotic to the ones with real internal masses in the limit $widths\to 0^+$. For the scalar loop integrals, the results are compared to the ones calculated numerically in the limit of large widths, {\it e.g.} $\Gamma_{t,W}=100$GeV. Furthermore, for the scalar box integrals the results can be checked by using the segmentation technique described in \cite{Boudjema:2005hb}. The idea is the following. At the boundary of the phase space, the $4$- point functions can be written as a sum of $4$ three point functions. The $3$-point functions with complex masses can be calculated by using LoopTools. In this way, we have verified that the results of the scalar $4$-point functions are correct at the boundary of phase space. We have also carried out a comparison with a purely
numerical approach based on an extension of the $\eps$-extrapolation technique \cite{yuasa:2004}. We have
found perfect agreement\footnote{We thank F. Yuasa for sending us the results of the extrapolation technique.}. 

For the whole calculation, we have performed two checks at the amplitude level. First, by taking the limit $widths\to 0^+$ we have observed that the results approach to the results calculated with real internal masses. This again is just a consistency check. Second and it is the most important check where we have verified that the results calculated with complex internal masses are QCD gauge invariant.     

As the LLS is integrable at interference level, the NLO calculation with $\la_{bbH}\neq 0$ performed in chapter~\ref{chapter_bbH1} can be trivially extended to the region of $M_H\ge 2M_W$ by using the same method without introducing widths for unstable internal particles. However, there is a small problem related to the universal correction $(\delta Z_H^{1/2}-\delta\upsilon)$ where $\delta Z_H^{1/2}$ related to the derivative of the Higgs
two-point function becomes singular when $M_H$ equal to $2M_W$ or $2M_Z$. This problem is separately treated by introducing the widths of the $W$ and the $Z$. To be complete, the results are given in section~\ref{section_nlo}.   

\section{Results in the limit of vanishing $\la_{bbH}$}
\label{section_nnlo}
The input parameters and kinematical cuts are the same as given in section~\ref{section_bbH_result1}. 
We write here additional parameters related to the widths of unstable particles appearing in the calculation: 
\bea \Gamma_{W}=2.1\GeV, \hs \Gamma_{Z}=2.4952\GeV, 
\eea
the top-quark width is calculated at the tree level in the SM
\bea
\Gamma_t=\fr{G_\mu(m_t^2-M_W^2)^2(m_t^2+2M_W^2)}{8\pi\sqrt{2}m_t^3}\approx 1.5\GeV
\label{eq_top-width}
\eea
where the bottom-quark mass has been neglected.  
\subsection{Total cross section}
\begin{figure}[htp]
\begin{center}
\mbox{\includegraphics[width=0.45\textwidth]{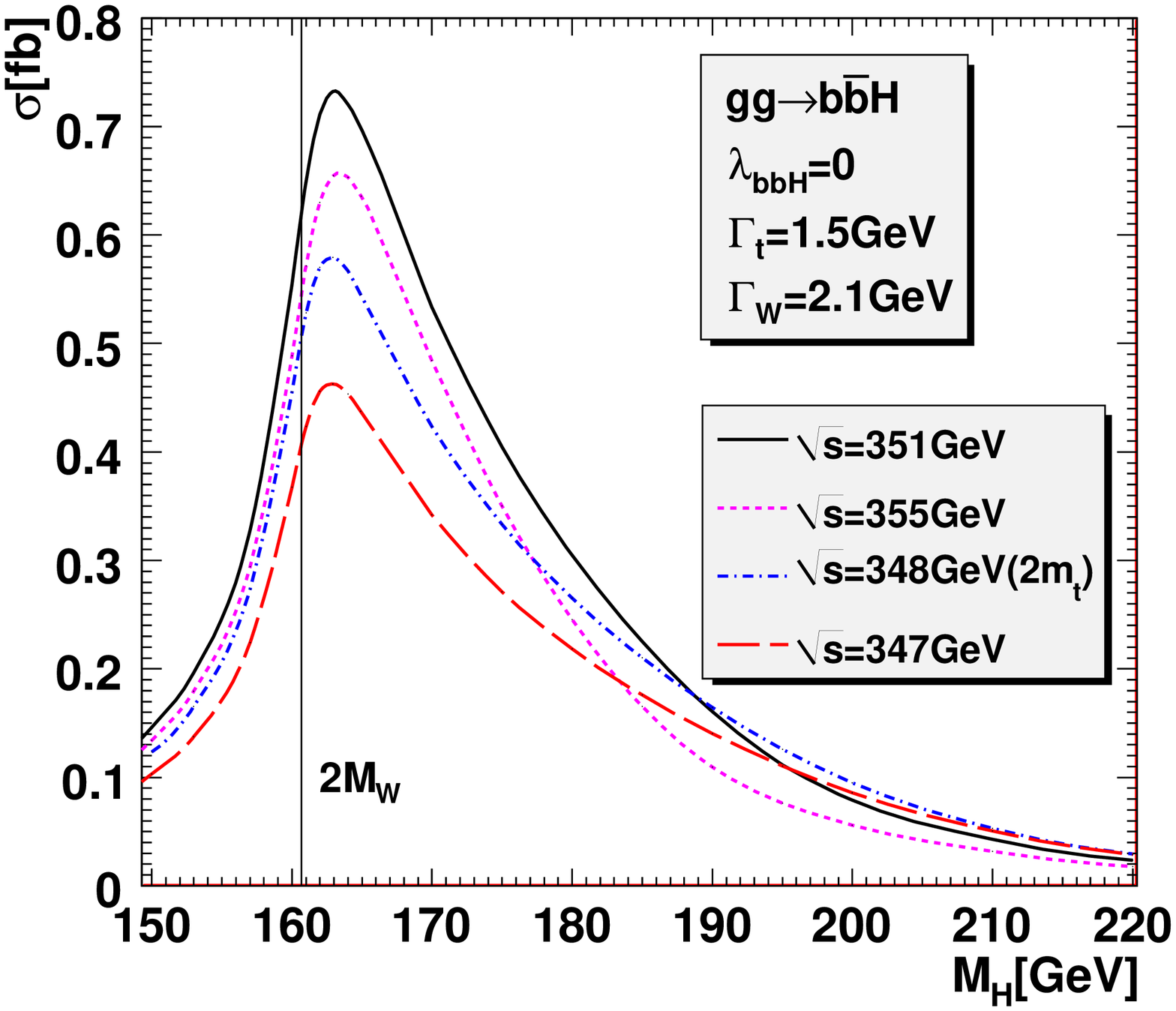}
\hspace*{0.075\textwidth}
\includegraphics[width=0.45\textwidth]{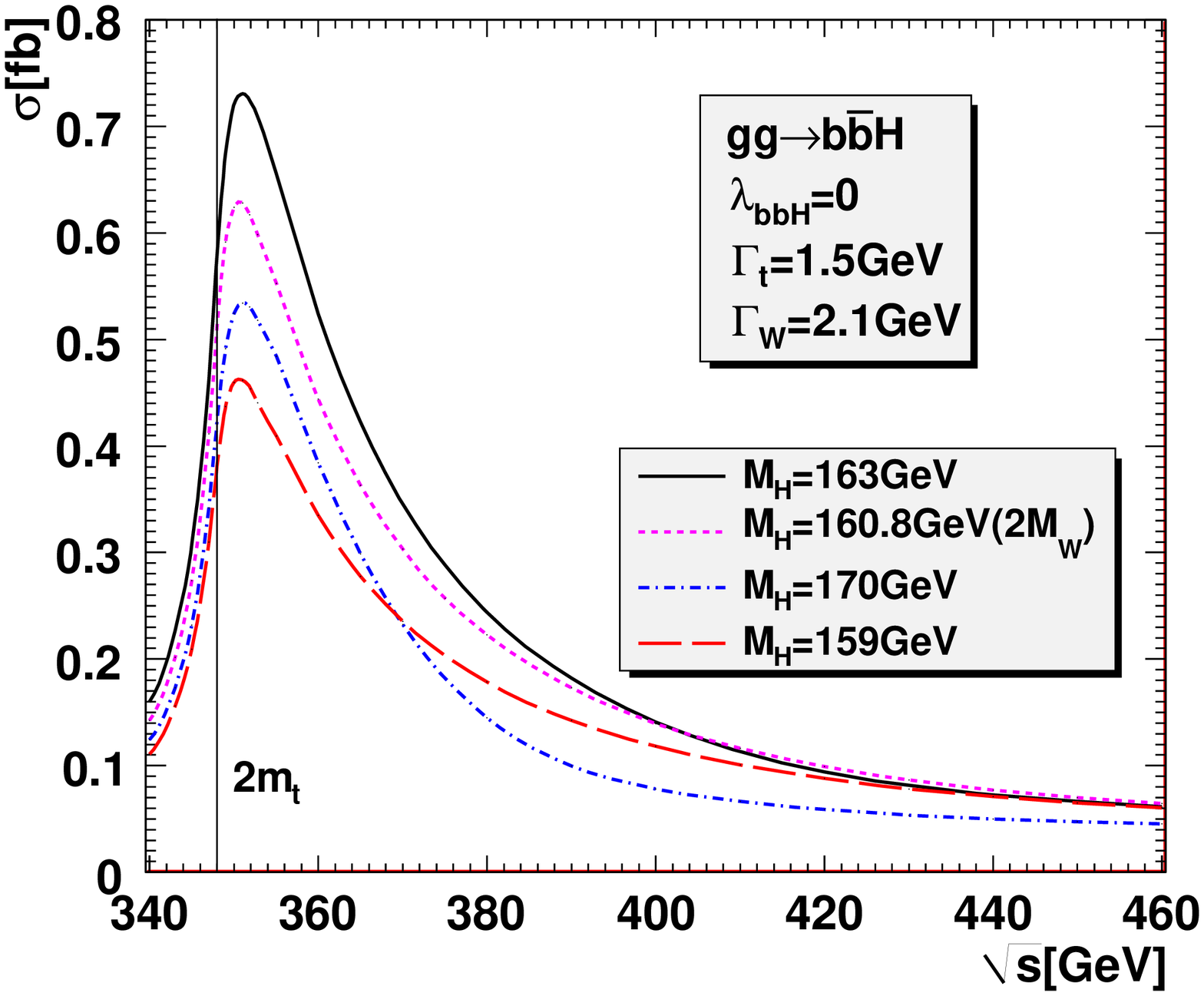}}
\caption{\label{gg_LL_rs}{\em Left: the cross section for the subprocess $gg\to b\bar{b}H$ as functions of $M_H$ for 
various values of $\sqrt{s}$ including the case $\sqrt{s}=2m_t=348\GeV$. Right: the cross section for the subprocess $gg\to b\bar{b}H$ as functions of $\sqrt{s}$ for 
various values of $M_H$ including the case $M_H=2M_W=160.7532\GeV$.}}
\end{center}
\end{figure}
We start with the cross section in the case where $\la_{bbH} = 0$. In section \ref{section_bbH_result1} we reported on results
up to $M_H = 150$GeV that showed that this cross section was rising fast as one approached
the threshold $M_H = 2M_W$. Beyond this threshold our integrated cross sections showed
large instabilities. As discussed in subsection \ref{bbH_landau_4pt} this is due to the appearance of a leading
singularity which as we have advocated can be cured by the introduction of a width for
the unstable top-quark and W gauge boson. Before convoluting with the gluon distribution function let us briefly look at the behaviour of the
partonic cross section $\ggbbH$ paying a particular attention to this leading Landau
singularity region shown in Fig. \ref{landau_range}, see also Eq. \ref{bounds_MH_s}. 

Figs \ref{gg_LL_rs} show that indeed the widths do regulate the cross section. Moreover it is within
the range of LLS that the cross section is largest. The (highest) peak
of the cross section occurs for a Higgs mass about $163$GeV (slightly higher than the normal threshold value $2M_W=160.7532$GeV) and 
for $\sqrt{s}\approx 351$GeV (slightly higher than the normal threshold value $2m_t=348$GeV). The peak positions are slightly shifted from 
the normal threshold values, this is due to the width effect. The reason for the peak to occur at the normal threshold position is that {\em all} Landau singularities (leading and sub-leading) start showing up at this point and the size of Landau singularity region is largest at the position of 
normal threshold, see Fig. \ref{landau_range}. To see the LLS effect, we look at the two curves $\sqrt{s}=347$GeV and $\sqrt{s}=2m_t=348$GeV of Fig. \ref{gg_LL_rs}(left). For the former, there is no LLS and the peak effect is due to sub-LLSs (the normal thresholds and $3$-point sub-LLSs). The later includes additional LLS contribution, which is significant even after being regulated. 
We emphasize that although the $4$-point LLS is special in the sense that it 
is not integrable at the level of one-loop amplitude squared one should not overlook this point and use it as an account for the bulk of the large correction around the normal threshold position. The $3$-point (sub-) LLSs and normal thresholds have also significant contributions.   
\begin{figure}[h]
\begin{center}
\mbox{\includegraphics[width=0.45\textwidth]{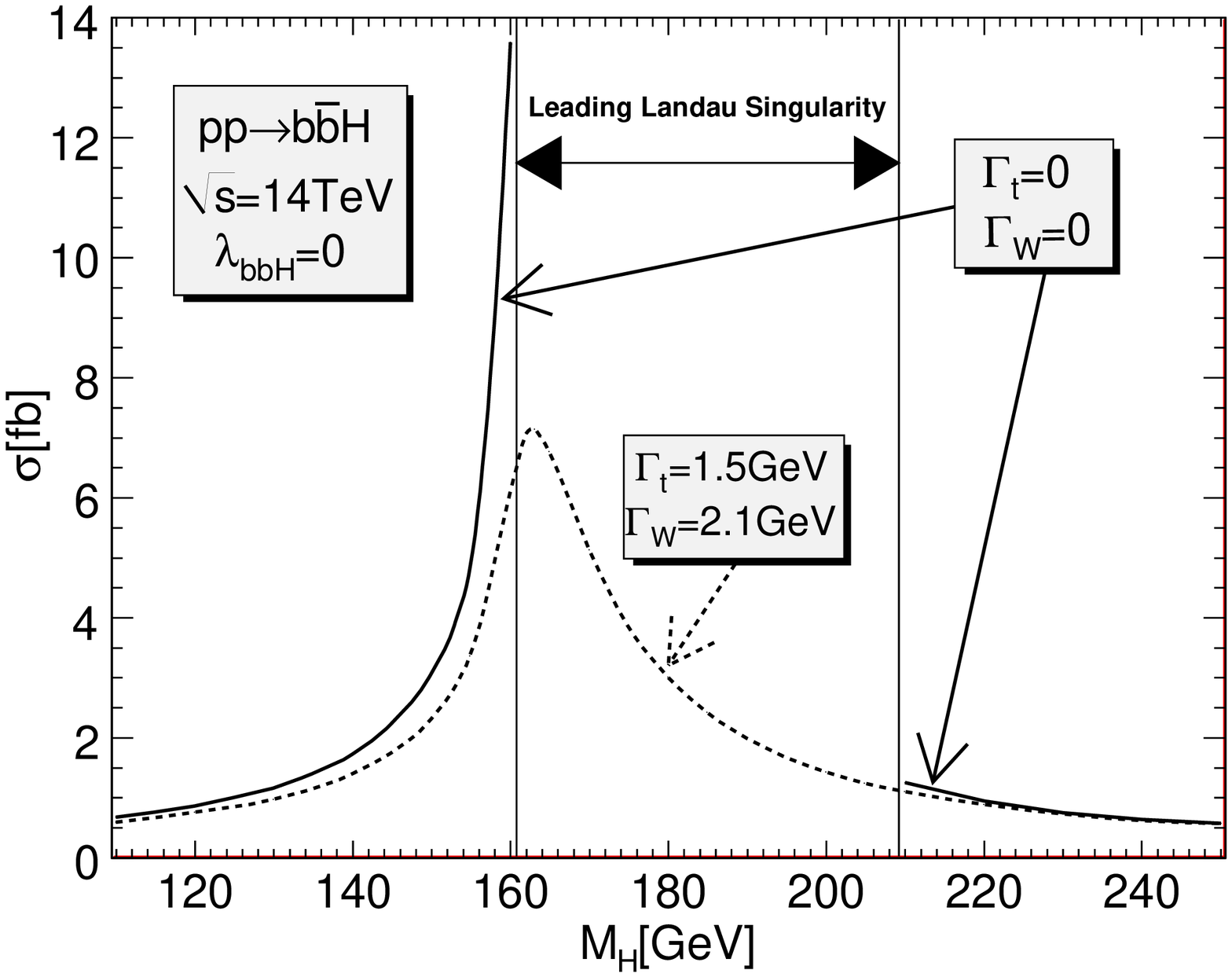}
\hspace*{0.075\textwidth}
\includegraphics[width=0.45\textwidth]{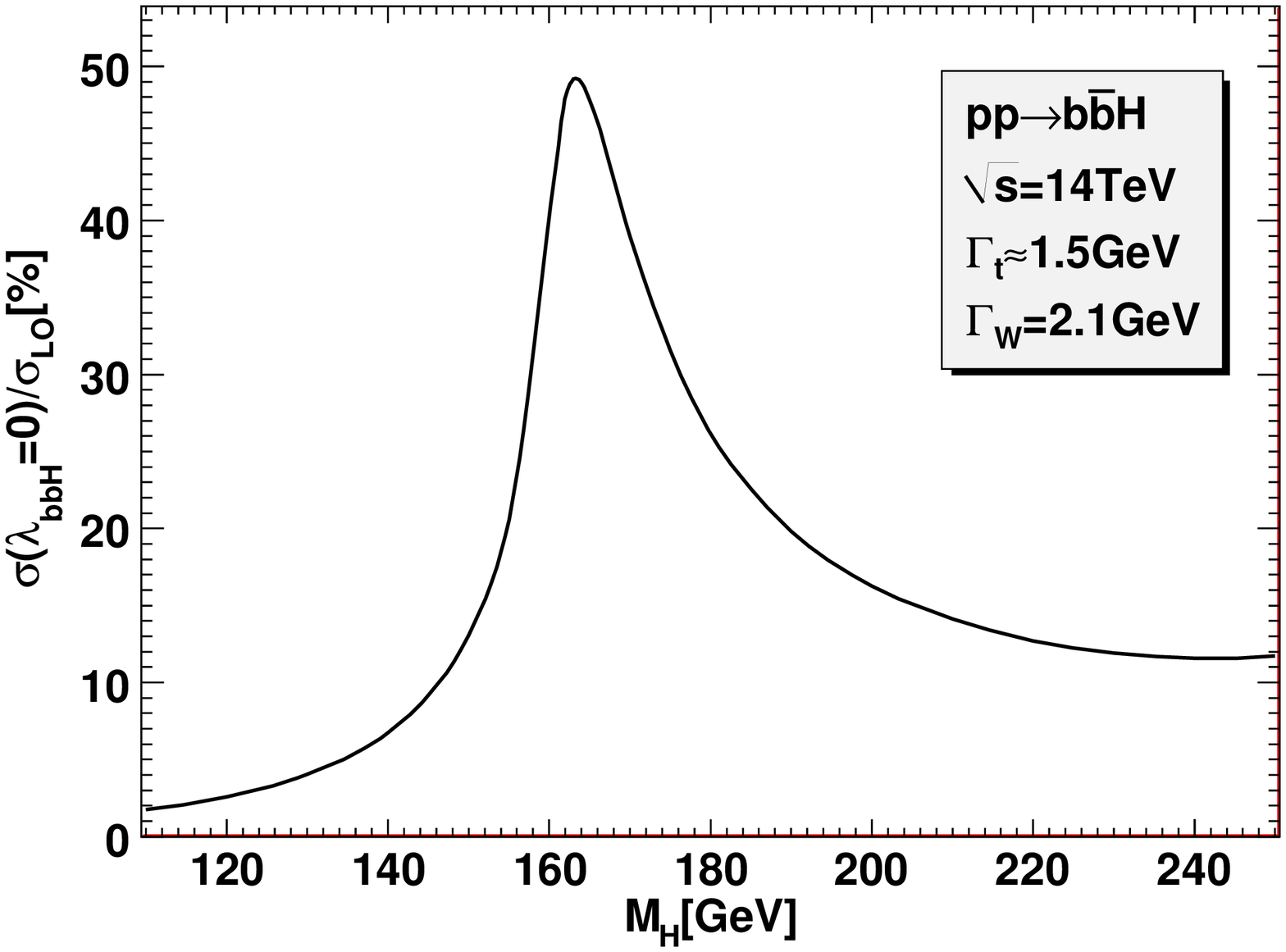}}
\caption{\label{p_LL_mH}{\em Left: the one-loop induced cross section as
a function of $M_H$ in the limit of vanishing bottom-Higgs Yukawa
coupling for two cases: with and without widths. Right: the percentage correction of
the contribution with widths relative to the tree level cross section calculated with
$\la_{bbH}\neq 0$.}}
\end{center}
\end{figure}

The cross section at the $pp$ level for the $14$TeV centre of mass energy (LHC) as a function of
the Higgs mass is shown in Fig. \ref{p_LL_mH} taking into account the width of the top-quark and the
W gauge boson. For comparison we also show the cross section without the width effect outside the
leading Landau singularity range of $M_H$. The sharp rise above $M_H > 150$GeV is nicely
tamed. On the other hand note that on leaving the leading Landau singularity region
around $M_H = 211$GeV, the width effect is much smaller and the figures suggest that
one could have entered this region from the right without having recourse to introducing
a width. Indeed our numerical integration routine over phase space with the default
LoopTools library does not show any bad behaviour until we venture around values of
$2M_W < M_H < 200$GeV. The reason for this can be understood by taking a glance at
Fig. \ref{landau_range}. For $200\text{GeV} < M_H < 211$GeV the singularity region is considerably shrunk so that one is
integrating over an almost zero measure. The effect of the widths outside the singularity
region is to reduce the cross section for $M_H =$ $120$GeV, $140$GeV and $150$GeV by\footnote{The relative difference is defined by $[\sigma(\Gamma=0)-\sigma(\Gamma\neq 0)]/\sigma(\Gamma\neq 0)$.} respectively
$15\%$, $24\%$ and $33\%$ while for $M_H =$ $210$GeV, $230$GeV and $250$GeV the reduction is comparatively
more modest with respectively $15\%$, $5\%$ and $2\%$.

The relative correction to the tree level cross section is shown in Fig. \ref{p_LL_mH} (right). 
It amounts to $2.6\%$ for 
$M_H=120$GeV, increases to as much as $49\%$ when $M_H=163$GeV and finally becomes almost constant at about $10\%$ for large values of $M_H$. Large contribution is due to the effect of Landau singularities.
\subsection{Distributions} 
\begin{figure}[hp]
\begin{center}
\mbox{\includegraphics[width=0.45\textwidth]{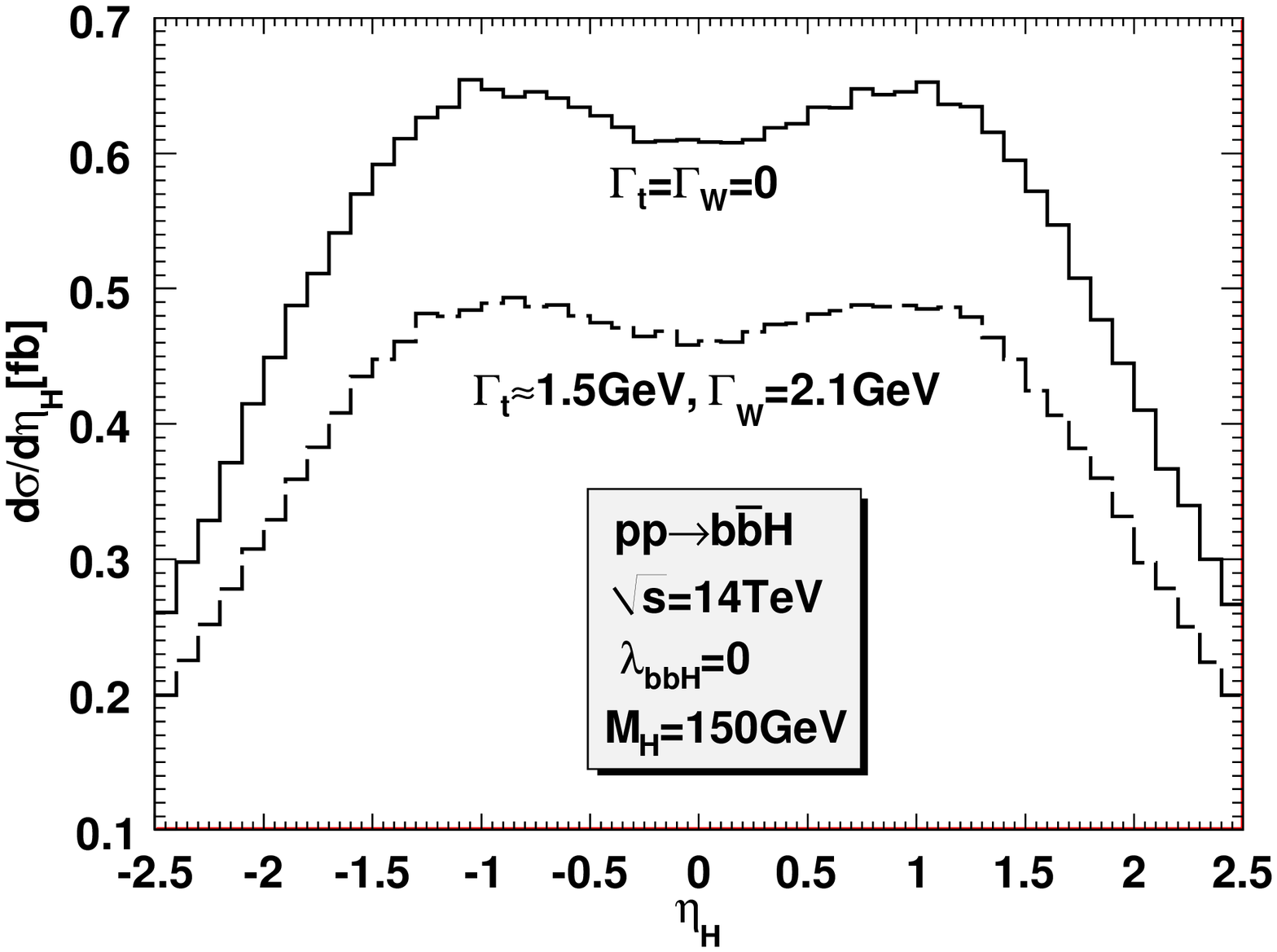}
\hspace*{0.075\textwidth}
\includegraphics[width=0.45\textwidth]{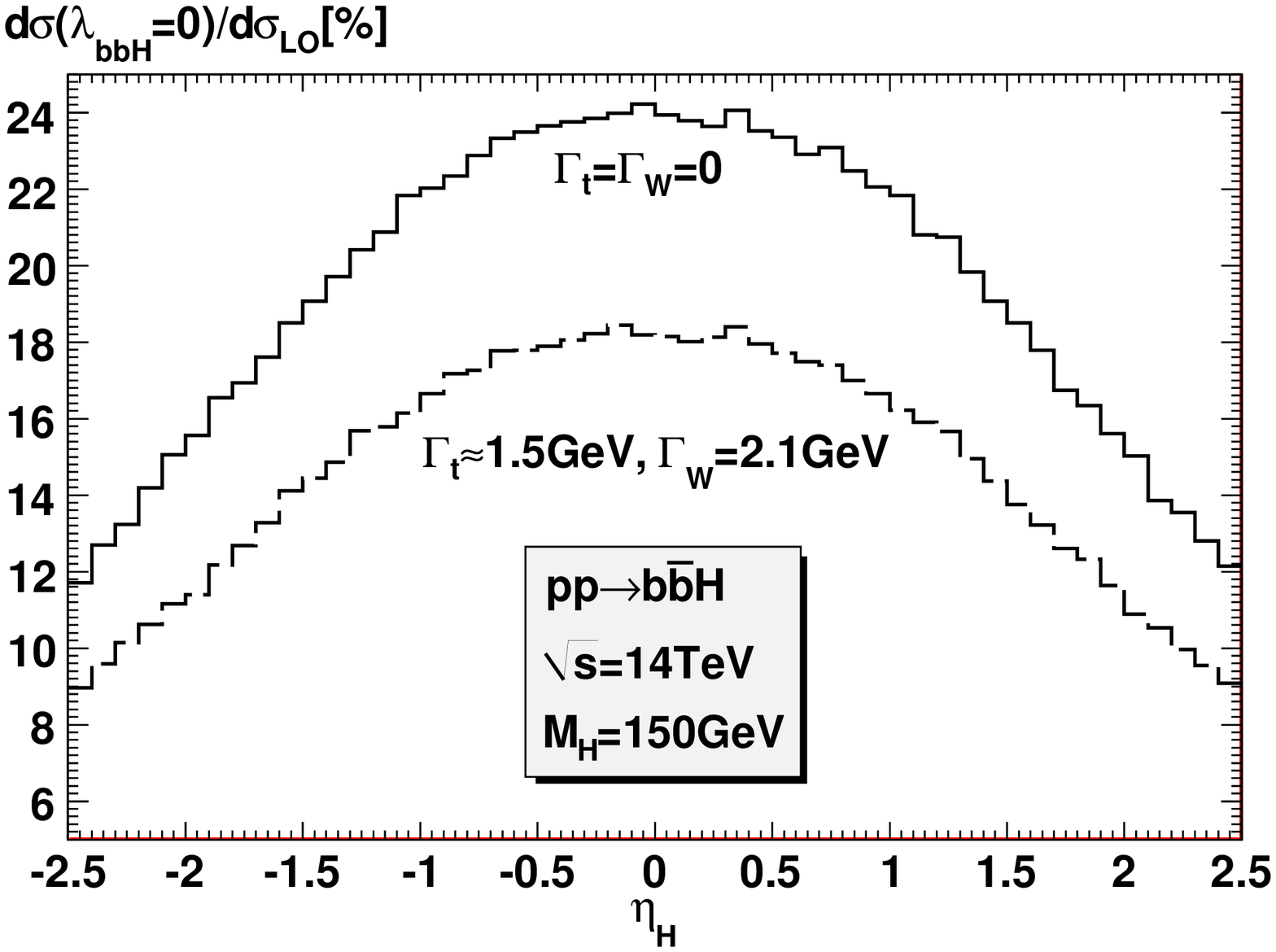}}
\mbox{\includegraphics[width=0.45\textwidth]{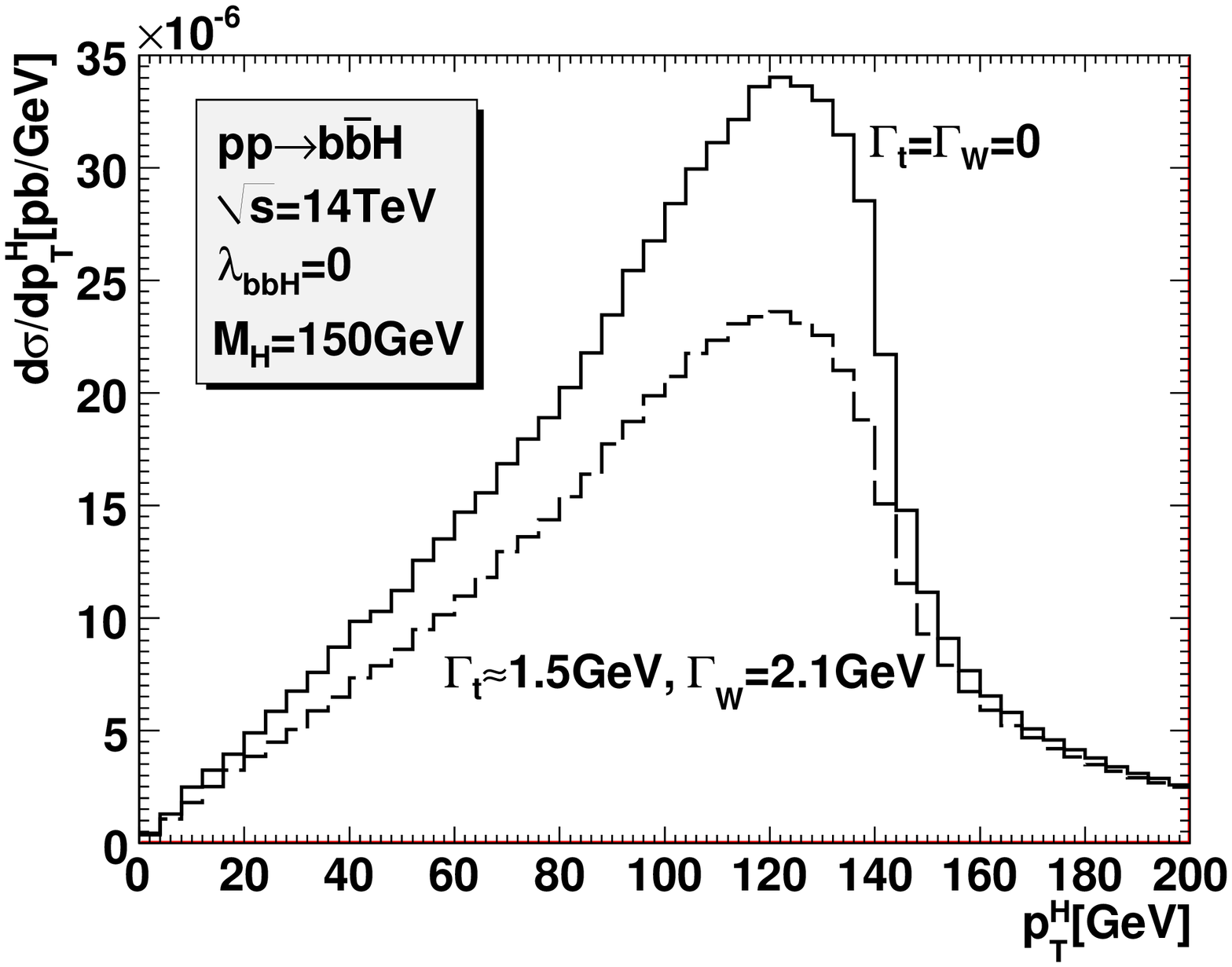}
\hspace*{0.075\textwidth}
\includegraphics[width=0.45\textwidth]{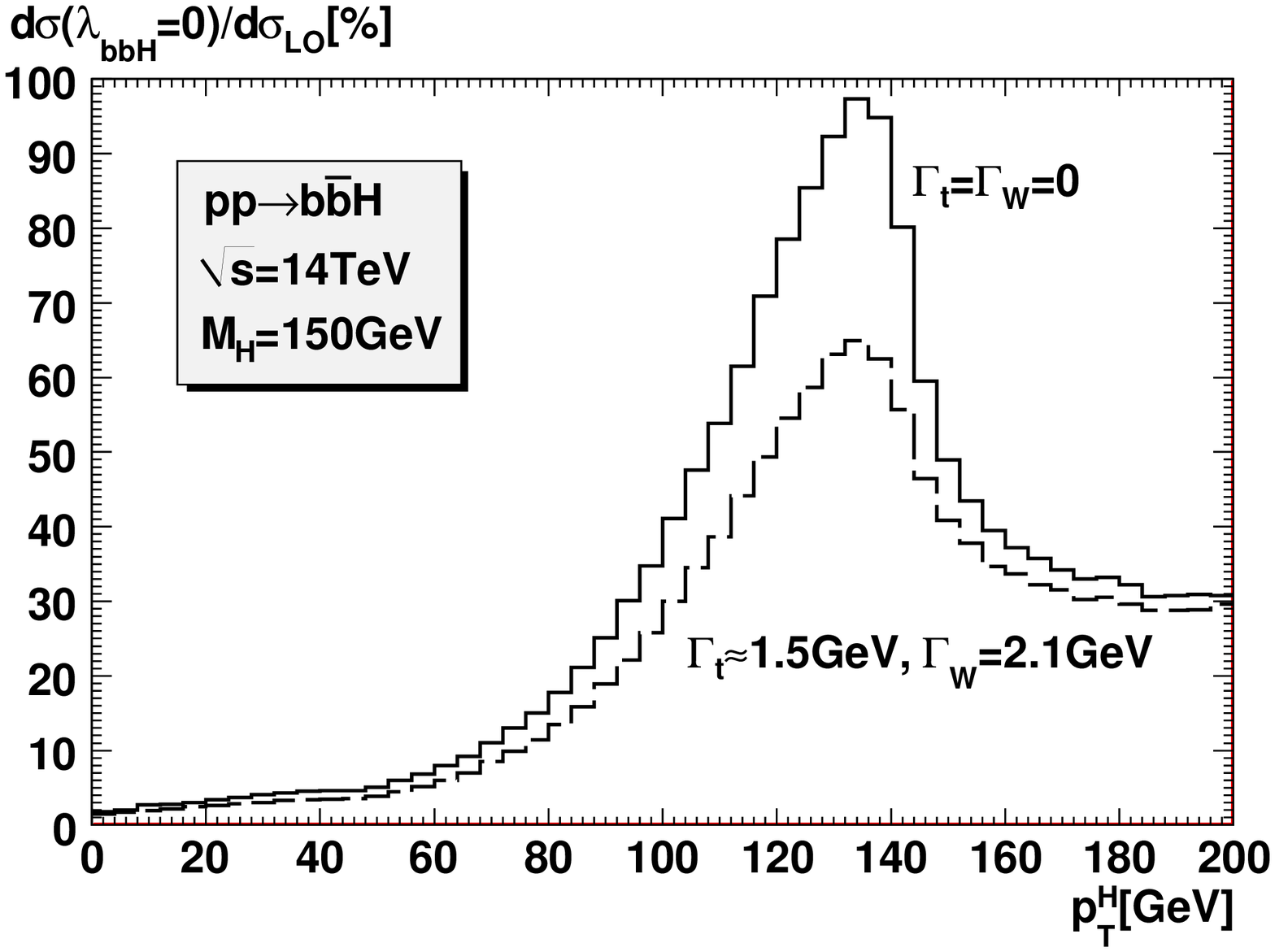}}
\mbox{\includegraphics[width=0.45\textwidth]{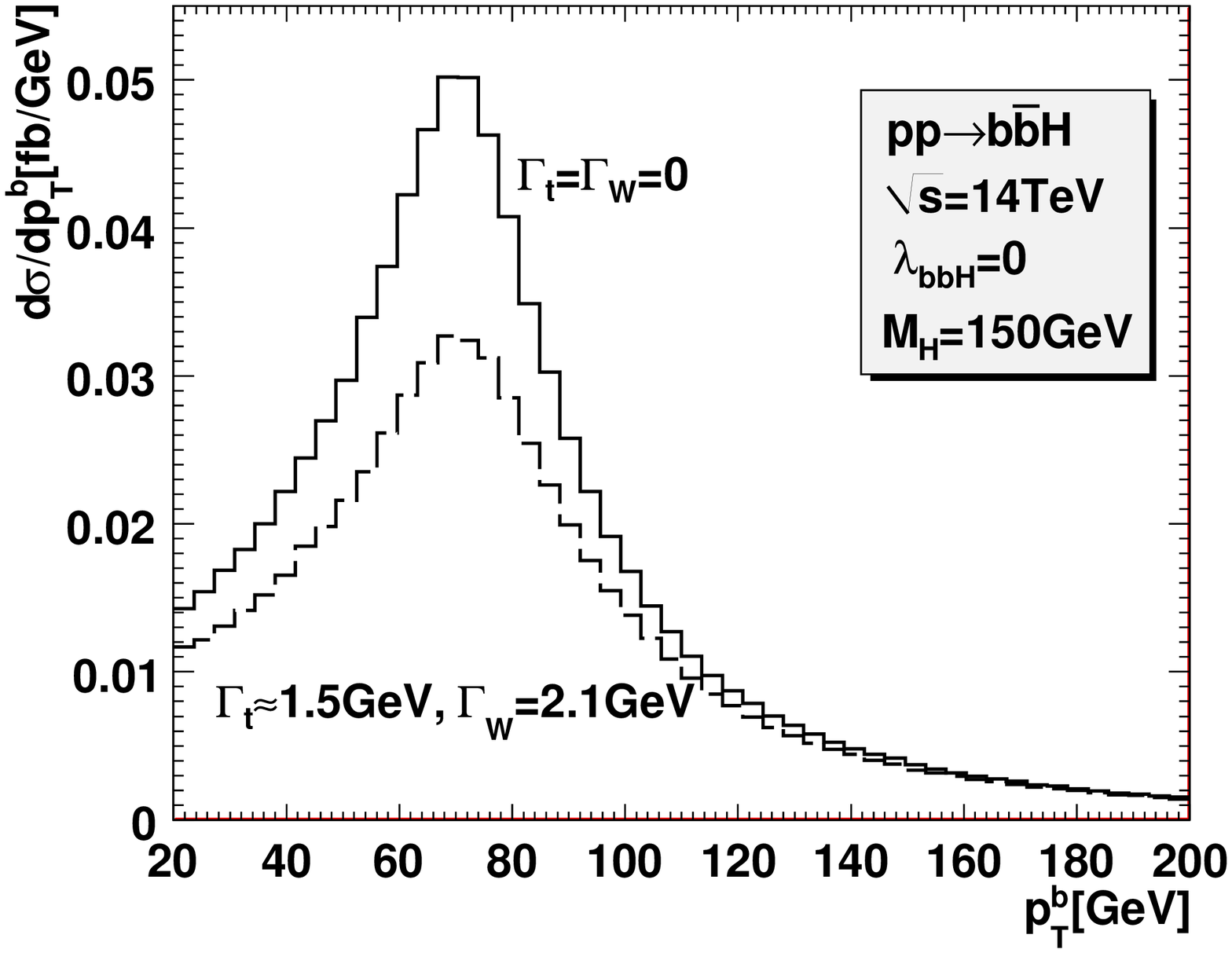}
\hspace*{0.075\textwidth}
\includegraphics[width=0.45\textwidth]{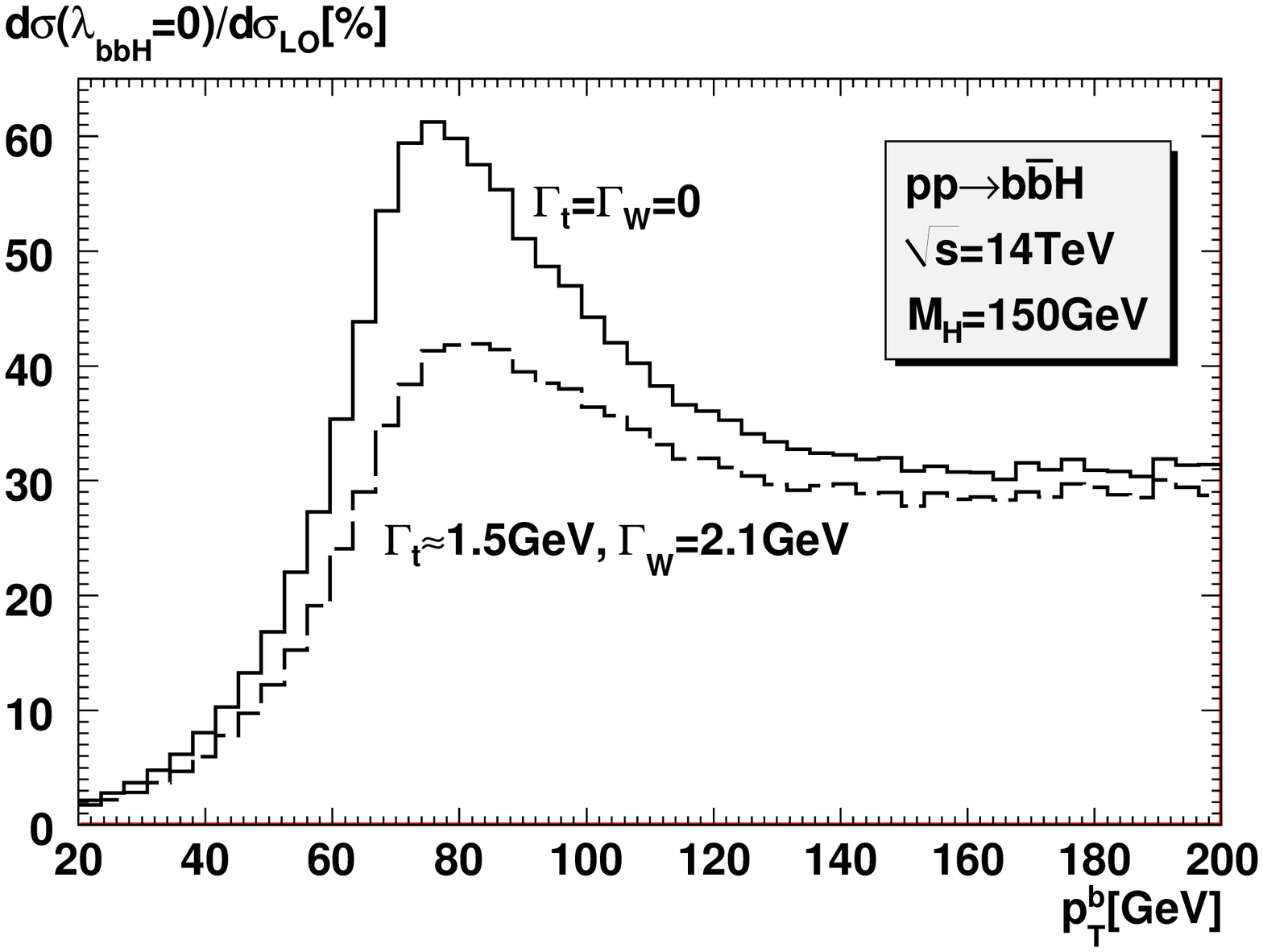}}
\caption{\label{p_LL_histo_mH150}{\em Left: The pseudo-rapidity of the Higgs and
transverse momentum distributions of the Higgs and the bottom for
$M_H=150$GeV arising from the purely one-loop contribution in
the limit of vanishing LO ($\la_{bbH}=0$) for two cases: with and without widths. 
Right: Its relative percentage contribution $d\sigma(\lambda_{bbH}=0)/d\sigma_{LO}$ is also
shown.}}
\end{center}
\end{figure}
\begin{figure}[h]
\begin{center}
\mbox{\includegraphics[width=0.45\textwidth]{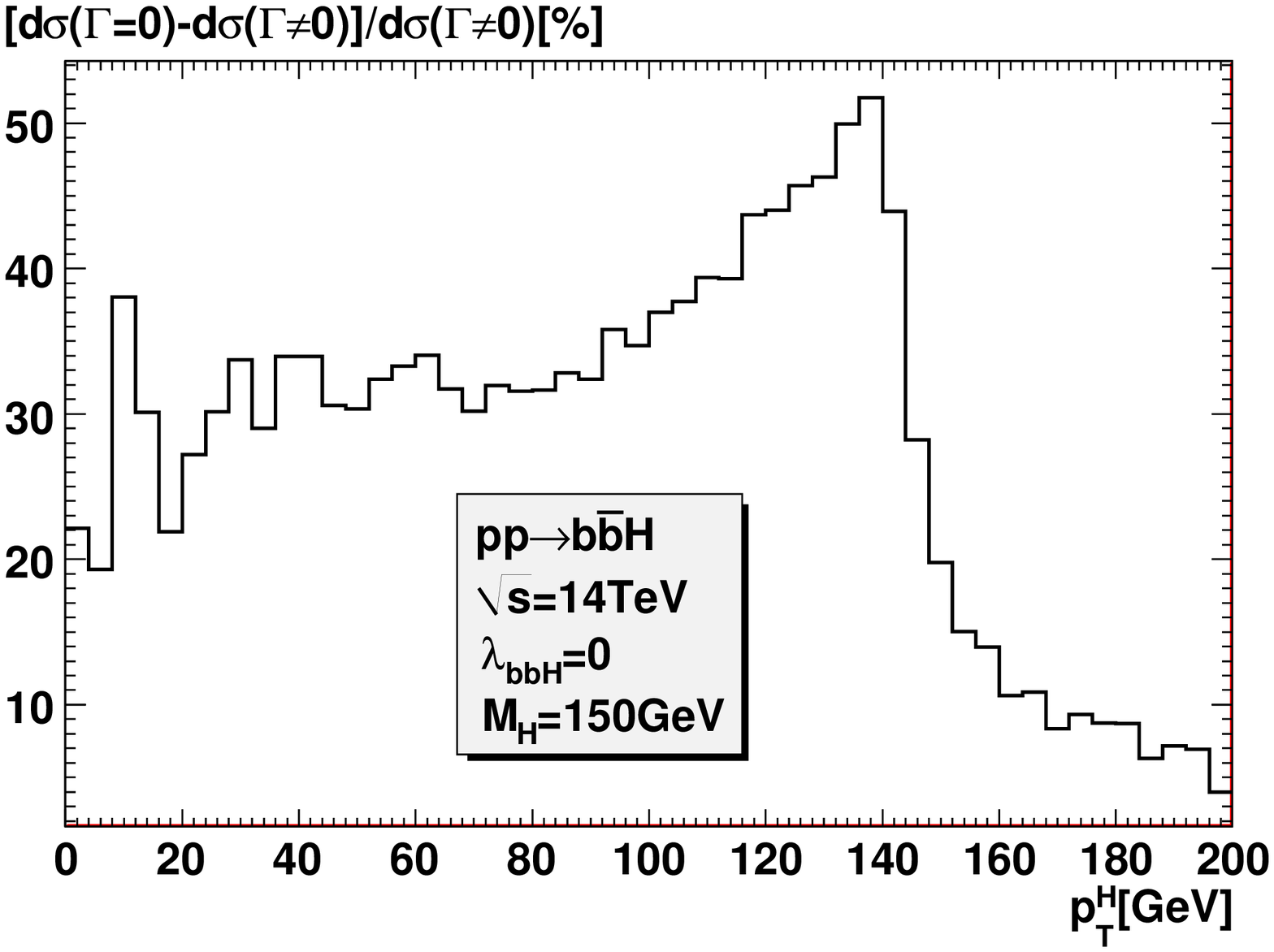}
\hspace*{0.075\textwidth}
\includegraphics[width=0.45\textwidth]{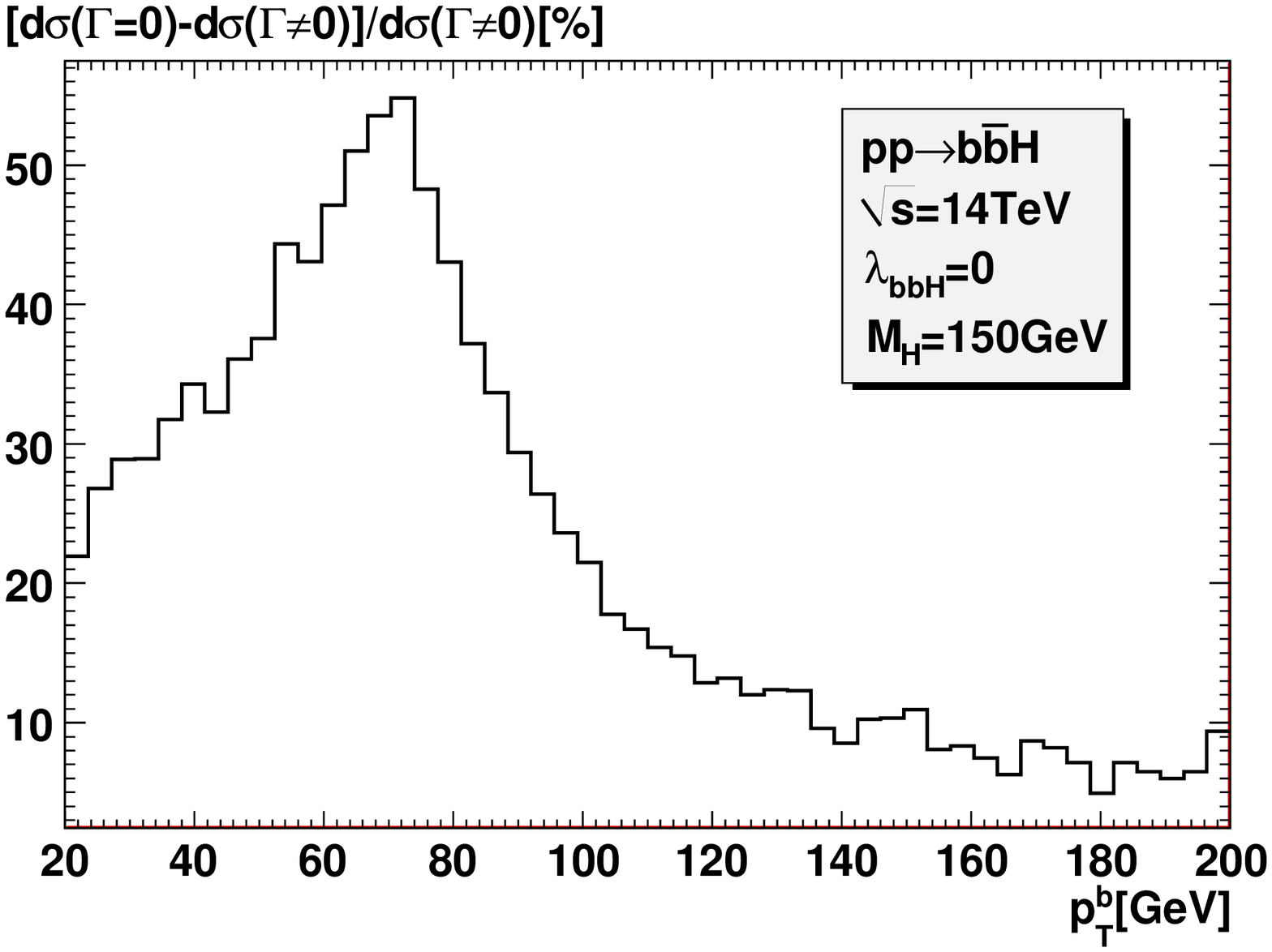}}
\caption{\label{p_LL_histo_mH150a}{\em The relative difference between two cases: with and without widths, 
defined by $[d\sigma(\Gamma=0)-d\sigma(\Gamma\neq0)]/d\sigma(\Gamma\neq0)[\%]$, for the 
transverse momentum distributions of the Higgs and the bottom quark.}}
\end{center}
\end{figure}
\newpage
\begin{figure}[hp]
\begin{center}
\mbox{\includegraphics[width=0.45\textwidth]{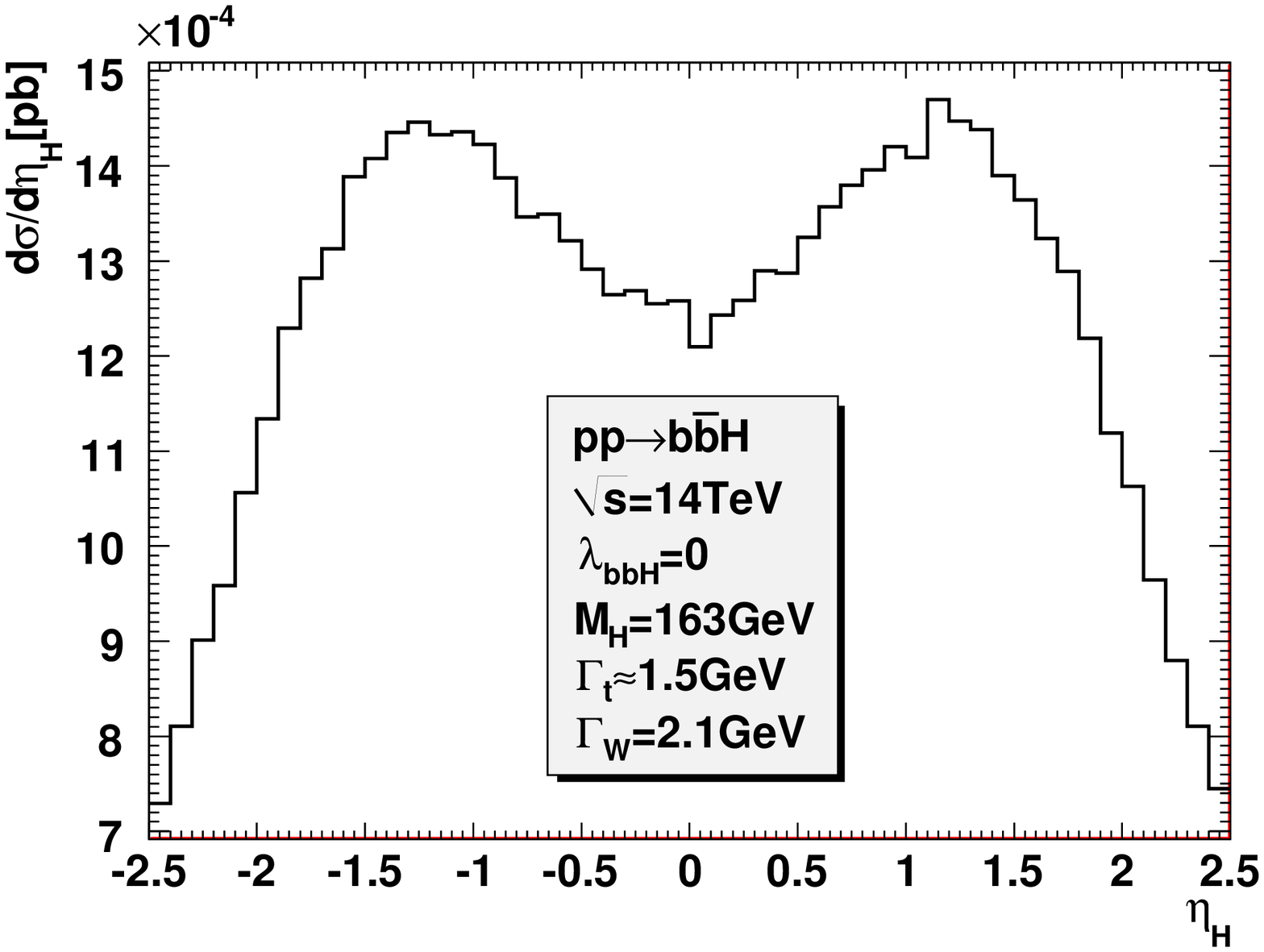}
\hspace*{0.075\textwidth}
\includegraphics[width=0.45\textwidth]{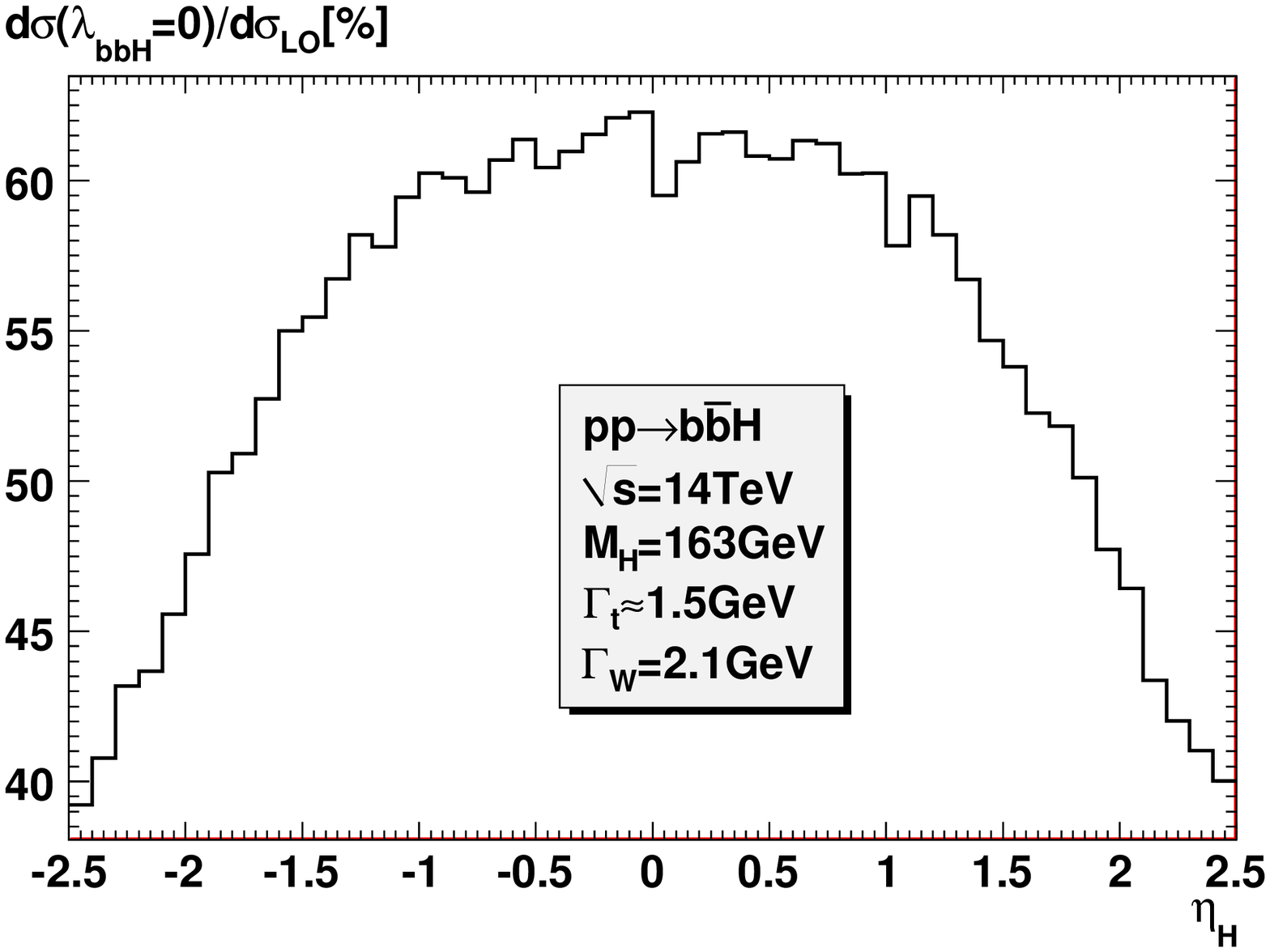}}
\mbox{\includegraphics[width=0.45\textwidth]{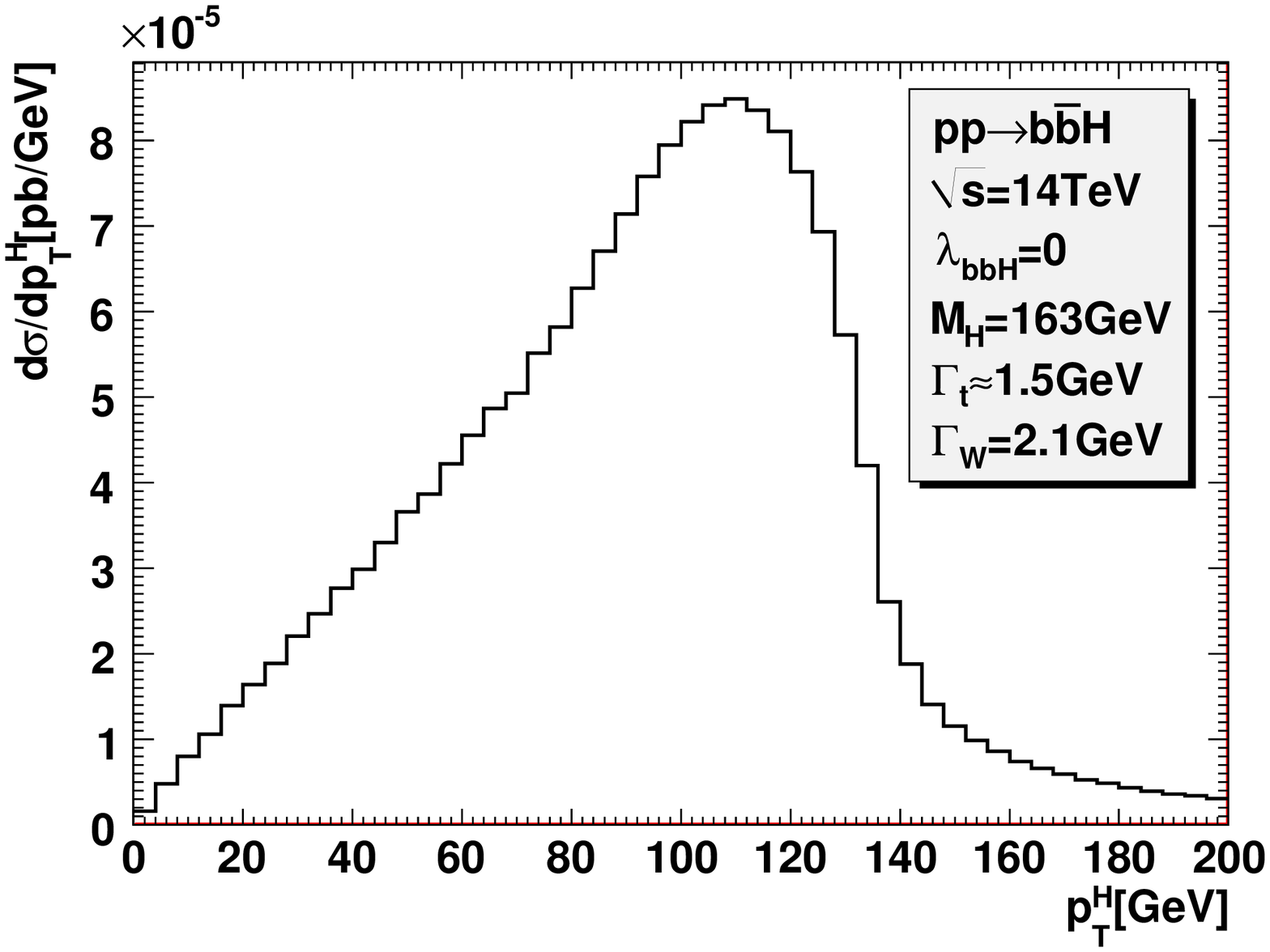}
\hspace*{0.075\textwidth}
\includegraphics[width=0.45\textwidth]{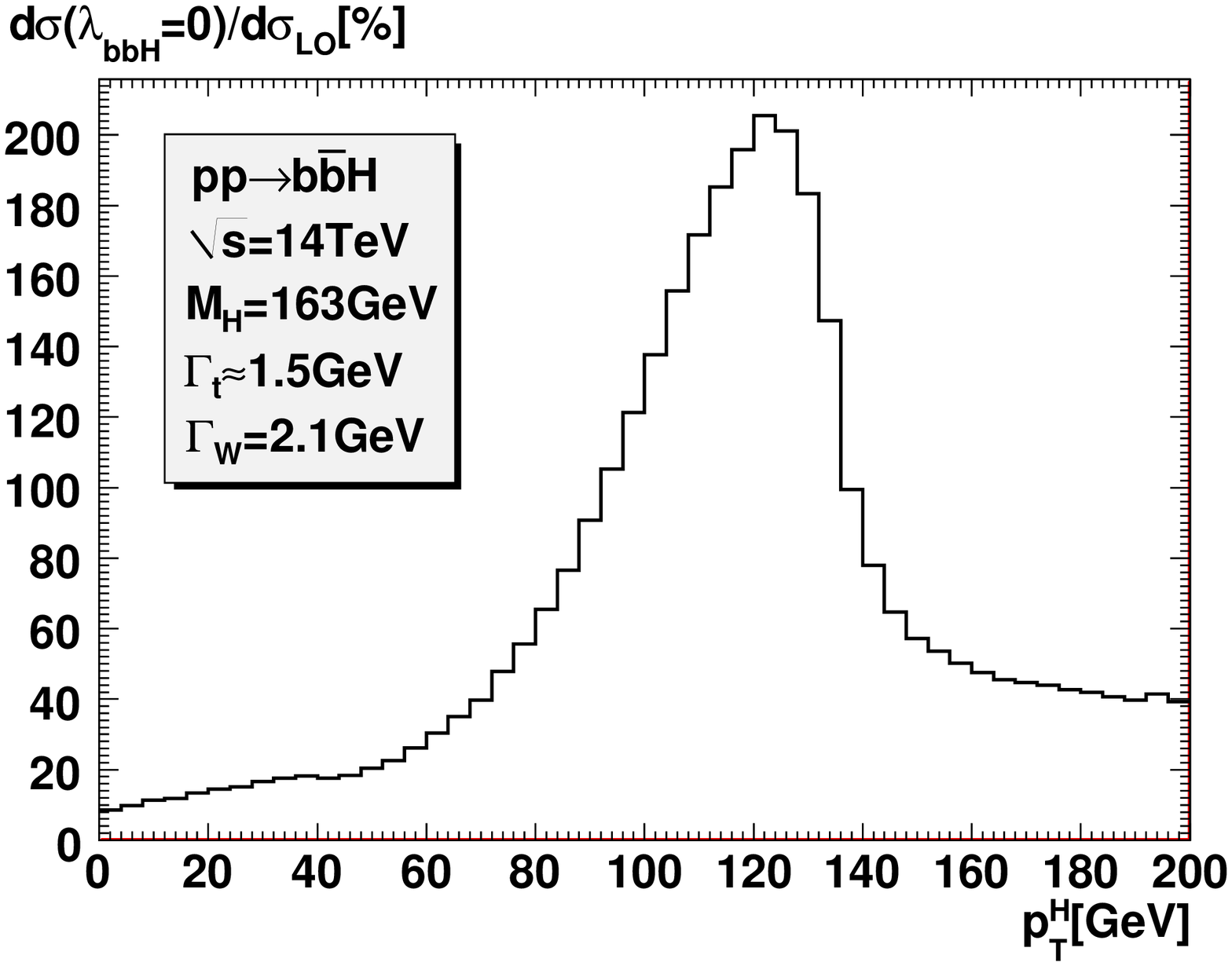}}
\mbox{\includegraphics[width=0.45\textwidth]{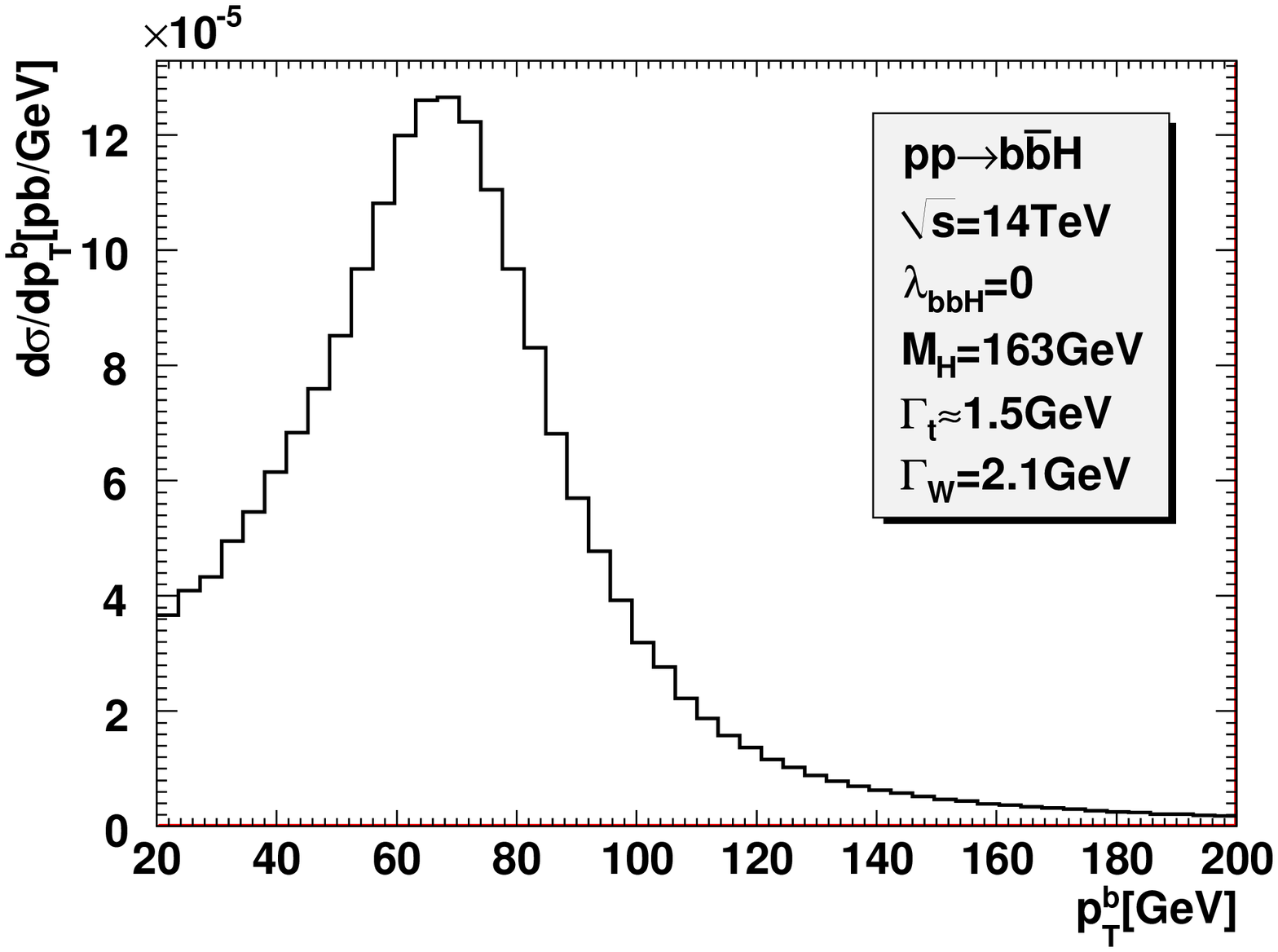}
\hspace*{0.075\textwidth}
\includegraphics[width=0.45\textwidth]{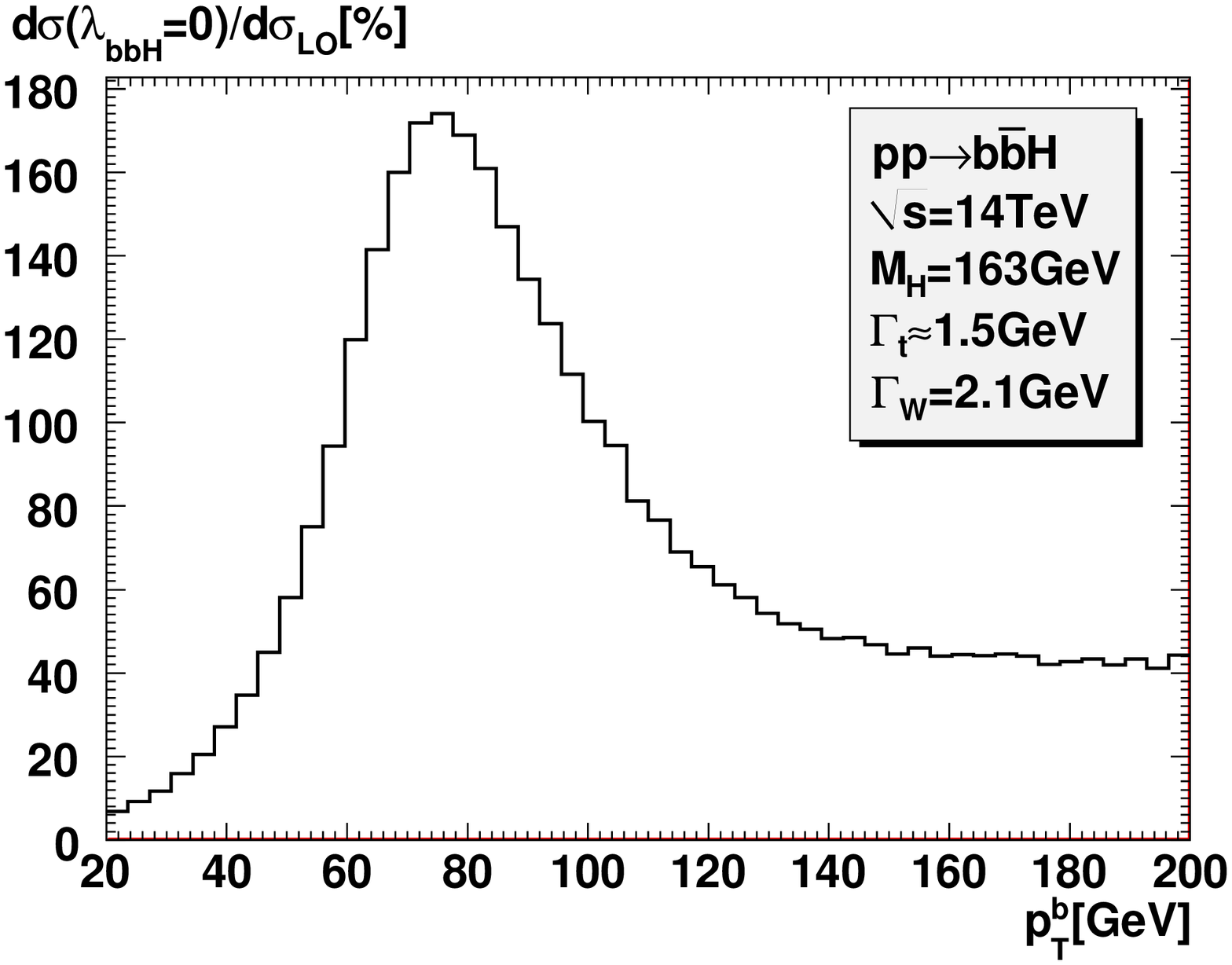}}
\caption{\label{p_LL_histo_mH163}{\em Left: The pseudo-rapidity of the Higgs and
transverse momentum distributions of the Higgs and the bottom for
$M_H=163$GeV arising from the purely one-loop contribution in
the limit of vanishing LO ($\la_{bbH}=0$). 
Right: Its relative percentage contribution $d\sigma(\lambda_{bbH}=0)/d\sigma_{LO}$ is also
shown}}
\end{center}
\end{figure}

In order to see the width effect on distributions, we first consider the case where $M_H=150$GeV $<2M_W$ (no LLS in this case). Figures \ref{p_LL_histo_mH150} and 
\ref{p_LL_histo_mH150a} show the difference between the two cases: with and without widths. The relative difference is rather uniform, about $33\%$, on the Higgs pseudorapidity distribution. 
For the transverse momentum distributions, the relative difference is not uniform but has structure as shown in Figure 
\ref{p_LL_histo_mH150a}. We remark that the peaks in the transverse momentum distributions occur at the position where 
the width effect is largest, hence are related to the opening of sub-leading Landau singularities as discussed in the previous subsection.   

The largest relative corrections to the tree level distributions are shown in Fig. \ref{p_LL_histo_mH163} for the case $M_H=163$GeV $>2M_W$ (LLS does occur here) which corresponds to the maximum value of the cross section in the limit $\la_{bbH}=0$ as displayed in Fig.~\ref{p_LL_mH} (right). The correction to the Higgs pseudorapidity distribution is about $60\%$ around the center region. The corrections to the $p_T$ distributions can be enormous in some region of phase space, up to $200\%$ for the Higgs and about $170\%$ for the bottom-quark case. These huge corrections to the distributions in some region of phase space are again due to the effect of Landau singularities.

\section{Results at NLO with $\la_{bbH}\neq 0$}
\label{section_nlo}
The purpose of this section is to complete the study of subsection~\ref{bbH_result_nlo_1}, to cover higher values of Higgs mass. 
Moreover, we would like to know the width effect at NLO in the presence of Landau singularities in various one-loop diagrams. 
We recall that the LLS is integrable at NLO. 

As discussed in subsection \ref{section_3classes} the NLO Yukawa corrections consist of $3$ gauge invariant
classes, see Fig. \ref{diag_3group}. The study of subsection \ref{bbH_result_nlo_1} revealed that 
class (a) gives a totally negligible correction below $0.1\%$. 
We will not discuss this contribution any further
here. Moreover, leading Landau singularities we have discussed only show up in class
(c). We will therefore study separately the NLO correction due to class (c) and weigh the
effect of implementing the width of the internal particles. Class (b) does not develop any
leading Landau singularity and therefore the width effect will be marginal. 
\subsection{Width effect at NLO}
\begin{figure}[htp]
\begin{center}
\mbox{\includegraphics[width=0.45\textwidth]{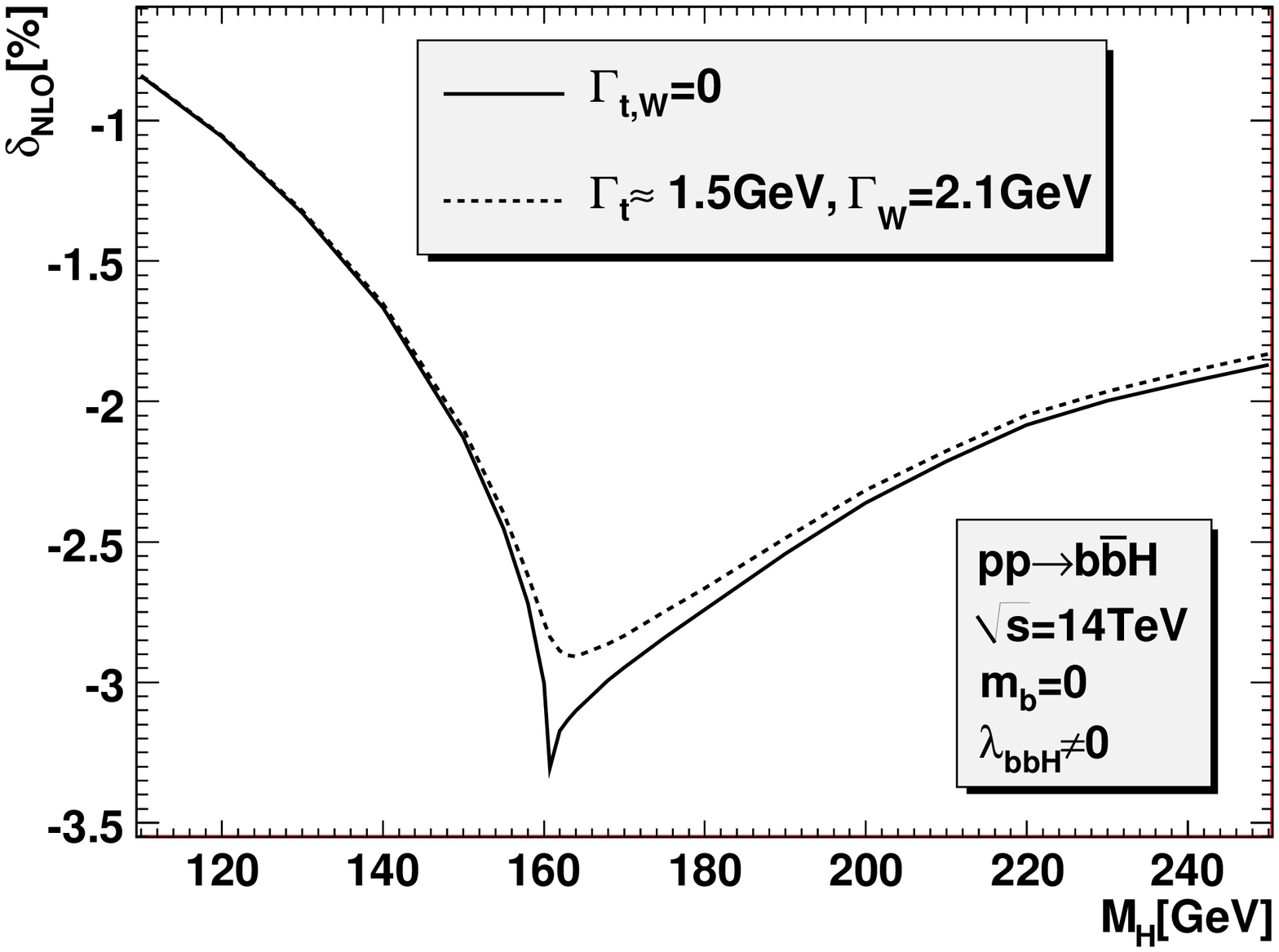}
\hspace*{0.075\textwidth}
\includegraphics[width=0.45\textwidth]{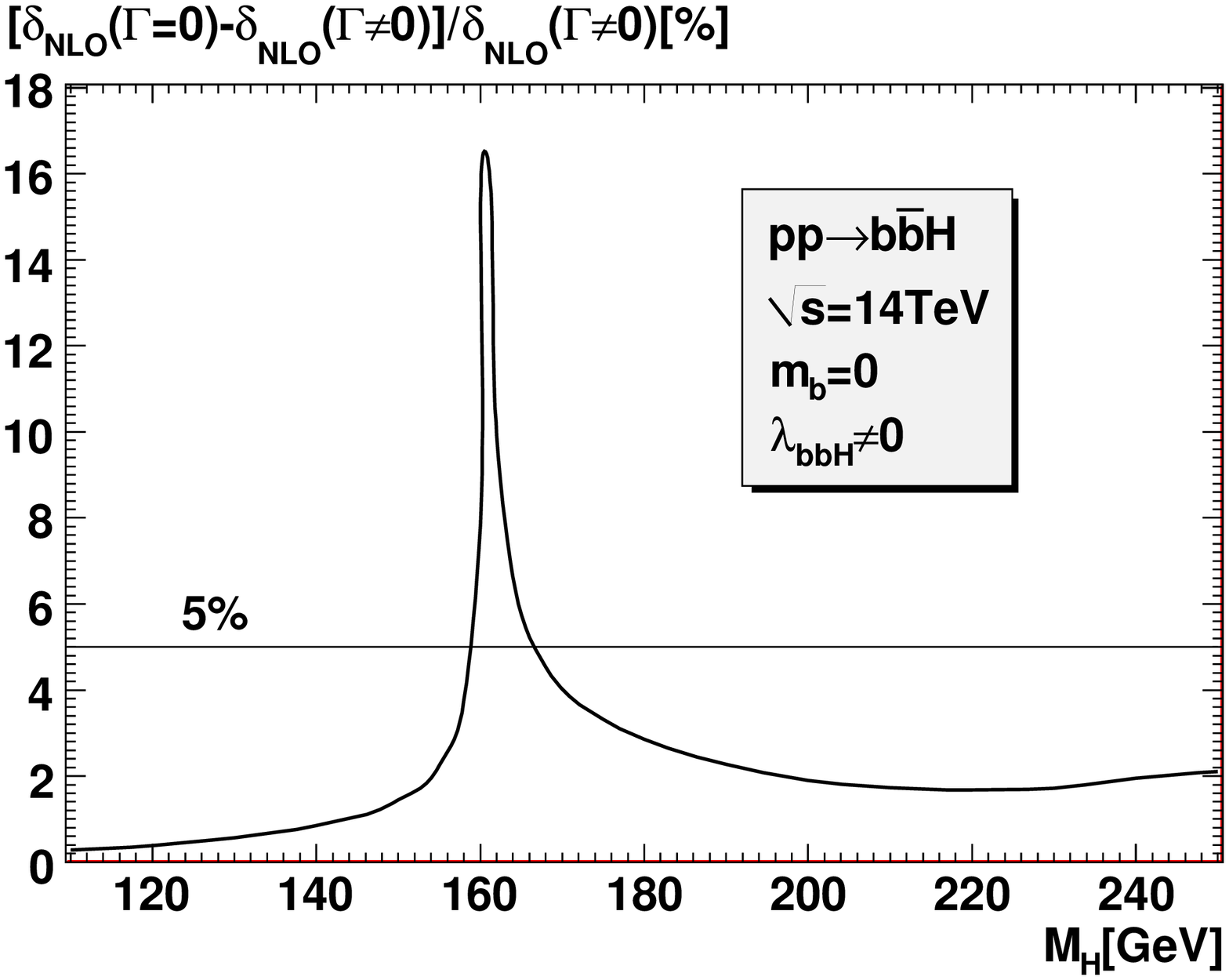}}
\caption{\label{pp_Loop_mb0_mH}{\em The relative difference between two cases: with and without widths, 
defined by $[d\sigma(\Gamma=0)-d\sigma(\Gamma\neq0)]/d\sigma(\Gamma\neq0)[\%]$, for the 
NLO correction of the class (c) where the LLS occurs. The tree-level amplitude is calculated with massive bottom quark. The one-loop amplitude is calculated by keeping $m_b$ only in the $\la_{bbH}$ coupling.}}
\end{center}
\end{figure}
The class (c) has problem with the $4$-point LLS. However, 
this LLS is integrable at NLO. Thus one can calculate the NLO cross section without introducing widths for unstable particles in loops. 
However, one can still expect some effect of Landau singularities (leading and sub-leading) and wonder if the width effect is significant in this case? 
The answer to this question is given in Fig.~\ref{pp_Loop_mb0_mH}. 

The results in this plot are calculated by setting $m_b=0$ in the kinematics (spinors and propagators) and loop integrals while keeping $m_b=4.62$GeV only in the $\la_{bbH}$ coupling which can be regarded as an independent parameter. The first remark is that if $M_H< 158$GeV or $M_H>165$GeV then the NLO width effect is below $5\%$. For $M_H<158$GeV the W-Goldstone bosons in the loops can never be on-shell and thus the width effect is completely small. For $M_H>165$GeV the W-Goldstone bosons can be on-shell and thus the width effect is a bit bigger than in the former case. Indeed even at the peak $M_H=2M_W$ the width effect is just about $17\%$. From this analysis we conclude that the width effect at the NLO is small and can be neglected. 

\subsection{NLO corrections with $m_b\neq 0$}
\begin{figure}[th]
\begin{center}
\mbox{\includegraphics[width=0.45\textwidth]{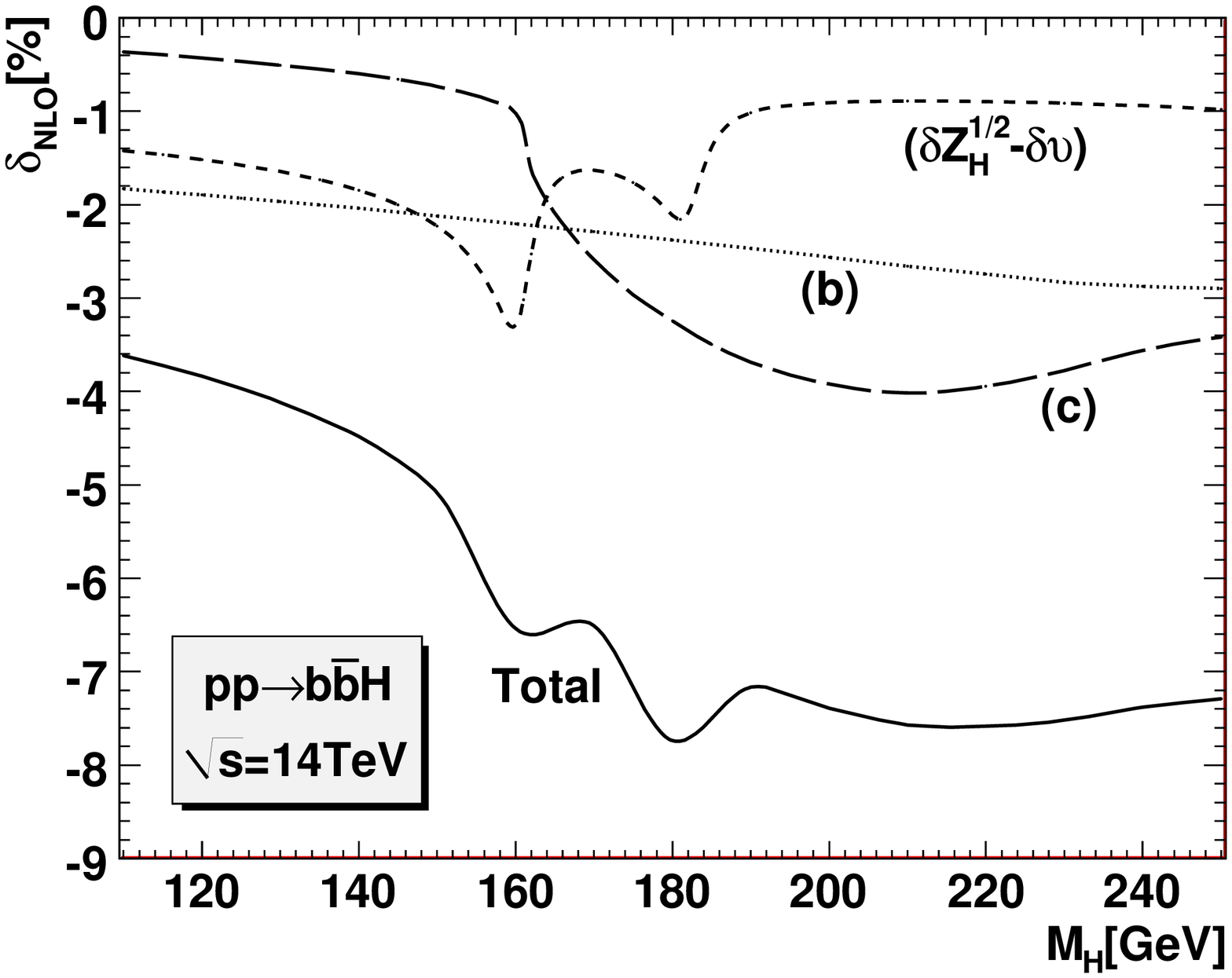}
\hspace*{0.075\textwidth}
\includegraphics[width=0.45\textwidth]{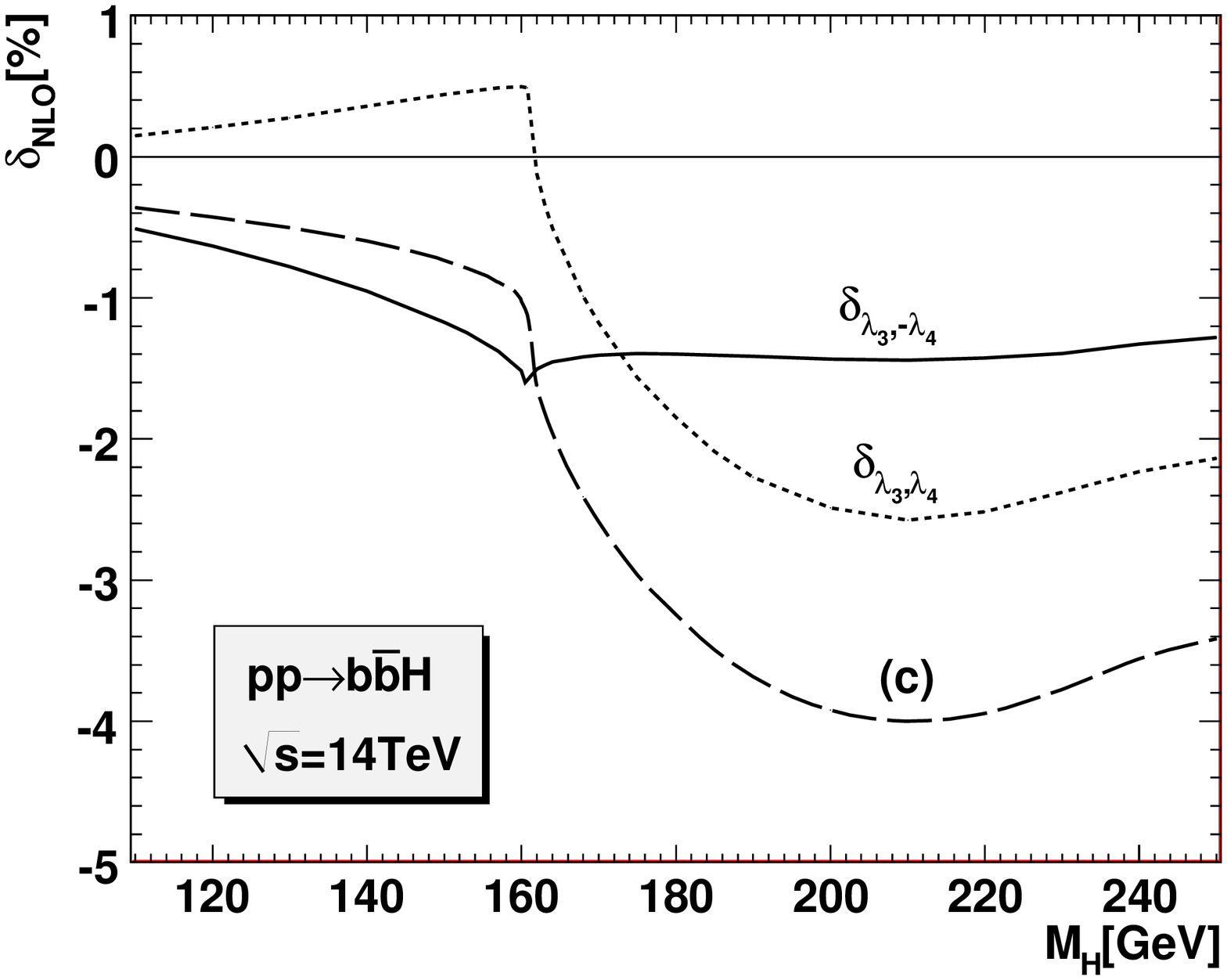}}
\caption{\label{pp_Loop_mb_mH}{\em Left: The relative NLO EW corrections normalized to the tree-level cross section. (b) and (c) correspond to 
the two classes of diagrams displayed in Fig.~\ref{diag_3group}. $(\delta Z_{H}^{1/2}-\delta\upsilon)$ is the universal correction contained in the renormalization of the $b\bar{b}H$ vertex. "Total" refers to the sum of those $3$ corrections. 
$\delta Z_{H}^{1/2}$ is calculated by taking into account the widths of $W$, $Z$ and the top quark. Right: the structure of (c)-correction which is a sum of two independent helicity configurations.}} 
\end{center}
\end{figure}

The results for the NLO corrections are shown in Fig.~\ref{pp_Loop_mb_mH} (left). For classes (b) and (c) the widths of unstable particles are neglected. 
For the universal correction, $(\delta Z_{H}^{1/2}-\delta\upsilon)$ where $\delta Z_{H}^{1/2}$ involving the derivative of the two-point function Higgs self-energy diverges when $M_H$ is equal to $2M_W$ or $2M_Z$, all the widths of unstable particles are kept \footnote{Note that $\delta Z_{H}^{1/2}$ does not diverge when $M_H=2m_t$ and the top-quark width thus has a marginal effect.}. 

The effect of Landau singularities is obvious if one compares the (c)-curve to the (b)-curve. 
The contribution from class (b) where the Higgs couples to the internal top decreases
very slowly as the Higgs mass increases from $110$GeV to $250$GeV, as expected there is
no structure as would be the case if this contribution were sensitive to any threshold or
singularity. Class (c) on the other hand does, as expected, reveal some structure around
$M_H = 2M_W$ where we see a fall in the relative correction. The correction is however,
despite this fall, quite modest ranging from about $-1\%$ for $M_H = 160$GeV to about $-4\%$ for $M_H = 210$GeV. 

The universal correction, $(\delta Z_{H}^{1/2}-\delta\upsilon)$, contributes from $-1\%$ to $-3\%$, where the highest correction is at the $H\to WW$ threshold ($M_H=2M_W$).

The detailed structure of the class (c) is shown in Fig.~\ref{pp_Loop_mb_mH} (right). It consists of two independent helicity structures where the helicities of two bottom quarks in the final state are the same or different. We call them $\delta_{\la_3,\la_4}$ and $\delta_{\la_3,-\la_4}$ structures. In the massless bottom limit whose result is displayed in Fig.~\ref{pp_Loop_mb0_mH}, only the $\delta_{\la_3,-\la_4}$ structure survives. It is remarked that the behaviors of two different helicity 
structures as functions of $M_H$ are very different despite the fact that they both have a common denominator.

\section{Summary}
We have extended the study of section \ref{section_bbH_result1} to cover higher values of $M_H$. This study is important and highly nontrivial due to the 
appearance of Landau singularities in various one-loop Feynman diagrams when $M_H\ge 2M_W$. 

We have applied the general study of Landau singularities, presented 
in chapter \ref{section_landau_introduction}, to the specific case of $\ggbbH$. Namely, we have studied in detail the LLS in the case of one-loop $3$-point and $4$-point functions. The nature of those LLS is carefully explained by using two different methods. On one hand, we used the general formulae (which are for the singular part only) obtained in section~\ref{subsection_landau_nature}. On the other hand, we used explicit results obtained by performing loop integrals using the traditional method of 't Hooft and Veltman. The latter results in various plots of real and imaginary parts of scalar loop integrals. This indeed helped us to understand much better the various structures of Landau singularities. 

We have performed a detailed study to understand how the LLS terminates. From this, we got the upper bounds of external parameters.

We have argued that the problem of $4$-point LLS, which is not integrable at the level of one-loop amplitude squared, is solved by introducing a width for 
internal unstable particles. In order to do so, we have applied the loop calculation method of 't Hooft and Veltman to write down explicitly two formulae to calculate scalar box integrals with complex 
internal masses (see Appendix~\ref{appendix-box-integral}). The restriction is that at least two external momenta are lightlike. We have implemented those two formulae into the library LoopTools. 

We have studied the width effect in the presence of Landau singularities in various one-loop diagrams. At NLO, the width effect is negligible. At NNLO, 
the width effect is extremely important around the normal threshold position ($M_H=2M_W$). 

We have shown various results of one-loop corrections to the cross section as well as important distributions. NNLO corrections, calculated in the limit 
$\la_{bbH}=0$, can be very large in some region of phase space or parameter space where Landau singularities show up. For the $p_T$ distribution, those corrections can be enormous. NLO corrections remain small although some structure related to Landau singularities does show up.

\chapter{Conclusions}
\label{chapter_conclusions}
In the previous chapters, we have explained how to calculate one-loop Yukawa corrections to the process of 
SM Higgs production associated with two high $p_T$ bottom quarks at the LHC and their physical content. 
In particular, we have studied one-loop Landau 
singularities and shown how to handle them in practice for our process. The properties and effect of the Landau singularities in the 
process $\ppbbH$ were also carefully investigated. 

The entirely dominant contribution to the process $\ppbbH$ is the sub-process $\ggbbH$. 
The physics of one-loop Yukawa corrections to this sub-process is very rich due to the fact that the bottom-quark and 
the top-quark are in the same $SU(2)_L$ doublet with a large mass splitting. At tree level, the Higgs boson couples directly 
to bottom quarks with a small Yukawa coupling $\la_{bbH}$. At one-loop level, the Higgs can couple to heavy particles like 
the top-quark or W Goldstone bosons. Thus, it is easy to see that, if we do the expansion in the small coupling $\la_{bbH}$, the one-loop 
corrections can still give some contribution when this coupling vanishes. The salient property of the one-loop corrections is related to the 
smallness of $m_b$ and is best seen if we take the limit $m_b\to 0$ while keeping $\la_{bbH}$ (which can be regarded as a different parameter) unchanged. In this limit, the tree level contribution contains only {\em even} helicity configurations ("even" means that both bottom-quarks have the same helicity) while the one-loop contribution includes both even and {\em odd} helicity configurations ("odd" means that the two bottom-quarks have opposite helicities). The even one-loop contribution is proportional to the small $\la_{bbH}$ while the odd one, which comes from $m_t$ insertion in loops, is independent of $\la_{bbH}$ and proportional to large couplings like $\la_{ttH}$ and $\la_{HHH}$. We remark immediately that the NLO correction is proportional to $\la_{bbH}^2$ hence must be small. However, the most interesting thing is that the purely one-loop contribution which can be extracted by setting $\la_{bbH}=0$ consists of only odd helicity configurations. Thus, this new loop induced contribution can be very different from the tree level or NLO contributions. 
Moreover, the best way to understand Landau singularities, which are intrinsic analytic properties of loop integral, is to look at purely loop corrections. 

The numerical results of chapters \ref{chapter_bbH1} and \ref{chapter_bbH2} confirm those qualitative conclusions. The NLO correction is small and changes from $-4\%$
for $M_H = 120$GeV to $-8\%$ at $M_H = 2M_Z$ stabilising to around $-7\%$ for larger values of $M_H$ up to $250$GeV, despite the appearance of various Landau singularities. The purely one-loop contribution (NNLO), calculated by taking into account the widths of the top-quark and W gauge boson, amounts to $2.6\%$ for 
$M_H=120$GeV, increases to as much as $49\%$ when $M_H=163$GeV and finally becomes almost constant at about $10\%$ for large values of $M_H$.  The difference between the two corrections becomes clearest when one looks at the $p_T$ distributions. The NNLO correction to $p_T$ distributions can be enormous in some region of phase space for $M_H$ about $2M_W$ while the NLO correction is modest. 
Large NNLO corrections are due to the effect of Landau singularities and occur around the normal threshold position $M_H=2M_W$. 

Calculating one-loop corrections contains a lot of technical features. The amplitude squared was calculated by using 
the helicity amplitude method (HAM) where each helicity amplitude, which is a complex number, is numerically calculated. In order 
to do so, we have reduced all loop integrals (up to five-point function) in terms of Passarino-Veltman loop functions. 
This was easily done with the help of LoopTools. 
After doing so, we obtained a huge algebraic expression for each helicity configuration which takes a lot of computer time to calculate. 
The problem was how to optimise the calculation. We found a systematic way to do this by using FORM based on the fact that each helicity 
amplitude can be factorised in terms of independent blocks. The advantage is that some complicated blocks which appear several times in the computation can 
be put in a common sector hence just need to be calculated once. 

In our calculation with two gluons as external particles, the HAM leads to a very easy and 
convenient way to check the QCD gauge invariance which means that the amplitude squared is independent of the reference vectors needed to define the gluon polarization vectors. Indeed, this is a very powerful way to check the correctness of the results and can be used for other processes with at least one external gluon/photon. 

Another advantage of HAM is that since the tree-level and one-loop helicity amplitudes are calculated separately, the 
one-loop amplitude squared is immediately obtained when the NLO calculation is done.  

There is a special class of one-loop $\ggbbH$ diagrams where the Higgs boson is produced by W gauge boson fusion. If the Higgs mass is heavy enough for 
this normal threshold to open ($M_H\ge 2M_W$), then all Landau singularities of two-, three-, four- point functions show up. In particular, the four-point leading Landau singularity leads to severe numerical instabilities when we calculate the cross section involving the square of a one-loop amplitude. We have solved this problem by taking into account the fact that W gauge boson and the top-quark are unstable particles hence have a width. For this, we have followed the standard technique of 't Hooft and Veltman to calculate scalar four-point functions with complex internal masses. The restriction is that at least two external momenta are lightlike. We have implemented those formulae into the library LoopTools. Various checks have been performed to make sure that this implementation is correct. We have also observed that the same implementation can be done for the case of one lightlike external momentum. However, the calculation of the scalar four-point functions with complex internal masses in the most general case with no restriction on the external momenta 
is not tractable if one uses the method of 't Hooft and Veltman. Another disadvantage of this method is that the results, even in some special cases with massless external particle, contain many Spence functions. It may be better to find another way. 

Although the main calculation of this thesis is for a very specific process, it is quite obvious for us that some of the results discussed above can be used or generalised for other complicated one-loop calculations. 


\begin{appendix}
\chapter{The helicity amplitude method }
\label{appendix-helicity}
\section{The method}
We use a combination of helicity amplitude methods as  described
in \cite{kleiss_stirling,maina} to calculate the total cross
section. In the following we only want to highlight some key
features that were most useful for our calculation, for details of
the method we refer to\cite{kleiss_stirling,maina}. For our
process $g(p_1,\la_1)+g(p_2,\la_2)\rightarrow
b(p_3,\la_3)+\bar{b}(p_4,\la_4)+H(p_5)$ where the particles are
denoted by their momentum $p_i$ and  helicity $\lambda_i$ we write
the corresponding helicity amplitude as $\ali$.
\def\slashepi{\epsilon_i\kern -.720em {/}}
\def\slashpi{p_i\kern -.670em {/}}
\bea \ali&=&\eps_{\mu}(p_1,\la_1;q_1)\eps_{\nu}(p_2,\la_2;q_2) {\cal
M}^{\mu \nu}(\la_3,\la_4), \crn
 {\cal M}^{\mu\nu}(\la_3,\la_4)&=&
\bar{u}(p_3,\la_3)\Gamma^{\mu\nu} v(p_4,\la_4). \label{amp_form}
\eea
$\Gamma^{\mu\nu}$ is a string of Dirac $\gamma$ matrices. These $\gamma$
matrices represent either interaction vertices or momenta from the
fermion propagators. In our case the interaction vertices are the
vectorial gluon vertices in which case they represent $\slashepi
\;$, the scalar Higgs vertex and at one-loop  the pseudo-scalar
Goldstone coupling. For the momenta, in our implementation we
re-express them in terms of the independent external momenta
$p_1,p_2,p_3,p_4$. This applies also to the loop momenta after the
reduction formalism of the tensor integrals has been performed.
The first step in the idea of the helicity formalism we follow is
to  turn  each of these $\gamma$ matrices (apart from the
pseudo-scalar and the trivial scalar) into a combination of spinor
function $u \bar u$. We therefore transform our helicity
amplitude into products of spinors such as the helicity amplitude
could be written like a product $\bar u\; u
 \bar u\; ...u \bar u \; v$ with the possible insertion of $\gamma_5$'s in the string.
The different $u$, $\bar u$, $v$ in the string we have written have
of course, in general, different arguments. Nonetheless one can
turn each spinor product of two adjacent $\bar u u$, etc into a
complex number written in terms of the momenta in our problem as
we will see.

\noi In the first step, for the momentum $\slashpi$ with
$p_i^2=m_i^2$ we use \bea \slashpi=u(p_i,-)\bar
u(p_i,-)+u(p_i,+)\bar u(p_i,+)-m_i. \eea \noindent The transverse
polarization vector of the initial gluon $i$,
$\eps_{\mu}(p_i,\la_i;q_i)$, is also first expressed in terms of
spinors such as
\bea\eps_\mu(p_i,\la_i;q_i)&=&\fr{\bar{u}(p_i,\la_i)\ga_\mu
u(q_i,\la_i)}{[4(p_i.q_i)]^{1/2}}, \label{eps12_def}
\eea
\noindent where $q_{i}$ is an  {\em arbitrary} reference vector
satisfying the following conditions
\bea q_i^2=0, \quad
p_i.q_i\neq 0.\label{cond_refer_vector}
\eea
Using the trace technique one can easily prove that definition (\ref{eps12_def}) indeed satisfies the 
physical polarisation sum identity given in Eq. (\ref{gluon_axial}), namely
\bea
\sum_{\la=\pm}\eps_\mu(p,\la;q)\eps_\nu(p,\la;q)^*=-g_{\mu\nu}+\fr{p_\mu q_\nu+p_\nu q_\mu}{p.q}.  
\eea
Gauge invariance (transversality condition) requires that the
cross sections are independent of the choice of the reference
vector as we will see later. This acts as an important check of
the calculation, see later.  It is not difficult to prove that the
choice (\ref{eps12_def}) satisfies all the conditions for a
transverse polarization vector. In particular,
\bea
p_i.\eps(p_i,\la_i)&=&0, \quad  \eps(p_i,\la_i).\eps(p_i,\la_i)=0,
\crn  \eps_{\mu}(p_i,-\la_i)&=&\eps_{\mu}(p_i,\la_i)^*, \quad
\eps(p_i,\la_i).\eps(p_i,-\la_i)=-1,
\eea
where the reference vector is not written down explicitly. $i=1,2$
and no sum over $i$ must be understood. Then for
$\slashepi=\epsilon_\mu \gamma^\mu$ one uses  the so-called
Chisholm identity
\bea\bar{u}(p,\la)\ga_\mu
u(q,\la)\ga^\mu=2[u(p,-\la)
\bar u(q,-\la)+u(q,\la)\bar u(p,\la)],  \label{chisholm}
\eea
where all the spinors in Eq.~(\ref{chisholm}) are for massless
states in view of the lightlike condition on the reference frame
vector and of course the momentum of the real gluon.

With $U(p_i,\la_i)$ representing
either $u(p_i,\la_i)$ or $v(p_i,\la_i)$ one uses the general
formulae
\bea
\bar
U(p_i,\la_i)\gamma_5U(p_j,\la_j)&=&-\la_i
\frac{A_{\la_i\la_j}(p_i,p_j)-M_i B_{\la_i\la_j}(p_i,p_j)+M_j
C_{\la_i\la_j}(p_i,p_j)} {\sqrt{(p_i.k_0)(p_j.k_0)}},\crn
\bar U(p_i,\la_i)U(p_j,\la_j)&=&\frac{A_{\la_i\la_j}(p_i,p_j)+M_i
B_{\la_i\la_j}(p_i,p_j)+M_j
C_{\la_i\la_j}(p_i,p_j)}{\sqrt{(p_i.k_0)(p_j.k_0)}},\label{def_U5U}
\eea
where
\bea M_i&=&+m_i \,\,\,\,\, \text{if}\,\,\,\,\,
U(p_i,\la_i)=u(p_i,\la_i),\crn M_i&=&-m_i \,\,\,\,\,
\text{if}\,\,\,\,\, U(p_i,\la_i)=v(p_i,\la_i),\crn
A_{\la_i\la_j}&=&\delta_{\la_i -\la_j} \la_i
\left((k_0.p_i)(k_1.p_j)-(k_0.p_j)(k_1.p_i)-i \la_i
\epsilon_{\mu\nu\rho\sigma}k_0^\mu k_1^\nu p_i^\rho p_j^\sigma
\right), \crn B_{\la_i\la_j}&=&\delta_{\la_i\la_j}(k_0.p_j),\,\,
C_{\la_i\la_j}=\delta_{\la_i\la_j}(k_0.p_i),
\eea
with $k_{0,1}$ being auxiliary vectors such that $k_0^2=0$,
$k_1^2=-1$ and $k_0.k_1=0$. No sum over repeated indices must be
understood. For instance, we can choose $k_0=(1,0,1,0)$
and $k_1=(0,1,0,0)$. With this choice, it is obvious to see that
the denominator in Eq. (\ref{def_U5U}) can never vanish if the bottom mass is kept.
If one would like to neglect $m_b$, that choice can bring $p_3.k_0$ or
$p_4.k_0$ to zero in some cases. If this happens, one can tell the code to
choose $k_0=(1,0,-1,0)$ instead of the above choice. In fact, that is what we did
in our codes.

In the case of spinors representing a massless state, the helicity
formalism simplifies considerably. Only $A_{\la_i\la_j}$ is
needed. Traditionally we introduce the $C$-numbers $s(p,q)$ and
$t(p,q)$,
\bea 
s(p,q)&\equiv&\bar{u}(p,+)u(q,-)=A_{+-}(p,q),\hs t(p,q)\equiv\bar{u}(p,-)u(q,+)=-s(p,q)^*,\crn
|s(p_1,p_2)|^2&=&s(p_1,p_2)t(p_2,p_1)=2p_1.p_2\,.\eea
These are the functions that appear in our code for the massless
$b$ quark. The massless case is also used when expressing the
gluon polarisation vector to which we now turn.
\section{Transversality and gauge invariance}
The reference vector used for the polarisation of the gluon can be
changed at will. Changing the reference vector from $q$ to
$q^\prime$ amounts essentially to a gauge transformation. Indeed
one has \cite{kleiss_stirling}
\bea
\eps^\mu(p,\la;q^\prime)=e^{i\phi(q^\prime,q)}\eps^\mu(p,\la;q)+\beta(q^\prime,q)p^\mu,\,\,
\label{relation_eps}
\eea where
\bea
e^{i\phi(q^\prime,q)}&=&\left[\fr{s(p,q)}{t(p,q)}\fr{t(p,q^\prime)}{s(p,q^\prime)}\right]^{1/2},\crn
\beta(q^\prime,q)&=&\fr{2}{[4(q^\prime.p)]^{1/2}}\fr{t(q,q^\prime)}{t(q,p)}.
\eea
Therefore up to the phase factor, the difference is contained in
the momentum vector of the gluon. QCD gauge invariance for our
process leads to the important identity
%
%
\bea
|\aliq|^2=|\aliqp|^2,\label{check_gauge}\eea as long as
$q^\prime_{1,2}$ satisfy condition (\ref{cond_refer_vector}).
We have carefully checked that the numerical result for the norm
of each helicity amplitude at various point in  phase space is
independent of the reference vectors $q_{1,2}$ up to 12 digits using double precision.
By default, our numerical evaluation is based on the
use of $q_{1,2}=(p_2,p_1)$. For the checks in the case of massive
$b$ quarks the result with $q_{1,2}=(p_2,p_1)$ is compared with
the one using any $q_{1,2}$ such as conditions (\ref{cond_refer_vector}) are obeyed.
In the case of massless $b$ quarks it is simplest to take
$q_{1,2}=(p_3,p_4)$.

This check is a an important check on many ingredients that enter
the calculation: the Dirac spinors, the gluon polarization
vectors, the propagators, the Lorentz indices, the tensorial loop integrals.
It has been used extensively in our numerical calculation.

\chapter{Optimization with FORM}
\label{optimisation}
\section{Optimization}
Each helicity amplitude ${\cal A}(\la_1,\la_2;\la_3\la_4) \equiv
\alio$, a C-number, is calculated numerically in the Fortran code.
The price to pay is that the number of helicity amplitudes to be
calculated can be large, $16$ in our case for the electroweak loop
part. Some optimisation is necessary. The categorisation of the
full set of diagrams into three  gauge invariant classes as shown
in section~\ref{section_3classes} is a first step. We have sought
to write each diagram as a compact product of blocks and
structures containing different properties of the amplitude. We
write the amplitude according to a colour ordering pattern that
defines three channels. The ordering is in a one-to-one
correspondance with the three channels or diagrams shown in
Fig.~\ref{diag_gg_LO}. The $T$-type is the direct channel, the
$U$-type is the  crossed one obtained from the $T$-type by
interchanging the two gluons and the $S$-type is the one involving
the triple gluon vertex. The helicity amplitude for each diagram
can thus be represented as\footnote{The method we use here is very similar to the one described by Denner in \cite{Denner:1991kt}.}
\bea
\alio^{T,U,S}=CME(a,b)\times Cc\times \sum [FFE\times SME(\la_i)],
\eea
where
\begin{itemize}
\item
$CME(a,b)$ is the colour matrix element. $a,b$ are the colour
indices of the two initial gluons\footnote{Other colour indices of
the bottom quarks are omitted here for simplicity}. The colour
products  can be $(T^aT^b)$, $(T^bT^a)$ or $[T^a,T^b]$
corresponding to the three $T$, $U$, $S$ channels respectively
\item
$Cc$ contains all the common coefficients like the strong coupling constant
$g_s$ or factors common to all diagrams and amplitudes such as the
normalisation factor entering the representation of the
polarisation vector of the gluon, see Eq.~(\ref{eps12_def}). $Cc$ is the same for all diagrams and 
is included at the very end of the numerical evaluation stage.
\item
$FFE$, form factor element, contains all the denominators of
propagators, loop functions as well as various scalar products of
external momenta $\{p_1,p_2,p_3,p_4\}$ {\it i.e.} all the scalar
objects which do not depend on the helicity $\la_i$
\item
$SME(\hat \la)$, standard matrix element, is a product of the
scalar spinor functions $A_{\la_i\la_j}$, $B_{\la_i\la_j}$ and
$C_{\la_i\la_j}$ defined in Appendix~\ref{appendix-helicity}.
\end{itemize}

For each channel, say $\alio^T$, the most complicated and
time-consuming part is the $FFE$. That is why we want to factorise
it out and put it in a common block so that in order to calculate
all the $16$ helicity configurations of $\alio^T$ we just need to
calculate $FFE$ once. This is done at every point in  phase space.
This kind of factorisation can be easily carried out in FORM (see section~\ref{section_form_details}). 

$SME(\hat \la)$ is also complicated  because the bottom quark
is massive and $\gamma_5$ occurs in the ``helicity strings". Thus
we have to optimize this part as well. The way we do it for all
the $3$ groups is as follows. In FORM, we have to find out all the
generic expressions of $SME(\hat \la)$. There are $12$ of them at
tree level and $68$ at one-loop if we choose $q_{1,2}=p_{2,1}$ for
the reference vectors. For instance,
\bea
SME_1&=&[\bar{u}(\la_3,p_3)v(\la_4,p_4)]\times [\eps_{\mu}(\la_1,p_1,p_2)p_4^{\mu}]\times [\eps_{\nu}(\la_2,p_2,p_1)p_4^{\nu}],\crn
&=&BME_1(\la_3,\la_4)\times BME_2(\la_1)\times BME_3(\la_2),
\eea
can be expressed in terms of  $3$ basic matrix elements ($BME$).
Each $BME$ occurs several times when calculating all the $SME(\hat
\la)$. The number of $BME$ is $31$. Each $BME$ is written in terms
of scalar spinor functions $A_{\la_i\la_j}$, $B_{\la_i\la_j}$,
$C_{\la_i\la_j}$. All the $SME$ or $BME$ can be found and
abbreviated in FORM. As an alternative, we can use Perl for such
an operation. The FORM output is converted directly into a Fortran
code
for numerical evaluation. Needless to say, all the abbreviations of $SME$ or $BME$ must be put in common blocks.

To get the final result, we have to sum over all the channels. The
grouping can be re-arranged in terms of an Abelian part and a
non-Abelian part according to
\bea \alio&=&
\alio^{T}+\alio^{U}+\alio^{S}\,,\crn
&\equiv&\{T^a,T^b\}\aliot^{Abel}+[T^a,T^b]\aliot^{NAbel}\,,\eea
where \bea \aliot^{Abel}&=&\fr{1}{2}(\aliot^T+\aliot^U)\,,\crn
\aliot^{NAbel}&=&\aliot^S+\fr{1}{2}(\aliot^T-\aliot^U)\,
\label{A_abel_nabel}
\eea
corresponding to the Abelian and Non-Abelian parts respectively. Amplitudes without color factor are denoted 
by a tilde. The amplitude squared then contains no interference term between
the Abelian and Non-Abelian parts: \bea\mid
\alio\mid^2=\fr{1}{256}\left(\fr{28}{3}\mid
\aliot^{Abel}\mid^2+12\mid \aliot^{NAbel}\mid^2\right)\,
\label{A2_abel_nabel}
\eea where
$\fr{1}{256}=\fr{1}{4}\times\fr{1}{8}\times\fr{1}{8}$ is the spin-
and colour- averaging factor.

The task of our FORM code is to calculate $\sum_{T-diagrams}\aliot^T$, $\sum_{U-diagrams}\aliot^U$ and $\sum_{S-diagrams}\aliot^S$ as functions of $FFE$s and $SME$s. These algebraic expressions are fed into a Fortran code which uses Eqs. (\ref{A_abel_nabel}) and (\ref{A2_abel_nabel}) to calculate the total amplitude squared.  
\section{Technical details}
\label{section_form_details}
What we actually do 
in our FORM code to find all $FFE$ and $SME(\la_i)$ is the following. Consider, for example, a pentagon diagram in 
the class (b) of Fig.~\ref{diag_3group}. In the FORM code, we have to write a subroutine to calculate the loop integral given by
\bea
&&E^{\mu\nu}_r(k_i,m_i)=\int d^Dq\crn
&\times&\fr{N^{\mu\nu}}{(q^2-m_0^2)[(q+k_1)^2-m_1^2][(q+k_2)^2-m_2^2][(q+k_3)^2-m_3^2]
[(q+k_4)^2-m_4^2]},\crn
N^{\mu\nu}&=&(1+A\gamma_5)(m_1+\slashq+\slashk_1)(m_2+\slashq+\slashk_2)\crn
&\times&\gamma^\mu
(m_3+\slashq+\slashk_3)\gamma^\nu(m_4+\slashq+\slashk_4)(1-A\gamma_5),
\label{Emunu_b}
\eea
where $A$ is just some coupling constant; $k_1=p_5$, $k_2=p_5+p_3$, $k_3=k_2-p_1$, $k_4=k_3-p_2$, $m_0=M_W$, 
$m_1=m_2=m_3=m_4=m_t$. 
In order to write $E^{\mu\nu}_r$ in terms of Passarino-Veltman functions $E_{ijk\ldots}$, 
$D_{ij\ldots}$ and $C_{i\ldots}$ we have to expand the numerator $N^{\mu\nu}$. We then use Dirac algebra to re-organise $N^{\mu\nu}$ 
such that each term has the form $\gamma_5(\ldots)^\nu\times q^\mu$ or $\gamma_5(\ldots)^{\mu\nu}\times \slashq$. 
For terms with $q^2$, $q^4$ or $q.k_i$, we use
\bea
q^2&=&(q^2-m_0^2)+m_0^2,\crn
q.k_1&=&\fr{1}{2}\left\{[(q+k_1)^2-m_1^2]-(q^2-m_0^2)+(m_1^2-m_0^2-k_1^2)\right\},\crn
q^2&=&[(q+k_1)^2-m_1^2]+(m_1^2-k_1^2)-2k_1^\eta q_\eta.
\label{id_reduce}
\eea
$D$ and $C$ functions appear when terms in the right-hand-side cancel with denominator factors. 
The advantage of using Eq.~(\ref{id_reduce}) 
is that the rank of tensorial functions is reduced. In the present example, Eq.~(\ref{id_reduce}) helps us avoid tensorial 
five-point functions with rank $4$, which are very complicated. Obviously, numerical evaluation is much faster 
this way. However, there may be a problem when using Eq.~(\ref{id_reduce}) if the library of scalar loop integrals is not complete. This is the situation of section~\ref{section_nnlo}. In that calculation, we have to deal with complex internal masses and our library for scalar four-point functions includes only special cases with $2$ lightlike external momenta. If we use 
Eq.~(\ref{id_reduce}), it will create four-point functions with $3$ or $4$ massive external particles since the momenta in the denominator of Eq.~(\ref{Emunu_b}) are shifted: $k_i\to k_i-k_1$, $i=2,3,4$ and $(k_i-k_1)^2$ are not necessarily zero. We then have a problem with these four-point functions. In this situation, one should not use Eq.~(\ref{id_reduce}) but simply write 
$q^2=g_{\eta\rho}q^\eta q^\rho$, $q.k_i=k_i^\eta q_\eta$ instead. The first aim is to write $E_r^{\mu\nu}$ as the following 
generic expression
\bea
E_r^{\mu\nu}(k_i,m_i)=\sum_{s}f_s^{\mu\nu}(\gamma_5,k_i)T_s(k_i^2,m_i^2),
\label{id_reduce_generic}
\eea
where $T_s$ are Passarino-Veltman loop functions, $f_s^{\mu\nu}(\gamma_5,k_i)$ are complicated functions of $g^{\mu\nu}$, 
$\gamma^\mu$, $k_i^\mu$ and $\gamma_5$. 

In order to write helicity amplitudes in simple forms, we have to choose a good momentum basis. In the calculation 
of $\ppbbH$ our basis is $\left\{p_1,p_2,p_3,p_4\right\}$. 
We have to replace $k_i$ by $p_i$ to get $E_r^{\mu\nu}(p_i,m_i)$. One can then use the on-shell condition to simplify $E_r^{\mu\nu}$. In practice, one knows that each $T_s(p_i^2,m_i^2)$ appears several times in the calculation and has a very lengthy 
expression. For optimization and having compact expression of 
$E_r^{\mu\nu}(p_i,m_i)$ we introduce abbreviations for $T_s(k_i^2,m_i^2)$. Those abbreviations serve as the library of Passarino-Veltman loop functions for the present calculation. The way to do this in FORM is as follows.
\begin{quote}
local $F=\sum E_r^{\mu\nu}(p_i,m_i)$;
bracket $T_s$;
.sort\\
collect cf1;
.sort\\
polyfun cf1;
.sort\\
polyfun;
id cf1(x?)=1;
.sort\\
print +s $F$;
.end
\end{quote} 
The result of this simple FORM script is that $F$ is a sum of independent Passarino-Veltman loop functions $T_s(k_i^2,m_i^2)$. One can use the command "\#write" to produce an output file if one wishes. 
Now one can use Perl \cite{perl} to introduce an abbreviation for each term: $T_s(k_i^2,m_i^2)\to fT_s$. We have
\bea
E_r^{\mu\nu}(p_i,m_i)=\sum_{s}f_s^{\mu\nu}(\gamma_5,p_i)\times fT_s.
\label{id_reduce_generic_abbr}
\eea  
Apart from $CME$ and $Cc$, a generic $T$-channel helicity amplitude is calculated in FORM as
\bea
\aliot^T=\bar{u}(\la_3,p_3)(\sum_{r} E_r^{\mu\nu})\eps_\mu(p_1,q_1)\eps_\nu(p_2,q_2)v(\la_4,p_4).
\eea
We can now simplify this expression by using the transversality condition for the gluon polarization vectors and Dirac equation. For the latter, we have to re-organise $f_s^{\mu\nu}(\gamma_5,p_i)$ to have the form: 
$\slashp_3\gamma_5 \slashp_1\gamma^\mu \gamma^\nu \slashp_2 \slashp_4$.
\bea
p_1^\mu\eps_\mu(p_1,q_1)&=&0, \hs p_2^\nu\eps_\nu(p_2,q_2)=0,\crn
\bar{u}(\la_3,p_3)\slashp_3&=&m_b\bar{u}(\la_3,p_3), \hs \slashp_4v(\la_4,p_4)=-m_bv(\la_4,p_4).
\eea 
Notice that if one chooses $q_1=p_2$ and $q_2=p_1$ then $\aliot$ is further simplified by using 
\bea p_2^\mu \eps_\mu(p_1,p_2)=0, \hs p_1^\nu \eps_\nu(p_2,p_1)=0.\eea 
We are now in the position to factorize each term of $\alio^T$ as a product of $FFE$ and $SME(\la_i)$. The trick to find all $SME(\la_i)$ is the same as above: using the combination (bracket, collect, polyfun).
\begin{quote}
local $FT=\aliot^T$;
local $FU=\aliot^U$;
local $FS=\aliot^S$;\\
bracket $\bar{u},v,\eps,\gamma$;
.sort\\
collect cf0;
.sort\\
local $FTUS=FT+FU+FS$;
.sort\\
polyfun cf0;
.sort\\
polyfun;
id cf0(x?)=1;
.sort\\
print +s $FTUS$;
.end
\end{quote}
The result of this simple FORM script is that $FTUS$ is a sum of independent helicity structures 
$\bar{u}(\la_3,p_3)f_s^{\mu\nu}(\gamma_5,p_i)\eps_\mu(p_1,q_1)\eps_\nu(p_2,q_2)v(\la_4,p_4)$. Now one can use Perl to give each term a name $SME(\la_i)$. For $FFE$, it is just slightly more complicated 
\begin{quote}
local $FT=\aliot^T$;
local $FU=\aliot^U$;
local $FS=\aliot^S$;\\
bracket $\bar{u},v,\eps,\gamma$;
.sort\\
collect cf0;
normalize cf0;
.sort\\
local $FTUS=FT+FU+FS$;
bracket cf0;
.sort\\
collect cf1;
.sort\\
polyfun cf1;
.sort\\
polyfun;
id cf1(x?)=1;
.sort\\
print +s FTUS;
.end
\end{quote}
where the command "normalize cf0;" is very important. The result of this FORM script is that $FTUS$ is a sum of cf0$[\sum(X_s(p_i.p_j)\times fT_s)]$ where $X_s(p_i.p_j)$ are just simple algebra expressions, $fT_s$ are loop functions. 
 We can then use Perl to give the argument of each function cf0 a name $FFE$. Thus $FFE\equiv \sum(X_s(p_i.p_j)\times fT_s)$. 

To sum up, the working stream of our FORM code is the following. Input: all expressions of helicity amplitudes for all Feynman diagrams. This is just simply applying the Feynman rules. Output: $\sum_{T-diagrams}\aliot^T$, $\sum_{U-diagrams}\aliot^U$ and $\sum_{S-diagrams}\aliot^S$ as functions of $FFE$s and $SME$s. The major source of bugs is at the beginning when we type in the Feynman-rule-amplitude-expressions. The rest is almost automatic. One might wonder about the connection between FORM and Perl. This is semi-automatic in our code, \ie we have to run FORM and Perl separately. The three FORM scripts described above generate their output files. We write three very simple Perl scripts to read those files and introduce abbreviations. Those Perl scripts also prepare three output files to be read by FORM. Indeed if one does not like using Perl and wants to do everything automatically within FORM, this is possible. 
\section{Automation with FORM}
In this section, we would like to show that the working stream described in the previous section can be automatized in FORM without invoking Perl. We have not done this in the $\ppbbH$ calculation. However, the implementation is straight forward. The only difficulty is "How to introduce abbreviations?". The answer is in the following FORM example\footnote{We have learnt these tricks from a private communication with Vermaseren.}.
\begin{quote}
Symbol	a,b,c,d,e,x,y,n;
CFunction	f1,f2,f3;\\
Local	F = x*(1+a+b+c)\^\ \hspace*{-1.5mm}3+y*(1+a+b+c)\^\ \hspace*{-1.5mm}2;\\
AntiBracket	x,y;
.sort\\
Collect f1;
Makeinteger f1;
Bracket	f1;
.sort

*** which terms to be abbreviated? ***\\ 
Keep Brackets;\\
id	f1(x?) = f1(nterms\_(x),x);\\
id	f1(1,x?) = x;\\
id	f1(n?,x?) = f1(-termsinbracket\_(0),x);\\
id	f1(-1,x?) = x;\\
Bracket	f1;
.sort

*** give it a name and store it by using \$ variable ***\\
Keep Brackets;\\
\#\$cou = 0;\\
if ( count(f1,1)!=0 );
	\$cou = \$cou + 1;
	id f1(n?,x?) = f2(\$cou)*f3(\$cou,x);
endif;\\
Bracket	f2,f3;
.sort\\
\#do i = 1,`\$cou'
	id	f3(`i',x?\$t`i') = 1;
\#enddo
.sort

*** using temporary expressions for writing output file ***\\
\#do i = 1,`\$cou'
local XX`i'=\$t`i';
\#enddo
.sort

*** for fortran output files ***\\
format doublefortran;\\
\#write $<$abbrf2.F$>$ "      Subroutine abbreviation(x,y)"\\
\#write $<$abbrf2.F$>$ "      IMPLICIT DOUBLE PRECISION (A-H, O-Z)"\\
\#write $<$abbrf2.F$>$ "      DOUBLE PRECISION f2(`\$cou')"\\
\#write $<$abbrf2.F$>$ "      Common/abbr/f2"\\
\#do i = 1,`\$cou'\\
*	\#write $<$abbrf2.F$>$ "      f2(`i') = `\$t`i''"\\
	\#write $<$abbrf2.F$>$ "      f2(`i') = \%e",XX`i'\\
\#enddo\\
\#write $<$abbrf2.F$>$ "      End"\\
\#write $<$funct.F$>$ "      Function fun(x,y,a,b,c)"\\
\#write $<$funct.F$>$ "      IMPLICIT DOUBLE PRECISION (A-H, O-Z)"\\
\#write $<$funct.F$>$ "      DOUBLE PRECISION f2(`\$cou')"\\
\#write $<$funct.F$>$ "      Common/abbr/f2"\\
\#write $<$funct.F$>$ "      fun=\%e",F(fun)\\
\#write $<$funct.F$>$ "      Return"\\
\#write $<$funct.F$>$ "      End"\\
.end
\end{quote}        
 The working stream of this example is the following. Input: an algebraic expression named F. Output: two Fortran files to calculate 
F: "funct.F" and "abbrf2.F". The latter computes all abbreviations which are "complicated" functions of (x,y) and appear several times in the final result. This is nothing but the idea of optimization.

\chapter{Phase space integral}
\label{appendix_integral_phase}
\section{$2\to 3$ phase space integral}
\begin{figure}[hpt]
\begin{center}
\includegraphics[width=0.6\textwidth]{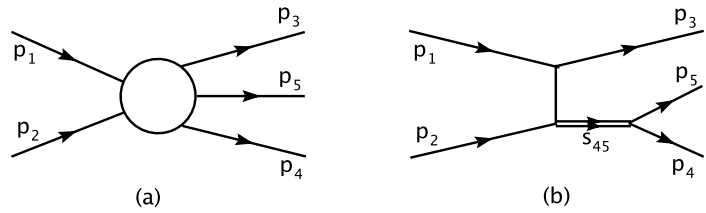}
\caption{\label{R3_R_2}{\em A typical $2\to 3$ Feynman diagram. The arrow gives the momentum direction.}}
\end{center}
\end{figure}
The phase space integral is given by 
\bea
R_3=\int\fr{d^3\pp_3}{2e_3}\fr{d^3\pp_4}{2e_4}\fr{d^3\pp_5}{2e_5}\delta^4(p_1+p_2-p_3-p_4-p_5).\label{int_r3}\eea
There are $9$ integration variables with $4$ constraints from the Dirac delta function. All the interactions we consider are spherically symmetric, it means that there is one trivial variable $\phi$ corresponding to rotation around the $z$-axis. Integration over $\phi$ gives a factor of $2\pi$. Thus, there are $4$ essential final state variables\footnote{For a $2\to n$ process, the number of essential final state variables is $3n-5$.}. We define the kinematical function
\bea \lambda (x,y,z)&\equiv &x^2+y^2+z^2-2xy-2xz-2yz. \eea
$R_3$ can be factorized into $2\to 2$ and $1\to 2$ processes (see Fig. \ref{R3_R_2}b)
\bea R_3=\int ds_{45}R_2(s,s_{45},m_b^2)R_2(s_{45},m_b^2,m_H^2)\,,\label{int_r3_2r2}\eea 
with \bea R_2(s_{ij},m_i^2,m_j^2)=\fr{\lambda^{1/2}(s_{ij},m_i^2,m_j^2)}{8s_{ij}}\int
d\cos\theta_i^{(ij)}d\phi_i^{(ij)}\,,\eea 
where $\theta
(\phi)_i^{(ij)}$ are the angles determined in the rest frame of
$(i+j)$, $s_{ij}=(p_i+p_j)^2$ with $i,j=3,4,5$. Clearly, formula (\ref{int_r3_2r2}) is just one way of factorizing the phase space integral. We can replace $s_{45}$ by $s_{34}$ or $s_{35}$. In practice, choosing a good set of integration variables makes the integral convergent much faster. For a complicated calculation it is very difficult to know which choice is the best. In that case, we should always start with the tree level and try all the possibilities of phase space factorization with the same number of Monte Carlo points (if one uses the Monte Carlo method) and compare the integration errors to judge the best choice. That is what we did to find out that Eq. (\ref{int_r3_2r2}) is the best way to parameterise the phase space in the case of $\ggbbH$ calculation. 

\begin{figure}[hpt]
\begin{center}
\includegraphics[width=0.6\textwidth]{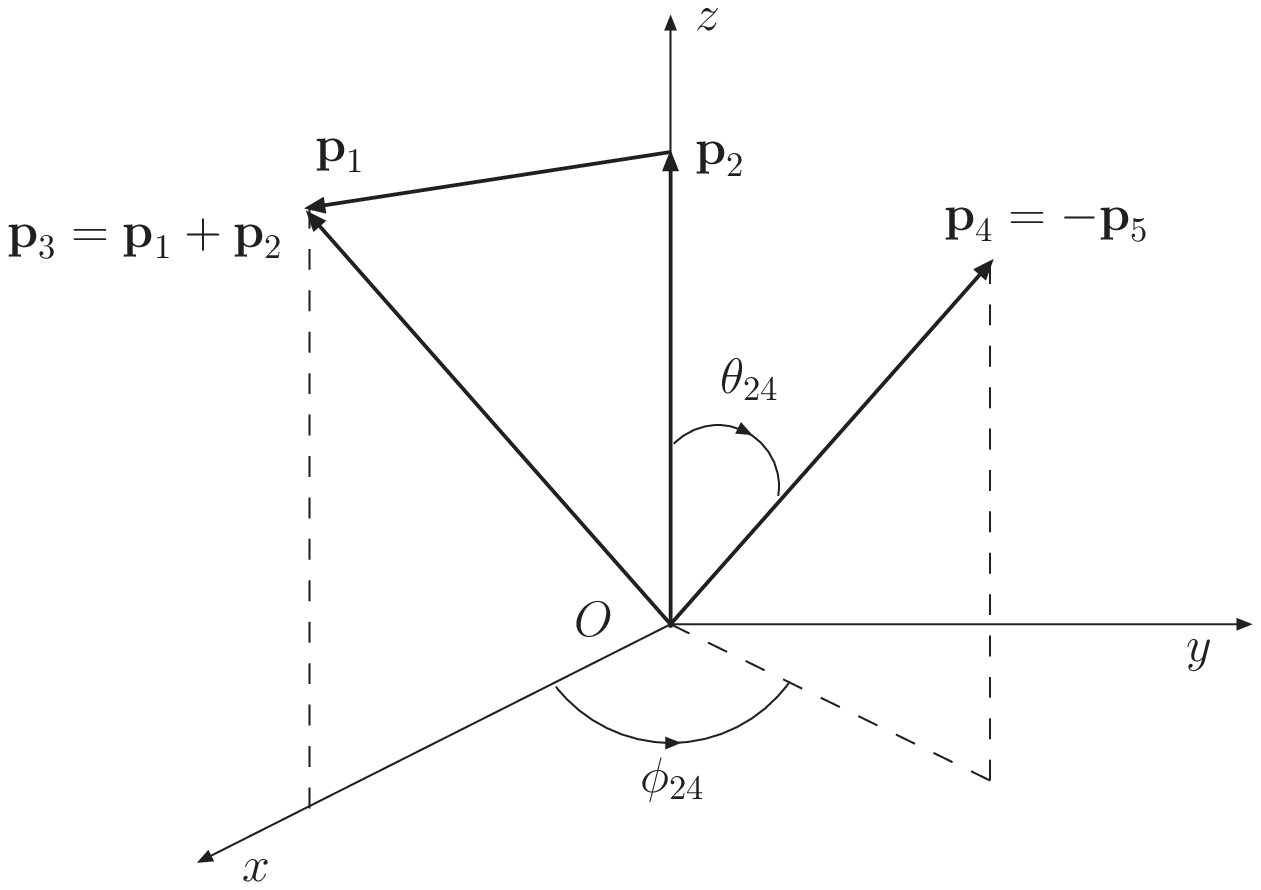}
\caption{\label{R3_Oxyz}{\em Coordinate system in the frame $\pp_4+\pp_5=0$.}}
\end{center}
\end{figure}
We choose $Oz\parallel p_2$. In the center-of-mass system (CMS) of $(4+5)$, we call this the CMS45 hereafter, one has $\pp_4+\pp_5=\pp_1+\pp_2-\pp_3=0$. Thus $\pp_1$, $\pp_2$ and $\pp_3$ define a plane chosen to be $Oxz$ (see Fig. \ref{R3_Oxyz}). $\pp_4$ is defined by two angles $\theta_{24}$ and $\phi_{24}$. 
\bea R_2(s_2,m_b^2,M_H^2)=\fr{\lambda^{1/2}(s_{2},m_b^2,M_H^2)}{8s_{2}}
\int_{-1}^1d\cos\theta_{24}\int_0^{2\pi}d\phi_{24}.\eea
In the CMS of $(1+2)$, we call this the CMSgg hereafter, one gets
\bea
R_2(s,s_2,m_b^2)=2\pi\fr{\lambda^{1/2}(s,s_2,m_b^2)}{8s}
\int_{-1}^1d\cos\theta_{23}\,,
\eea
where the factor $2\pi$ comes from the trivial integration over $\phi_{23}$. The range of $s_2$ is given by
\bea
s\ge &s_2&\ge (m_b+M_H)^2,\crn
s_0\ge s&=&x_1x_2s_0 \ge (2m_b+M_H)^2,\label{range_s2_s} 
\eea
where $s_0$ is the invariant mass of the initial protons, $x_{1,2}\in [0,1]$ are the momentum fractions carried by the initial gluons. The second equation in (\ref{range_s2_s}) implies that
\bea
1\ge x_1\ge \fr{(2m_b+M_H)^2}{s_0},\hs 1\ge x_2 \ge \fr{(2m_b+M_H)^2}{s_0x_1}.
\eea
The integration formula we actually use in the Fortran code, taking into account the convolution of the gluon structure functions, reads
\bea
\hat{R}_3&=&2\pi\int_0^1dx_1^\prime\int_0^1dx_2^\prime\int_0^1ds_2^\prime\int_{-1}^1d\cos\theta_{23}
\int_{-1}^1d\cos\theta_{24}\int_0^{2\pi}d\phi_{24},\crn
&\times&J\fr{\lambda^{1/2}(s,s_2,m_b^2)}{8s}\fr{\lambda^{1/2}(s_{2},m_b^2,M_H^2)}{8s_{2}}
\label{R3_final}\eea
where we have changed the integration variables as follows
\bea
x_1&=&x_1^\prime\left[1-\fr{(2m_b+M_H)^2}{s_0}\right]+\fr{(2m_b+M_H)^2}{s_0},\crn
x_2&=&x_2^\prime\left[1-\fr{(2m_b+M_H)^2}{s_0x_1}\right]+\fr{(2m_b+M_H)^2}{s_0x_1},\crn
s_2&=&s_2^\prime[s-(m_b+M_H)^2]+(m_b+M_H)^2,
\eea 
and $J$ is the Jacobian
\bea
J=\left[1-\fr{(2m_b+M_H)^2}{s_0}\right]\left[1-\fr{(2m_b+M_H)^2}{s_0x_1}\right][s-(m_b+M_H)^2].
 \eea
We should stress again that $\theta_{24}$ and $\phi_{24}$ are defined in the CMS45 while $\cos\theta_{23}$ is defined in the CMSgg. 
In order to calculate the helicity amplitudes, which are Lorentz invariant, one has to reconstruct from $\{s_2,\cos\theta_{23},\cos\theta_{24},\phi_{24}\}$ all the components of $5$ external momenta in some reference frame. The way we do this for the CMSgg is as follows. First, the components of $p_4$ and $p_5$ can be easily calculated in the CMS45 
\bea
\pp_5&=&-\pp_4, \hs \mid\pp_5\mid=\mid\pp_4\mid=\fr{\lambda^{1/2}(s_2,m_b^2,M_H^2)}{2\sqrt{s_2}}, \crn
\pp_{4z}&=&\mid\pp_4\mid\cos\theta_{24}, \crn
\pp_{4x}&=&\mid\pp_4\mid\sin\theta_{24}\cos\phi_{24},\crn 
\pp_{4y}&=&\mid\pp_4\mid\sin\theta_{24}\sin\phi_{24}.
\eea 
In the CMSgg we get 
\bea
p_1&=&(\sqrt{s},0,0,-\sqrt{s}), \hs p_2=(\sqrt{s},0,0,\sqrt{s}),\crn
\pp_{45}&=&\pp_4+\pp_5=-\pp_3, \hs \mid\pp_3\mid=\fr{\lambda^{1/2}(s,s_2,m_b^2)}{2\sqrt{s}}, \hs e_{45}=\sqrt{s_2+\mid\pp_3\mid^2},\crn
\pp_{3z}&=&\mid\pp_3\mid\cos\theta_{23}, \hs \pp_{3x}=\mid\pp_3\mid\sin\theta_{23}, \hs \pp_{3y}=0.
\eea
In the CMSgg one sees that the CMS45 is moving with the $4$-component momentum $p_{45}=(e_{45},-\pp_3)$. We then boost $p_4$ calculated above in the 
CMS45 to be in the CMSgg by using the following general Lorentz transformation
\bea
e^\prime&=&\gamma_0 e-\gamma_0\vv_0.\pp,\crn
\pp^\prime&=&\pp+\gamma_0\vv_0(\fr{\gamma_0\vv_0.\pp}{\gamma_0+1}-e),\label{trans_lorentz}
\eea        
with 
\bea
\gamma_0=\fr{e_0}{m_0}, \hs m_0=\sqrt{e_0^2-\pp_0^2}, \vv_0=\fr{1}{e_0}(\pp_{0x},\pp_{0y},\pp_{0z}),\label{trans_lorentz1}
\eea
where $p_0$ is the $4$-component momentum of a reference frame $K^\prime$ observed in $K$, $p$ is any momentum observed in $K$ and 
$p^\prime$ is the same momentum observed in $K^\prime$. The inverse of equations (\ref{trans_lorentz}) and (\ref{trans_lorentz1}) are obtained by changing 
$\vv_0$ to $-\vv_0$ and by interchanging primed and unprimed variables. Now we have all the components of $p_1$, $p_2$, $p_3$, $p_4$, $p_5=p_1+p_2-p_3-p_4$ in one 
reference frame, CMSgg. 

In the CMS of
two initial protons (CMSPP), the proton and gluon momenta are \bea
P_2^\mu&=&(\sqrt{s_0},0,0,\sqrt{s_0})\,\,,P_1^\mu=(\sqrt{s_0},0,0,-\sqrt{s_0})\,,\crn
p_1&=&x_1P_1\,\,,p_2=x_2P_2\,,\eea where we have neglected the
proton mass\footnote{The proton mass $m_p=0.9383$GeV.}. All the kinematical cuts are defined in the CMSPP. 
One can move from the CMSgg to CMSPP by using the following boost matrix
along the $z$ axis
\bea\Lambda_{gg\rightarrow PP}=\left(%
\begin{array}{cccc}
  \gamma & 0 & 0 & -\beta\gamma \\
  0 & 1 & 0 & 0 \\
  0 & 0 & 1 & 0 \\
  -\beta\gamma & 0 & 0 & \gamma \\
\end{array}%
\right)\,,\eea with $\gamma=\fr{x_1+x_2}{2\sqrt{x_1x_2}}$,
$\beta=\fr{x_1-x_2}{x_1+x_2}$. This boost matrix can be easily
found from the Lorentz transformation of the four vector
$(p_1+p_2)$ from the CMSgg to CMSPP. 
The helicity amplitudes can be calculated in the CMSgg or CMSPP as one wishes.   

For experimental purpose, 
one has to impose cuts on the transverse momenta and pseudorapidities of the bottom and
anti-bottom in CMSPP
\bea |\textbf{p}_{3T}|\,,|\textbf{p}_{4T}|\ge  
|\textbf{p}_{T}|_{min}\hs \textrm{and}\hs \vert \eta_{3,4}\vert \le \eta_{max}
\eea
where the values of $|\textbf{p}_{T}|_{min}$ and $\eta_{max}$ depend on the experiment, the definition of pseudorapidity is 
\bea
\eta=-\ln\left[\tan\left(\fr{\theta}{2}\right)\right].
\eea 
Those kinematical cuts also help to avoid some possible zero poles associated with some bottom-quark propagators in the massless limit and with some Gram determinants related to the tensorial reduction of loop integrals.
\section{Numerical integration with BASES}
\label{section_bases}
BASES is a Monte Carlo integrator which functions by means of the importance and stratified sampling method \cite{bases}. 
Executions of BASES consists of the grid optimization and integration steps. Both steps are done by performing a number of {\em iterations}. An iteration is the process of computing the estimate of an integral and its variance. Each iteration is a Monte Carlo integration with $N_{call}$ sample points and is realised as follows. The full multi-dimensional integral volume is covered by a grid of $N_{cube}$, the number of hypercubes (each hypercube is divided into many subregions). In each hypercube, the integral and its variance are evaluated with $N_{trial}=N_{call}/N_{cube}$ sample points. The results of each iteration are obtained by summing up results of all hypercubes. $N_{cube}$ is calculated as folows
\bea
N_{cube}=N_{region}^{N_{wild}}<32768, \hs N_{region}=\left(\fr{N_{call}}{2}\right)^{1/N_{wild}}\le 25,
\eea
with $N_{wild}$ is the number of {\em wild variables} on which the integrand depends strongly or exhibits singular behavior. $N_{call}$ and $N_{wild}$ are BASES input parameters. Clearly $N_{wild}\le N_{dim}$, $N_{dim}$ the number of integration variables, and the maximum number of wild variables is $15$. As input information for BASES, one has to decide which are the "wild" variables and place them at the beginning of the integration variable array. 
\begin{itemize}
\item Grid optimization step: at the first iteration the grid is uniformly defined for each variable axis. After each iteration the grid is adjusted so as to make the size of the subregions narrower at the parts with larger function value and wider at the parts with the smaller one. In this way a suited grid to the integrand is obtained. The number of iterations for this step is denoted $ITMX1$ (default $15$), a BASES input parameter.
\item Integration step: the probability to select each hypercube and the maximum value of the function in it are calculated as well as the estimate of integral with the frozen grid determined in the former step. The number of iterations for 
this step is denoted $ITMX2$ (default $100$), a BASES input parameter.  
\end{itemize}
The typical BASES input parameters for calculating (\ref{R3_final}) are: $N_{dim}=6$, $N_{wild}=2$, $N_{call}=10^5$, $ITMX1=20$ and $ITMX2=130$. The two wild variables are $\cos\theta_{23}$ and $\cos\theta_{24}$ placed at the beginning of the integration variable array. With those input parameters, the typical error we obtained for the $\ppbbH$ calculation is $0.08\%$.

\chapter{Mathematics}
\label{appendix_math}
\section{Logarithms and Powers}
The natural logarithm $\ln(z)$ is defined as
\bea
\ln(z)=\ln(|z|)+i\,\arg(z),
\eea 
with $-\pi\le \arg(z)\le \pi$. The logarithm $\ln z$ has a branch cut along the negative real axis. 
The general power $w=z^\alpha$ ($\alpha$ is a complex constant) is defined with the aid 
of the exponential function 
\bea
z^\alpha=(e^{\ln z})^\alpha=e^{\alpha\ln z}.
\eea
With those definitions, one has the following rules
\bea
\ln(z_1z_2)&=&\ln(z_1)+\ln(z_2)+\eta(z_1,z_2),\label{log_z1z2}\\
\eta(z_1,z_2)&=&2\pi i[\theta(-\Img\,z_1)\theta(-\Img\,z_2)\theta(\Img\,z_1z_2)-
\theta(\Img\,z_1)\theta(\Img\,z_2)\theta(-\Img\,z_1z_2)],\crn
(z_1z_2)^\alpha&=&e^{\alpha\ln(z_1z_2)}=e^{\alpha[\ln(z_1)+\ln(z_1)+\eta(z_1,z_2)]}=e^{\alpha\eta(z_1,z_2)}z_1^\alpha z_2^\alpha,
\label{power_z1z2}
\eea
which have important consequences
\bea
\ln(z_1z_2)&=&\ln(z_1)+\ln(z_2) \,\,\, \text{if $\Img\,z_1$ and $\Img\,z_2$ have different sign}\crn
\ln\fr{z_1}{z_2}&=&\ln(z_1)-\ln(z_2) \,\,\, \text{if $\Img\,z_1$ and $\Img\,z_2$ have the same sign}\crn
(z_1z_2)^\alpha&=&z_1^\alpha z_2^\alpha \,\,\, \text{if $\Img\,z_1$ and $\Img\,z_2$ have different sign}.
\label{app_math_lnab}
\eea
For $-z=a-i\rho$ with $a$ real and $\rho\to 0^+$ we have
\bea
\ln(-z)&=&\left\{ \begin{array}{ll}\ln\vert a\vert & \textrm{if}\hs a>0\\ 
                                  \ln\vert a\vert-i\pi & \textrm{if}\hs a<0 \end{array}\right. \crn
\arg(-z)&=&\arg(z)-\pi,\crn 
\ln(-z)&=&\ln(|z|)+i\,\arg(-z)=\ln(|z|)+i\,\arg(z)-i\pi=\ln(z)-i\pi,\crn
(-z)^\alpha&=&e^{-i\pi\alpha}e^{\alpha\ln(z)}=e^{-i\pi\alpha}z^\alpha.\label{rel_pmz}
\eea
If $A$ and $B$ are real then 
\bea
\ln(AB-i\rho)=\ln(A-i\rho^\prime)+\ln(B-i\rho/A),
\label{log_AB}
\eea
where $\rho^\prime$ is infinitesimal and has the same sign as $\rho$. From this we get
\bea
(AB-i\rho)^\alpha=e^{\alpha\ln(AB-i\rho)}=e^{\alpha[\ln(A-i\rho^\prime)+\ln(B-i\rho/A)]}=(A-i\rho^\prime)^\alpha(B-i\rho/A)^\alpha.
\eea
\section{Dilogarithms}
The dilogarithm or Spence function is defined by \cite{hooft_velt, Maximon2003, Lewin81}
\bea
\Sp(z)=-\int_0^1dt\fr{\ln(1-zt)}{t},
\eea
where $z$ may be complex. The logarithm has a branch cut along the negative real axis, implying for the Spence function a cut along the positive 
real axis from $1$ to $+\infty$. When one is in a problematic situation, the following transformation formulae may be helpful
\bea
\Sp(z)&=&-\Sp(\fr{1}{z})-\fr{1}{6}\pi^2-\fr{1}{2}\ln^2(-z),\label{spence_inverse}\\
\Sp(z)&=&-\Sp(1-z)+\fr{1}{6}\pi^2-\ln(1-z)\ln(z)\label{spence_one_minus}.
\eea 
More transformation formulae can be found in \cite{Maximon2003, Lewin81}.
\section{Gamma and Beta functions}
The gamma function $\Gamma(z)$ is a function of the complex variable $z$. For $\Rel z>0$ it is defined by
\bea
\Gamma(z)=\int_0^\infty dt e^{-t}t^{z-1}
\label{gamma1}
\eea
where the principal value of $t^{z-1}$ is to be taken. 
For $\Rel z\le 0$ the alternative definition reads \cite{book_carrier}
\bea
\Gamma(z)=\fr{1}{2i\sin\pi z}\int_C dt e^t t^{z-1},
\label{gamma2}
\eea
where the path of integration $C$ starts at $-\infty$ on the real axis, circles the origin once in the positive direction, and returns to $-\infty$; the initial and final arguments of $t$ are to be $-\pi$ and $\pi$, respectively. The latter defines an analytic function for all $z$ other then $0$, $\pm 1$, 
$\pm 2$, \ldots . For positive integers $n$ definition (\ref{gamma1}) gives
\bea
\Gamma(n+1)=n\Gamma(n)=n!.
\eea
$\Gamma(z)$ is analytic everywhere, except at the points $z=0$, $-1$, $-2$, \ldots . 
The following properties of the gamma function are very useful
\bea
\Gamma(z+1)&=&z\Gamma(z),\crn
\Gamma(z)\Gamma(1-z)&=&\fr{\pi}{\sin\pi z},\crn
\Gamma(\eps)&=&\fr{1}{\eps}-\gamma_E+\OO(\eps), \hs \eps\to 0,
\eea
where $\gamma_E$ is Euler constant.

The beta function is defined by
\bea
B(p,q)=\int_0^1 dt t^{p-1}(1-t)^{q-1}
\eea
where $\Rel p>0$ and $\Rel q>0$; the principal values of the various powers are to be taken. The analytic continuation of $B(p,q)$ onto 
the left halves of the $p$ and $q$ planes is achieved by using
\bea
B(p,q)=\fr{\Gamma(p)\Gamma(q)}{\Gamma(p+q)}.
\eea
\section{Integrals}
Formulae to move to spherical coordinates:
\bea
\int dt_1\cdots dt_{K}=\int r^{K-1}drd\Omega_{K-1},\,\,\, \int d\Omega_{K-1}=\fr{2\pi^{K/2}}{\Gamma(K/2)}.
\label{int_spherical}
\eea
The following integral formula is very useful in many cases
\bea
\int_0^{\infty}ds\fr{s^{\alpha-1}}{(z+s)^\beta}=z^{(\alpha-\beta)}
\frac{\Gamma(\beta-\alpha)\Gamma(\alpha)}{\Gamma(\beta)},
\label{int_master_z}
\eea
where $z$ can be complex.\\
The following integral usually appears in loop calculation \cite{hooft_velt}
\bea
S_3=\int_0^1dy\fr{\ln(ay^2+by+c)-\ln(ay_0^2+by_0+c)}{y-y_0},
\eea
where $a$ is real, while $b$, $c$ and $y_0$ may be complex, with the restriction that $\Img(ay^2+by+c)$ has the same sign for $0\le y\le 1$. 

Let $\eps$ and $\rho$ be infinitesimally real quantities having the opposite sign to $\Img(ay^2+by+c)$ and $\Img(ay_0^2+by_0+c)$ respectively. 
Using Eq. (\ref{log_AB}) we get
\bea
\ln(ay^2+by+c)&=&\ln(a-i\eps)+\ln[(y-y_1)(y-y_2)],\crn
\ln(ay_0^2+by_0+c)&=&\ln(a-i\rho)+\ln[(y_0-y_1)(y_0-y_2)],
\eea
where $y_{1,2}$ are two roots of equation
\bea
y^2+\fr{b}{a}y+\fr{c}{a}=0.
\eea
We then use Eq. (\ref{log_z1z2}) to get
\bea
S_3&=&\int_0^1dy\fr{\ln[(y-y_1)(y-y_2)]-\ln[(y_0-y_1)(y_0-y_2)]}{y-y_0}\crn
&-&\eta(a-i\eps,\fr{1}{a-i\rho})\ln\fr{y_0-1}{y_0}.
\eea
We write
\bea
\ln[(y_0-y_1)(y_0-y_2)]&=&\ln(y_0-y_1)+\ln(y_0-y_2)+\eta(y_0-y_1,y_0-y_2)\crn
\ln[(y-y_1)(y-y_2)]&=&\ln(y-y_1)+\ln(y-y_2)+\eta(y-y_1,y-y_2)
\eea
with 
\bea
\eta(y-y_1,y-y_2)=\eta(-y_1,-y_2)
\eea
since $y$ is real and $\Img[(y-y_1)(y-y_2)]=\Img(\fr{b}{a}y+\fr{c}{a})=\Img(\fr{c}{a})=\Img(y_1y_2)$ as we assumed at the beginning. We have
\bea
S_3&=&\int_0^1dy\fr{\ln(y-y_1)-\ln(y_0-y_1)+\ln(y-y_2)-\ln(y_0-y_2)}{y-y_0}\crn
&+&\left[\eta(-y_1,-y_2)-\eta(y_0-y_1,y_0-y_2)-\eta(a-i\eps,\fr{1}{a-i\rho})\right]\ln\fr{y_0-1}{y_0}.\;\;\;
\eea
For this, we have to calculate
\bea
R(y_1,y_0)=\int_0^1dy\fr{\ln(y-y_1)-\ln(y_0-y_1)}{y-y_0}.
\eea
We change the integration variable $y=y^\prime+y_1$ to get
\bea
R=\int_{-y_1}^{1-y_1}dy\fr{\ln y-\ln(y_0-y_1)}{y-y_0+y_1}.
\eea
Since the residue of the pole is zero and the logarithmic cut along the negative real axis is outside the triangle $[0,-y_1,1-y_1]$, we can write
\bea
\int_{-y_1}^{1-y_1}=\int_{0}^{1-y_1}-\int_{0}^{-y_1}.
\eea   
We then make the substitutions $y=(1-y_1)y^\prime$ and $y=-y_1y^\prime$ to get
\bea
R&=&\int_0^1dy\fr{d}{dy}\left[\ln\left(1+y\fr{1-y_1}{y_1-y_0}\right)\right]\left\{\ln[(1-y_1)y]-\ln(y_0-y_1)\right\}\crn
&-&\int_0^1dy\fr{d}{dy}\left[\ln\left(1-y\fr{y_1}{y_1-y_0}\right)\right]\left\{\ln(-y_1y)-\ln(y_0-y_1)\right\}.
\eea
Since $y$ is real and positive the logarithmic arguments never cross the cut along the negative real axis. We do partial integration to get
\bea
R&=&\ln\left(\fr{1-y_0}{y_1-y_0}\right)\left[\ln(1-y_1)-\ln(y_0-y_1)\right]-\ln\left(\fr{-y_0}{y_1-y_0}\right)\left[\ln(-y_1)-\ln(y_0-y_1)\right]\crn
&-&\Sp\left(\fr{y_1}{y_1-y_0}\right)+\Sp\left(\fr{1-y_1}{y_0-y_1}\right).
\eea
One uses Eq. (\ref{spence_one_minus}) to obtain
\bea
R(y_1,y_0)&=&\Sp\left(\fr{y_0}{y_0-y_1}\right)-\Sp\left(\fr{y_0-1}{y_0-y_1}\right)+\ln\left(\fr{y_0}{y_0-y_1}\right)\eta(-y_1,\fr{1}{y_0-y_1})\crn
&-&\ln\left(\fr{1-y_0}{y_1-y_0}\right)\eta(1-y_1,\fr{1}{y_0-y_1}).
\eea
The result for $S_3$ reads
\bea
S_3&=&R(y_1,y_0)+R(y_2,y_0)\crn
&+&\left[\eta(-y_1,-y_2)-\eta(y_0-y_1,y_0-y_2)-\eta(a-i\eps,\fr{1}{a-i\rho})\right]\ln\fr{y_0-1}{y_0}\hs
\label{result_S3}
\eea
which contains $4$ Spence functions.

\chapter{Scalar box integrals with complex masses }
\label{appendix-box-integral}
The calculation of the scalar one-loop function for the box (N = 4) with imaginary
internal masses in the most general case with no restriction on the external invariants is
not tractable. The standard technique of 't Hooft and Veltman \cite{hooft_velt} (see also \cite{Denner:1991qq}) has some restriction on the values of external momenta. In particular, the method works if at least one of the external momenta is lightlike. In our present calculation, there are at least $2$ lightlike external momenta in all boxes. 
We explain here our derivation based on the method given in \cite{hooft_velt} for this special case. 

With $N=D=4$, from Eq.~(\ref{eq_TN0_3}) we get
\bea
D_0&\equiv& (4\pi)^2T_0^4=\int_0^1dx\int_0^xdy\int_0^ydz\crn
&\times&\fr{1}{(ax^2+by^2+gz^2+cxy+hxz+jyz+dx+ey+kz+f)^2},\label{d0_xyz}
\eea 
where we have changed the integration variables as $t=\sum_{i=1}^4x_i$, $x=\sum_{i=1}^3x_i$, $y=x_1+x_2$, $z=x_1$; and 
\bea
a&=&\fr{1}{2}(Q_{33}+Q_{44}-2Q_{34})=p_3^2,\hspace*{3mm} b=\fr{1}{2}(Q_{22}+Q_{33}-2Q_{23})=p_2^2,\crn
g&=&\fr{1}{2}(Q_{11}+Q_{22}-2Q_{12})=p_1^2,\hspace*{3mm} c=Q_{23}+Q_{34}-Q_{33}-Q_{24}=2p_2.p_3,\crn
h&=&Q_{13}+Q_{24}-Q_{14}-Q_{23}=2p_1.p_3,\hspace*{3mm} j=Q_{12}+Q_{23}-Q_{22}-Q_{13}=2p_1.p_2,\crn
d&=&Q_{34}-Q_{44}=m_3^2-m_4^2-p_3^2,\hspace*{3mm} e=Q_{24}-Q_{34}=m_2^2-m_3^2-p_2^2-2p_2.p_3,\crn
k&=&Q_{14}-Q_{24}=m_1^2-m_2^2+p_1^2+2p_1.p_4,\hspace*{3mm} f=\fr{Q_{44}}{2}-i\eps=m_4^2-i\eps,
\eea    
with $Q_{ij}$ is defined in Eq.~(\ref{def_Qij}). $d$, $e$, $k$, $f$ are complex while other parameters are real. 
There are two cases corresponding to the fact that the positions of two lightlike momenta are opposite or adjacent. 
\section{Integral with two opposite lightlike external momenta}
\begin{figure}[htb]
\begin{center}
\includegraphics[width=0.3\textwidth]{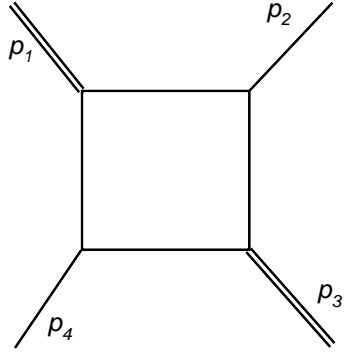}
\caption{\label{box_p1p3}{\em A box diagram with two opposite lightlike external momenta $p_1$ and $p_3$. Double line means massless.}}
\end{center}
\end{figure}
For the box shown in Fig.~\ref{box_p1p3} with $p_1^2=p_3^2=0$ one gets $a=g=0$ and writes
\bea
D_0^{(13)}=\int_0^1dx\int_0^xdy\int_0^ydz\fr{1}{(by^2+cxy+hxz+jyz+dx+ey+kz+f)^2}.\label{d0_xyz_13}
\eea
Integrating over $z$ to get
\bea
D_0^{(13)}=\int_0^1dx\int_0^xdy\fr{y}{(Ax+B)(Cx+D)}.
\eea
with 
\bea
A&=&cy+d, \hs B=by^2+ey+f,\crn
C&=&(c+h)y+d, \hs D=(b+j)y^2+(e+k)y+f.\label{ABC_d0_13} 
\eea
One changes the integration order as
\bea
\int_0^1dx\int_0^xdy=\int_0^1dy\int_y^1dx.
\eea 
We get
\bea
D_0^{(13)}=\int_0^1dyy\int_y^1dx\fr{1}{(Ax+B)(Cx+D)},\label{d0_xy_13}
\eea
where $A$, $B$, $C$, $D$ are complex. Integrating over $x$ as follows
\bea
\int_y^1dx\fr{1}{(Ax+B)(Cx+D)}&=&\fr{1}{AC}\int_y^1\fr{dx}{(x+\fr{B}{A})(x+\fr{D}{C})}\crn
&=&\fr{1}{AD-BC}\int_y^1\left(\fr{1}{x+\fr{B}{A}}-\fr{1}{x+\fr{D}{C}}\right)dx\crn
&=&\fr{1}{AD-BC}\left(\ln\fr{1+\fr{B}{A}}{y+\fr{B}{A}}-\ln\fr{1+\fr{D}{C}}{y+\fr{D}{C}}\right)\crn
&=&\fr{1}{AD-BC}\left(\ln\fr{A+B}{Ay+B}-\ln\fr{C+D}{Cy+D}\right),\label{int_x_ABCD}
\eea
where we have made sure that the arguments of the logarithms never cross the cut along the negative real axis. One easily gets 
\bea
D_0^{(13)}&=&\int_0^1dy\crn
&\times&\fr{1}{(cj-bh)y^2+(dj+ck-eh)y+dk-fh}\left(\ln\fr{A+B}{Ay+B}-\ln\fr{C+D}{Cy+D}\right)\hs\hs\hs
\label{eq_d013_ac}
\eea
where the discriminant of the quadratic denominator in the prefactor is nothing but the Landau determinant
\bea
\det(Q_4)=(dj+ck-eh)^2-4(cj-bh)(dk-fh).
\eea  
We write
\bea
D_0^{(13)}=\fr{1}{(cj-bh)(y_{2}-y_{1})}\int_0^1\left(\fr{1}{y-y_2}-\fr{1}{y-y_1}\right)\left(\ln\fr{A+B}{Ay+B}-\ln\fr{C+D}{Cy+D}\right)\hs\label{d0_y_13}
\eea
with 
\bea
y_{1,2}=\fr{-(dj+ck-eh)\mp\sqrt{\det(Q_4)}}{2(cj-bh)},
\eea
where the indices $1$, $2$ correspond to $-$ and $+$ signs respectively. 

Now we have to look at the imaginary parts of the arguments of the logarithms in Eq.~(\ref{d0_y_13}). We write them explicitly
\bea
A+B&=&by^2+(c+e)y+d+f,\crn
Ay+B&=&(b+c)y^2+(e+d)y+f,\crn
C+D&=&(b+j)y^2+(e+k+c+h)y+d+f,\crn
Cy+D&=&(b+j+c+h)y^2+(e+k+d)y+f.
\eea
Imaginary parts read
\bea
\Img(A+B)&=&\Img(ey+d+f)=\Img[ym_2^2+(1-y)m_3^2-i\eps]<0,\crn
\Img(Ay+B)&=&\Img(ey+dy+f)=\Img[ym_2^2+(1-y)m_4^2-i\eps]<0,\crn
\Img(C+D)&=&\Img[(e+k)y+d+f]=\Img[ym_1^2+(1-y)m_3^2-i\eps]<0,\crn
\Img(Cy+D)&=&\Img[(e+k)y+dy+f]=\Img[ym_1^2+(1-y)m_4^2-i\eps]<0.\hs\hs \label{sign_img_deno}
\eea
Using formula $\ln(a/b)=\ln a-\ln b$ for $\Img(a)\Img(b)>0$, we rewrite Eq.~(\ref{d0_y_13}) as
\bea
D_0^{(13)}=\fr{1}{\sqrt{\det(Q_4)}}\sum_{i=1}^2\sum_{j=1}^4(-1)^{i+j}\int_0^1dy\fr{1}{y-y_i}\ln(A_jy^2+B_jy+C_j)\label{d0_y_13_sumij}
\eea
with
\bea
A_1&=&b+c, \hs B_1=e+d, \hs C_1=f,\crn
A_2&=&b, \hs B_2=c+e, \hs C_2=d+f,\crn
A_3&=&b+j, \hs B_3=e+k+c+h, \hs C_3=d+f,\crn
A_4&=&b+j+c+h, \hs B_4=e+k+d, \hs C_4=f.
\eea
We would like to make an important remark here. From Eq.~(\ref{sign_img_deno}) we can re-write Eq.~(\ref{d0_y_13}) in the form 
\bea
D_0^{(13)}=\fr{1}{(cj-bh)(y_{2}-y_{1})}\int_0^1\left(\fr{1}{y-y_2}-\fr{1}{y-y_1}\right)\left(\ln\fr{A+B}{C+D}-\ln\fr{Ay+B}{Cy+D}\right).\hs\hs \label{d0_y_13_ab}
\eea
We notice that if $y=y_{1,2}$ then $AD=BC$ which means
\bea
\fr{A+B}{C+D}\Big\vert_{y=y_{1,2}}=\fr{Ay+B}{Cy+D}\Big\vert_{y=y_{1,2}}=\fr{B}{D}\Big\vert_{y=y_{1,2}}.
\eea
Thus, we get
\bea
\int_0^1\left(\fr{1}{y-y_2}-\fr{1}{y-y_1}\right)\left(\ln\fr{A+B}{C+D}\Big\vert_{y=y_{1,2}}-\ln\fr{Ay+B}{Cy+D}\Big\vert_{y=y_{1,2}}\right)=0.
\eea
Subtracting this zero contribution from Eq.~(\ref{d0_y_13_sumij}) we get another form
\bea
D_0^{(13)}&=&\fr{1}{\sqrt{\det(Q_4)}}\sum_{i=1}^2\sum_{j=1}^4(-1)^{i+j}\crn
&\times&\int_0^1dy\fr{\ln(A_jy^2+B_jy+C_j)-\ln(A_jy_i^2+B_jy_i+C_j)}{y-y_i}\crn
&+&\fr{1}{\sqrt{\det(Q_4)}}\sum_{i,j=1}^2(-1)^{i+j}\eta_{ij}\ln\fr{y_i-1}{y_i},
\label{d0_y_13_sumij_extra}
\eea
where $\eta_{i1}=\eta(A+B,1/(C+D))\vert_{y=y_i}$ and $\eta_{i2}=\eta(Ay+B,1/(Cy+D))\vert_{y=y_i}$ with $i=1,2$. 
This
representation is more convenient for the evaluation in
terms of Spence functions.

Each integral in Eq.~(\ref{d0_y_13_sumij}) or (\ref{d0_y_13_sumij_extra}) can be written in terms of $4$ Spence functions as given in Eq.~(\ref{result_S3}). 
Thus $D_0^{(13)}$ can be written in terms of $32$ Spence functions. 
\section{Integral with two adjacent lightlike external momenta}
\begin{figure}[htb]
\begin{center}
\includegraphics[width=0.3\textwidth]{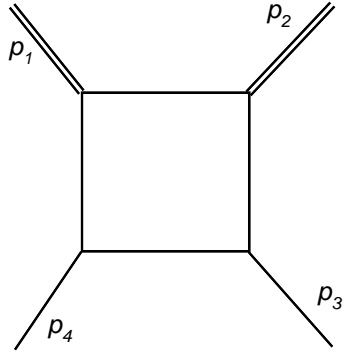}
\caption{\label{box_p1p2}{\em A box diagram with two adjacent lightlike external momenta $p_1$ and $p_2$. Double line means massless.}}
\end{center}
\end{figure}
For the box shown in Fig.~\ref{box_p1p2} with $p_1^2=p_2^2=0$ one gets $b=g=0$ and writes
\bea
D_0^{(12)}=\int_0^1dx\int_0^xdy\int_0^ydz\fr{1}{(ax^2+cxy+hxz+jyz+dx+ey+kz+f)^2}.\label{d0_xyz_12}
\eea
As in the case of $D_0^{(13)}$, 
integrating over $z$ gives
\bea
D_0^{(12)}=\underbrace{\int_0^1dx\int_0^xdy\fr{1}{a_1b_1}}_{I_1}+s_k\underbrace{\int_0^1dx\int_0^xdy\fr{1}{-s_ka_1(a_1y+b_1)}}_{I_2},
\eea
with 
\bea
s_k&=&sign(\Img(k)),\hs -s_ka_1=-s_k(hx+jy+k)-i\eps^\prime,\crn
b_1&=&ax^2+cxy+dx+ey+f,\crn
a_1y+b_1&=&ax^2+jy^2+(c+h)xy+dx+(e+k)y+f-i\eps,
\eea
where we have used the fact that $\Img(a_1y+b_1)=\Img[dx+(e+k)y+f]=\Img[(x-y)m_3^2+(1-x)m_4^2+ym_1^2-i\eps]<0$ because 
$0\le y\le x \le1$. $\eps$ and $\eps^\prime$ are infinitesmally positive and carry the sign of the imaginary parts of $-s_ka_1$ and $a_1y+b_1$. For $I_1$, we integrate over $y$, similar to Eq.~(\ref{int_x_ABCD}), to get
\bea
I_1&=&\int_0^1dy\fr{1}{(ja-hc)y^2+(jd-he-kc)y+jf-ke}\crn
&\times&\left[\ln\fr{(j+h)y+k-i\eps^\prime}{hy+k-i\eps^\prime}-\ln\fr{(a+c)y^2+(d+e)y+f}{ay^2+dy+f}\right].
\eea
Consider the prefactor
\bea
det(Q_4)&=&(jd-he-kc)^2-4(ja-hc)(jf-ke),\crn
y_{11(12)}&=&\fr{(he+kc-jd)\mp\sqrt{det(Q_4)}}{2(ja-hc)},
\eea
where the indices $11$, $12$ correspond to $-$ and $+$ signs respectively.
We rewrite $I_1$ as
\bea
I_1&=&\fr{1}{\sqrt{det(Q_4)}}\sum_{i=1}^2(-1)^i\crn
&&\int_0^1dy\fr{1}{y-y_{1i}}\left[\ln\fr{(j+h)y+k-i\eps^\prime}{hy+k-i\eps^\prime}-\ln\fr{(a+c)y^2+(d+e)y+f}{ay^2+dy+f}\right]\crn
&=&\fr{1}{\sqrt{det(Q_4)}}\sum_{i=1}^2\sum_{j=1}^4(-1)^{i+j}\int_0^1dy\fr{1}{y-y_{1i}}\ln(A_{1j}y^2+B_{1j}y+C_{1j})\label{I1_result}
\eea
with 
\bea
A_{11}&=&0,\hs B_{11}=h,\hs C_{11}=k,\crn
A_{12}&=&0,\hs B_{12}=j+h,\hs C_{12}=k,\crn
A_{13}&=&a+c,\hs B_{13}=d+e,\hs C_{13}=f,\crn
A_{14}&=&a,\hs B_{14}=d,\hs C_{14}=f.
\eea
Thus $I_1$ can be written in terms of $24$ Spence functions. For $I_2$ we shift $y=y+\alpha x$, $\alpha$ such that
\bea
j\alpha^2+(c+h)\alpha+a=0.
\eea
There are, in general, two values of $\alpha$. The final result does not depend on which value of $\alpha$ we take. 
We have used this freedom to find bugs in the numerical calculation and it turns out to be a very powerful method to check the correctness of the imaginary part which can be very tricky for the case of equal masses.
One gets
\bea
I_2=\int_0^1dx\int_{-\alpha x}^{(1-\alpha)x}dy\fr{1}{(Gx+H-i\eps^\prime)(Ex+F-i\eps)},\label{int_xy_EF}
\eea
with 
\bea
G&=&-s_kh-s_kj\alpha,\hs H=-s_kjy-s_kk,\crn
E&=&(2j\alpha+c+h)y+d+\alpha(e+k),\hs F=jy^2+(e+k)y+f.
\eea
For real $\alpha$ we have
\bea
\int_0^1dx\int_{-\alpha x}^{(1-\alpha)x}dy&=&\int_0^1dx\int_0^{(1-\alpha)x}dy-\int_0^1dx\int_{0}^{-\alpha x}dy\crn
&=&\int_0^{1-\alpha}dy\int_{y/(1-\alpha)}^{1}dx-\int_0^{-\alpha}dy\int_{-y/\alpha}^{1}dx.
\eea
We write
\bea
\fr{1}{(Gx+H-i\eps^\prime)(Ex+F-i\eps)}=\fr{1}{GF-HE}\left(\fr{G}{Gx+H-i\eps^\prime}-\fr{E}{Ex+F-i\eps}\right).\hs\hs
\eea
Integrating over $x$, we get
\bea
I_2&=&\int_{-\alpha}^{1-\alpha}\fr{dy}{GF-HE}\ln\fr{G+H}{E+F}-\int_0^{1-\alpha}\fr{dy}{GF-HE}\ln\fr{\fr{Gy}{1-\alpha}+H}{\fr{Ey}{1-\alpha}+F}\crn
&+&\int_0^{-\alpha}\fr{dy}{GF-HE}\ln\fr{\fr{Gy}{-\alpha}+H}{\fr{Ey}{-\alpha}+F}.
\eea
The prefactor
\bea
\fr{GF-HE}{s_k}&=&j(j\alpha+c)y^2+(2\alpha jk+jd-he+kc)y+\alpha(ke+k^2-jf)+kd-hf\crn
&=&j(j\alpha+c)(y-y_{21})(y-y_{22}),
\eea
with 
\bea
y_{21(22)}=\fr{-(2\alpha jk+jd-he+kc)\mp\sqrt{det(Q_4)}}{2j(j\alpha+c)},
\eea
where the indices $21$, $22$ correspond to $-$ and $+$ signs respectively.
We rewrite $I_2$ as
\bea
I_2&=&\fr{1}{s_k\sqrt{det(Q_4)}}\sum_{i=1}^2(-1)^iI_2^{(i)},\crn
I_2^{(i)}&=&\int_{-\alpha}^{1-\alpha}\fr{dy}{y-y_{2i}}\ln\fr{G+H}{E+F}-\int_0^{1-\alpha}\fr{dy}{y-y_{2i}}\ln\fr{\fr{Gy}{1-\alpha}+H}{\fr{Ey}{1-\alpha}+F}\crn
&+&\int_0^{-\alpha}\fr{dy}{y-y_{2i}}\ln\fr{\fr{Gy}{-\alpha}+H}{\fr{Ey}{-\alpha}+F}.\label{I2_alpha}
\eea
We make the substitutions $y=y-\alpha$ for the first integral, $y=(1-\alpha)y$ for the second integral and $y=-\alpha y$ for the third integral to get
\bea
I_2^{(i)}&=&\int_{0}^{1}\fr{dy}{y-\alpha-y_{2i}}\ln\fr{-s_kjy-s_kh-s_kk-i\eps^\prime}{jy^2+(c+h+e+k)y+a+d+f-i\eps}\crn
&-&\int_0^{1}\fr{(1-\alpha)dy}{(1-\alpha)y-y_{2i}}\ln\fr{-s_k(j+h)y-s_kk-i\eps^\prime}{(a+c+j+h)y^2+(d+e+k)y+f-i\eps}\crn
&+&\int_0^{1}\fr{-\alpha dy}{-\alpha y-y_{2i}}\ln\fr{-s_khy-s_kk-i\eps^\prime}{ay^2+dy+f-i\eps}.\label{I2_01}
\eea
Consider the arguments of the three logarithms, as demonstrated in Eq.~(\ref{sign_img_deno}), it is easy to see that the sign of the imaginary parts of the denominators is negative as indicated by $-i\eps$. The derivation is for real $\alpha$. However, this result can be easily
generalized to cover the case of complex $\alpha$ as shown below. We can now rewrite $I_2$ as
\bea
I_2=\fr{1}{s_k\sqrt{det(Q_4)}}\sum_{i=1}^2\sum_{j=1}^6(-1)^{i}\int_0^1dy\fr{c_j}{a_jy-b_j-y_{2i}}\ln(A_{2j}y^2+B_{2j}y+C_{2j})\label{I2_result}
\eea
with
\bea
c_1&=&1,\hs a_1=1,\hs b_1=\alpha,\crn
c_2&=&-(1-\alpha),\hs a_2=1-\alpha,\hs b_2=0,\crn
c_3&=&-\alpha,\hs a_3=-\alpha,\hs b_3=0,\crn
c_4&=&-1,\hs a_4=1,\hs b_4=\alpha,\crn
c_5&=&1-\alpha,\hs a_5=1-\alpha,\hs b_5=0,\crn
c_6&=&\alpha,\hs a_6=-\alpha,\hs b_6=0,\crn 
A_{21}&=&0,\hs B_{21}=-s_kj,\hs C_{21}=-s_kk-s_kh,\crn
A_{22}&=&0,\hs B_{22}=-s_k(j+h),\hs C_{22}=-s_kk,\crn
A_{23}&=&0,\hs B_{23}=-s_kh,\hs C_{23}=-s_kk,\crn
A_{24}&=&j,\hs B_{24}=c+h+e+k,\hs C_{24}=a+d+f,\crn
A_{25}&=&a+c+j+h,\hs B_{25}=d+e+k,\hs C_{25}=f,\crn
A_{26}&=&a,\hs B_{26}=d,\hs C_{26}=f.
\eea
$I_2$ can be written in terms of $36$ Spence functions. Thus 
\bea 
D_0^{(12)}=I_1+s_kI_2
\label{box_adj}
\eea
contains $60$ Spence functions. For the evaluation of $D_0^{(12)}$ in terms of Spence functions 
and to generalize Eq.~(\ref{I2_result}) for
complex $\alpha$, we have to do the following replacement for
each logarithm in $I_{1,2}$: 
\bea
\ln(A_{1j}y^2+B_{1j}y+C_{1j})&\to& \ln(A_{1j}y^2+B_{1j}y+C_{1j})-\ln(A_{1j}y_{1i}^2+B_{1j}y_{1i}+C_{1j}),\crn
\ln(A_{2j}y^2+B_{2j}y+C_{2j})&\to& \ln(A_{2j}y^2+B_{2j}y+C_{2j})-\ln(A_{2j}\hat{y}_{2i}^2+B_{2j}\hat{y}_{2i}+C_{2j}),\nn
\eea
with $\hat{y}_{2i}=(y_{2i}+b_j)/a_j$ and add the corresponding extra
terms related to the eta functions. The argument for this is similar to that explained in the previous section, see Eq.~(\ref{d0_y_13_sumij_extra}).
 
For the boxes with one lightlike external momentum, the result is written in terms of $72$ Spence functions by 
using exactly the same method.

\end{appendix}

\bibliographystyle{hunsrt}
\bibliography{Thesis}

\end{document}